%% file: main.tex
\pdfoutput=1
\documentclass[final,10pt, 3p, twocolumn, times]{elsarticle}

\input{packages}
\input{macros}

\input{citation_blocks}

\makeatletter
\def\bstctlcite{\@ifnextchar[{\@bstctlcite}{\@bstctlcite[@auxout]}}
\def\@bstctlcite[#1]#2{\@bsphack
  \@for\@citeb:=#2\do{%
    \edef\@citeb{\expandafter\@firstofone\@citeb}%
    \if@filesw\immediate\write\csname #1\endcsname{\string\citation{\@citeb}}\fi}%
  \@esphack}
\makeatother

\begin{document}

\bstctlcite{IEEEexample:BSTcontrol}

\begin{frontmatter}

\title{\vspace{-15 pt}\huge \textbf{A Modern Primer on \gf{Processing-In-Memory}}}

\author[1]{Onur Mutlu}
\author[2]{Saugata Ghose}
\author[3]{Juan G\'omez-Luna} 
\author[4]{Rachata Ausavarungnirun} 
\author[1]{\\Mohammad Sadrosadati}
\author[1]{Geraldo F. Oliveira}

\address{SAFARI Research Group}
\address[1]{ETH Z\"urich}
\address[2]{University of Illinois Urbana-Champaign}
\address[3]{NVIDIA Research}
\address[4]{MangoBoost Inc.\vspace{-10pt}}

\begin{abstract}
\input{sections/00-abstract}

\end{abstract}

\end{frontmatter}

\ifcameraready
\ifversionfour
\else
  \thispagestyle{firststyle}
\fi
\else
\fi

\clearpage

\setcounter{tocdepth}{2}
\renewcommand{\baselinestretch}{0.80}\normalsize
\tableofcontents
\renewcommand{\baselinestretch}{1.0}\normalsize

\input{sections/01-introduction}

\input{sections/02-trends-main-memory}
\input{sections/03-intelligent-controllers}

\input{sections/04-perils-processor-centric}

\input{sections/05-PIM-approaches}
\input{sections/06-PUM}
\input{sections/07-PNM}

\input{sections/08-enabling-adoption}

\input{sections/09-other-resources}
\input{sections/10-conclusion}

\section*{\sg{Acknowledgments}}

This \omvii{article} is a drastically revised and extended version of an earlier article published in 2019~\cite{mutlu2019} \omv{and \omvi{a largely} extended and revised version of our 2022 version of the \omvi{article}~\cite{mutlu2020modern,mutlu2020modernarxiv}}. 
This \omiv{paper} also incorporates revised material from \omiv{other} earlier \omiv{articles} published in 2019~\cite{ghose2019processing} \omvii{and 2020--2024}~\omvii{\cite{mutlu2020modern,mutlu2020modernarxiv,mcciede2024}}. 
The shorter, initial version of this work~\cite{mutlu2019} is based on a keynote talk delivered by Onur Mutlu at the
3rd Mobile System Technologies (MST) Workshop in Milan, Italy on 27
October 2017\sg{~\cite{mutlu.msttalk17}}. 

The mentioned keynote talk is similar to a series of talks given by Onur Mutlu in a wide variety of venues since 2015 until now. This talk has evolved significantly
over time with the accumulation of new works and feedback received from
many audiences. \omiv{Recent versions of the talk were delivered at \gfv{several places, including \omvii{at} the \omvii{first} EFCL Summer School~\cite{mutlu.efcl24.talk}\omvii{, the 20th} \omvi{HiPEAC ACACES} Summer School~\cite{mutlu.access24}}}\omvii{, and the 31st IEEE International Conference on Electronics Circuits and Systems on 19 November 2024~\cite{mutlu.icecs24.talk}}. Earlier versions of the talk were delivered as a
distinguished lecture at George Washington
University in February 2019~\cite{mutlu.gwutalk19}, as an Invited Talk at ISSCC Special Forum on ``\emph{Intelligence at the Edge: How Can We Make Machine Learning More Energy Efficient?}'', as part of the 2019 International Solid State Circuits Conference in February 2019~\cite{mutlu.isscctalk19}, as a keynote talk at the 29th ACM Great Lakes Symposium on VLSI~\cite{mutlu.glvlsitalk19}, as a keynote talk at the  International Symposium on Advanced Parallel Processing Technology in August 2019~\cite{mutlu.appttalk19}, and as a keynote talk at the 37th IEEE International Conference on Computer Design in November 2019~\cite{mutlu.iccdtalk19}.
\omvi{This article also benefited from  a recent invited paper~\cite{mcciede2024} presented at The ``\emph{AI Memory}'' focus session of the IEDM 2024 conference.}
\omvii{We also acknowledge two interviews published \omix{in \emph{HiPEAC info}} in 2018 and 2024 on the progress in memory systems and memory-centric computing~\cite{hipeacinterview18,hipeacinterview24}. }

This article and the associated talks are based on research done over
the course of the past \omx{16} years in the SAFARI Research Group on the
topic of \sg{processing-in-memory} (PIM) \omx{in particular, and memory \& storage systems and computer architecture in general}. We thank all of the members
of the SAFARI Research Group, and our collaborators at Carnegie
Mellon, ETH Zürich, and other universities, who have contributed to
the various works we describe in this paper. Thanks also goes to our
research group's industrial sponsors over the past \omx{16} years,
especially Alibaba, ASML, \omiv{Facebook,} Google, Huawei, Intel, Microsoft, NVIDIA, Samsung, Seagate, VMware\omiv{, and Xilinx}. 
\omvi{Part of the research that is described in this article} was also partially supported by the \omiv{ETH Future Computing Laboratory,} \omvi{HKUST} \omv{ACCESS Center,} Intel Science and Technology Center for Cloud Computing, the
Semiconductor Research Corporation, the Data Storage Systems Center at
Carnegie Mellon University, various NSF and NIH grants, and various awards,
including the NSF CAREER Award, the Intel Faculty Honor Program Award, \omx{Huawei OlympusMons Award in Storage Systems,}
and a number of Google \omx{Faculty Awards} and IBM Faculty Research Awards to Onur Mutlu.

{
\bibliographystyle{IEEEtran}
\bibliography{refs}
}

\end{document}

%% file: packages.tex
\usepackage{amssymb}
\biboptions{sort&compress}

\usepackage[dvipsnames]{xcolor}
\usepackage{fancyhdr}
\usepackage[utf8]{inputenc}
\usepackage{datetime2}
\usepackage{xurl}
\usepackage{microtype}
\usepackage{flushend}
\usepackage[binary-units]{siunitx}
\usepackage{setspace}
\usepackage{multirow}
\usepackage{booktabs}
\usepackage{caption}
\usepackage[compact]{titlesec}

\usepackage{subcaption}
\usepackage[acronym,nonumberlist,nowarn]{glossaries}
    \glsdisablehyper
     \loadglsentries{acronyms}

\usepackage{tikz}
\usepackage{xcolor}
\usepackage{csquotes}
\usepackage{dblfloatfix}
\usepackage{amsmath}
\usepackage{enumitem}
\usepackage{xargs} 
\usepackage{xspace}
\usepackage{todonotes}
\usepackage{hyperref} 
\usepackage{soul}

\hypersetup{
    colorlinks,
    linkcolor={red!50!black},
    citecolor={blue!50!black},
    urlcolor={blue!80!black}
}

%% file: acronyms.tex
\newacronym{ADI}{ADI}{Alternating Direction Implicit}
\newacronym{AGU}{AGU}{Address Generation Unit}
\newacronym[longplural={Access History Tables}]{AHT}{AHT}{Access History Table}
\newacronym{AHT-R}{AHT-R}{Access History Table - Read}
\newacronym{AHT-W}{AHT-W}{Access History Table - Write-Back}
\newacronym{AIP}{AIP}{Access Interval Predictor}
\newacronym{AL}{AL}{Added Latency for column accesses}
\newacronym{ASIC}{ASIC}{Application Specific Integrated Circuit}
\newacronym{COTS}{COTS}{commercial off-the-shelf}
\newacronym{AVX}{AVX}{Advanced Vector Extensions}
\newacronym[longplural={Basic Block Vectors}]{BBV}{BBV}{Basic Block Vector}
\newacronym{BHT}{BHT}{Branch History Table}
\newacronym{HBM}{HBM}{Hybrid Bandwidth Memory}
\newacronym{DDR4}{DDR4}{Double-Data Rate 4}
\newacronym{MIMD}{MIMD}{multiple-instruction multiple-data}
\newacronym{SIMT}{SIMT}{single-instruction multiple-thread}
\newacronym{SLP}{SALP}{subarray-level parallelism}

\newacronym{NN}{NN}{Neural Network}
\newacronym{BTB}{BTB}{Branch Target Buffer}
\newacronym{CAS}{CAS}{Column Address Strobe}
\newacronym{AMAT}{AMAT}{Average Memory Access Time}
\newacronym{CCD}{CCD}{Column to Column Delay}
\newacronym{AI}{AI}{Arithmetic Intensity}
\newacronym[longplural={Columnar Database Systems}]{CDBS}{CDBS}{Columnar Database System}
\newacronym[longplural={Chip Multiprocessors}]{CMP}{CMP}{Chip Multiprocessor}
\newacronym{CMT}{CMT}{Chip Multithreading}
\newacronym[longplural={Coarse-Grain Reconfigurable Arrays}]{CGRA}{CGRA}{Coarse-Grain Reconfigurable Array}
\newacronym{C-RAM}{C-RAM}{Computational-RAM}
\newacronym{CWD}{CWD}{Column Write Delay}
\newacronym{DBI}{DBI}{Dynamic Binary Instrumentation}
\newacronym[longplural={Database Management Systems}]{DBMS}{DBMS}{Database Management System}
\newacronym{DDR}{DDR}{Double Data Rate}
\newacronym{DEWP}{DEWP}{Dead Line and Early Write-Back Predictor}
\newacronym{DRAM}{DRAM}{dynamic random access memory}
\newacronym{DSBP}{DSBP}{Dead Sub-Block Predictor}
\newacronym{DSM}{DSM}{Decomposed Storage Manage}
\newacronym{FAW}{FAW}{Four row Activation Window}
\newacronym{FP}{FP}{Floating-Point}
\newacronym{EDP}{EDP}{Energy-Delay Product}
\newacronym{FCFS}{FCFS}{First-Come First-Serve}
\newacronym{FIFO}{FIFO}{Fist In, First Out}
\newacronym{FSM}{FSM}{Finite State Machine}
\newacronym{FPGA}{FPGA}{Field-Programmable Gate Array}
\newacronym[longplural={Functional Units}]{FU}{FU}{Functional Unit}
\newacronym{GAg}{GAg}{Global Adaptive branch prediction using one Global PHT}
\newacronym{GAs}{GAs}{Global Adaptive branch prediction using per-Set PHT}
\newacronym{GCC}{GCC}{GNU Compiler Collection}
\newacronym{GEMS}{GEMS}{General Execution-driven Multiprocessor Simulator}
\newacronym[longplural={General Purpose Processors}]{GPP}{GPP}{General Purpose Processor}
\newacronym{HMC}{HMC}{Hybrid Memory Cube}
\newacronym{HIVE}{HIVE}{HMC Instruction Vector Extensions}
\newacronym{IATAC}{IATAC}{Inter-Access Time per Access Count}
\newacronym{ILP}{ILP}{Instruction Level Parallelism}
\newacronym{IPC}{IPC}{Instructions per Cycle}
\newacronym{ISA}{ISA}{Instruction Set Architecture}
\newacronym{IRAM}{IRAM}{Intelligent RAM}
\newacronym{KIPS}{KIPS}{Kilo Instructions per Second}
\newacronym{LDS}{LDS}{Linked Data Structure}
\newacronym{LFSR}{LFSR}{Linear Feedback Shift Register}
\newacronym{LFMR}{LFMR}{Last-to-First Miss-Ratio}
\newacronym{MLP}{MLP}{Memory-Level Parallelism}

\newacronym{LLC}{LLC}{last-level cache}
\newacronym{LRU}{LRU}{Least Recently Used}
\newacronym{LSU}{LSU}{Load Store Unit}
\newacronym{LTP}{LTP}{Last-Touch Predictor}
\newacronym{LvP}{LvP}{Live-time Predictor}
\newacronym{LWP}{LWP}{Last Write Predictor}
\newacronym{MARSS}{MARSS}{Micro Architectural and System Simulator}
\newacronym{McPAT}{McPAT}{Multi-core Power, Area, and Timing}
\newacronym{MIPS}{MIPS}{Microprocessor without Interlocked Pipeline Stages}
\newacronym{MOB}{MOB}{Memory Order Buffer}
\newacronym{MMU}{MMU}{memory management unit}
\newacronym{PUD}{PUD}{Processing-using-DRAM}
\newacronym{MPKI}{MPKI}{Misses per Kilo-Instruction}
\newacronym{MSHR}{MSHR}{Miss-Status Handling Registers}
\newacronym{NAS}{NAS}{Numerical Aerodynamic Simulation}
\newacronym{NDP}{NDP}{\underline{n}ear-\underline{d}ata \underline{p}rocessing}
\newacronym{NMP}{NMP}{Near Memory Processor}
\newacronym{NoC}{NoC}{Network-on-Chip}
\newacronym{NPB}{NPB}{NAS Parallel Benchmark}
\newacronym{NUCA}{NUCA}{Non-Uniform Cache Architecture}
\newacronym{NUMA}{NUMA}{Non-Uniform Memory Access}
\newacronym{OoO}{OoO}{Out-of-Order}
\newacronym{OpenMP}{OpenMP}{Open Multi-Processing}
\newacronym{OS}{OS}{Operating System}
\newacronym{PAg}{PAg}{Per-address Adaptive branch prediction using one Global PHT}
\newacronym{PAs}{PAs}{Per-address Adaptive branch prediction using per-Set PHT}
\newacronym{PC}{PC}{Program Counter}
\newacronym[longplural={Pointer-Chasing Engines}]{PCE}{PCE}{Pointer-Chasing Engine}
\newacronym{PCM}{PCM}{Performance Counter Monitor}
\newacronym{PIM}{PIM}{\underline{p}rocessing-\underline{i}n-\underline{m}emory}
\newacronym[longplural={Pattern History Tables}]{PHT}{PHT}{Pattern History Table}
\newacronym{RAPL}{RAPL}{Running Average Power Limit}
\newacronym{RAS}{RAS}{Row Address Strobe}
\newacronym{RAT}{RAT}{Registers Alias Table}
\newacronym{RC}{RC}{Row Cycle}
\newacronym{RCD}{RCD}{RAS to CAS Delay}
\newacronym[longplural={Row-based Database Systems}]{RDBS}{RDBS}{Row-based Database System}
\newacronym{ROB}{ROB}{Reorder Buffer}
\newacronym{RRD}{RRD}{Row to Row activation Delay}
\newacronym{DNN}{DNN}{Deep Neural Network}
\newacronym{ANN}{ANN}{Artificial Neural Network}
\newacronym{PNM}{PNM}{processing-near-memory}
\newacronym{PUM}{PUM}{processing-using-memory}
\newacronym{FFD}{FFD}{first-fit-decreasing}
\newacronym{HFF}{HFF}{helper flip-flop}

\newacronym{RP}{RP}{Row Precharge}
\newacronym{RTL}{RTL}{Register Transfer Level}
\newacronym{RTP}{RTP}{Read To Precharge}
\newacronym{RVU}{RVU}{Reconfigurable Vector Unit}
\newacronym{SDP}{SDP}{Skewed Dead-Block Predictor}
\newacronym{SESC}{SESC}{Superescalar Simulator}
\newacronym{SSE}{SSE}{Streaming SIMD Extensions}
\newacronym{SFP}{SFP}{Spatial Footprint Predictor}
\newacronym{SiNUCA}{SiNUCA}{Simulator of Non-Uniform Cache Architectures}
\newacronym{SMT}{SMT}{Simultaneous Multi-Threading}
\newacronym{SIMD}{SIMD}{single-instruction multiple-data}
\newacronym{SPEC}{SPEC}{Standard Performance Evaluation Corporation}
\newacronym{SoC}{SoC}{System-on-Chip}
\newacronym{SPP}{SPP}{Spatial Pattern Predictor}
\newacronym{SRAM}{SRAM}{Static Random Access Memory}
\newacronym{SSOR}{SSOR}{Symmetric Successive Over-Relaxation}
\newacronym{SSV}{SSV}{Search Set Vector}
\newacronym{TLB}{TLB}{Translation Look-aside Buffer}
\newacronym{TLP}{TLP}{Thread Level Parallelism}
\newacronym[longplural={through-silicon vias}]{TSV}{TSV}{through-silicon via}
\newacronym{VWQ}{VWQ}{Virtual Write Queue}
\newacronym{WTR}{WTR}{Write Recovery time}
\newacronym{WR}{WR}{Write To Read delay time}
\newacronym{TRA}{TRA}{triple row activation}

%% file: macros.tex
\newcommand\tempcommand[1]{\renewcommand{\arraystretch}{#1}}

\titlespacing\section{0pt}{5pt plus 2pt minus 2pt}{0pt plus 2pt minus 2pt}
\titlespacing\subsection{0pt}{5pt plus 2pt minus 2pt}{0pt plus 2pt minus 2pt}
\titlespacing\subsubsection{0pt}{5pt plus 2pt minus 2pt}{0pt plus 2pt minus 2pt}

\makeatletter
\g@addto@macro{\normalsize}{%
  \setlength{\abovedisplayskip}{8pt plus 0.5pt minus 1pt}
  \setlength{\belowdisplayskip}{8pt plus 0.5pt minus 1pt}
  \setlength{\abovedisplayshortskip}{0pt}
  \setlength{\belowdisplayshortskip}{0pt}
  \setlength{\intextsep}{5pt plus 1pt minus 1pt}
  \setlength{\textfloatsep}{5pt plus 1pt minus 1pt}
  \setlength{\skip\footins}{5pt plus 1pt minus 1pt}}
\makeatother

\newcommand{\paratitle}[1]{\noindent\textbf{#1.}}

\newcommand{\incircledd}[1]{%
  \tikz[baseline=(char.base)]{
    \node[shape=circle,minimum size=10.5pt,draw,inner sep=0.5pt,fill=white,text=black,font=\itshape\footnotesize] (char) {\fontfamily{lmss}\selectfont#1};}%
}%

\newcommand*\circled[1]{\tikz[baseline=(char.base)]{
            \node[shape=circle,fill,inner sep=0.8pt] (char) {\textcolor{white}{#1}};}}

\newif\ifcameraready
\camerareadytrue

\newif\ifrevision
\revisiontrue

\newif\ifversionthree
\versionthreetrue

\newif\ifversionfour
\versionfourtrue

\newcommand{\versionnum}[0]{11.1}

\definecolor{dblue}{rgb}{0.00, 0.00, 0.75}
\definecolor{electriccrimson}{rgb}{1.0, 0.0, 0.25}
\definecolor{airforceblue}{rgb}{0.36, 0.54, 0.66}
\definecolor{dodgerblue}{rgb}{0.12, 0.56, 1.0}
\definecolor{brandeisblue}{rgb}{0.0, 0.44, 1.0}
\definecolor{darkgoldenrod}{rgb}{0.72, 0.53, 0.04}

\ifcameraready
\ifversionfour
\else
  \fancypagestyle{firststyle}
  {
    \fancyhead[C]{\textcolor{airforceblue}{\emph{Version \versionnum~---~\today, \DTMcurrenttime}}
    }
  }
\fi
\else
\fi

\newcommand{\ch}[0]{}
\newcommand{\chii}[0]{}
\newcommand{\chiii}[0]{}

\newcommand{\chv}[0]{}
\newcommand{\chvi}[0]{}
\newcommand{\chvii}[0]{}

\ifcameraready

  \newcommand{\juan}[0]{}
  \newcommand{\sg}[0]{}
  \newcommand{\sgii}[0]{}
  
  \newcommand{\rachata}[0]{}
\else
  
  \newcommand{\juan}[0]{}

  \newcommand{\sg}[0]{}
  \newcommand{\rachata}[1]{\textcolor{Orange}{#1}}
  \newcommand{\sgii}[1]{\textcolor{black}{#1}}
  
\fi

\ifrevision
\newcommand{\juanr}[1]{\textcolor{black}{#1}}
\newcommand{\juanrr}[1]{\textcolor{black}{#1}}
\newcommand{\juanrri}[1]{\textcolor{black}{#1}}
\else
\newcommand{\juanr}[1]{\textcolor{ForestGreen}{#1}}
\newcommand{\juanrr}[1]{\textcolor{BrickRed}{#1}}
\newcommand{\juanrri}[1]{\textcolor{Orange}{#1}}
\fi

\ifversionthree
    \newcommand{\juanrrr}[1]{\textcolor{black}{#1}}
    \newcommand{\juanrrri}[1]{\textcolor{black}{#1}}
\else
    \newcommand{\juanrrr}[1]{\textcolor{brandeisblue}{#1}}
    \newcommand{\juanrrri}[1]{\textcolor{airforceblue}{#1}}
    
\fi

\ifversionfour
    \newcommand{\gf}[0]{}
    \newcommand{\gfv}[0]{}
    \newcommand{\gfvi}[0]{}
    \newcommand{\gfvii}[0]{}
    \newcommand{\gfviii}[0]{}
    \newcommand{\gfix}[0]{}
    \newcommand{\gfx}[0]{}

    \newcommand{\agy}[0]{}
    
    \newcommand{\omiv}[0]{}
    \newcommand{\omv}[0]{}
    \newcommand{\omvi}[0]{}
    \newcommand{\omvii}[0]{}
    \newcommand{\omviii}[0]{}
    \newcommand{\omix}[0]{}
    \newcommand{\omx}[0]{}
    \newcommand{\omxi}[0]{}

    \newcommand{\msvii}[0]{}
    \newcommand{\omcomment}[1][]{}
    \newcommand{\agycomment}[1][]{}

    \newcommandx{\unsure}[2][1=]{\todo[disable,#1]{#2}}
    \newcommandx{\change}[2][1=]{\todo[disable,#1]{#2}}
    \newcommandx{\feedback}[2][1=]{\todo[disable,#1]{#2}}
    \newcommandx{\improvement}[2][1=]{\todo[disable,#1]{#2}}
    \newcommandx{\thiswillnotshow}[2][1=]{\todo[disable,#1]{#2}}
    \newcommandx{\completedRevision}[2][1=]{\todo[disable,#1]{#2}}
    \newcommandx{\dataSource}[2][1=]{\todo[disable,#1]{#2}}
    \newcommandx{\info}[2][1=]{\todo[disable,#1]{#2}}
\else
    \newcommand{\gf}[0]{}
    \newcommand{\gfv}[0]{}
    \newcommand{\gfvi}[0]{}
    \newcommand{\gfvii}[0]{}
    \newcommand{\gfviii}[0]{}
    \newcommand{\gfix}[0]{}
    \newcommand{\gfx}[0]{}

    \newcommand{\agy}[1]{\textcolor{black}{#1}}
    
    \newcommand{\omiv}[0]{}
    \newcommand{\omv}[0]{}
    \newcommand{\omvi}[0]{}
    \newcommand{\omvii}[0]{}
    \newcommand{\omviii}[0]{}
    \newcommand{\omix}[0]{}
    \newcommand{\omx}[0]{}
    \newcommand{\omxi}[1]{\textcolor{BrickRed}{#1}}

    \newcommand\omcomment[1]{\noindent{\color{BrickRed} {\bf \fbox{OM}} {\it#1}}}
    \newcommand\agycomment[1]{\noindent{\color{BrickRed} {\bf \fbox{AGY:}} {\it#1}}}

    \newcommand{\msvii}[0]{}

    \newcommandx{\unsure}[2][1=]{\todo[linecolor=red,backgroundcolor=red!25,bordercolor=red,#1, size=\tiny,fancyline]{#2}}
    \newcommandx{\change}[2][1=]{\todo[linecolor=blue,backgroundcolor=blue!25,bordercolor=blue,#1,size=\tiny]{\textbf{#2}}}
    \newcommandx{\feedback}[2][1=]{\todo[linecolor=yellow,backgroundcolor=yellow!25,bordercolor=yellow,#1,size=\tiny]{#2}}
    \newcommandx{\improvement}[2][1=]{\todo[linecolor=Plum,backgroundcolor=Plum!25,bordercolor=Plum,#1]{#2}}
    \newcommandx{\thiswillnotshow}[2][1=]{\todo[disable,#1]{#2}}
    \newcommandx{\completedRevision}[2][1=]{\todo[disable,backgroundcolor=red,#1]{#2}}
    \newcommandx{\dataSource}[2][1=]{\todo[disable,backgroundcolor=red,#1]{#2}}
    \newcommandx{\info}[2][1=]{\todo[linecolor=green,backgroundcolor=green!25,bordercolor=green,#1, size=\tiny,fancyline]{#2}}

    \usepackage{marginnote} 
    
\fi

%% file: citation_blocks.tex
\newcommand\membottleneck{\cite{mutlu.imw13, mutlu.superfri15, dean2013tail, kanev.isca15, ferdman.asplos12, wang.hpca14, boroumand.asplos18, boroumand2021google, boroumand2021google_arxiv, mutlu2019enabling, mutlu2019, mutlu2020vlsi, ghose2019processing, alser2020accelerating, cali2020genasm, koppula2019eden, kanellopoulos2019smash, oliveira2021pimbench, oliveira2021pimbench_arxiv, oliveira2021.SLS, mutlu.msttalk17, mutlu.gwutalk19, mutlu.isscctalk19, mutlu.glvlsitalk19, mutlu.appttalk19, mutlu.iccdtalk19,narancic2014evaluating,ayers2018memory,jia2016understanding,Manegold_2000,gholami2020ai,stengel2015quantifying,chishti2019memory,burger1995declining,gupta2020architectural,sriraman2019softsku,zhao2022understanding,ruan2019insider,gan2018architectural,sriraman2020accelerometer,yuan2023rambda,delimitrou2018amdahl,hsia2020cross,lottarini2018vbench,wang2022characterizing,richins2020missing,mckee.cf04}\xspace}

\newcommand\memscalingissue{\cite{kang.memoryforum14, mutlu.imw13,
mutlu.superfri15, mckee.cf04,wilkes.can01,kim-isca2014,salp,yoongu-thesis,raidr,mutlu2017rowhammer,donghyuk-ddma,lee-isca2009,yoon2012row,yoon-taco2014,lim-isca09,
wulf1995hitting, chang.sigmetrics16, lee.hpca13, lee.hpca15,
chang.sigmetrics17, lee.sigmetrics17, luo2014characterizing,
luo.arxiv17,hassan2017softmc, chargecache, kevinchang-thesis, patel2017reaper, hasan2019crow, ghose2018vampire, kim2018solar, kim2020revisiting, mutlu2020retrospective, wang2018cal, mutlu2018recent, ghose2019demystifying,mutlu2015main,hong2010memory,sites1996,luo2023rowpress,yaglikci2024spatial,olgun2024abacus,bostanci2024comet,yauglikcci2022hira}}

\newcommand\srampum{\cite{1802.shs, aga.hpca17, agrawal2018x, al2020towards, eckert2018neural, fujiki2019duality, jeloka201628, jiang2020c3sram, kang.icassp14, kang2015energy, kim2021colonnade, nag2019gencache, si2019dual, simon2020blade, wang2019bit, wang2023infinity}\xspace}

\newcommand\flashpum{\cite{flashcosmos,gao2021parabit,choi2020flash,han2019novel,merrikh2017high,wang2018three,lue2019optimal,kim2021behemoth,wang2022memcore,han2021flash,kang2021s,lee2020neuromorphic,lee20223d}\xspace}

%% file: sections/00-abstract.tex

Modern computing systems are overwhelmingly \omv{processor-centric: they are} designed to move data to
computation. This design choice goes directly against at least three
key trends in computing that cause performance, scalability and energy
bottlenecks: 
(1)~data access is a key bottleneck
as many important applications \omvi{in various domains} \omv{(e.g., machine learning, genome analytics, databases, graph analytics\gfv{, high-performance computing, \omvi{and a wide variety of} mobile and server-class workloads})} are increasingly data-intensive, and memory bandwidth and energy do not scale well, 
(2)~energy consumption is a key limiter in almost all computing platforms, \omv{and}  (3)~data movement, especially off-chip to on-chip, is very expensive in terms of energy\omv{,} latency, \omv{and bandwidth}\omvi{,} much more so than
computation. These trends are especially severely-felt in the
data-intensive server and energy-constrained mobile systems of today.
\omv{As applications continue to become more data-intensive, processor-centric systems will likely increasingly waste more energy and performance, and they will hurt sustainability.}

At the same time, conventional memory technology is facing many
technology scaling challenges in terms of \omv{robustness}, energy, and
performance. As a result, \omv{(}memory\omv{)} system architects are open to
organizing memory in different ways and making it more intelligent, at
the expense of higher cost. The emergence of 3D-stacked memory plus
logic, the adoption of error correcting codes inside DRAM
chips, 
\gfv{the} proliferation of different main memory standards and chips specialized for different purposes (e.g., low-power, high bandwidth, \omv{graphics,} low latency), and the necessity of designing new solutions to serious \omv{robustness (i.e.,} reliability\gf{,} security\gf{, safety}\omv{)} issues, such as the RowHammer \gf{and RowPress phenomena},
are evidence of this trend.
 
This \gf{paper} discusses recent research that aims to enable computation close to data, an approach we \omv{broadly} call \juanrrr{\emph{\underline{p}rocessing-\underline{i}n-\underline{m}emory} (\emph{PIM})}. PIM places computation
mechanisms in or near where the data is stored (i.e., inside
memory chips \gf{or modules}, in the logic layer of 3D-stacked memory, in the memory controllers\omv{, in storage devices or chips}), so that data movement between the
computation units and memory\gfv{/storage} \omvi{units} is reduced or eliminated. While
the general idea of PIM is not new, we discuss motivating trends in
applications as well as memory circuits \omv{and} technology that greatly exacerbate the need for enabling it in modern computing systems. We examine at least two promising new approaches to designing PIM systems to accelerate
important data-intensive applications: 
(1)~{\em processing\gf{-}using\gf{-}memory}\omv{, which exploits fundamental} analog operational \omv{principles} of \omv{memory} chips to perform
massively-parallel operations \omv{in-situ} in memory,
(2)~{\em processing{-}near{-}memory}\omv{, which exploits} \gf{different logic and memory integration technologies (e.g., 3D-stacked memory technology)} \omv{to place computation logic close to memory circuitry, and thereby enable high-bandwidth, low-energy, and low-latency access to data}. In both approaches, we describe and tackle relevant cross-layer research, design, and adoption
challenges in devices, architecture, systems, \omvi{compilers,} programming models\omvi{, and applications}. Our focus is {on} the development of
\omv{PIM} designs that can be adopted in real computing platforms at low cost. We conclude by discussing work on solving key challenges to the practical adoption of PIM.
\gfv{We believe that the shift from a processor-centric to a memory-centric mindset (and infrastructure) remains the largest adoption challenge for PIM, which, once overcome, can unleash a fundamentally energy-efficient, high-performance, and sustainable new way of designing, using, and programming computing systems.}
\vspace{-5pt}

%% file: sections/01-introduction.tex
\section{Introduction}
\label{sec:introduction}

Main memory, \omv{prominently} built using the \gf{\gls{DRAM}} technology~\cite{dennard1968field}, is a major component in nearly all computing
systems, including servers, cloud platforms, mobile/embedded devices, and sensor systems. 
Across all of these systems, the data working set sizes of modern applications are rapidly growing, while the need for fast analysis of such data is increasing. Thus, main memory is becoming an increasingly significant bottleneck across a wide variety of computing systems and \omvi{application domains} \omv{(including machine learning, databases, graph analytics, genome analysis, \gfv{high-performance computing, security, data manipulation, \omvi{and a wide variety of} mobile and server-class workloads})}~\gf{\membottleneck}. 
Alleviating the main memory bottleneck requires the memory capacity,
energy, cost, and performance to all scale in an efficient manner across technology generations. 
Unfortunately, it has become increasingly difficult in recent years, especially the past decade, to scale all of these dimensions~\memscalingissue, and thus the main memory bottleneck has been worsening.

A major reason for the main memory bottleneck is the high energy and latency cost
associated with \emph{data movement}. In modern computers, to
perform any operation on data that resides in main memory, the processor must retrieve
the data from main memory. This requires the memory controller
to issue commands to a DRAM module across a relatively slow and power-hungry off-chip bus (known as the \emph{memory channel}).  The DRAM module sends the requested data across the memory
channel, after which the data is placed in the caches and registers.
The CPU can perform computation on the data once the data is
in its registers.  Data movement from
the DRAM to the CPU incurs long latency and consumes a significant
amount of energy~\cite{hashemi.isca16,cont-runahead, ahn.tesseract.isca15, ahn.pei.isca15,boroumand.asplos18, boroumand2021google, boroumand2021google_arxiv, keckler2011gpus}. 
These costs are often exacerbated by the fact that much of the data brought
into the caches is \emph{not reused} by the
CPU~\omv{\cite{qureshi.isca07,qureshi-hpca07, ahn.tesseract.isca15, ahn.pei.isca15,oliveira2021pimbench,tyson1995modified,johnson1997run,johnson1999run,seshadri2012evicted}} \omiv{or accelerators~\gfv{\cite{boroumand2021google,boroumand2021google_arxiv,amiraliphd,rhu2013locality,chen2014adaptive,chatterjee2017architecting,koo2017access}},}
providing little benefit in
return for the high latency and energy cost. 

The cost of data movement is a fundamental issue with the
\emph{processor-centric} nature of contemporary computer systems.
The CPU is considered to be the master in the system, and
computation is performed only in the processor (and accelerators).  In contrast, data storage and communication units, including the main memory, are
treated as unintelligent workers that are incapable of computation. 
As a result of this processor-centric design paradigm, data moves
a lot in the system \omv{(back and forth} between the computation units and communication/storage units\omv{)} so that computation can be done on it. With the increasingly \emph{data-centric}
nature of contemporary and emerging applications, the
processor-centric design paradigm leads to great
inefficiency in performance, energy\gf{,} and cost.
For
example, most of
the real estate within a single compute node is already dedicated to
handling data movement and storage (e.g., large caches, memory
controllers, interconnects, \omv{communication interfaces and associated circuitry,} main memory)~\cite{kumar.isscc2009,howard201048core,jowani2010x8664,gillespie2014steam,singh2017zen},  
and our recent \omiv{works show} that 
\omiv{(1)}~more than 62\% of the entire system energy of a mobile device is spent on data movement between the processor and the memory
hierarchy for widely-used mobile workloads~\gf{\cite{boroumand.asplos18}};
\omiv{(2)~more than 90\% of the entire system energy is spent on memory when executing large \omv{commercial edge} neural network models on modern edge machine learning accelerator\omv{s}~\cite{boroumand2021google,boroumand2021google_arxiv}}.

The large overhead of data movement in modern systems along
with technology advances that enable better integration of memory
and logic have recently prompted the re-examination of an old idea~\omv{\cite{Kautz1969,stone1970logic}}
that we will \omv{broadly} call {\emph{\gls{PIM}}. The key
idea is to place computation mechanisms in or near where the data
is stored (i.e., inside the memory chips, in the logic layer of 3D-stacked
memory, in the memory controllers,  inside large caches\omv{, inside storage \omvi{units} or \omvi{inside} sensing units}),
so that data movement between where the computation is done
and where the data is stored is reduced or eliminated, compared to
contemporary processor-centric systems. \gf{\gls{PIM}} 
enables the ability to perform operations and execute software tasks either using 
(1)~the \omv{operational properties of the} memory \omv{circuitry}
itself, or 
(2)~some form of processing logic (e.g., accelerators,
simple cores, reconfigurable logic) \omvi{added} inside the memory subsystem \omvi{\omvii{close to} the memory circuitry}.

The idea of \gls{PIM} has been around for at least \omv{six}
decades \gf{~\cite{Kautz1969, stone1970logic,shaw1981non, elliott1992computational, kogge1994execube, gokhale1995processing, patterson1997case, oskin1998active, kang2012flexram, fraguela.ppopp03, Draper:2002:ADP:514191.514197, Mai:2000:SMM:339647.339673, elliott.dt99, riedel.1998, keeton.1998, kaxiras.1997, acharya.1998, jino1978magnetic, doty1980magnetic, bongiovanni1980magnetic, kim1999assessing, gebis2000trends,saulsbury1996missing,murphy2001characterization}}. 
However, past efforts were {\em not} widely adopted for various reasons, including 
\gf{(}1)~the difficulty of integrating processing elements with DRAM~\gf{\cite{kim1999assessing}}, 
\gf{(}2)~the lack of critical memory-related scaling challenges that current technology and applications face today~\memscalingissue, and 
\gf{(}3)~the \gfv{\omvi{large effort expended to tolerate} data movement bottlenecks (via processor-centric techniques) \omvi{at the cost of more complexity in software and hardware}}~\omvii{\membottleneck}. 
As a result of advances in modern memory architectures, {e.g., the integration of logic and memory in a 3D-stacked manner}, various recent works explore a range of \gls{PIM} architectures for multiple different \gfv{use cases} (e.g., \cite{zhu2013accelerating, pugsley2014ndc, zhang.hpdc14, farmahini2015nda, ahn.tesseract.isca15, ahn.pei.isca15,   loh2013processing, hsieh.isca16, pattnaik.pact16,   DBLP:conf/isca/AkinFH15, impica, DBLP:conf/sigmod/BabarinsaI15, DBLP:conf/IEEEpact/LeeSK15, DBLP:conf/hpca/GaoK16, chi2016prime, gu.isca16, kim.isca16, asghari-moghaddam.micro16, boroumand2016pim, hashemi.isca16, cont-runahead, GS-DRAM, liu-spaa17, gao.pact15, guo2014wondp, sura.cf15, morad.taco15, hassan.memsys15, li.dac16, kang.icassp14, aga.hpca17, shafiee2016isaac, seshadri2013rowclone, Seshadri:2015:ANDOR, chang.hpca16, seshadri.arxiv16,  seshadri.micro17, hajinazarsimdram, nai2017graphpim,kim.arxiv17,kim.bmc18, li.micro17, kim.sc17, boroumand.asplos18, boroumand2021google, boroumand2021google_arxiv,cali2020genasm,fernandez2020natsa,singh2019napel,seshadri2020indram, wang2020figaro, gao2020computedram, olgun2021pidram, rezaei2020nom,herruzo2021enabling, boroumand2021polynesia, boroumand2022icde, syncron, besta2021sisa_micro, besta2021sisa, asgarifafnir,denzler2021casper, li2018scope}). 
We believe it is crucial to re-examine \gls{PIM} today with a fresh perspective (i.e., with novel approaches and ideas), by exploiting new memory technologies, with realistic workloads and systems, and with a mindset to ease adoption and feasibility.

In this \omv{article}, we explore two new approaches to enabling
processing-in-memory in modern systems. 
\omv{The first approach exploits the analog operational properties of the memory circuitry to perform simple yet powerful common operations that the chip is inherently efficient at or could be made efficient at performing~\cite{seshadri.bookchapter17, seshadri2013rowclone,chang.hpca16,kevinchang-thesis,seshadri.thesis16,Seshadri:2015:ANDOR, seshadri.arxiv16, seshadri.micro17, hajinazarsimdram, li.micro17,GS-DRAM, ghose.bookchapter19, ghose.bookchapter19.arxiv,deng.dac2018, seshadri2020indram, wang2020figaro, gao2020computedram, olgun2021pidram, rezaei2020nom, ghose2019processing,xin2020elp2im,aga.hpca17,eckert2018neural,fujiki2019duality,kang.icassp14,li.dac16,angizi2018pima,angizi2018cmp,angizi2019dna,levy.microelec14,kvatinsky.tcasii14,shafiee2016isaac,kvatinsky.iccd11,kvatinsky.tvlsi14,gaillardon2016plim,bhattacharjee2017revamp,hamdioui2015memristor,xie2015fast,hamdioui2017myth,yu2018memristive,yavits2021giraf,li2018scope}.
We call this approach {\emph{\gls{PUM}}}~\cite{ghose2019processing, seshadri.bookchapter17, seshadri.bookchapter17.arxiv, seshadri2020indram}.
This approach has the potential to provide large performance and energy gains with minimal changes to memory chips and circuitry.}
Some solutions that fall under
this approach take advantage of the existing DRAM design to cleverly
and efficiently perform \emph{bulk operations} (i.e., operations on an
entire row of DRAM cells), such as bulk copy \gf{and} data initialization~\gf{\cite{seshadri.bookchapter17, seshadri2013rowclone,seshadri2018rowclone,chang.hpca16,seshadri.thesis16,seshadri.bookchapter17.arxiv,seshadri2020indram,rezaei2020nom}}, \gf{bitwise Boolean operations~\gf{\cite{seshadri.micro17, gao2020computedram, xin2020elp2im, besta2021sisa, li.micro17,Seshadri:2015:ANDOR,seshadri.thesis16,seshadri.arxiv16,seshadri.bookchapter17.arxiv,seshadri2020indram}}, arithmetic operations~\gf{\cite{deng.dac2018, gao2020computedram,li.micro17,angizi2019graphide, hajinazarsimdram,li2018scope,mimdram}}, and lookup table {based} operations~\gf{\cite{ferreira2021pluto,ferreira2022pluto,deng2019lacc,sutradhar2021look,sutradhar2020ppim}}.}
Other solutions take advantage of the analog operational principles of \gf{SRAM}~\srampum, \gf{NAND flash~\flashpum, and} emerging non-volatile memory technologies \gf{(e.g., 
phase-change memory, PCM~\cite{le202364,joshi2020accurate}\omv{,} 
spin-transfer torque magnetic RAM, STT-MRAM~\cite{jain2017computing,roy2019towards,kang2017memory}\omv{,}
metal-oxide resistive RAM, ReRAM~\cite{song2018graphr,song2017pipelayer,marinella2018multiscale,yuan2021forms,ankit2020panther,yang2020retransformer,chen2018regan,truong2022adapting,truong2021racer})} to perform similar bulk operations~\gfvi{\cite{li.dac16,
aga.hpca17,eckert2018neural,si2019dual,simon2020blade,nag2019gencache,wang2019bit,al2020towards,kang.icassp14,kim2021colonnade,jiang2020c3sram,jeloka201628,wang2023infinity,kang2015energy,
flashcosmos,gao2021parabit,choi2020flash,han2019novel,merrikh2017high,wang2018three,lue2019optimal,kim2021behemoth,wang2022memcore,han2021flash,kang2021s,lee2020neuromorphic,lee20223d,fernandez2024matsa}} or other specialized computations like convolutions and matrix multiplications~\gfvi{\cite{angizi2018pima,angizi2018cmp,angizi2019dna,levy.microelec14,kvatinsky.tcasii14,shafiee2016isaac,kvatinsky.iccd11,kvatinsky.tvlsi14,gaillardon2016plim,bhattacharjee2017revamp,hamdioui2015memristor,xie2015fast,hamdioui2017myth,yu2018memristive,ankit2019puma,mao2022genpip,shahroodi2023swordfish}}.

The second approach enables \gls{PIM} in
a \omv{potentially} more general-purpose \omv{and flexible} manner by \omiv{adding} computation capability \omiv{to}
conventional memory controllers~\cite{hashemi.isca16,cont-runahead}\gf{,
\omiv{memory chips~\gfv{\cite{gupta2023evaluating,gomez2023evaluating,oliveira2023transpimlib,chen2023simplepim,oliveira2023dappa,gomez2022machine,giannoula2022towards,gomezluna2022ieeeaccess,giannoula2022sparsep,giannoula2022sigmetrics,kaxiras.1997,patterson1997case,devaux2019,CASES_MVX,samsunghc23,lee2022isscc,rhyner2024pimopt,hyun2024pathfinding,abecassis2023gapim}},
\omiv{memory modules~\cite{DBLP:conf/sigmod/BabarinsaI15,asghari-moghaddam.micro16,sun2021abc,lee2022improving,dai2022dimmining,zhou2023dimm,feng2022menda,chen2023metanmp,zhou2022gnnear,yun2023grande,patel2023xfm,chen2024bridge}}}}
or 
the logic layer(s) of the relatively new \emph{3D-stacked memory technologies}~\gf{\cite{
fernandez2020natsa, cali2020genasm, kim.bmc18, ahn.pei.isca15, ahn.tesseract.isca15, boroumand.asplos18, boroumand2021google, boroumand2021google_arxiv, boroumand2019conda, boroumand2016pim, boroumand.arxiv17, singh2019napel, asghari-moghaddam.micro16, JAFAR, chi2016prime, farmahini2015nda, gao.pact15, DBLP:conf/hpca/GaoK16, gu.isca16, guo2014wondp, hashemi.isca16, cont-runahead, hassan.memsys15, hsieh.isca16, kim.isca16, kim.sc17, DBLP:conf/IEEEpact/LeeSK15, liu-spaa17, morad.taco15, nai2017graphpim, pattnaik.pact16, zhang.hpdc14, zhu2013accelerating, DBLP:conf/isca/AkinFH15, gao2017tetris, drumond2017mondrian, dai2018graphh, huang2020heterogeneous, zhuo2019graphq, herruzo2021enabling, boroumand2021polynesia, boroumand2022icde, syncron, besta2021sisa_micro, besta2021sisa, asgarifafnir, upmem2018, devaux2019, shin2018mcdram, cho2020mcdram, denzler2021casper, gomez2022machine, giannoula2022towards, fernandez2022exploiting, oliveira2022heterogeneous, balasubramonian2014near, jacob2016compiling, nair2015active, lloyd2018dse, gokhale2015rearr, lloyd2015memory, rodrigues2016scattergather, lloyd2017keyvalue, landgraf2021combining, nair2015evolution, kim2021aquabolt, ke2021near, lee2022improving, loh2013processing, pugsley2014ndc, DBLP:conf/sigmod/BabarinsaI15, impica, kim.arxiv17, sura.cf15, singh2020nero, singh2021fpga, singh2021accelerating, RVU, NIM, gu2016leveraging, amiraliphd, kwon202125, lee2021hardware, niu2022isscc, azarkhish2016logic, azarkhish2018neurostream, de2018design, akin2014hamlet, liu2018processing, tsai:micro:2018:ams, gu2020ipim, DRAMA_CAL_2014, Asghari-Moghaddam_2016, huang2019active, kersey2017lightweight, li2019pims, zhang2018graphp, lim2017triple, smc_sim, HIVE, jang2019charon, hadidi2017cairo, santos2018processing, hadidi2017demystifying, gu2020dlux, asgari2020mahasim, baskaran2020decentralized, ahmed2019compiler, picorel2017near}}. 
We call this general approach {\em{\gls{PNM}}}~\cite{ghose2019processing}.\footnote{\omv{This approach is also called \gls{NDP} in some of the literature~\cite{balasubramonian2014near,fernandez2022exploiting,fernandez2020natsa,DBLP:conf/sigmod/BabarinsaI15,DBLP:conf/hpca/GaoK16,gu.isca16,hsieh.isca16,gao.pact15,huang2019active,JAFAR,kim.sc17,yun2023grande,lee2016application,liang2019ins,Kim2018HowMC,hong2016accelerating,xu2015scaling,oliveira2021pimbench}.}} 
This approach is especially catalyzed by recent advancements in 3D-stacked memory technologies that include a logic processing layer underneath memory layers~\gfv{\cite{ramulator,lee.taco16,HBM, jeddeloh2012hybrid, jedec.hbm.spec, hmc.spec.1.1, hmc.spec.2.0, loh2008stacked}} \omiv{and recent prototypes that map the computing capability inside DRAM chips~\cite{gupta2023evaluating,gomez2023evaluating,oliveira2023transpimlib,chen2023simplepim,oliveira2023dappa,gomez2022machine,giannoula2022towards,gomezluna2022ieeeaccess,giannoula2022sparsep,giannoula2022sigmetrics,kaxiras.1997,patterson1997case,devaux2019,CASES_MVX} and DRAM modules~\cite{DBLP:conf/sigmod/BabarinsaI15,asghari-moghaddam.micro16,sun2021abc,lee2022improving,dai2022dimmining,zhou2023dimm,feng2022menda,chen2023metanmp,zhou2022gnnear,yun2023grande,patel2023xfm,chen2024bridge}.}
In order to stack multiple layers of memory, 3D-stacked chips use vertical \glspl{TSV} to connect the layers to each other, and to the I/O drivers of the chip~\cite{lee.taco16}. 
The \glspl{TSV} provide much greater \emph{internal} bandwidth within the 3D stack layers than is available externally on the memory
channel.  
Several such 3D-stacked memory architectures, such as the
Hybrid Memory Cube~\cite{hmc.spec.1.1, hmc.spec.2.0} and
High-Bandwidth Memory~\cite{jedec.hbm.spec,lee.taco16}, include a
\emph{logic layer}, where designers can add some processing
logic (e.g., accelerators, simple cores, reconfigurable logic) to take advantage of this high internal bandwidth. \omv{Emerging} die-stacking \omvi{and packaging} technologies, like \omv{\emph{hybrid bonding}~\cite{kagawa2016novel,niu2022isscc,schmidt1998wafer} and {\em monolithic 3D \omvi{integration}}~\omiv{\cite{gopireddy2019m3d,mitra2018vlse,hwang2018cmos,mitra2015nano,rich2020nano,sabry2015abundant,sabry2019n3xt,ghiasi2022revamp3d}}},
can amplify the benefits of this approach by greatly improving internal bandwidth \omv{across layers} and \omv{potentially adding} logic layers between memory layers. 

Regardless of the approach taken to \gls{PIM}, there are key practical
adoption challenges that system architects and \omv{system} programmers must
address to enable the widespread adoption of \gls{PIM} across the computing
landscape and in different domains of workloads\gf{,} including 
\gfv{(1)~programming models and code generation support (via compilers, high-level APIs, and software frameworks) for \gls{PIM};
(2)~runtime engines for adaptive code and data scheduling, data mapping, access/sharing control;
(3)~memory coherence mechanisms that allow for collaborative host--\gls{PIM} execution;
(4)~virtual memory support for \omvi{a} unified memory space between host and \gls{PIM} main memory;
(5)~data structures that inherently take into account the concurrent execution model of a many-core \gls{PIM} system; and 
(6)~infrastructures to assess the benefits and feasibility of \gls{PIM} systems, including benchmarks and simulation \omvi{infrastructures} for \gls{PIM} prototyping.} In addition to describing \omv{research and development} along
the two key approaches \omv{to \gls{PIM}}, we also discuss these challenges  
\juanrrr{in this \omv{article}}\omvi{,} along with existing work that addresses these challenges.

Before we describe in detail the two modern approaches to \gls{PIM} in Section~\ref{sec:pim}, we first describe major trends affecting main memory (Section~\ref{sec:majortrends}), then demonstrate  many reasons why we need to have intelligent memory controllers to enhance memory scaling into the future (Section~\ref{sec:intmemcont}), followed by an analysis of the major shortcomings of the processor-centric computing paradigm \omv{that} \gls{PIM} intends to augment, disrupt, and perhaps in some cases displace (Section~\ref{sec:processorcentric}).

%% file: sections/02-trends-main-memory.tex
\glsresetall

\section{Major Trends Affecting Main Memory}
\label{sec:majortrends}

Main memory is a major, critical component of all computing
systems, including cloud and server platforms, desktop computers,
mobile and embedded devices, and sensors. It constitutes one of the main pillars of any computing platform, together with 
(1)~the processing
elements (or computational elements), which can include CPU cores, GPU cores, accelerators, or reconfigurable devices, and 
(2)~the communication elements, which can include interconnects, network interfaces, and network processing units.

Due to its relatively low cost and low latency, \gls{DRAM}~\cite{dennard1968field}
is the predominant data storage technology that is used to build main memory. The growing data working set sizes of modern applications~\cite{mutlu.imw13, mutlu.superfri15,dean2013tail, kanev.isca15, ferdman.asplos12,wang.hpca14, boroumand.asplos18, boroumand2021google, boroumand2021google_arxiv,pekhimenko2012lcp,pekhimenko2013lcp,churin1988camac,abali2001mxt,friedrich2014power8}
impose an ever-increasing demand for higher DRAM capacity and performance. \omv{For example, memory capacity requirements of large machine learning models increased by \omvii{more than} 10,000 \omvi{times} in the past \omvi{five} years~\cite{gholami2020ai}.} Unfortunately, DRAM technology scaling is becoming increasingly challenging: it is increasingly difficult to enlarge DRAM chip capacity at low cost while also maintaining DRAM performance, energy efficiency, and reliability~\gf{\cite{liu.isca13,kim-isca2014,mutlu2017rowhammer, yauglikcci2021blockhammer, hassan2021uncovering, orosa2021deeper,frigo2020trr,kang.memoryforum14, mutlu.imw13,wilkes.can01,salp,yoongu-thesis,raidr,donghyuk-ddma,lee-isca2009,yoon2012row,yoon-taco2014,lim-isca09,
wulf1995hitting, chang.sigmetrics16, lee.hpca13, lee.hpca15,chang.sigmetrics17, lee.sigmetrics17, luo2014characterizing,luo.arxiv17,hassan2017softmc, chargecache, kevinchang-thesis, patel2017reaper, hasan2019crow, ghose2018vampire, kim2018solar, kim2020revisiting, mutlu2020retrospective, wang2018cal, mutlu2018recent, ghose2019demystifying,mutlu2015main,hong2010memory,sites1996,luo2023rowpress,yaglikci2024spatial,olgun2024abacus,bostanci2024comet,yauglikcci2022hira, mutlu.superfri15, dean2013tail, kanev.isca15, ferdman.asplos12, wang.hpca14, boroumand.asplos18, boroumand2021google, boroumand2021google_arxiv, mutlu2019enabling, mutlu2019, mutlu2020vlsi, ghose2019processing, alser2020accelerating, cali2020genasm, koppula2019eden, kanellopoulos2019smash, oliveira2021pimbench, oliveira2021pimbench_arxiv, oliveira2021.SLS, mutlu.msttalk17, mutlu.gwutalk19, mutlu.isscctalk19, mutlu.glvlsitalk19, mutlu.appttalk19,mutlu.iccdtalk19,narancic2014evaluating,jia2016understanding,Manegold_2000,gholami2020ai,stengel2015quantifying,chishti2019memory,burger1995declining,gupta2020architectural,sriraman2019softsku,zhao2022understanding,ruan2019insider,gan2018architectural,sriraman2020accelerometer,ayers2018memory,yuan2023rambda,delimitrou2018amdahl,hsia2020cross,lottarini2018vbench,wang2022characterizing,richins2020missing,mckee.cf04}}\omv{.}
Thus, fulfilling the increasing memory needs of modern workloads is becoming
increasingly costly and difficult~\cite{dean2013tail,kanev.isca15,mckee.cf04,mutlu.superfri15,wilkes.can01,salp,kang.memoryforum14,yoongu-thesis,raidr,ahn.tesseract.isca15,ahn.pei.isca15,hsieh.isca16,donghyuk-ddma,lee-isca2009,yoon2012row,yoon-taco2014,lim-isca09,wulf1995hitting,chang.sigmetrics16,lee.hpca13,lee.hpca15,chang.sigmetrics17,lee.sigmetrics17,luo2014characterizing,luo.arxiv17,hassan2017softmc,chargecache,wang2018cal,das.dac18,kim2018solar, kevinchang-thesis, wang2020figaro, kim2020revisiting, luo2020clr, koppula2019eden, mutlu2020retrospective, hasan2019crow, patel2019understanding, patel2020bit, patel2021harp, ghose2019demystifying, kim.hpca19}.

If CMOS technology scaling is coming to an end~\cite{denning.2016},
the projections are significantly worse for DRAM technology
scaling~\cite{itrs.2009}. DRAM technology scaling affects all
  major characteristics of DRAM, including capacity,
bandwidth, latency, reliability, energy, and cost.  
We next describe the key issues and trends in DRAM technology scaling 
and discuss how these trends motivate the need for {\em intelligent memory controllers}, i.e., controllers that have intelligence and computation capability to enable better scaling of main memory in terms of all metrics of interest. Such intelligent memory controllers can also more easily pave the way to \omv{\gls{PIM}} and be used as a starting substrate for \omv{\gls{PIM}}.

\begin{figure}[ht]
\centering
\begin{subfigure}[h]{0.46\textwidth}
    \includegraphics[width=1.0\linewidth]{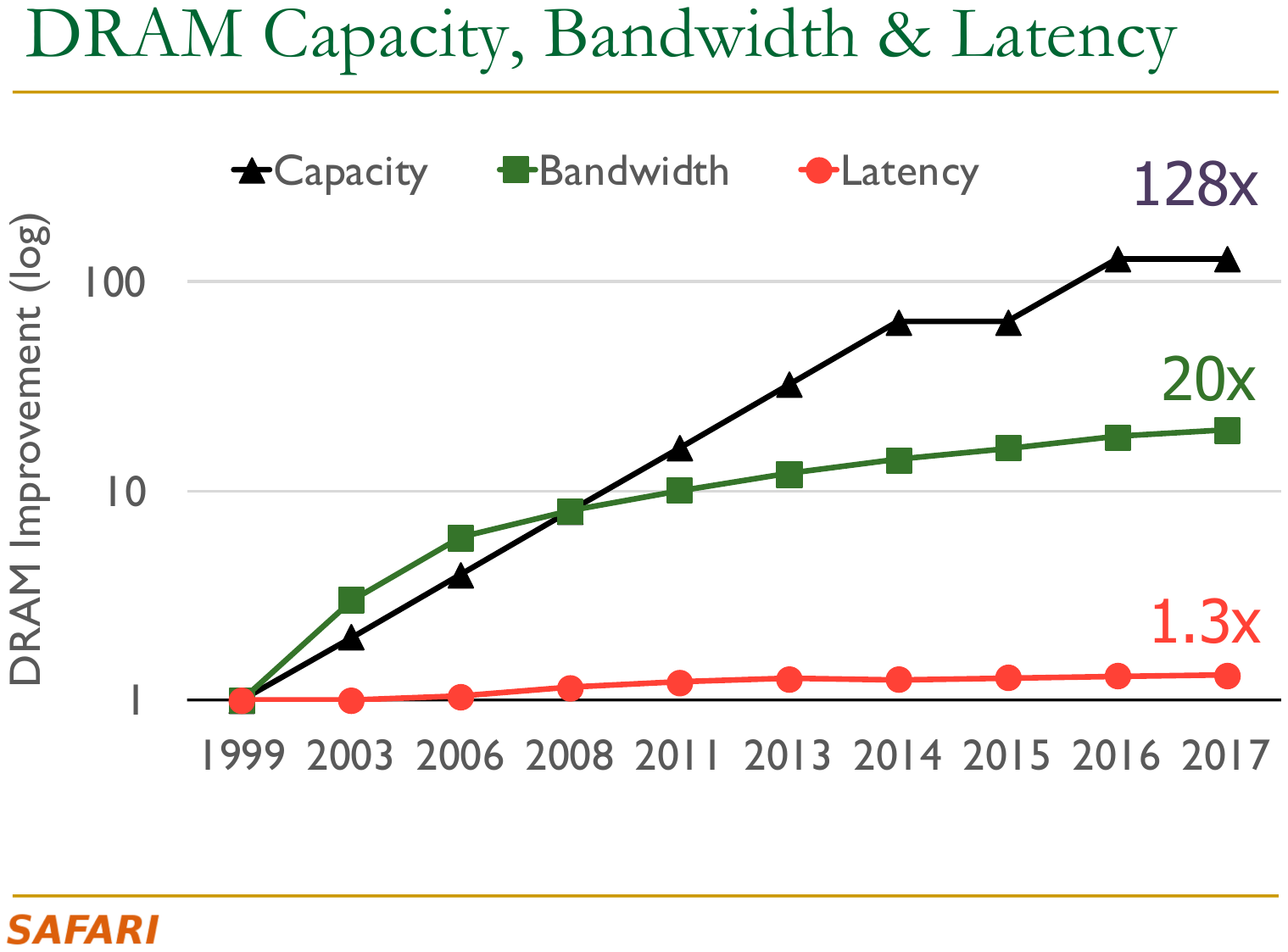}
    \caption{Scaling of DRAM capacity, bandwidth and latency between 1999 and 2017, normalized to the value in 2017. Data depicted for the most common type of DDRx chip of each year. Reproduced from~\cite{mutlu.iccdtalk19}. Originally presented in~\cite{chang.sigmetrics16,chang.sigmetrics16talk,kevinchang-thesis}.}
    \label{fig:dram-lat-band-cap}
\end{subfigure}
\par\bigskip 
\begin{subfigure}[h]{0.46\textwidth}
    \includegraphics[width=1.0\linewidth]{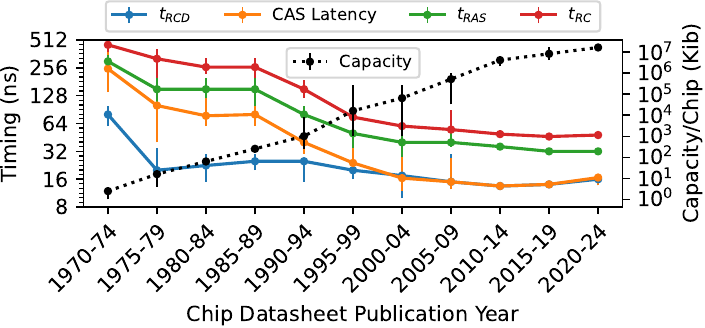}
    \caption{\gfvii{\omix{Capacity and latency data} from DRAM chips with more \omiv{expanded coverage of dates (1970--202\omv{4}).
    The semi-log plot shows the evolution of key DRAM access timings (left) and per-chip storage capability (right) across each 5-year period of time. The figure depicts four JEDEC-standardized parameters~\cite{jedec1994synchronous} found in DRAM chip datasheets: (i)~$t_{RCD}$, the minimum row activation to column operation delay; 
    (ii)~$CAS Latency$, the read operation to data access latency; 
    (iii)~$t_{RAS}$, the minimum row activation to precharge delay; 
    (iv)~$t_{RC}$, the minimum delay between accesses to different rows.
    Reproduced from~\cite{patel2024rethinking,patel2024rethinkingieee,patel2022acase}.}}}
    \label{fig:dram-lat-cap}
\end{subfigure}
\caption{\gfvii{Overview of DRAM \omx{technology} scaling over \omix{more than five} \omviii{decades}.}}
\label{fig:memscaling}
\end{figure}

The first key {concern is the difficulty} of scaling DRAM
capacity (i.e., density, or cost per bit), bandwidth, and latency {\em
  at the same time}.  
While the processing core count doubles every
two years, the DRAM capacity doubles only every three
years, as shown by~\cite{lim-isca09}, and the latter is slowing down. 
This trend causes the \emph{memory capacity per
  core} to drop by approximately 30\% every two
years~\cite{lim-isca09}. The trend is even worse for \emph{memory
  bandwidth per core} -- in the approximately two decades between 1999 and 2017, {DRAM chip storage capacity (for the most commonly-used DDRx chip of the time)} has improved \omv{by}
approximately 128$\times$ while DRAM bandwidth has improved only \omv{by} approximately
20$\times$~\cite{kevinchang-thesis,chang.sigmetrics16,lee.hpca13}, as shown in \omvii{Figure~\ref{fig:dram-lat-band-cap}}. In
the same period of about two decades, DRAM latency (as measured by the row
cycling time~\omv{\cite{salp,lee.hpca13}}) has remained almost constant (i.e., reduced by only
30\%, as shown in \omviii{Figure~\ref{fig:dram-lat-band-cap}}), making it a significant performance bottleneck for {many
  modern workloads, including} in-memory
databases~\gf{\cite{ailamaki1999dbmss,boncz.1999,clapp.2015,JAFAR,boroumand2016pim,GS-DRAM,seshadri.micro17,boroumand2019conda,boroumand2022icde,RVU,drumond2017mondrian}},
graph processing~\gf{\cite{ahn.tesseract.isca15,umuroglu.2015,xu.2014,ahn.pei.isca15,kanellopoulos2019smash,besta2021sisa_micro,besta2021sisa,huang2020heterogeneous,zhuo2019graphq,dai2018graphh,zhang2018graphp,angizi2019graphide,santos2018processing}},
data analytics~\gf{\cite{awan.2015,awan.2016,clapp.2015,yasin.2014,boroumand2022icde}}, datacenter workloads~\cite{kanev.isca15}, neural networks~\gf{\cite{koppula2019eden, chi2016prime, desa2018high, han2016eie, long2018reram, schuiki2018scalable, boroumand.asplos18, boroumand2021google, boroumand2021google_arxiv,yang2020retransformer,samsunghc23,zhou2022transpim}},
\omv{large language models~\gfvi{\cite{gholami2020ai,zhou2022transpim,park2024attacc,yang2020retransformer,park2024lpddr,samsunghc23}},}
and consumer workloads~\cite{boroumand.asplos18}. 
As low-latency computing is becoming ever more important~\cite{mutlu.imw13, mutlu.superfri15, cali2020genasm, cali2018nano, alser2020accelerating, alser.bioinformatics17, alser2020sneakysnake, alser2019shouji,dean2013tail},
e.g., due to the ever-increasing need to process large amounts of data at real time, and predictable performance continues to be a critical concern in the design of modern computing systems~\cite{moscibroda-usenix2007,mutlu-micro2007,mutlu-isca2008,mutlu.superfri15,lavanya-thesis,donghyuk-ddma,subramanian-hpca2013,usui-taco2016,subramanian-micro2015,kim2010thread,kim2014bounding,kim2016bounding},
it is increasingly \omv{important} to design low-latency main memory chips.
\omviii{
Figure~\ref{fig:dram-lat-cap} provides extended data from DRAM chips from 1970--2024, demonstrating that memory latency has been improved much less than storage capacity\omix{:} over the course of more than 50 years, DRAM storage capacity has \omix{improved} by more than \gfviii{1~million} times whereas \omix{DRAM} latency \omix{has} reduced by \omix{approximately} only \gfviii{8~times}, and \omix{DRAM} latency has remained almost constant in the past 20 years or so. 
Unfortunately, this scaling trend that favors capacity over latency has \omix{also caused} to DRAM latency to be \omix{an even larger} performance bottleneck, as described earlier.}

\begin{figure}[ht]
\centering
\includegraphics[width=1.0\linewidth]{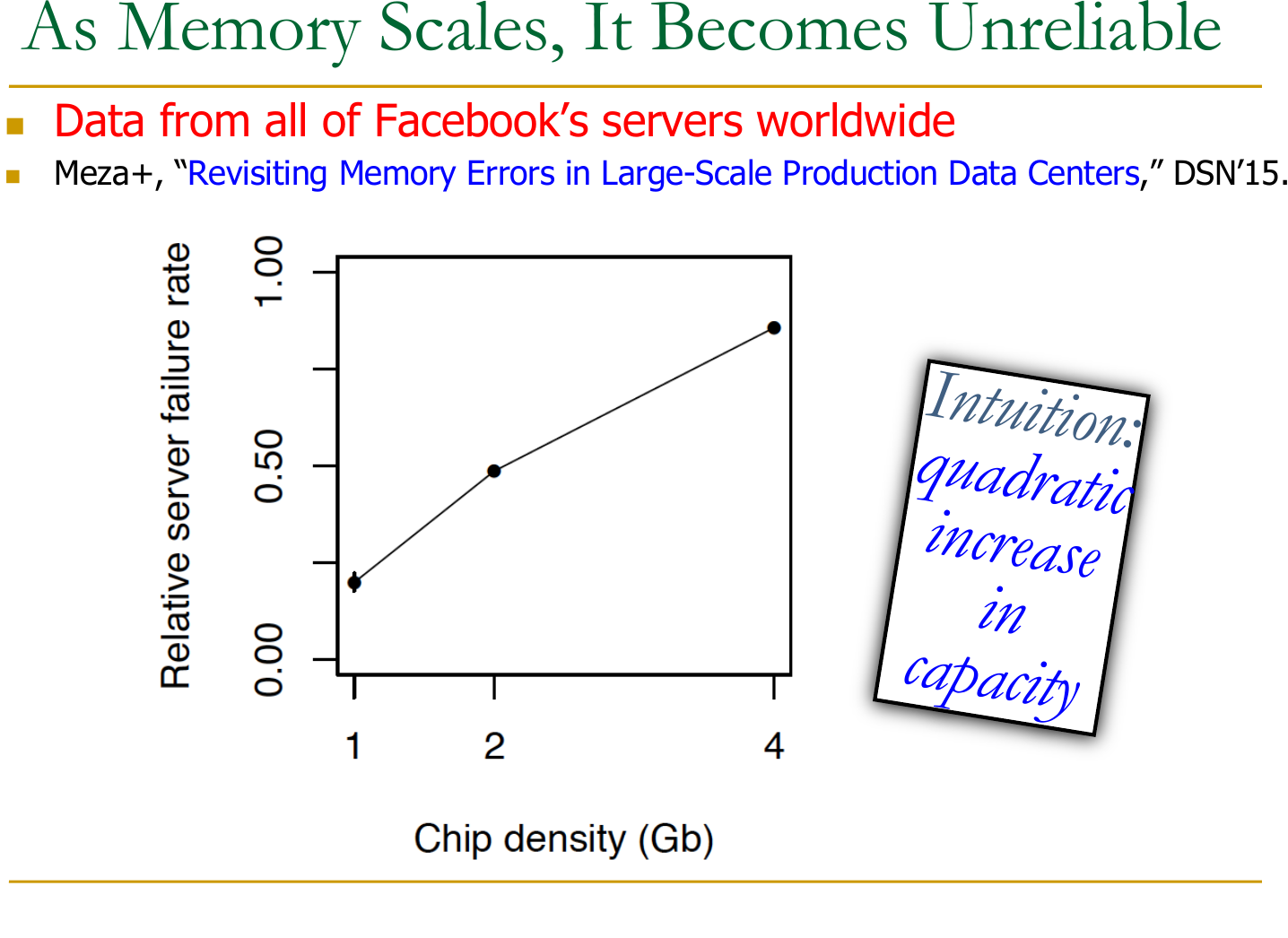}
\caption{\gf{The relative failure rate
for servers \omv{using DRAM chips with different} densities. \omv{Higher density chips} (related to
newer technology nodes) \omv{correlate with} higher \omv{server} failure rates. Reproduced from~\cite{mutlu.accml23.talk}. Originally presented in~\cite{meza.dsn15}.}
}
\label{fig:error_density_facebook}
\end{figure}

The second key concern is that DRAM technology scaling to
  smaller nodes adversely affects DRAM reliability.  A DRAM cell
stores one bit of data in the form of charge in a capacitor, which is
accessed via an access transistor and peripheral circuitry. For a DRAM
cell to operate correctly, both the capacitor and the access
transistor (as well as the peripheral circuitry) need to operate
reliably. As the size of the DRAM cell reduces, both the capacitor and
the access transistor become less reliable, more leaky, and generally more vulnerable to electrical noise and disturbance.  As a result,
{reducing the size of the DRAM cell} increases the difficulty of
correctly storing and detecting the desired original value in the DRAM, as shown in various recent works that study DRAM reliability by analyzing data retention and other reliability issues of modern DRAM chips
cell~\gf{\cite{liu.isca13,mutlu.imw13,kim-isca2014,mutlu2017rowhammer,khan.sigmetrics14, khan.dsn16, khan.cal16, khan.micro17, qureshi.dsn15, patel2017reaper, hassan2017softmc, hasan2019crow, raidr, patel2020bit, patel2021harp, patel2019understanding, kim2020revisiting, yauglikcci2021blockhammer, hassan2021uncovering, orosa2021deeper, mutlu2020retrospective, frigo2020trr, cojocar2020susceptible,luo2023rowpress,yaglikci2024spatial,olgun2024abacus,bostanci2024comet,yauglikcci2022hira}}.
Hence, memory technology scaling causes memory errors to appear more frequently. For example, a study of Facebook's entire production datacenter servers
showed that memory errors, and thus the server failure rate, are strongly positively  correlated with the density of the chips employed in the
servers~\cite{meza.dsn15}: the higher the density of the chip used in a server, the more likely the server is to experience a memory error and server failure\omiv{, as Figure~\ref{fig:error_density_facebook} depicts}. Thus, it is critical to make the main memory
system more reliable to build reliable computing systems on top of it.

The third key {issue} is that the reliability problems caused by
aggressive DRAM technology scaling can lead to new \omv{robustness-related (including security \omv{and safety \omvi{problems}) vulnerabilities}.}
\omv{For instance, many works demonstrated that the \emph{read disturbance} \omvi{phenomenon} in DRAM can be exploited to cause security and safety problems that enable attackers to 
take over a system~\agy{\cite{aga2017good, bosman.2016, cojocar19exploiting, frigo2020trr, gruss2016rowhammerjs, anotherflip, qiao2016new, tatar2018defeating, drammer-github, vanderveen.2016, vanderveen2018guardion, zhang2020pthammer,wang2022research, kurmus2017from,seaborn.2015,seaborn.2016,gruss.2015,glitch-vu,nethammer,throwhammer}}, 
read data they do \emph{not} have access to~\agy{\cite{ cojocar19exploiting, carre2018openssl, cohen2022hammerscope, frigo2020trr, ji2019pinpoint, kwong2020rambleed, qiao2016new, tobah2022spechammer, kaur2022workinprogress, li2023fphammer,vanderveen.2016}}, break out of virtual machine sandboxes~\cite{razavi.2016,cloudflops}, 
corrupt important data (even rendering machine learning inference useless)~\cite{kim-isca2014,yao2020deephammer,tol2023don,hong2019terminal, cojocar2020susceptible, deridder2021smash, jattke2022blacksmith, glitch-vu, frigo2020trr, hassan2021uncovering, kogler2022halfdouble, pessl2016drama, qiao2016new, razavi.2016, zhang2018triggering}, 
steal secret data (e.g., cryptographic keys~\agy{\cite{bhattacharya2016curious,bhattacharya2018advanced,carre2018openssl, cohen2022hammerscope, cojocar19exploiting, fahrjr2022when, frigo2020trr, ji2019pinpoint, kwong2020rambleed, poddebniak2018attacking, tobah2022spechammer, weissman2020jackhammer, mus2022jolt, fahr2022theeffects, islam2022signature, tomita2022extracting,razavi.2016}} and \omvii{steal or} \agy{alter} machine learning model parameters~\agy{\cite{hong2019terminal, liu2022generating, rakin2022deepsteal, tol2023don, zheng2022trojvit, cai2022onthe, roohi2022efficient, staudigl2022neurohammer,yang2022sociallyaware,rakin2019bit}}).} 

\begin{figure}[ht]
\centering
\includegraphics[width=1.0\linewidth]{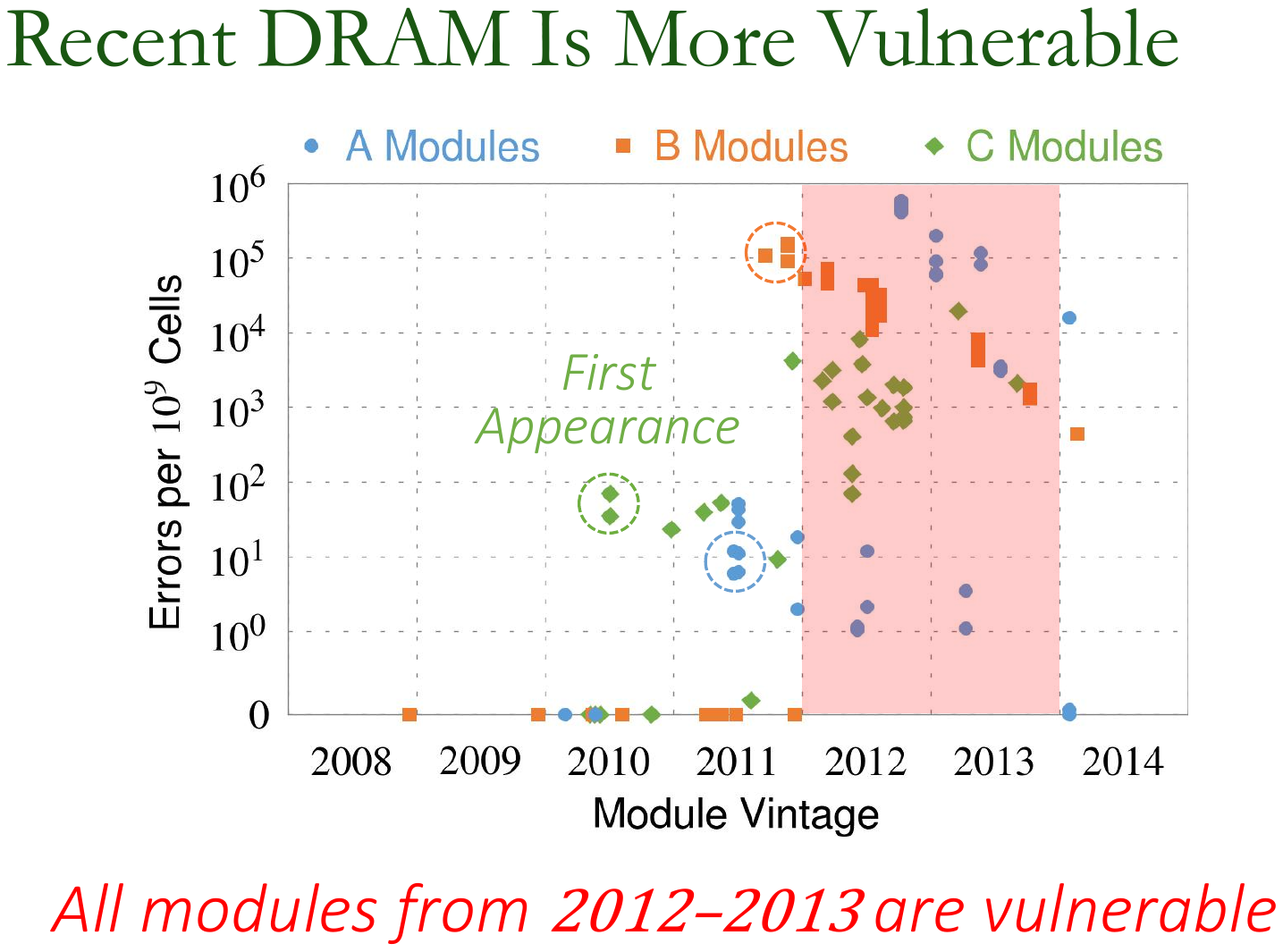}
\caption{RowHammer vulnerability for DRAM modules manufactured between 2008 and 2014. Reproduced from~\cite{mutlu.nsfpim20}. Originally presented in~\cite{kim-isca2014,kim.isca2014talk}.
}
\label{fig:rowhammer}
\end{figure}

\gf{Read disturbance is the phenomenon that reading data from a memory device causes physical disturbance (e.g., voltage deviation, electron injection, electron trapping) on another piece of data that is \emph{not} accessed but physically located near the accessed data. Two prime examples of read disturbance in modern DRAM chips are RowHammer~\cite{kim-isca2014} and RowPress~\gf{\cite{luo2023rowpress,luo2024rowpress}}.}
{The RowHammer}~\omv{\cite{kim-isca2014,mutlu2017rowhammer,mutlu2020retrospective,kim2020revisiting,mutlu2023fundamentally,mutlu2023retrospective}} \gf{and RowPress~\gf{\cite{luo2023rowpress,luo2024rowpress}} phenomena} show that it is possible to {predictably} induce errors {(bit flips)} in most {modern} DRAM chips.  {Repeatedly reading \gf{or keeping active a} row in DRAM can corrupt data in physically-adjacent rows. Specifically, \gf{in RowHammer}, when a DRAM row is opened (i.e., activated) and closed (i.e., precharged) repeatedly (i.e., hammered), enough times within a DRAM refresh interval, one or more bits in physically-adjacent DRAM rows can be flipped to the wrong value.
\gf{Similarly, in RowPress, keeping a DRAM row open for a long period of time (i.e., pressing a DRAM row) amplifies the effects of read disturbance and induces RowPress bit flips, without \omv{requiring} many repeated DRAM row activations. }
A very simple user-level program~\cite{rowhammer.github} can reliably and consistently induce RowHammer \gf{and RowPress} errors in vulnerable DRAM modules.}  {The seminal paper that introduced RowHammer~\cite{kim-isca2014} \gf{shows} that more than 85\% of the chips tested, built by three major vendors between 2010 and 2014, were vulnerable to RowHammer-induced errors}. In particular, \emph{all} DRAM modules from 2012 and 2013 are vulnerable, as shown by Figure~\ref{fig:rowhammer}\omv{,} which depicts the observed RowHammer error vulnerability of DRAM modules manufactured between 2008 and 2014 by all three major DRAM manufacturers\omv{, called} A, B, C~\cite{kim-isca2014}. 
A \omv{more} recent technology scaling study~\cite{kim2020revisiting} 
of 1580 DRAM chips belonging to three different DRAM types and various different technology node sizes experimentally demonstrated that the RowHammer vulnerability is getting much worse at the circuit level: fewer number of activates to a row can cause bit flips in the most recent chips and recent chips experience higher bit flip rates due to RowHammer. The same work~\cite{kim2020revisiting} 
also showed that existing RowHammer mitigation mechanisms will not be effective in future DRAM chips that will be much more vulnerable to RowHammer, and thus RowHammer remains to be an open vulnerability to securely protect against. \omiv{RowPress exacerbates the read disturbance problem by 1--2 orders of magnitude and makes the read disturbance vulnerability of modern DRAM chips \omv{much} more acute.
\omv{Other recent works~\omvi{\cite{olgun2024read,olgun2023anexperimental}} demonstrate that the RowHammer vulnerability is similarly present in \gls{HBM}~\omvi{\cite{HBM, jedec.hbm.spec,lee.taco16}} chips, which are commonly used in machine learning and high-performance computing infrastructures of today.}}

\begin{figure}[ht]
\centering
\includegraphics[clip, trim=0.0cm 2.5cm 0.0cm 0.0cm, width=1.0\linewidth]{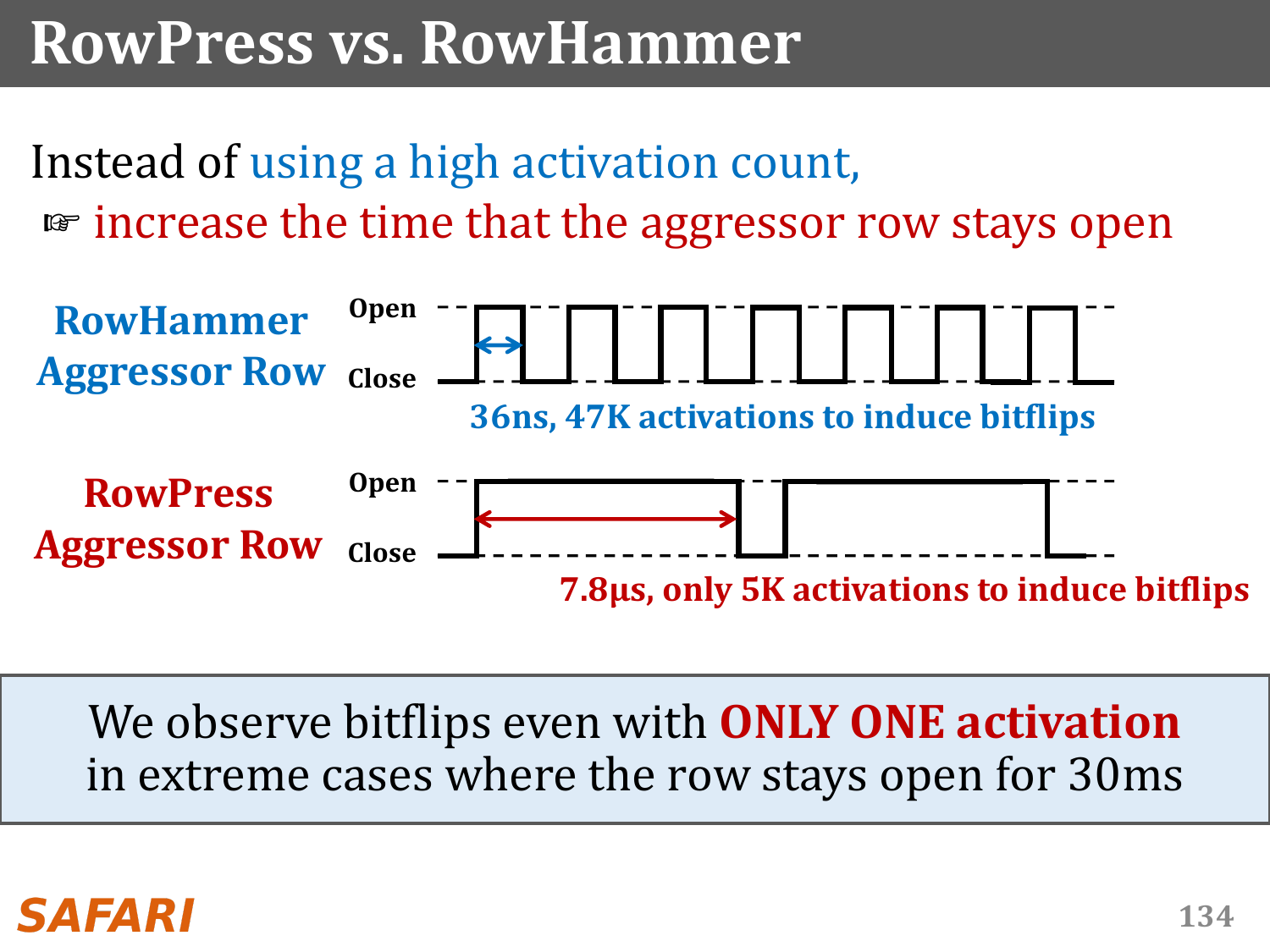}
\caption{\gfix{Fundamental difference between RowHammer and RowPress \omviii{(i.e., how long an aggressor row is kept open)} and the resulting effect on the number of activations required to induce a bitflip. Reproduced from~\cite{mutlu.njit2023talk}.}}
\label{fig:rowhammer_vs_rowpress}
\end{figure}

\begin{figure}[ht]
\centering
\includegraphics[width=1.0\linewidth]{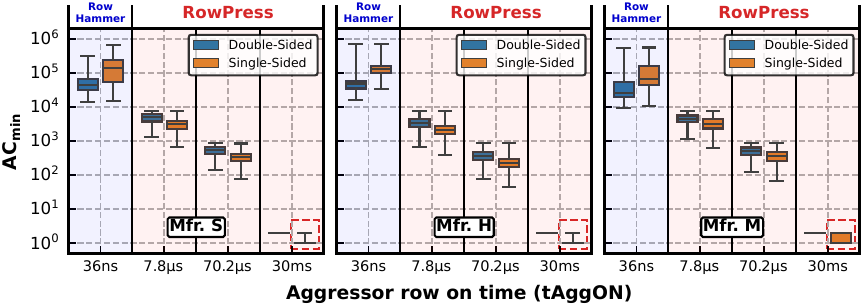}
\caption{\gfvii{The minimum number of total aggressor row activations to cause at least one bitflip ($AC_{min}$) distributions of conventional RowHammer and three representative cases of RowPress at \SI{80}{\degreeCelsius} with one
(single-sided) and two (double-sided) aggressor row(s) across 164 DDR4 chips from manufacturers S, H, and M \omviii{(i.e., Samsung, SK Hynix, Micron)}. Reproduced from~\cite{luo2023rowpress}.}}
\label{fig:rowpress}
\end{figure}

\gf{The \omv{original} paper \omv{that introduced RowPress}~\cite{luo2023rowpress} shows, by characterizing 164 real DDR4 DRAM chips from three major DRAM manufacturers, that RowPress 
(1)~\omv{is prevalent in} \emph{all} chips
from \emph{all} three major DRAM manufacturers, 
(2)~gets worse as DRAM
technology scales down to smaller node sizes, and 
(3)~affects a different set of DRAM cells from RowHammer and behaves differently
from RowHammer as temperature and access pattern changes.}
\omv{Figure~\ref{fig:rowhammer_vs_rowpress} shows the \omvi{fundamental difference between RowHammer and RowPress: RowPress keeps the row open larger after each activation. Doing so leads to a large reduction in number of activations required to induce a bitflip, as shown in the example in the figure}. 
\omvii{Under realistic conditions, RowPress leads to 1--2 orders of magnitude reduction in the number of activations required to induce a bitflip, as Figure~\ref{fig:rowpress} demonstrates.}
In RowPress, we observe bitflips even with \emph{only one} activation in extreme cases (i.e., when the aggressor row stays open for 30~ms). }

The RowHammer \gf{and RowPress} \gf{phenomena} entail a real \omv{robustness problem that stems from unwanted data corruption.
Such a robustness problem is not only a threat to DRAM reliability and technology scaling, but it also causes a real and prevalent security and safety problem.} 
\omv{RowHammer and RowPress}
break physical {memory} isolation between two addresses, one of
the fundamental building blocks of \omv{the} memory \omv{abstraction}, on top of which system
security principles are built. With RowHammer, accesses to one row
(e.g., an application page) can modify data stored in another memory
row (e.g., an \omv{operating system} page).  
\omv{Due to unwanted bitflips caused on other locations, the original RowHammer paper predicted that the RowHammer bitflips can be used to \omvi{inject} errors into other programs, \omvi{to crash} the system, \omvi{and to even hijack} control of the system~\cite{kim-isca2014}.}
This was confirmed in 2015 by researchers from Google Project Zero, who developed a user-level attack that uses RowHammer to
gain kernel privileges~\cite{seaborn.2015,seaborn.2016}.  Other
researchers have shown how RowHammer vulnerabilities can be exploited
in various ways to gain privileged access to various systems: in a
remote server RowHammer can be used to remotely take over the server
via the use of JavaScript~\cite{gruss.2015}; a virtual machine can
take over another virtual machine by inducing errors in the victim
virtual machine's memory space~\cite{razavi.2016}; a malicious
application without permissions can take control of an Android mobile
device~\cite{vanderveen.2016}; an attacker can gain arbitrary
read/write access in a web browser on a Microsoft Windows 10
system~\cite{bosman.2016}\omv{;
or \omvi{a user-level} attacker can read \omvi{secret cryptographic} keys from the root\omvi{-}level SSH daemon~\cite{kwong2020rambleed}}. 
Over the past \omv{ten} years, many security attacks \gf{and defenses} were developed to exploit~\gf{\cite{cloudflops,bosman.2016,anotherflip,qiao2016new,bhattacharya2016curious,jang2017sgx,aga2017good,pessl2016drama,gruss.2015,razavi.2016,vanderveen.2016,glitch-vu,fournaris2017exploiting,poddebniak2018attacking,nethammer,throwhammer,tatar2018defeating,carre2018openssl,barenghi2018software,zhang2018triggering,bhattacharya2018advanced,cojocar19exploiting,seaborn.2015,seaborn.2016,burleson2016invited,brasser2017cant,mutlu2017rowhammer,vanderveen2018guardion,ji2019pinpoint,mutlu2020retrospective,hong2019terminal,kwong2020rambleed,frigo2020trr,cojocar2020susceptible,weissman2020jackhammer,zhang2020pthammer,yao2020deephammer,deridder2021smash,hassan2021uncovering,jattke2022blacksmith,tol2022toward,kogler2022halfdouble,orosa2022spyhammer,zhang2022implicit,liu2022generating,cohen2022hammerscope,zheng2022trojvit,fahrjr2022when,tobah2022spechammer,rakin2022deepsteal,aydin2022cyber,mus2022jolt,wang2022research,lefforge2023reverse,fahr2022theeffects,kaur2022workinprogress,cai2022onthe,li2022cyberradar,roohi2022efficient,staudigl2022neurohammer,yang2022sociallyaware,islam2022signature,tomita2022extracting,france2022modeling}} \gf{and mitigate~\cite{apple2015about, enterprise2015hpmoonshot,lenovo2015rowhammer,greenfield2012throttling, kim-isca2014, kim2014architectural,  bains2015method, bains2016rowhammer, bains2016distributed, aichinger2015ddrmemory, aweke2016anvil, gomez2016dram_rowhammer, yang2016suppression, son2017making, seyedzadeh2017counterbased, seyedzadeh2017mitigating, seyedzadeh2018mitigating, irazoqui2016mascat, ryu2017overcoming, yang2017scanning, you2019mrloc, lee2019twice, park2020graphene, yaglikci2021security, yauglikcci2021blockhammer, canpolat2024breakhammer, frigo2020trr, kang2020cattwo, hassan2021uncovering, qureshi2022hydra, saileshwar2022randomized, brasser2017cant, konoth2018zebram, vanderveen2018guardion, vig2018rapid, hasan2019crow, gautam2019rowhammering, kim2022mithril, lee2021cryoguard, marazzi2023protrr, zhang2022softtrr, joardar2022learning, juffinger2023csirowhammercryptographic, yauglikcci2022hira, saxena2022aqua, manzhosov2022revisiting, ajorpaz2022evax, naseredini2022alarm, joardar2022machine, hassan2022acase, zhang2020leveraging,loughlin2021stop, devaux2021method, han2021surround, fakhrzadehgan2022safeguard, saroiu2022theprice, saroiu2022howto, loughlin2022moesiprime, zhou2022ltpim, hong2023dsac, mutlu2023fundamentally, marazzi2022rega, didio2023copyonflip, sharma2022areview, woo2023scalable, park2022rowhammer_reduction, wi2023shadow, kim2023a11v, guderamarao2023defending, guha2022criticality, france2022modeling, france2022reducing, bennett2021panopticon, enomoto2022efficient, arikan2022processor, tomita2022extracting, saxena2023ptguard, zhou2023dnndefender, bostanci2024comet, olgun2024abacus} RowHammer}.
\omiv{More} recently, the TRRespass attack~\cite{frigo2020trr} 
showed that existing \omiv{DDR4} DRAM chips that are advertised to be RowHammer-\omv{free}, as described by various DRAM vendors~\cite{lee2014green,micron2016ddr4},
are actually vulnerable because these mitigation mechanisms can be circumvented with a new type of RowHammer attack called {\em many-sided hammering}. 
\omv{Uncovering TRR (U-TRR)~\cite{2021utrr} shows that one can reverse-engineer DRAM chips that implement target row refresh (TRR) RowHammer mitigation mechanisms and craft specialized access patterns that essentially induce large numbers of bitflips on DRAM chips, even in the presence of TRR.}
\omiv{SMASH~\cite{deridder2021smash} and Blacksmith~\cite{jattke2022blacksmith} works \omvi{demonstrate} automated attacks that are successful against DDR4 chips. 
\omvi{Both Google and Microsoft recently \msvii{also} developed RowHammer attacks and solutions~\cite{kogler2022halfdouble,loughlin2022moesiprime,bennett2021panopticon,cojocar2020susceptible,loughlin2021stop,saroiu2022howto,saroiu2022theprice}.}
Even though the DRAM industry \omvi{is} finally \omv{writing} papers about the problem \omv{and suggesting different solutions} \gfvi{(such as~\cite{kim2023a11v} and \cite{hong2023dsac})}, \omv{as well as modifying the DRAM standards to incorporate better solutions \omvi{than} before~\omvi{\cite{canpolat2024understanding,jedec2024jesd795c,canpolat2025chronus}},} the problem remains to be securely and provably solved \omv{at low performance and energy overheads}~\omv{\cite{mutlu2023fundamentally,canpolat2024understanding}}.}
For a more detailed treatment of the
RowHammer problem and its consequences, as well as its root causes, modeling, and analyses, we refer the reader
to \omiv{various overview papers~\agy{\cite{mutlu2023fundamentally,mutlu2017rowhammer,mutlu2020retrospective,mutlu2019rowhammer,fournaris2017exploiting, mutlu2015themain,burleson2016whois, burleson2016invited, aydin2022cyber, giray-thesis}} and recent research papers~}~\gf{\cite{kim-isca2014,mutlu2017rowhammer,rowhammer-topinhes18,kim2020revisiting, yauglikcci2021blockhammer, hassan2021uncovering, orosa2021deeper,frigo2020trr,cojocar2020susceptible,redeker2002aninvestigation, park2014activeprecharge, park2016statistical, yang2016suppression, park2016experiments,lim2017active, ryu2017overcoming, yang2017scanning, lim2018study, yun2018study, yang2019trapassisted, gautam2019rowhammering, walker2021ondram, jiang2021quantifying, orosa2022spyhammer, cohen2022hammerscope, khan2018analysis, agarwal2018rowhammer_for, li2014write, ni2018write, genssler2022onthe, he2023whistleblower, baeg2022estimation, olgun2023anexperimental, olgun2023drambender, zhou2023doublesided, luo2023rowpress,giray-thesis}} \omiv{on the topic}. 

The fourth key {issue} is the power and energy consumption of
main memory. DRAM is inherently a power and energy hog, as it consumes
energy even when it is not used (e.g., it requires periodic memory
refresh~\cite{raidr}), due to its charge-based nature. And, energy
consumption of main memory is becoming worse due to three major
reasons. First, main memory's capacity, bandwidth, parallelism, and complexity are all
increasing, causing energy consumption to naturally increase due to higher amount of dynamic activity and higher overall static power consumption. Second, main memory has remained off the main processing
chip\omiv{(s)} and thus did not benefit from many energy reduction mechanisms that come with better integration, even though many other platform components have been integrated into the processing chip and have benefited from the aggressive energy/voltage scaling mechanisms~\gf{\cite{david2011memdvfs,haj2020sysscale,kim2008system,mallik2006user,wu2005voltage,haj2020flexwatts,gonzalez1996energy,rotem2011power,yahya2022darkgates}} and the low-energy communication substrate on-chip. Third, the difficulties in DRAM technology scaling are making DRAM energy reduction very difficult with technology generations. In fact, some of the mechanisms that are added to DRAM chips to compensate for reliability problems in smaller technology generations, e.g., in-DRAM error correcting codes~\gf{\cite{patel2017reaper,patel2019understanding,patel2020bit, patel2021harp,kang.memoryforum14,oh2014ecc,jedec2020ddr5,nair2016xedexposing,micron2017ecc,gong2017dram_scaling,son2015cidra,kwon2014understanding,mineshphd}}\omv{,} higher refresh rates~\cite{patel2020bit, patel2021harp,kwak2017refresh,kwon2017wearable,apple2015about}, 
\omv{and RowHammer-prevention mechanisms~\omvi{\cite{canpolat2024understanding,kim2023a11v,jedec2024jesd795c,canpolat2025chronus}},}
directly increase energy consumption. As a result of these three major issues that make main memory a larger energy bottleneck, the fraction of the entire system power consumed by main memory \omv{has been} increasing over the last two decades. 
In 2003,
{Lefurgy et al.~\cite{lefurgy.2003} showed that, in
  large commercial servers designed by IBM, the off-chip memory
  hierarchy (including, at that time, DRAM, interconnects, memory
  controller, and off-chip caches) consumed between 40\% and 50\% of
  the total system energy. The trend has become even worse over the
  course of the one-to-two decades.}  In recent computing systems with
CPUs or GPUs, {\em only DRAM itself} is shown to account for more
than 40\% of the total system power~\cite{ware.2010,paul.2015,chang.sigmetrics17,ghose2018vampire}.
Hence, the power and energy consumption of main memory is increasing relative
to that of other components in computing platform. As energy
efficiency and sustainability are critical necessities in computing
platforms today, it is critical to reduce the energy and power
consumption of main memory~\cite{david2011memdvfs,deng2011memscale,chang.sigmetrics17,ghose2018vampire,haj2020sysscale,haj2020techniques,ghose2019demystifying}.


%% file: sections/03-intelligent-controllers.tex
\section{The Need for Intelligent Memory Controllers to Enhance Memory \omvii{Technology} Scaling}
\label{sec:intmemcont}

A key promising approach to solving the four major issues \omv{described in Section~\ref{sec:majortrends}} is to
design {\em intelligent memory controllers} that can manage main
memory better~\omv{\cite{mutlu.imw13}}. If the memory controller is designed to be more
intelligent and more programmable, it can, for example, incorporate
flexible mechanisms to overcome various types of \omv{robustness} issues
(including RowHammer \omiv{and RowPress}), manage latencies and energy/power consumption better
based on a deep understanding of the DRAM chip and application
characteristics, provide enough support for programmability to prevent
security\omiv{,} reliability\omiv{, and safety} vulnerabilities that are discovered in the
field, and manage various types of memory technologies that are put
together as a hybrid main memory to enhance the scaling of the main
memory system. We provide a few examples of how an intelligent memory
controller can help overcome circuit- and device-level issues modern computing systems are
facing at the main memory level. We believe having intelligent memory
controllers can greatly alleviate the scaling issues encountered with
main memory today, as we have described in an earlier position
paper~\cite{mutlu.imw13}. This is a direction that is also supported
by key hardware manufacturers in computing industry today, as described in an informative paper written
collaboratively by Intel and Samsung engineers on DRAM technology
scaling issues\sg{~\cite{kang.memoryforum14}} \omiv{as well as more recent developments \omv{that} modify memory chips \omv{to include relatively intelligent \msvii{control} techniques that can} mitigate RowHammer~\omvi{\cite{kim2023a11v,hong2023dsac,bennett2021panopticon,canpolat2024understanding,jedec2024jesd795c,canpolat2025chronus}}}.

In this section, we give several examples of how an intelligent memory controller can help overcome major scaling challenges of modern DRAM.   
First, a slightly more intelligent memory controller than today's controllers can prevent the RowHammer vulnerability by
probabilistically refreshing rows that are physically adjacent to an activated row, with a very low probability. This solution, called PARA
(\gf{p}robabilistic \gf{a}djacent \gf{r}ow \gf{a}ctivation)~\cite{kim-isca2014} was shown
to provide strong, programmable, robust guarantees against RowHammer, with
very little power, performance and chip area
overheads~\cite{kim-isca2014}. It requires a slightly more intelligent
memory controller that 
(1)~knows (or that can figure out) the physical
adjacency of rows in a DRAM chip, 
(2)~is programmable enough to
adjust the probability of adjacent row activation depending on the vulnerability of a chip, and 
(3)~can issue refresh
requests to physically-adjacent rows accordingly to the probability supplied by
the system or discovered online. As described by prior
work~\cite{kim-isca2014,mutlu2017rowhammer,rowhammer-topinhes18,mutlu2020retrospective,kim2020revisiting},
this solution has much lower performance and energy overheads than \sg{increasing the} refresh rate 
for the entire main memory, which \omv{was the \omvi{first}} RowHammer solution
employed by systems~\omvi{\cite{apple2015about,mutlu2020retrospective}} \omv{when RowHammer was observed} in the field \omv{due to} simple and rigid
memory controllers. 
\omv{After this initial solution, \omvi{which was} proposed along with multiple other solutions in the original RowHammer paper~\cite{kim-isca2014}, many other solutions to RowHammer have been developed~\cite{apple2015about, enterprise2015hpmoonshot,lenovo2015rowhammer,greenfield2012throttling, kim-isca2014, kim2014architectural,  bains2015method, bains2016rowhammer, bains2016distributed, aichinger2015ddrmemory, aweke2016anvil, gomez2016dram_rowhammer, yang2016suppression, son2017making, seyedzadeh2017counterbased, seyedzadeh2017mitigating, seyedzadeh2018mitigating, irazoqui2016mascat, ryu2017overcoming, yang2017scanning, you2019mrloc, lee2019twice, park2020graphene, yaglikci2021security, yauglikcci2021blockhammer, canpolat2024breakhammer, frigo2020trr, kang2020cattwo, hassan2021uncovering, qureshi2022hydra, saileshwar2022randomized, brasser2017cant, konoth2018zebram, vanderveen2018guardion, vig2018rapid, hasan2019crow, gautam2019rowhammering, kim2022mithril, lee2021cryoguard, marazzi2023protrr, zhang2022softtrr, joardar2022learning, juffinger2023csirowhammercryptographic, yauglikcci2022hira, saxena2022aqua, manzhosov2022revisiting, ajorpaz2022evax, naseredini2022alarm, joardar2022machine, hassan2022acase, zhang2020leveraging,loughlin2021stop, devaux2021method, han2021surround, fakhrzadehgan2022safeguard, saroiu2022theprice, saroiu2022howto, loughlin2022moesiprime, zhou2022ltpim, hong2023dsac, mutlu2023fundamentally, marazzi2022rega, didio2023copyonflip, sharma2022areview, woo2023scalable, park2022rowhammer_reduction, wi2023shadow, kim2023a11v, guderamarao2023defending, guha2022criticality, france2022modeling, france2022reducing, bennett2021panopticon, enomoto2022efficient, arikan2022processor, tomita2022extracting, saxena2023ptguard, zhou2023dnndefender, bostanci2024comet, olgun2024abacus}.
Almost all of these solutions require an intelligent controller to mitigate the RowHammer vulnerability. Similarly, intelligent memory controllers were also designed to mitigate the RowPress vulnerability~\cite{qureshi2024impress,luo2023rowpress,luo2024rowpress}.}

\begin{figure}[ht]
\centering
\includegraphics[width=1.0\linewidth]{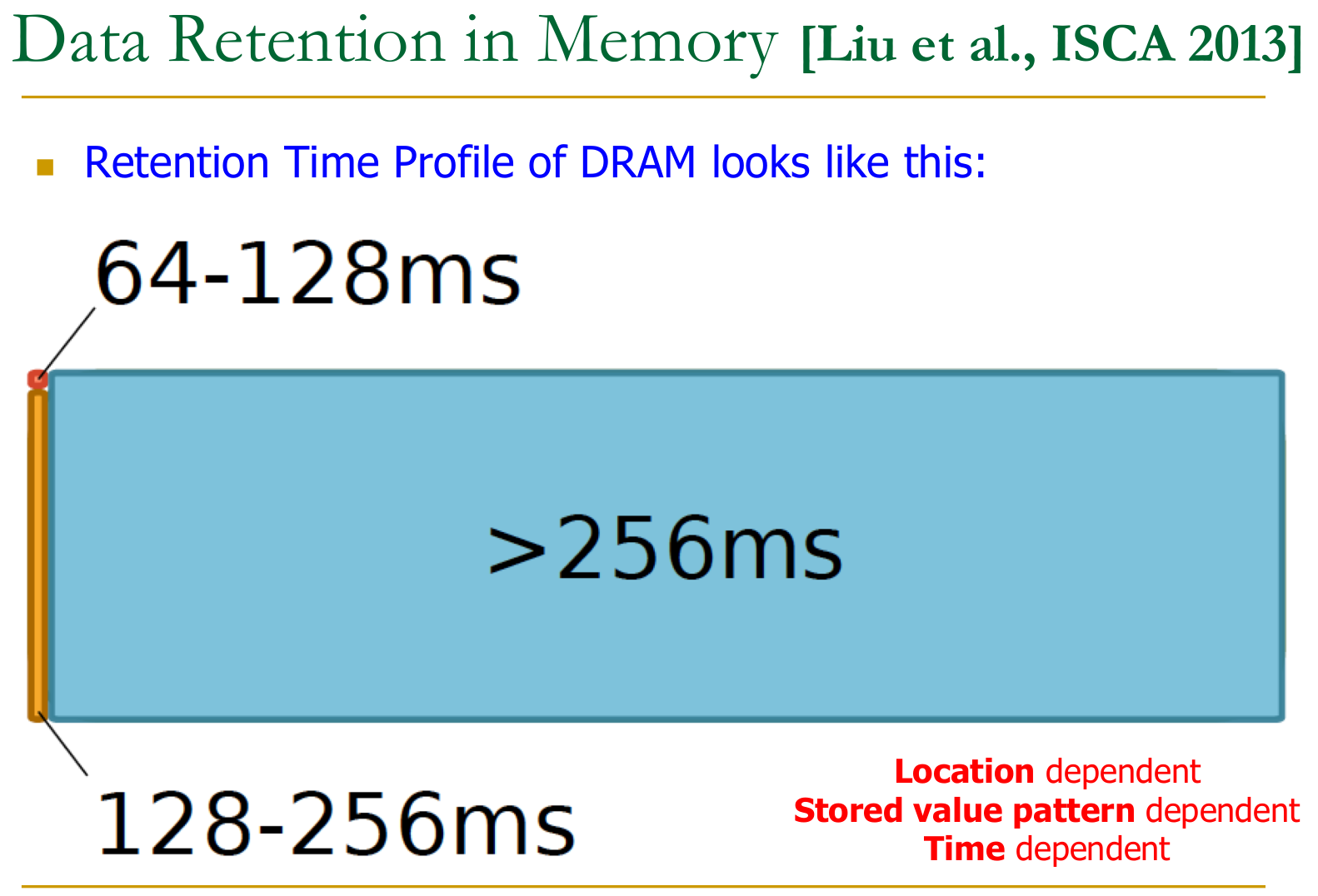}
\caption{Data retention times of different DRAM cells, represented as a \omv{cartoonish picture} based on experimental data obtained from real DRAM chips~\cite{kim2009new}.
Reproduced from~\cite{mutlu.nsfpim20}. Originally presented in~\cite{raidr.isca2012talk,mutlu.isca2013talk}.
}
\label{fig:dram-refresh}
\end{figure}

Second, an intelligent memory controller can greatly alleviate the
refresh problem in DRAM, and hence its negative consequences on
energy, performance, predictability, and technology scaling, by
understanding the retention time characteristics of different rows
well. It is well known that the retention time of different cells in
DRAM are widely different due to process manufacturing
variation~\cite{raidr,liu.isca13}. A very large fraction of all DRAM cells are strong (i.e., they
can retain data for hundreds of seconds), whereas only a small fraction of DRAM cells are weak
(i.e., they can retain data for only \SI{64}{\milli\second}~\omv{\cite{raidr,liu.isca13}}), as demonstrated in Figure~\ref{fig:dram-refresh}.
Yet, today's memory
controllers treat \sg{every cell} as equal and refresh all rows every \SI{64}{\milli\second} \omv{(in 2012; now every \SI{32}{\milli\second} or even lower~\omvi{\cite{2018ddr5,jedec2020ddr5}})},
which is the \sg{worst}-case retention time that is allowed. This
worst-case refresh rate leads to a large number of unnecessary
refreshes, and thus great energy waste and performance loss. Refresh
is also shown to be the key technology scaling limiter of
DRAM~\omv{\cite{kang.memoryforum14,raidr}}, and\omv{,} as such\omv{,} refreshing all DRAM cells
at the worst case rates is likely to make DRAM technology scaling \omv{more}
difficult.  An intelligent memory controller can overcome the refresh
problem by 
1)~identifying the minimum data retention time of each row
(during online operation)\omv{,} 
2)~refreshing each row at the rate it
really requires to be refreshed at or 
3)~by \sg{decommissioning} weak rows
such that data is not stored in them. 
As shown by a recent body of
work whose aim is to design such an intelligent memory controller that
can perform online profiling of DRAM cell retention times and online
adjustment of refresh rate on a per-row
basis~\omv{\cite{raidr,liu.isca13,khan.sigmetrics14,qureshi.dsn15,khan.dsn16,khan.cal16,patel2017reaper,khan.micro17,patel2021harp,patel2020bit,patel2019understanding,patel2024rethinking,patel2024rethinkingieee}},
including the works on RAIDR~\cite{raidr,liu.isca13},
AVATAR~\cite{qureshi.dsn15} and REAPER~\cite{patel2017reaper}, such an
intelligent memory controller can eliminate more than 75\% of all
refreshes at very low cost, leading to \omv{large system-level} energy reduction,
performance improvement, and quality of service benefits, all \omv{obtained} at the
same time. Thus, the downsides of DRAM refresh \omv{and the technology scaling challenges related to handling data retention issues in DRAM} can potentially be \omv{efficiently}
overcome with the design of intelligent memory controllers.

\sg{Third, an intelligent memory controller can enable performance
  improvements that can overcome the limitations of memory \omvi{technology} scaling.
  As we discuss in Section~\ref{sec:majortrends}, DRAM latency has
  remained almost constant over the last decades, despite the
  fact that low-latency computing has become even more important during
  that time. Similar to how intelligent memory controllers handle the
  refresh problem, the controllers can exploit the fact that not all
  cells in DRAM need the same amount of time to be accessed.
  Modern DRAM specifications require worst-case timing parameters that define the amount of
  time required to perform a memory access.  In order to guarantee
  correct operation, the timing parameters are chosen to ensure that
  the \emph{worst-case} cell in any DRAM chip that is acceptable (to satisfy a yield rate) can still
  be accessed correctly at \emph{worst-case operating
    temperatures}~\cite{chang.sigmetrics16, lee.hpca15,lee.sigmetrics17, kim2018solar,wang2018cal,hasan2019crow,luo2020clr}.
    However, we find that access
  latency to cells is very heterogeneous due to variation in 
  operating conditions (e.g., across different temperatures and
  operating voltage levels), manufacturing process (e.g., across different
  chips and different parts of a chip), and access patterns (e.g., based on
  whether or not the cell was recently accessed). We give eight examples
  of how an intelligent memory controller can exploit the various
  different types of heterogeneity in access latency.

  \sgii{(1)~At} low temperature, DRAM cells contain more charge, and
  as a result, can be accessed much faster than at high
  temperatures. We find that, averaged across 115 real DRAM modules
  from three major manufacturers, read and write latencies of DRAM can
  be reduced by 33\% and \sgii{55\%, respectively,} when operating at
  relatively low temperature (\sgii{\SI{55}{\celsius}}) compared to
  operating at worst-case temperature
  (\sgii{\SI{85}{\celsius}})~\cite{lee.hpca15, lee.thesis16}. Thus, a slightly
  intelligent memory controller can greatly reduce memory latency by
  adapting the access latency to operating temperature.

  \sgii{(2)~Due} to manufacturing process variation, we find that the
  majority of cells in DRAM (across different chips or within the same
  chip) can be accessed much faster than the manufacturer-provided
  timing parameters~\omv{\cite{chang.sigmetrics16, lee.hpca15,
    lee.sigmetrics17, kim2018solar, kevinchang-thesis, lee.thesis16,kim2020improving}}.  
  An intelligent memory controller
  can profile the DRAM chip and identify which cells can be accessed
  reliably at low latency, and use this information to reduce access
  latencies by as much as 57\%~\cite{chang.sigmetrics16,
    lee.sigmetrics17, kim2018solar}.

  \sgii{(3)~In} a similar fashion, an intelligent memory controller
  can use similar properties of manufacturing process variation to
  reduce the energy consumption of a computer system, by exploiting
  the minimum voltage required for \omv{reliable} operation of different parts
  of a DRAM chip~\cite{chang.sigmetrics17, kevinchang-thesis}. 
  The key idea is to reduce
  the operating voltage of a DRAM chip from the standard specification
  and tolerate the resulting errors by increasing access latency on a
  per-bank basis, while keeping performance degradation in check.

  \sgii{(4)~Bank} conflict latencies can be dramatically reduced by
  making modifications in the DRAM chip such that different subarrays
  in a bank can be accessed mostly independently\sgii{,} and designing
  an intelligent memory controller that can take \sgii{advantage} of
  requests that require data from different subarrays (i.e., exploit
  subarray-level parallelism)~\omv{\cite{salp, yoongu-thesis,yauglikcci2022hira}}. A similar approach is also shown to reduce the performance impact of refresh by enabling parallelization of refresh and access operations to a bank~\cite{chang.hpca14,yauglikcci2022hira}. 
  \omv{Recent work fascinatingly demonstrates that refresh access parallelization across different subarrays in a bank is possible in real \gls{COTS} DRAM chips by violating DRAM timing parameters~\cite{yauglikcci2022hira}.}

  \sgii{(5)~Access} latency to a portion of the DRAM bank can be
  greatly reduced by partitioning the DRAM array such that a subset of
  rows can be accessed much faster than the other rows and having an
  intelligent memory controller that decides what data should be
  placed in fast rows versus slow rows~\cite{lee.hpca13, lee.thesis16,luo2020clr,hasan2019crow,chang.hpca16,wang2020figaro}.
  \omv{Recent work~\cite{luo2020clr} makes the capacity--latency trade-off in DRAM configurable at the row granularity: the CLR-DRAM architecture enables each DRAM row to be \emph{dynamically} or \emph{statically} configured to operate in a \emph{low latency} mode or \emph{high capacity} mode\gfvi{, as Figure~\ref{fig:clrdram} illustrates}. 
  In low latency mode, many major timing parameters can be reduced between 35\% to 64\%\omvii{, as described and evaluated in detail in the CLR-DRAM paper}~\cite{luo2020clr}.}

\begin{figure}[ht]
    \centering
    \includegraphics[width=1.0\linewidth]{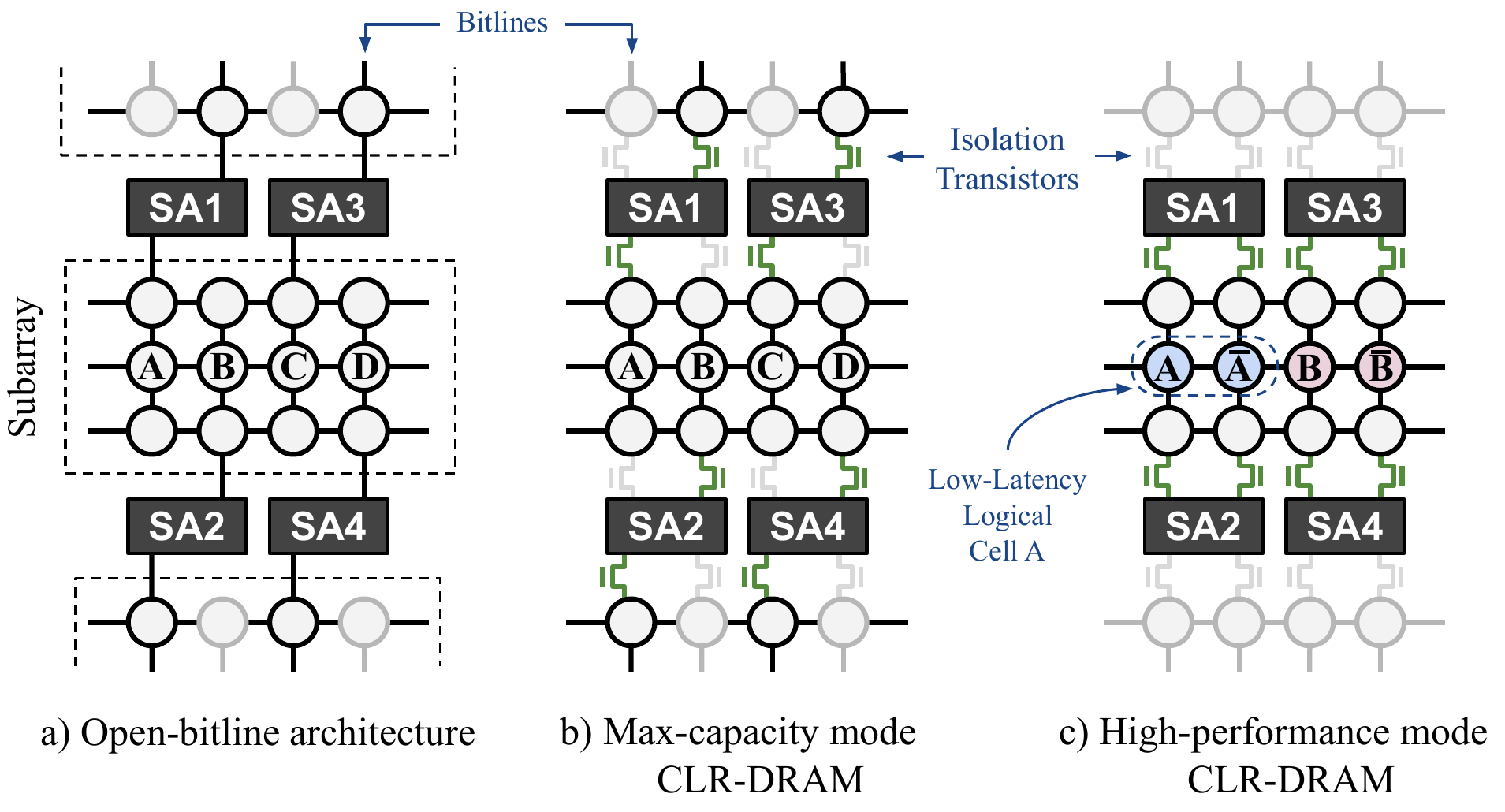}
    \caption{\gfvi{Comparison of a)~the conventional open-bitline architecture and CLR-DRAM architecture operating in b)~max-capacity mode and c)~high-performance (i.e., low latency) mode. The newly-added isolation transistors (highlighted in green) allow dynamic reconfiguration of \omvii{\emph{any}} DRAM row to switch between \omvii{the two} modes. Reproduced from~\cite{luo2020clr}.}}
    \label{fig:clrdram}
\end{figure}

  \sgii{(6)~We} find that a \sgii{recently-accessed or
    recently-refreshed} memory row can be accessed much more quickly
  than the standard latency if it needs to be accessed again soon,
  since the recent access and refresh to the row has replenished the
  charge of the cells in the row. An intelligent memory controller can
  thus keep track of the charge level of recently-accessed/refreshed
  rows and use the appropriate access latency that corresponds to the
  stored charge level~\cite{chargecache,wang2018cal,das.dac18}, leading to
  significant reductions in both access and refresh latencies.
  \sgii{Thus,} the poor scaling of DRAM latency and energy can
  potentially be overcome with the design of intelligent memory
  controllers that can facilitate a large number of effective latency
  and energy reduction techniques.}
  
  (7)~Two works \omv{that evaluate hundreds of real \gls{COTS} DRAM chips}~\cite{kim.hpca18,kim.hpca19} observe that the latency--reliability trade-off in modern DRAM devices can be exploited by an intelligent memory controller to 1) generate true random numbers at low latency and high throughput~\cite{kim.hpca19}, and 2) to evaluate physical unclonable functions quickly using a DRAM device~\cite{kim.hpca18}. These works exploit the heterogeneity in the latency-reliability trade-off of different cells: some cells fail truly randomly and some cells fail very consistently, when accessed with a low latency that violates the timing parameters. The former type of cells are used as true random number generator cells and the latter type of cells can be used as part of the challenge-response space of a DRAM-based physical unclonable function (PUF). An intelligent controller would determine the different types of cells using profiling mechanisms and enable the generation of true random numbers or PUF responses.
  
\begin{figure}[ht]
\centering
\includegraphics[width=1.0\linewidth]{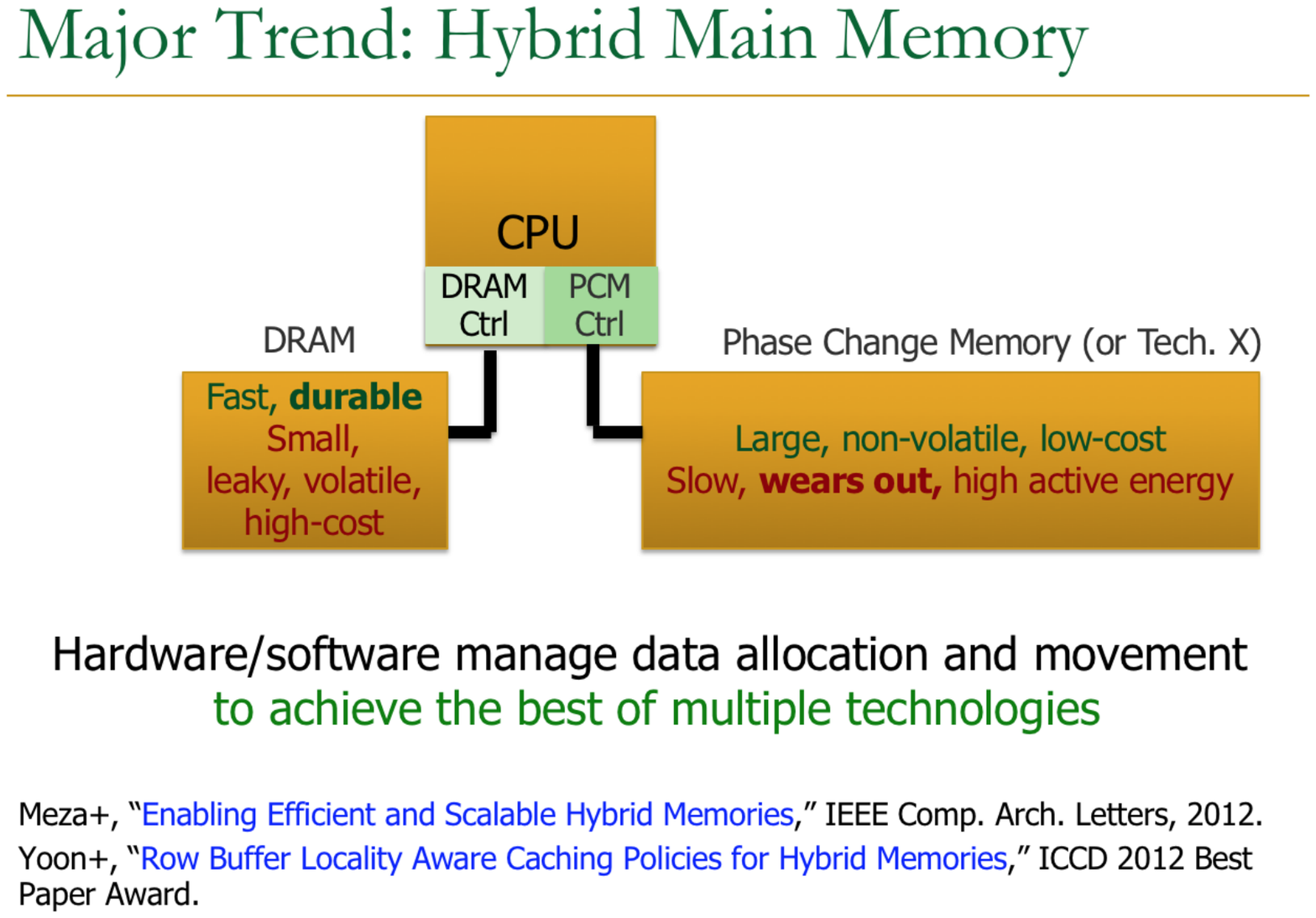}
\caption{Hybrid main memory. Reproduced from~\cite{mutlu.iccdtalk19}. Originally presented in~\cite{yoon2012row}.
}
\label{fig:hybrid}
\end{figure}

  (8)~An intelligent controller can use application and data characteristics to carefully map data across hybrid memories that consist of multiple different types of memories with different characteristics to maximize the benefits obtained from each memory type while avoiding the downsides of each memory type.
  Figure~\ref{fig:hybrid} depicts an example of such hybrid main memory composed of DRAM and PCM memories, as described by several works~\gfv{\cite{meza2012enabling,yoon2012row,qureshi.isca09,li2017utility,su2015hpmc,dhiman2009pdram,lee2013clock,ren.micro15,salkhordeh2019analytical,wang2019panthera,raybuck2021hemem}}.


  Many proposals exist for such intelligent controllers that manage hybrid memories, e.g.,~\gfvii{\cite{luo2014characterizing,ramos2011page,yu2017banshee,meza2012enabling,li2017utility,yoon2012row,qureshi.isca09,zhang2009exploring,song2020improving,su2015hpmc,dhiman2009pdram,lee2013clock,ren.micro15,wang2019panthera,raybuck2021hemem,Liu:2017:HCC:3079079.3079089}},
  indicating that such an intelligent controller can enhance memory scaling by enabling the best of multiple technologies. For example, the idea of {\em Heterogeneous Reliability Memory}~\cite{luo2014characterizing} 
  uses an intelligent memory controller that can communicate with both applications and memory devices to map each data element to different types of memories depending on the error vulnerability characteristics of the data element, thereby reducing memory cost. Similarly, EDEN~\cite{koppula2019eden} 
  uses a memory controller that can communicate with a neural network application to map different neural network layers to different DRAM partitions with different access latency and voltage parameters, depending on the error tolerance characteristics of each layer, thereby greatly improving energy efficiency and performance of neural network inference tasks. With increasing reliance on hybrid memories as well as increasing heterogeneity within each memory type to solve key memory scaling issues, it has become necessary to have intelligent controllers to manage data allocation, migration, and movement across the different heterogeneous parts. 
  


\begin{figure}[ht]
\centering
\includegraphics[width=1.0\linewidth]{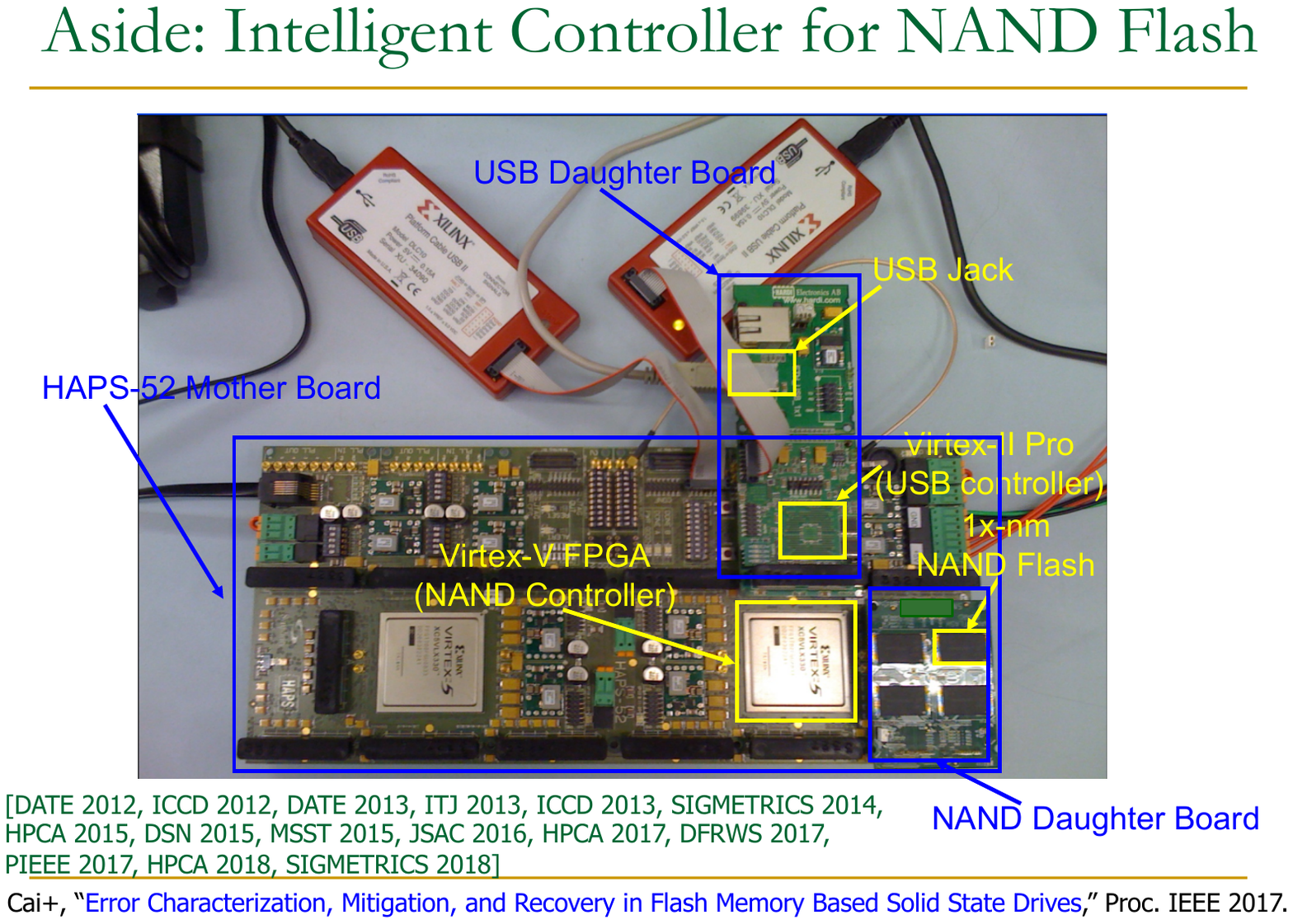}
\caption{Example of an intelligent flash memory controllers. The figure depicts a picture of one of our real experimental infrastructures (from~\cite{cai2012flash}) used for the design and evaluation of intelligent flash memory controllers. Reproduced from~\cite{mutlu.iccdtalk19}.}
\label{fig:flashI}
\end{figure}

Intelligent controllers are already in widespread use in another key part of a modern computing system.  In solid-state drives (SSDs) consisting of NAND flash memory, the flash memory controllers that manage the SSDs are designed to incorporate a significant level of intelligence in order to improve both performance and reliability~\omv{\cite{yucai.bookchapter18,cai.bookchapter18.arxiv, yucai-thesis,luo.thesis18,tavakkol.fast18,tavakkol.isca18,cai.procieee17,cai2012error,cai2012flash,cai2013program,cai2013threshold,cai2014neighbor,cai2015data,cai2015read,cai2017vulnerabilities,luo2018heatwatch,luo2018improving,luo2014characterizing,luo2015warm,luo2016enabling,cai2013error,kim2020evanesco,cho2024aero,nadig2023venice,park2021reducing}}.
Figure~\ref{fig:flashI} shows one of our real experimental infrastructures (from~\cite{cai2012flash}) used for the design and evaluation of intelligent flash memory controllers.

\begin{figure}[ht]
\centering
\includegraphics[width=1.0\linewidth]{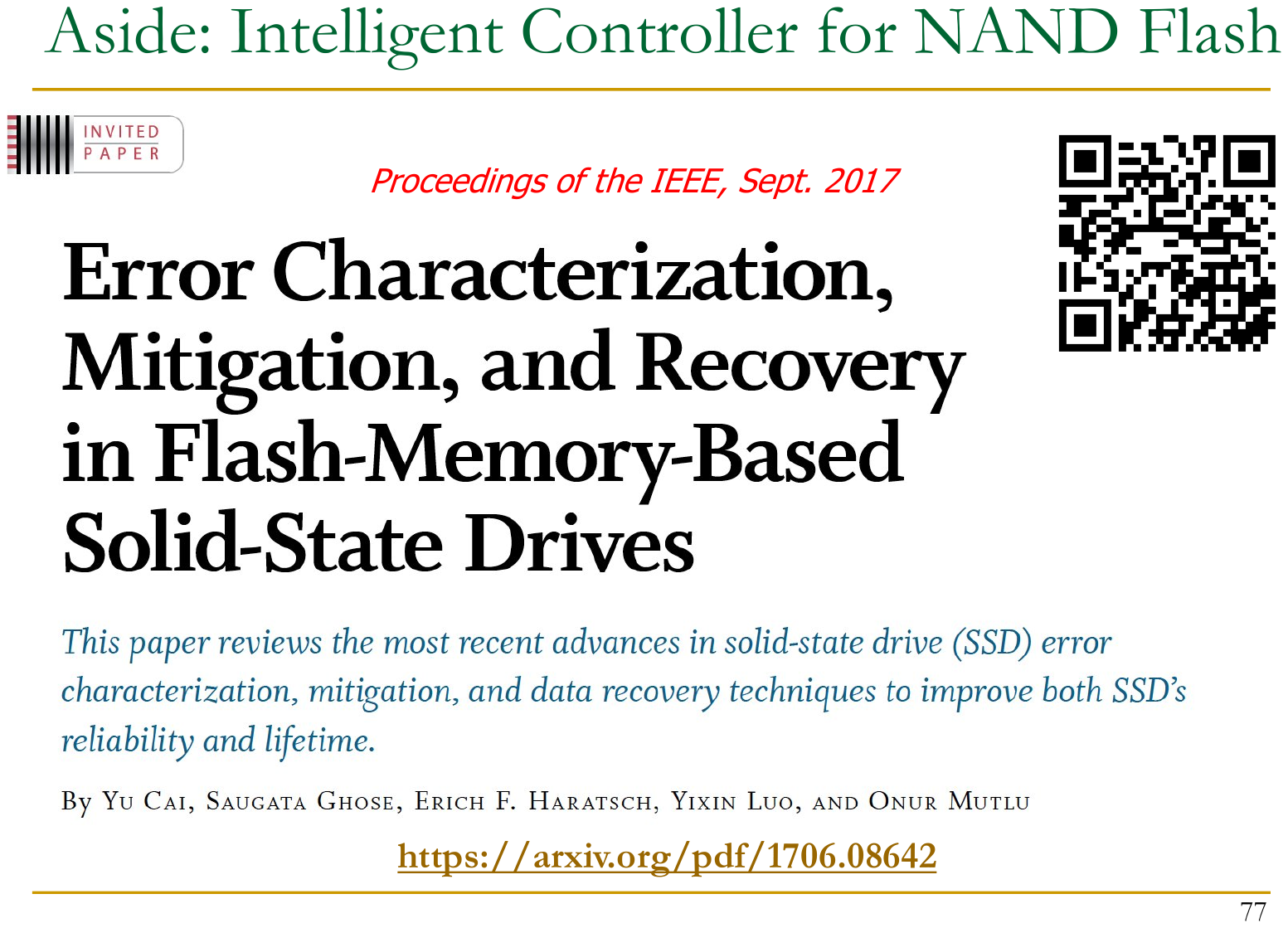}
\caption{A comprehensive review article on scaling issues of NAND flash memory and related mitigation techniques~\cite{cai.procieee17}\omv{.} Reproduced from~\cite{mutlu.iccdtalk19}.}
\label{fig:flashII}
\end{figure}

Modern flash \omvi{memory} controllers need to take into account a wide variety of issues such as remapping
data, performing wear leveling to mitigate the limited lifetime of
NAND flash memory devices, refreshing data based on the current
wearout of each NAND flash cell, optimizing voltage levels to
maximize memory lifetime, employing sophisticated error correction and recovery techniques to maximize lifetime and minimize error rates, and enforcing fairness across different
applications accessing the SSD.  Much of the complexity in flash
controllers is a result of mitigating issues related to the scaling
of NAND flash memory~\omvi{\cite{yucai.bookchapter18,cai.bookchapter18.arxiv,yucai-thesis,luo.thesis18,cai.procieee17,cai2012error,cai2012flash,cai2013program,cai2013threshold,cai2014neighbor,cai2015data,cai2015read,cai2017vulnerabilities,luo2018heatwatch,luo2018improving,luo2014characterizing,luo2015warm,luo2016enabling,cai2013error,kim2020evanesco,park2022deepsketch,park2021reducing,cho2024aero}}.
A comprehensive review of scaling issues of NAND flash memory and related mitigation techniques can be found in~\cite{cai.procieee17} (Figure~\ref{fig:flashII}) and~\cite{yucai.bookchapter18,cai.bookchapter18.arxiv}.

We argue that in order to overcome scaling
issues in main memory (DRAM), the time has come for main memory controllers to
also incorporate significant intelligence \omv{(just like flash controllers have been doing for decades)}. Yet, incorporating sophisticated intelligence in the DRAM controller is more challenging than doing so in a flash controller due to the much lower access latency and much higher access bandwidth of modern DRAM devices.


\sg{As we \omvii{have described} above, introducing intelligence into the memory
  controller can help us overcome a number of key challenges in memory
  scaling.  In particular, a significant body of work has
  demonstrated that the key reliability, refresh\sgii{,} and latency/energy
  issues in memory can be mitigated effectively with an intelligent
  memory controller that intelligently and meticulously manages the many different characteristics of underlying memory chips, which may consist of different types of memory technology. As we discuss in \omv{later sections}, such
  intelligence can go even further, by enabling the memory controllers
  (and the broader memory system) to perform \omv{special-purpose or general-purpose} computation
  in order to overcome the significant data movement bottleneck in
  modern and future computing systems.}

%% file: sections/04-perils-processor-centric.tex
\section{Perils of Processor-Centric Design}
\label{sec:processorcentric}


As described earlier, a major reason for performance and energy degradation in modern computing systems is the large amount of \emph{data movement} present in the systems. Such data movement is a natural consequence of the {\em processor-centric} execution model and design paradigm~\cite{burks.1946}, which creates a dichotomy between computation and memory/storage. The processor-centric design paradigm separates computation capability and memory/storage capability into two completely-\omv{disparate} system components (i.e., the computing unit versus the memory/storage unit) that are connected by long and energy-hungry interconnects: processing is done only in the computing unit, while data {is} stored in a completely {separate} place. As a result, data has to continuously move back and forth between the memory/storage unit and {the computing unit (e.g., CPU cores or accelerators)}, for any computation to be performed.

In order to perform an operation on data that is stored within memory,
a costly process is invoked. First, the CPU {(or an
  accelerator)} must issue a request to the memory controller, which
in turn sends a series of commands across the off-chip bus to the DRAM
module. Second, the data is read from the DRAM module and returned to
the memory controller. Third, the data is placed in the CPU cache
{and registers}, where it is accessible by the CPU
cores. Finally, the CPU can operate (i.e., perform computation) on the
data. All these steps consume substantial time and energy in order to
bring the data into the CPU
chip~\omv{\cite{kestor.iiswc2013,pandiyan.iiswc2014,kanev.isca15,boroumand.asplos18, boroumand2021google, boroumand2021google_arxiv,sites1996,DBLP:conf/hpca/MutluSWP03,DBLP:journals/micro/MutluSWP03,DBLP:conf/isca/MutluKP05,DBLP:journals/micro/MutluKP06}}.

{In current computing systems, the CPU (or any accelerator) is the only system
  component that is able to perform computation on data. \omv{Other}
  system components are devoted to only data storage (memory, caches,
  disks) and data movement (interconnects); they are incapable of
  performing computation.} As a result, current computing systems are
\emph{grossly imbalanced}, which leads to large amounts of energy
inefficiency\omv{,} lost performance\omv{, and system complexity}. As empirical evidence to the gross
imbalance caused by the processor--memory dichotomy in the design of
computing systems today, we have observed that
\gfv{(1)}~\omvi{more than} 62\%
of the entire system energy consumed by four major commonly-used
mobile consumer workloads (including the Chrome browser, TensorFlow
machine learning inference engine, and the VP9 video encoder and
decoder)~\cite{boroumand.asplos18} and 
\gfv{(2)~\omvi{more than} 90\% of the entire system energy consumed by large commercial edge neural network models~\cite{boroumand2021google,boroumand2021google_arxiv} is spent on data \omvi{access from main memory and data} movement between \omvii{a state-of-the-art} edge machine learning accelerator and the memory hierarchy}. 
Thus, the fact that current
systems \omv{are capable of performing} computation only in the computing unit (\omvii{e.g.,} CPU cores
and hardware accelerators) is causing significant waste in terms of
energy by necessitating \omvii{large amounts of} data movement across the entire system.

At least five factors contribute to the performance loss and the
energy waste associated with data movement between processor and memory. 
We briefly describe these next, to demonstrate the sweeping negative impact of data movement in modern computing systems.

First, the width of the off-chip bus between the memory controller and
the main memory is narrow, due to pin count and \sgii{cost} constraints,
leading to relatively low bandwidth and high latency to/from main memory. This makes {it} difficult to send a large number of requests to memory in
parallel to enable higher levels of parallelism and to tolerate the long main memory latency. As a result, systems that require higher levels of concurrency and lower latency require much higher cost because they require wider processor-memory interconnects or more processor-memory channels, both of which lead to higher power consumption and higher hardware area overheads~\omv{\cite{ghose2019demystifying,ghose2018vampire,mutlu.superfri15,HBM,jedec.hbm.spec,lee.taco16}}.

Second, current computing systems employ many sophisticated mechanisms to tolerate the data access from main memory. These mechanisms include complex multi-level cache
hierarchies with sophisticated insertion/promotion/eviction policies and sophisticated latency tolerance/hiding mechanisms (e.g.,
{sophisticated caching algorithms at many different caching
  levels~\gfv{\cite{seshadri-taco2015,samira2014improving,seshadri2012evicted,lee2010dramaware_lastlevel,MutluKP05,qureshi2006case,mutlu2005using,mutlu2004cache}},} multiple complex prefetching techniques~\gfv{\cite{bera2021pythia,bera2019dspatch,cont-runahead,DBLP:conf/hpca/MutluSWP03,DBLP:journals/micro/MutluSWP03,DBLP:conf/isca/MutluKP05,DBLP:journals/micro/MutluKP06,lee1987data,jnesbit2004acdc,roth1998}}, {high amounts of multithreading~\gfv{\cite{chen2013guided,borkar2007thousand}}, complex and power-hungry out-of-order execution mechanisms \omv{with large instruction windows~\gfv{\cite{lebeck2002alarge,akkary2003checkpoint,michaud2001dataflow}}}}). These components, while sometimes effective
at improving performance, are costly in terms of both die area and
energy consumption, {as well as the additional latency required
  to access/manage them}. When these components are not effective at improving performance, they result in a net energy waste and latency overheads that hurt the very performance that they are designed to improve~\omv{\cite{bera2022hermes}}. These components significantly increase the complexity of the system. Hence, the architectural and microarchitectural
techniques used in modern systems to tolerate the consequences of the
dichotomy between processing unit and main memory, lead to significant
energy waste and additional system complexity. As such, we are in a vicious cycle in system design due to the processor-centric design paradigm: 1) data movement between the processor and memory already causes significant energy waste and latency; 2) to tolerate the latency of such data movement, existing systems employ many complex mechanisms whose effectiveness varies depending on the workloads; 3) these complex mechanisms in turn cause additional energy waste and latency overheads. The fundamental cause of this vicious cycle is the processor-centric execution model and design paradigm, and hence breaking out of this vicious cycle requires tackling this fundamental cause by changing the paradigm (to a data-centric one).

Third, the many caches employed in computing systems are not always effective or efficient. Much of the data brought into the caches is \emph{not} reused by the
CPU~\omv{\cite{qureshi.isca07,qureshi-hpca07,ahn.pei.isca15,ahn.tesseract.isca15,seshadri2012evicted,
oliveira2021pimbench,tyson1995modified,johnson1997run,johnson1999run}},
resulting in a large waste of hardware area and memory bandwidth. For example, 1) random access to memory leads to poor locality, rendering caches almost completely ineffective, 2) strided access to memory where stride is greater than a cache block also renders caches ineffective, 3) even streaming access to memory where all elements in a cache block are used in a consecutive manner is inefficient to handle with large caches because the block is not reused again. There are many such access patterns in a wide variety of modern workloads~\omv{\cite{qureshi.isca07,qureshi-hpca07,ahn.tesseract.isca15,ahn.pei.isca15,seshadri2012evicted,ghose2019demystifying,khan2014improving,qureshi2006case,DBLP:journals/tc/MutluKP06,subramanian-micro2015,oliveira2021pimbench,oliveira2021pimbench_arxiv,ayers2020classifying}} 
that render caches either very inefficient or \omv{in extreme cases} unnecessary, exacerbating the energy waste due to data movement in processor-centric systems.

\begin{figure}[ht]
\centering
\includegraphics[width=1.0\linewidth]{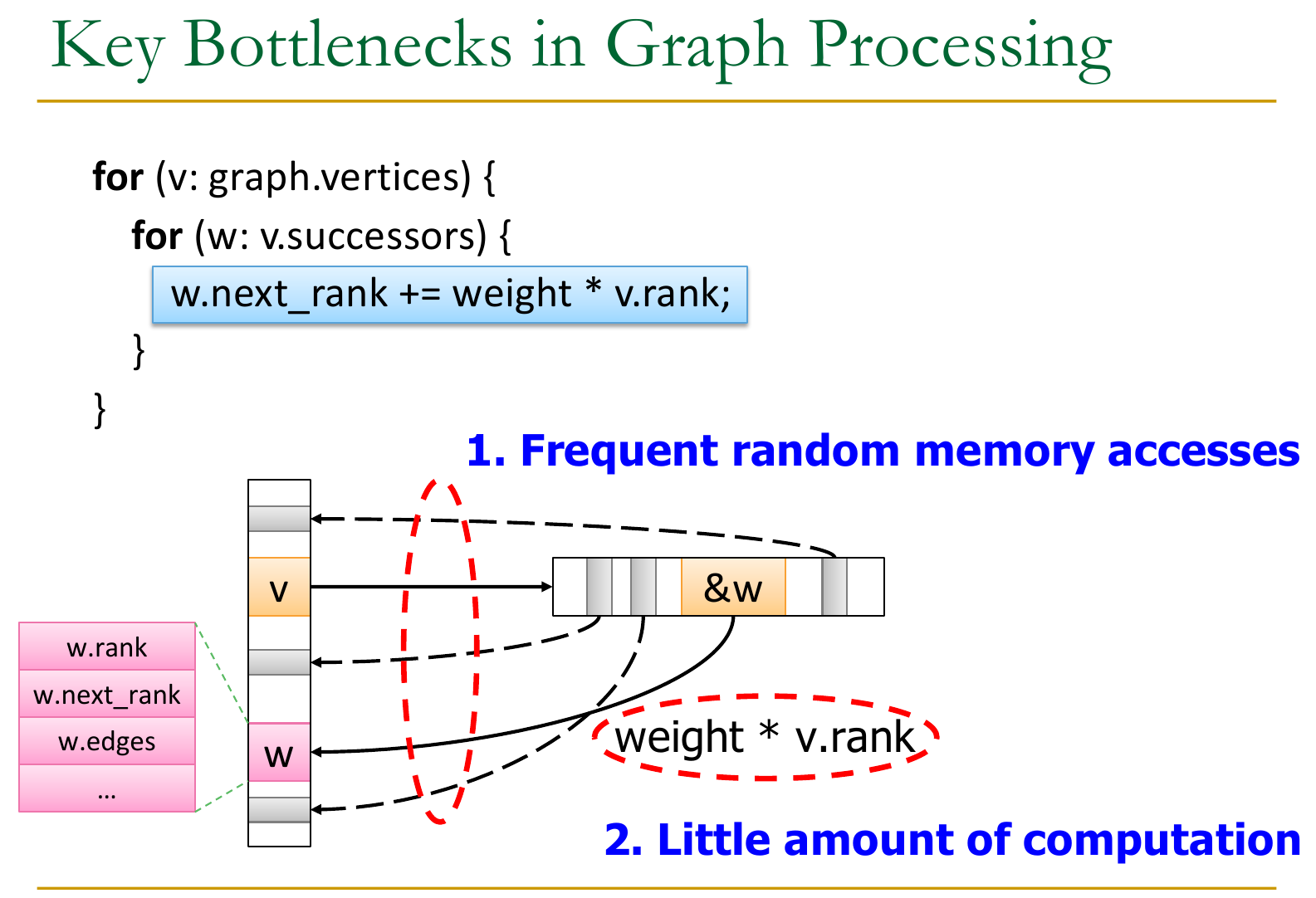}
\caption{Random memory accesses in the PageRank graph processing algorithm~\omv{\cite{page1999pagerank,CHO1998161}}. Reproduced from~\cite{mutlu.nsfpim20}. Originally depicted in~\cite{ahn.tesseract.isca15,ahn.tesseract.isca15talk}.
}
\label{fig:apps-graph}
\end{figure}

Fourth, many modern applications, such as graph
processing~\omv{\cite{ahn.tesseract.isca15,ahn.pei.isca15,besta2021sisa,besta2021sisa_micro,besta2021graphminesuite}} and workloads that operate on sparse data structures, such as sparse linear algebra~\cite{kanellopoulos2019smash,sadi2019efficient} and sparse neural networks~\cite{gondimalla2019sparten,hegde2019extensor,zhu2019sparse}, 
produce random memory access patterns. 
Figure~\ref{fig:apps-graph} shows the example of PageRank~\omv{\cite{page1999pagerank,CHO1998161}}, a graph processing algorithm with frequent random memory accesses and little amount of computation. 
With such random access patterns, not only the caches but also the main memory bus and the main memory itself are very inefficient, since only a small part of each memory row and cache line retrieved all the way from main memory is actually used by the CPU~\omv{\cite{olgun2022sectored,olgun2024sectored,cooper2010fine,Ani2010,zhang2014half,ha2016improving,lee2017partial,o2021energy,oconnor2017fine}}. Such random accesses are fundamentally difficult to prefetch, rendering prefetchers ineffective.
{This example demonstrates that modern memory hierarchies are not designed to work well for random memory access \sg{patterns} that are found in many modern workloads.}

Fifth, the processor (as well as accelerators) and the main memory are connected to each other via long, power-hungry interconnects. {These interconnects impose
  significant additional latency to every data access and represent a
  significant fraction of the energy spent on moving data to/from the
  DRAM memory.} In fact, off-chip interconnect latency and energy
consumption is a key limiter of performance and energy in modern
systems~\omv{\cite{lee.hpca13,donghyuk-ddma,seshadri2013rowclone,GS-DRAM,ahn.tesseract.isca15,boroumand.asplos18, boroumand2021google, boroumand2021google_arxiv}}, 
as it greatly exacerbates the cost of data movement. Unfortunately, off-chip interconnect latency and energy are not scaling (i.e., reducing) well with the scaling of technology node generations, which mainly benefits logic~\cite{dennard1974design}. 

\begin{figure}[ht]
\centering
\includegraphics[width=1.0\linewidth]{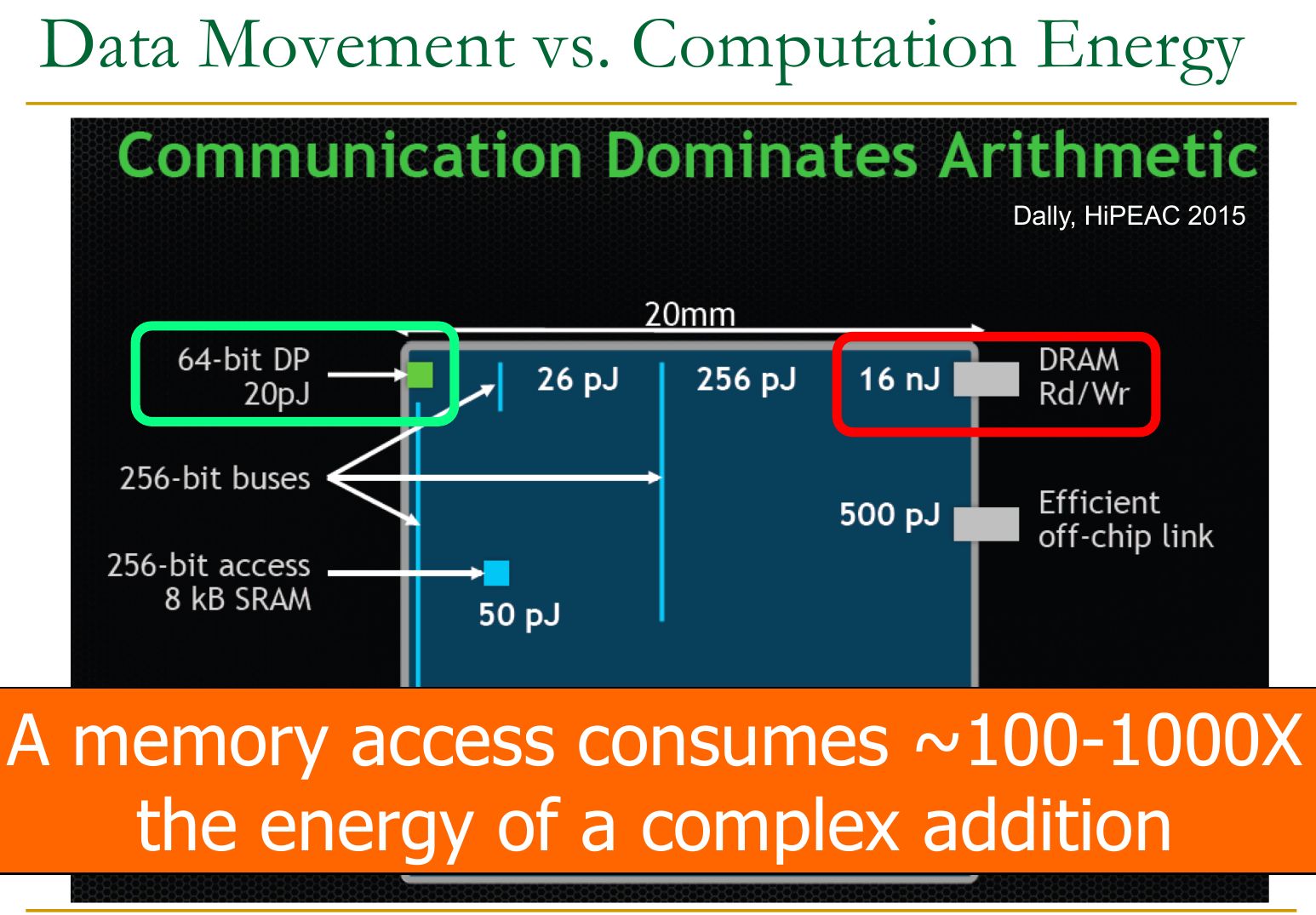}
\caption{Data movement versus computation energy.  The figure depicts the absolute amount of energy spent on various arithmetic and data movement operations, including a double-precision floating point addition and a single DRAM access. Reproduced from~\cite{mutlu.nsfpim20}, based on 
\juanr{a slide provided} in~\cite{dally.2015}.}
\label{fig:datamov-energy}
\end{figure}

The increasing disparity between processing technology and
memory/communication technology has resulted in systems in which
communication (\omv{including} data movement) costs dominate computation costs in
terms of energy consumption. The energy consumption of a main memory
access is between two to three orders of magnitude the energy
consumption of an addition operation today.  
For example,~\cite{pandiyan.iiswc2014} reports that the energy consumption
of a memory access is \gf{$\sim$$115\times$} the energy consumption of an
addition operation. 
Similarly, Figure~\ref{fig:datamov-energy} shows that the energy consumed by a DRAM access is \gf{$\sim$$800\times$} the energy consumption of a double precision addition operation, based on data reported by~\omv{\cite{dally.2015,han2016eie}}. 
As a result, data movement is empirically shown to account for 40\%~\cite{kestor.iiswc2013}, 35\%~\cite{pandiyan.iiswc2014}, 62\%~\cite{boroumand.asplos18}\omv{, and \omvi{more than} 90\%~\cite{boroumand2021google}} of the total system energy in
scientific, mobile, consumer applications, \omv{and edge machine learning,} respectively.  This
energy waste due to data movement is a huge burden that greatly limits
the efficiency and performance of all modern computing platforms, from
datacenters with a restricted power budget to mobile devices with
limited {battery life}.

Overcoming all the reasons that cause low performance and large energy
inefficiency (as well as high system design complexity) in current
computing systems first requires the realization that all of these reasons are caused by the processor-centric design paradigm employed by existing computing systems. As such, a fundamental solution to all of these reasons at the same time requires a paradigm shift~\omv{\cite{kuhn2012structure}}. 
We believe that future computing architectures should become \emph{data-centric}~\omv{\cite{mutlu2021intelligentdate}}: {they should (1) perform computation with} minimal data movement, and (2) {compute} where it makes sense (i.e., where
the data resides), as opposed to computing solely in the processor (i.e., CPU or accelerators). Thus, the traditional rigid dichotomy between the
computing units and the memory/communication units needs to be broken
and a new paradigm enabling computation where the data resides needs
to be invented and enabled. We refer to this general data-centric execution model and design paradigm as {\em Processing-in-Memory (PIM)}.

%% file: sections/05-PIM-approaches.tex
\section{Processing-in-Memory (PIM): Technology Enablers and Two Approaches}
\label{sec:pim}


Large amounts of data movement is a major result of the
  predominant processor-centric design paradigm of modern
  computers. Eliminating unnecessary data movement between memory 
  and the processor is essential to \omv{making} future computing architectures \omv{fundamentally}
  high performance, energy-efficient and sustainable. To this
  end, \emph{processing-in-memory} (PIM) equips the memory \sg{subsystem}
  with the ability to perform computation.

\omvi{PIM is part of a broader approach that advocates enabling data processing \emph{everywhere} in a computing system, including \gfv{(but not limited to)} sensors, caches, interconnects, main memory, storage, network controllers. We call this broader approach ``\emph{processing data where it makes sense to do so}'' and have written about it in prior works~\cite{mutlu2019,mutlu.msttalk17,mutlu.gwutalk19,mutlu.isscctalk19,mutlu.glvlsitalk19,mutlu.appttalk19,mutlu.iccdtalk19}. We believe a truly efficient and high performance system would have computation capability in \emph{all} its components and \emph{intelligently} use that computation capability according to workload, system, and user requirements.}

In this section, we first describe two new technology enablers for PIM: 
(1)~the emergence of 3D-stacked memories, and 
(2)~the use of \omv{relatively new} byte-addressable \omv{non-volatile} memories. These two relatively new main memory technologies provide new opportunities that can make it easier for modern computing systems to introduce and adopt PIM. 

Second, we introduce two promising approaches to implementing
PIM in modern architectures.  \omv{These approaches fundamentally differ in the \emph{nature} of computation they enable.}
The first approach, \gf{\em processing-using-memory} \omvi{(PUM)}, exploits the existing \omv{memory} architecture and the \omv{analog} \omvi{operational} principles of the \omv{memory} circuitry to enable (bulk) processing operations within the memory. This approach can \gf{be especially} powerful
in performing specialized computation in main memory by taking
advantage of what the main memory substrate is \omv{fundamentally} good at performing with \omv{small} changes to the existing memory chips.  
The second approach, \gf{\em processing-near-memory} \omvi{(PNM)}, exploits the ability to implement a wide variety of processing logic \omv{(i.e., computing capabilities)} \gf{near the memory arrays (e.g., in a DRAM chip\omv{, next to each memory bank \omvi{or subarray},}  at} the logic layer of 3D-stacked
memory\omvi{, in the memory controllers}\gf{)} and thus the high internal bandwidth {and low latency}
available \gf{inside the memory chip (e.g.,} {between the logic layer and the memory layers of}
3D-stacked memory\gf{)}.
This is a more general approach where the logic implemented \gf{near the memory arrays} can be \omv{more powerful (\omvi{and can be} specialized\omvi{, reconfigurable, or} general purpose)} and thus can benefit a wide variety of applications. 
%

\omvi{Both approaches, offering different \omvii{tradeoffs}, are important to exploit the full potential of PIM. 
PUM has two major advantages over PNM: 
(1)~PUM fundamentally reduces data movement by performing computation \emph{in-situ}, while data movement still occurs between computation units and memory arrays in PNM; 
(2)~PUM exploits the large internal bandwidth and parallelism available inside the memory arrays, while PNM is bottlenecked by the memory's internal \omvii{(and sometimes external)} data buses. 
In contrast, PNM can enable a wider set of functions (including complete processors) to be more easily implemented and exploited near memory due to its use of conventional logic. 
As such, both approaches to PIM should be seem as \emph{complementary} to each other and can be combined to exploit the maximum potential of a PIM system.}

Below, we provide a more detailed general overview of the two approaches, to \omvi{demonstrate} that the approaches are more general than what we will describe in more detail \omvi{in the rest of this article}. It is important for the reader to keep in mind that the two approaches can be applied to many different types of memory technologies, even though our major focus will be on DRAM, the predominant main memory technology for \omv{many} decades~\omvii{\cite{dennard1968field,mandelman.ibmjrd02,mutlu.imw13}}, in \omv{most of this article}. 


\subsection{New Technology Enablers: 3D-Stacked Memory and Non-Volatile Memory}
\label{sec:tech-enablers}

Memory manufacturers are actively developing  new approaches for main memory system design, due to the DRAM technology scaling issues we described in detail in Section~\ref{sec:majortrends}. Two promising \omvii{and relatively new} technologies are 3D-stacked memory and byte-addressable Non-Volatile Memory (NVM), both of which can be exploited to overcome prior barriers to introducing and implementing PIM architectures.

\subsubsection{3D-Stacked Memory Architectures}
\begin{figure}[ht]
\centering
\includegraphics[width=1.0\linewidth]{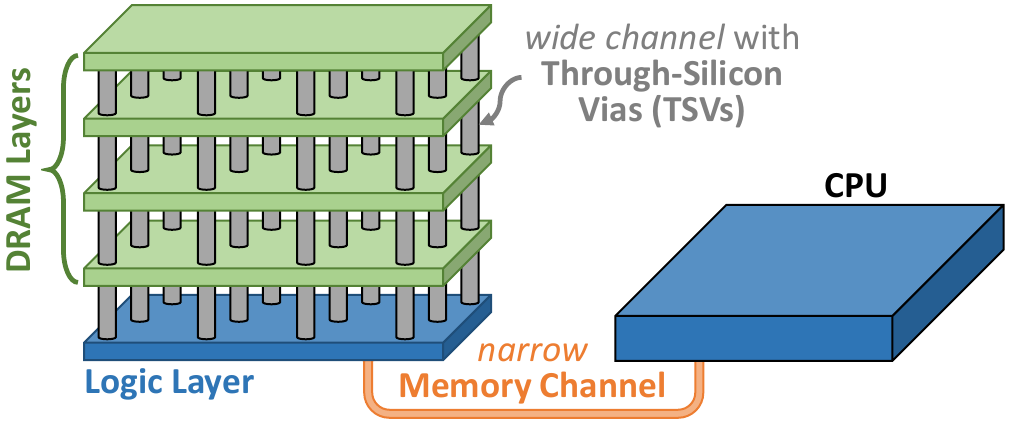}
\caption{High-level overview of a 3D-stacked DRAM architecture. Reproduced from~\cite{ghose2019arxiv}.}
\label{fig:3d-dram}
\end{figure}

The first major new approach to main memory design is 3D-stacked memory~\cite{ahn.tesseract.isca15,jedec.hbm.spec,lee.taco16,loh2008stacked,hmc.spec.1.1,hmc.spec.2.0, gokhale2015hmc}. 
In a 3D-stacked memory, multiple layers of memory (typically DRAM in already-existing systems) are stacked on top of each other, as shown in Figure~\ref{fig:3d-dram}.
These layers are connected together using vertical through-silicon vias (TSVs)~\cite{loh2008stacked,lee.taco16}. 
Using current manufacturing process technologies, thousands of TSVs can be placed within a single 3D-stacked memory chip. As such, the TSVs provide much greater internal memory bandwidth than the narrow memory channel. Examples of 3D-stacked DRAM available commercially include High-Bandwidth Memory (HBM)~\cite{jedec.hbm.spec,lee.taco16}, 
Wide I/O~\cite{wideio}, 
Wide I/O 2~\cite{wideio2}, 
and the Hybrid Memory Cube (HMC)~\cite{hmc.spec.2.0}. 
Detailed analysis of such 3D-stacked memories and their effects on modern workloads can be found in~\gfvi{\cite{lee.taco16,ghose2019demystifying,ramulator,loh2008stacked,luo2023ramulator,olgun2024read,oliveira2021pimbench}}.
\gfv{Emerging die-stacking \omvi{and packaging} technologies, like \emph{hybrid bonding}~\cite{kagawa2016novel,niu2022isscc,schmidt1998wafer} and \emph{monolithic 3D \omvi{integration}}~\cite{gopireddy2019m3d,mitra2018vlse,hwang2018cmos,mitra2015nano,rich2020nano,sabry2015abundant,sabry2019n3xt,ghiasi2022revamp3d}, can further amplify the benefits of conventional TSV-based 3D-stacked memory chips by greatly improving internal bandwidth across layers using \emph{high-density} inter-layer vias (ILVs)~\cite{gopireddy2019m3d,sabry2019n3xt} or \emph{direct} wafer-to-wafer \omvi{connections} via Cu--Cu bonding~\cite{lau2023recent,kagawa2016novel,niu2022isscc,schmidt1998wafer}, respectively. }

In addition to the multiple layers of DRAM, a number of prominent 3D-stacked DRAM architectures, including HBM and HMC, incorporate a logic layer inside the chip~\omv{\cite{lee.taco16,jedec.hbm.spec,hmc.spec.2.0,loh2008stacked,ahn.tesseract.isca15}}. 
The logic layer is typically the bottommost layer of the chip, and is connected to the same TSVs as the memory layers.\footnote{\gfv{\omvii{Logic} layer\omvi{(s)} \omvii{in monolithic 3D architectures} can also be added \emph{between} memory layers~\cite{ebrahimi2014monolithic,kim2023van}.}} 
The logic layer provides area inside the \omv{3D-stacked DRAM system} where architects can implement functionality that interacts with both the processor and the DRAM cells. Currently, manufacturers make limited use of the logic layer and there is significant amount of area the logic layer can provide\omv{, especially when manufactured with a \omvi{high-quality} logic fabrication process}. This presents a promising opportunity for architects to implement new \omvi{and efficient} PIM logic in the available area of the logic layer. We can potentially add a wide range of computational logic (e.g., 
general-purpose cores~\omv{\cite{boroumand.asplos18, ahn.tesseract.isca15,drumond2017mondrian,boroum2019conda,pugsley2014ndc,singh2019napel,oliveira2021pimbench,azarkhish2016logic,azarkhish2018neurostream}}, 
accelerators~\omv{\cite{zhang.hpdc14,RVU,NIM,gao2017tetris,hsieh.isca16,cali2020genasm,boroumand2021mitigating,boroumand2021google,boroumand2021polynesia,fernandez2020natsa,LiM_3D_FFT_MM,akin2014hamlet}}, 
reconfigurable \omv{logic}~\omv{\cite{DBLP:conf/hpca/GaoK16,farmahini2015nda}}, \omv{special-purpose functional units~\cite{boroumand.asplos18,nai2017graphpim,kim.bmc18,ahn.pei.isca15,kwon202125,lee2021hardware}}, or a combination of \omvi{different} \omv{such} types of logic) in the logic layer, as long as the added logic meets area, energy, and thermal dissipation constraints, which are important and potentially limiting constraints in 3D-stacked systems~\omv{\cite{boroumand.asplos18, ahn.tesseract.isca15}}.

\subsubsection{Non-Volatile Memory (NVM) Architectures}
The second major new approach to main memory design is the development of byte-addressable resistive nonvolatile memory (NVM). In order to circumvent DRAM scaling limitations, researchers and manufacturers have been developing new memory devices that can \omv{potentially} store data at much higher densities than the typical density available in existing DRAM manufacturing process technologies. 
Manufacturers are exploring at least \gfv{four} types of emerging NVMs to augment or replace DRAM: 
(1)~phase-change memory (PCM)~\cite{lee-isca2009,lee.cacm10,lee.ieeemicro10,qureshi.isca09,wong.procieee10,yoon-taco2014,zhou.isca09,atwood2018pcm,bock2011analyzing,burr2008overview,du2013bit,ferreira2010increasing,jiang2012fpb,jiang2013hardware,kannan2016energy,qureshi2011payasyougo,qureshi2010improving,qureshi2010morphable,sebastian2017temporal,wang2015exploit,yue2013accelerating,zhou2012writeback,zhou2013writeback,song2020improving,song2019enabling,yoon2013techniques,yoon2013techniques,dhiman2009pdram,meza2012acase,song2021aging}\gf{,}
(2)~magnetic RAM (MRAM)~\gf{\cite{kultursay.ispass13,naeimi.itj13, girard2020survey,wang2013low,chen2010advances,diao2007spin,hosomi2005novel,raychowdhury2009design,meza2012acase}}, 
(3)~metal-oxide resistive RAM (RRAM) or memristors~\gf{\cite{chua.tct71,strukov.nature08,wong.procieee12,akinaga2010resistive,yang2013memristive,chi2016prime,song2018graphr,song2017pipelayer,yao2017face,hu2016dot}, and} 
\gfv{(4)~ferroelectric RAM (FeRAM)~\cite{bondurant1990ferroelectronic,scott1989ferroelectric,scott2007applications,mikolajick2001feram}}.  
All \gfv{four} of these NVM types are expected to provide memory access latencies and energy usage that are competitive with or close enough to DRAM, while enabling much larger capacities per chip and nonvolatility in main memory.
Since they are emerging and their designs do not have the long-term ``baggage'' other main memories (DRAM) have accumulated, NVMs present architects with an opportunity to redesign how the main memory subsystem operates from the cell and chip levels all the way up to software and algorithms. While it can be relatively difficult to modify the design of DRAM arrays due to the delicacy of DRAM manufacturing process technologies as we approach scaling limitations, NVMs have yet to approach such scaling limitations. As a result, architects can potentially design NVM memory arrays that integrate PIM functionality from the getgo. A promising direction for this functionality is the ability to manipulate NVM cells at the circuit level in order to perform logic operations using the memory cells themselves. A number of recent works have demonstrated that NVM cells can be used to perform a complete family of Boolean logic operations~\gfvi{\cite{li.dac16,levy.microelec14,kvatinsky.tcasii14,angizi2018pima,angizi2018cmp,angizi2019dna, borghetti2010memristive, linn2012beyond,kvatinsky.iccd11,kvatinsky.tvlsi14,lehtonen2009stateful,kim2011field,lehtonen2012applications,mahmoudi2013implication,kim2019single,xie2017scouting,gaillardon2016plim,fernandez2024matsa,truong2021racer,leitersdorf2023aritpim,kang2017memory}}, 
similar to such operations that can be performed in DRAM cells~\gfvi{\cite{seshadri.micro17, hajinazarsimdram,seshadri.bookchapter17,Seshadri:2015:ANDOR,seshadri.arxiv16,seshadri2020indram,missingnot,xin2020elp2im,mimdram,gao2020computedram,besta2021sisa_micro,olgun2022pidram, missingnot}}. 
NVMs have also been shown to perform more sophisticated operations like \omv{matrix-vector} multiplication \omv{in the analog domain}~\gfvi{\cite{alibart2012high, angizi2019mrima, ankit2019puma, ankit2020panther, bojnordi2016memristive, cai2018training, challapalle2020gaas, chen2018regan, cheng2017time, chi2016prime, chou2019cascade, feinberg2018enabling, feinberg2018making, holmes1993use, hu2012hardware, huang2017highly, ielmini2018memory, imani2019floatpim, imani2020dual, jia2020programmable, joshi2020accurate, jung2022crossbar, le2018mixed, le202364, li2019long, li2020timely, long2018reram, mao2018lergan, marinella2018multiscale, nag2018newton, sebastian2020memory, shafiee2016isaac, song2017pipelayer, song2018graphr, tang2017aepe, tang2017binary, valavi201964, wan2022compute, wang2014energy, wang2018snrram, wen2019memristor, wen2020ckfo, xia2016switched, xia2017fault, yang2019sparse, yang2020retransformer, yuan2021forms, zhu2019configurable}}, 
which are more difficult to \omvi{naturally} implement in DRAM.

\subsection{Two Approaches: \gf{Processing-Using-Memory} (PUM) vs. \gf{Processing-Near-Memory} (PNM)}

Many recent works take advantage of the memory technology innovations that we discuss in Section~\ref{sec:tech-enablers} to enable and implement PIM. We find that these works generally take one of two approaches, which are categorized in Table~\ref{tab:pum-pnm}: 
(1) {\em \gf{processing-using-memory}} or (2) {\em \gf{processing-near-memory}}. We briefly describe each approach here. Sections~\ref{sec:PUM} and~\ref{sec:PNM} will provide example approaches and more detail for both.

\begin{table}[ht]
\tempcommand{0.9}
\caption{Summary of enabling technologies for the two approaches to PIM used by recent works. Adapted from~\cite{ghose2019arxiv} \juanrr{and extended}.}
\label{tab:pum-pnm}
\resizebox{\linewidth}{!}{
\begin{tabular}{ll}
\toprule
\textbf{Approach} & \textbf{\juanrr{Example} Enabling Technologies} \\
\midrule
\multirow{7}{*}{\gf{Processing-Using-Memory}} & SRAM  \\
                                        & DRAM  \\
                                        & \gf{NAND flash} \\ 
                                        & Phase-change memory (PCM)  \\
                                        & Magnetic RAM (MRAM)  \\
                                        & Resistive RAM (RRAM)/memristors  \\
                                        & \omvi{Ferroelectric RAM (FeRAM)} \\
\hline
\multirow{7}{*}{\gf{Processing-Near-Memory}} & Logic layers in 3D-stacked memory  \\
                                        & \omvi{(e.g., using} \gf{hybrid bonding}\omvi{, silicon \omvii{interposers},} \\ 
                                        & \omvi{or monolithic 3D integration)}  \\
                                        & Logic in memory controllers  \\
                                        & \juanr{Logic in memory chips (e.g., near bank\omvi{, near subarray})}  \\
                                        & \juanr{Logic in memory modules}  \\
                                        & \juanrr{Logic \omvi{in} caches}  \\
                                        & \juanrr{Logic in storage devices}  \\
                                        & \omv{Logic in sensors}  \\
\bottomrule
\end{tabular}
}
\end{table}

{\em \bf \gf{Processing-using-memory} (PUM)} exploits the existing
memory architecture and the operational principles of the memory circuitry
to enable operations within main memory\omvi{,} \omv{usually with small changes}. PUM makes use of intrinsic properties and operational principles of the memory cells and cell arrays themselves, by inducing interactions between cells such that the cells and/or cell arrays can perform useful computation. 
Prior works show that \gf{PUM} is possible using static RAM (SRAM)~\cite{aga.hpca17,eckert2018neural,fujiki2019duality,kang.icassp14},  
DRAM~\omv{\cite{chang.hpca16,seshadri.micro17, hajinazarsimdram,seshadri2013rowclone,seshadri2020indram,seshadri.bookchapter17.arxiv,seshadri.bookchapter17,seshadri.arxiv16,Seshadri:2015:ANDOR,angizi2019graphide, ferreira2021pluto,mimdram,missingnot,yuksel2024simultaneous,olgun2022pidram}}, 
PCM~\cite{li.dac16}, 
MRAM~\cite{angizi2018pima,angizi2018cmp,angizi2019dna},  RRAM/memristive~\cite{levy.microelec14,kvatinsky.tcasii14,shafiee2016isaac,kvatinsky.iccd11,kvatinsky.tvlsi14,gaillardon2016plim,bhattacharjee2017revamp,hamdioui2015memristor,xie2015fast,hamdioui2017myth,yu2018memristive,yavits2021giraf, xi2020memory, zheng2016tcam, truong2021racer, truong2022adapting}, \gf{
\gfvi{FeRAM~\cite{ma20232,slesazeck20192tnc,wang20211t2c}}, or 
NAND flash~\cite{flashcosmos,gao2021parabit,choi2020flash,han2019novel,merrikh2017high,wang2018three,lue2019optimal,kim2021behemoth,wang2022memcore,han2021flash,kang2021s,lee2020neuromorphic,lee20223d}}
devices. \gf{PUM} architectures enable a wide range of different functions, such as bulk as well as finer-grained data copy/initialization~\omv{\cite{chang.hpca16,seshadri2013rowclone,aga.hpca17,rezaei2020nom,wang2020figaro,olgun2022pidram,yuksel2024simultaneous}}, 
bulk bitwise operations (e.g., a complete set of Boolean logic operations)~\cite{seshadri.micro17,li.dac16,angizi2018pima,angizi2018cmp,angizi2019dna,Seshadri:2015:ANDOR,seshadri.arxiv16,seshadri2020indram,aga.hpca17,li.micro17,mutlu2020retrospective,mandelman.ibmjrd02,chang.sigmetrics16,xin2020elp2im,gao2020computedram, olgun2021pidram, olgun2022pidram, li2018scope}, 
simple arithmetic operations (e.g., addition, multiplication, implication)~\cite{levy.microelec14,kvatinsky.tcasii14,aga.hpca17,kang.icassp14,li.micro17,shafiee2016isaac,eckert2018neural,fujiki2019duality,kvatinsky.iccd11,kvatinsky.tvlsi14,gaillardon2016plim,bhattacharjee2017revamp,hamdioui2015memristor,xie2015fast,hamdioui2017myth,yu2018memristive,deng.dac2018,angizi2019graphide}, 
\juanr{and lookup table queries~\cite{ferreira2021pluto}.}
\juanr{\omv{Two} recent \omv{works}~\cite{hajinazarsimdram,mimdram} provide flexible \omv{frameworks} to enable user-defined PUM implementation of complex operations,} \juanrr{which improves both PUM performance and programmability.}

{\em \bf \gf{Processing-near-memory (PNM)}} involves adding or integrating PIM logic (e.g., accelerators, simple processing cores, \omv{functional units,} reconfigurable logic) close to or inside the memory (e.g.,~\cite{fernandez2020natsa,cali2020genasm,kim.bmc18,ahn.pei.isca15,ahn.tesseract.isca15,boroumand.asplos18, boroumand2021google, boroumand2021google_arxiv,boroumand2019conda,boroumand2016pim,boroumand.arxiv17,singh2019napel,asghari-moghaddam.micro16,JAFAR,chi2016prime,farmahini2015nda,gao.pact15,DBLP:conf/hpca/GaoK16,gu.isca16,guo2014wondp,hashemi.isca16,cont-runahead,hassan.memsys15,hsieh.isca16,kim.isca16,kim.sc17,DBLP:conf/IEEEpact/LeeSK15,liu-spaa17,morad.taco15,nai2017graphpim,pattnaik.pact16,pugsley2014ndc,zhang.hpdc14,zhu2013accelerating,DBLP:conf/isca/AkinFH15,gao2017tetris,drumond2017mondrian,dai2018graphh,zhang2018graphp,huang2020heterogeneous,zhuo2019graphq,herruzo2021enabling,boroumand2021polynesia, boroumand2022icde, syncron, besta2021sisa_micro, besta2021sisa, asgarifafnir, upmem2018, devaux2019, shin2018mcdram, cho2020mcdram, denzler2021casper, gomez2022machine, giannoula2022towards, fernandez2022exploiting, oliveira2022heterogeneous, balasubramonian2014near, jacob2016compiling, nair2015active, lloyd2018dse, gokhale2015rearr, lloyd2015memory, rodrigues2016scattergather, lloyd2017keyvalue, landgraf2021combining, nair2015evolution, kwon202125, lee2021hardware, kim2021aquabolt, ke2021near, lee2022improving,siddique2024architectural,jaiyeoba2023acts,lenjani2022pulley,lenjani2021supporting,zhou2021ultra,sadredini2021sunder,lenjani2020fulcrum,mosanu2022pimulator}). Many of these works place PIM logic inside the logic layer of 3D-stacked memories or at the memory controller, but recent advances in silicon interposers (\omv{i.e.,} in-package wires that connect directly to the through-silicon vias in a 3D-stacked chip)~\cite{jedec.hbm.spec, fernandez2020natsa, singh2020nero, singh2021fpga, singh2021accelerating}
also allow for separate logic chips to be placed in the same die package as a 3D-stacked memory while still taking advantage of the TSV bandwidth. 
\juanrr{Industry has recently developed commercial PNM systems and PNM prototypes that incorporate near-bank logic in DRAM chips~\cite{kwon202125, lee2021hardware, upmem2018, devaux2019, shin2018mcdram, cho2020mcdram, lee2022isscc, kim2021aquabolt}, logic in the logic layer of 3D-stacked DRAM~\cite{niu2022isscc}, and logic in memory modules~\cite{ke2021near, lee2022improving, kim2021aquabolt}.}

Note that more functionality can be potentially integrated into a memory chip using PNM than using PUM, but both approaches can be combined to get even higher benefit from PIM\omv{, e.g., as shown by the MIMDRAM work~\cite{mimdram}}. In Section~\ref{sec:PUM}, we provide a detailed overview of PUM\omv{, focusing especially on} the \omv{principles of the} commodity DRAM technology. In Section~\ref{sec:PNM}, we provide a detailed overview of PNM\omv{, focusing especially on} the 3D-stacked DRAM technology \omv{as well as planar DRAM technology}. \omvi{We focus especially on DRAM~\cite{dennard1968field} due to its dominance as the main memory technology and very large capacity that can house many data-intensive workloads at reasonably low access latency.}
\gfv{In both \omvi{sections}, we also briefly discuss the implementation of PUM and PNM architectures using non-volatile memories and SRAM devices, \omvi{as appropriate}.}
We note that the described approaches and techniques in Sections~\ref{sec:PUM} and~\ref{sec:PNM} are applicable to \omvi{various} other types of technologies as well, with  modifications. 


%% file: sections/06-PUM.tex
\section{\gf{Processing-Using-Memory} (PUM)}
\label{sec:PUM}
\label{sec:minimally}



The PUM approach to processing-in-memory \omv{has the potential to provide large energy and performance benefits with small modifications to memory chips.}
This approach takes advantage of the existing interconnects in and analog operational behavior of conventional \gf{memory} architectures (e.g., \gf{SRAM}, DDRx, LPDDRx, HBM\gf{, NVM, and NAND flash}), \omv{\emph{without}} \omvii{requiring} dedicated logic processing elements \omvii{or a dedicated logic layer}, and usually with low \omvii{additional area and power} overheads. 
Mechanisms that use this approach take advantage of the high internal bandwidth available within each \gf{memory} cell array. 
There are a number of example PIM architectures that make use of the PUM approach~
\gfv{
\cite{seshadri.bookchapter17,
seshadri2013rowclone,
chang.hpca16,
kevinchang-thesis,
seshadri.thesis16,
Seshadri:2015:ANDOR,
seshadri.arxiv16, 
seshadri.micro17, 
hajinazarsimdram, 
seshadri2020indram,
chi2016prime, 
shafiee2016isaac, 
li.micro17, 
seshadri.bookchapter17.arxiv, 
deng.dac2018, 
xin2020elp2im, 
song2018graphr, 
song2017pipelayer,
gao2020computedram, 
eckert2018neural, 
aga.hpca17,
fujiki2019duality,
besta2021sisa_micro,
li.dac16,
ferreira2022pluto,
imani2019floatpim,
he2020sparse,
flashcosmos,
truong2022adapting,
truong2021racer,
olgun2021quactrng,
kim.hpca19,
kim.hpca18,
bostanci2022dr,
olgun2022pidram,
ali2019memory,
angizi2019graphide,
li2018scope,
subramaniyan2017parallel,
zha2020hyper,
fujiki2018memory,
orosa2021codic,
sharad2013ultra,
gao2021parabit,
choi2020flash,
han2019novel,
merrikh2017high,
wang2018three,
lue2019optimal,
kim2021behemoth,
wang2022memcore,
han2021flash,
kang2021s,
lee2020neuromorphic,
lee20223d,
si2019dual,
simon2020blade,
nag2019gencache,
wang2019bit,
al2020towards,
kang.icassp14,
kim2021colonnade,
jiang2020c3sram,
jeloka201628,
wang2023infinity,
kang2015energy,
imani2020dual,
deng2019lacc,
sutradhar2021look,
sutradhar2020ppim,
peng2023chopper,
shahroodi2023swordfish,
mimdram,
missingnot,
yuksel2024simultaneous,
yavits2023drama,
jahshan2024majork,
khalifa2023clapim,
garzon2022aida,
hanhan2022edam,
morad2016resistive,
wu2022dramcam_generalpurpose,
sadredini2020flexamata,
sadredini2019eap,
angstadt2018aspen,
wang2016sequential,
wang2015association}}. 

\gf{In this section, we elaborate on the development of PUM architectures leveraging different memory technologies, including
(1)~processing-using-DRAM,
(2)~processing-using-NVM, and 
(3)~processing-using-SRAM.}

\subsection{\gf{Processing-Using-DRAM}}
\label{sec:pum:pud}

\gf{Processing-using-DRAM architectures leverage the analog operating principles of DRAM cells \omv{and DRAM circuitry} to implement different operations.} 
\gf{First, we} focus on two such designs: 
RowClone, which enables in-DRAM \omv{(}bulk\omv{)} data movement \omv{(e.g., copy and initialization)} operations~\cite{seshadri2013rowclone} and 
Ambit, which enables in-DRAM bulk bitwise operations~\cite{Seshadri:2015:ANDOR,seshadri.arxiv16,seshadri.micro17,seshadri2020indram}. 
\juanrrr{\gf{Second}, we introduce SIMDRAM~\cite{hajinazarsimdram, oliveira2022methodologies}, an end-to-end framework for bit-serial SIMD computing in DRAM\gf{, and MIMDRAM~\cite{mimdram}, \omv{an} architecture \omv{and system} for \omv{more efficient and easier-to-program }bit-serial MIMD computing in DRAM,} \omv{both of which} build upon RowClone and Ambit mechanisms. }
\gf{Third, we describe the implementation of lookup table-based operations using the pLUTo substrate~\cite{ferreira2021pluto, ferreira2022pluto}.}
\gf{Fourth,}
we describe a low-cost substrate that performs data reorganization for non-unit strided access patterns~\cite{GS-DRAM}. 
\juanrrr{\gf{Fifth}, we describe PUM-based security primitives that generate physical unclonable functions (PUFs)~\cite{kim.hpca18} and true random numbers (TRNs)~\cite{kim.hpca19, olgun2021quactrng}.}

\omvi{\omvii{Fascinatingly, many} operations envisioned by these PUD works can already be performed in \omvii{\emph{real}} unmodified commercial off-the-shelf (COTS) DRAM chips, by violating manufacturer-recommended DRAM timing parameters. 
Recent works show that COTS DRAM chips can perform 
(1)~data copy \omvii{and} initialization~\cite{gao2020computedram,olgun2022pidram} (as in RowClone~\cite{seshadri2013rowclone}), 
(2)~three-input bitwise MAJ and two-input \texttt{AND} \& \texttt{OR} operations~\cite{gao2020computedram,gao2022fracdram,olgun2023drambender} (as in Ambit~\cite{Seshadri:2015:ANDOR,seshadri.arxiv16,seshadri.micro17,seshadri2020indram}), 
\omvii{(3)~bitwise \texttt{NOT} operation~\cite{missingnot},
(4)~up to 16-input bitwise \texttt{AND}, \texttt{NAND}, \texttt{OR}, \& \texttt{NOR} operations~\cite{missingnot},}
\gfvii{(5)}~true random number generation \& physical unclonable functions~\cite{olgun2021quactrng,kim.hpca18,kim.hpca19}. We describe some of these works \omvii{on COTS DRAM chips} here and some others in Section~\ref{sec:adoption}.}

\subsubsection{RowClone}
\label{sec:rowclone}

Two important classes of bandwidth-intensive memory operations are
(1)~\emph{bulk data copy}, where a large quantity of data is copied
from one location in physical memory to another; and (2)~\emph{bulk
  data initialization}, where a large quantity of data is initialized
to a specific value. We refer to these two operations as \emph{bulk
  data movement operations}. Prior
research\ch{~\cite{os-hardware,arch-os, kanev.isca15}} has shown that
operating systems \ch{and data center workloads} spend a significant
portion of their time performing bulk data movement
operations. For example, a paper by Google shows that close to 5\% of the execution time in Google's data center workloads is spent on executing only two data movement function calls, {\em memset} and {\em memcopy}. Therefore, accelerating \omv{bulk data movement} operations will likely
improve system performance \sg{and energy} efficiency.

We have developed a mechanism called
\emph{RowClone}~\cite{seshadri2013rowclone}, which takes advantage of the fact that bulk data movement operations do \emph{not} require any computation \omv{(or involvement)} on the part of the processor \omv{and the processor-memory data bus}. 
RowClone exploits the
internal organization and operation of DRAM to perform bulk data
copy/initialization quickly and efficiently inside {a DRAM chip}.
A DRAM chip contains multiple banks, where the banks are connected
together and to external I/O circuitry by a shared internal bus. Each bank
is divided into multiple
\emph{subarrays}~\omv{\cite{salp,chang.hpca14,seshadri2013rowclone,yauglikcci2022hira}}.  Each
subarray contains {many} rows of DRAM cells, where each column of
DRAM cells is connected together across the multiple rows using
\emph{bitlines}.

\begin{figure}[!t]
\centering
\includegraphics[width=1.0\linewidth]{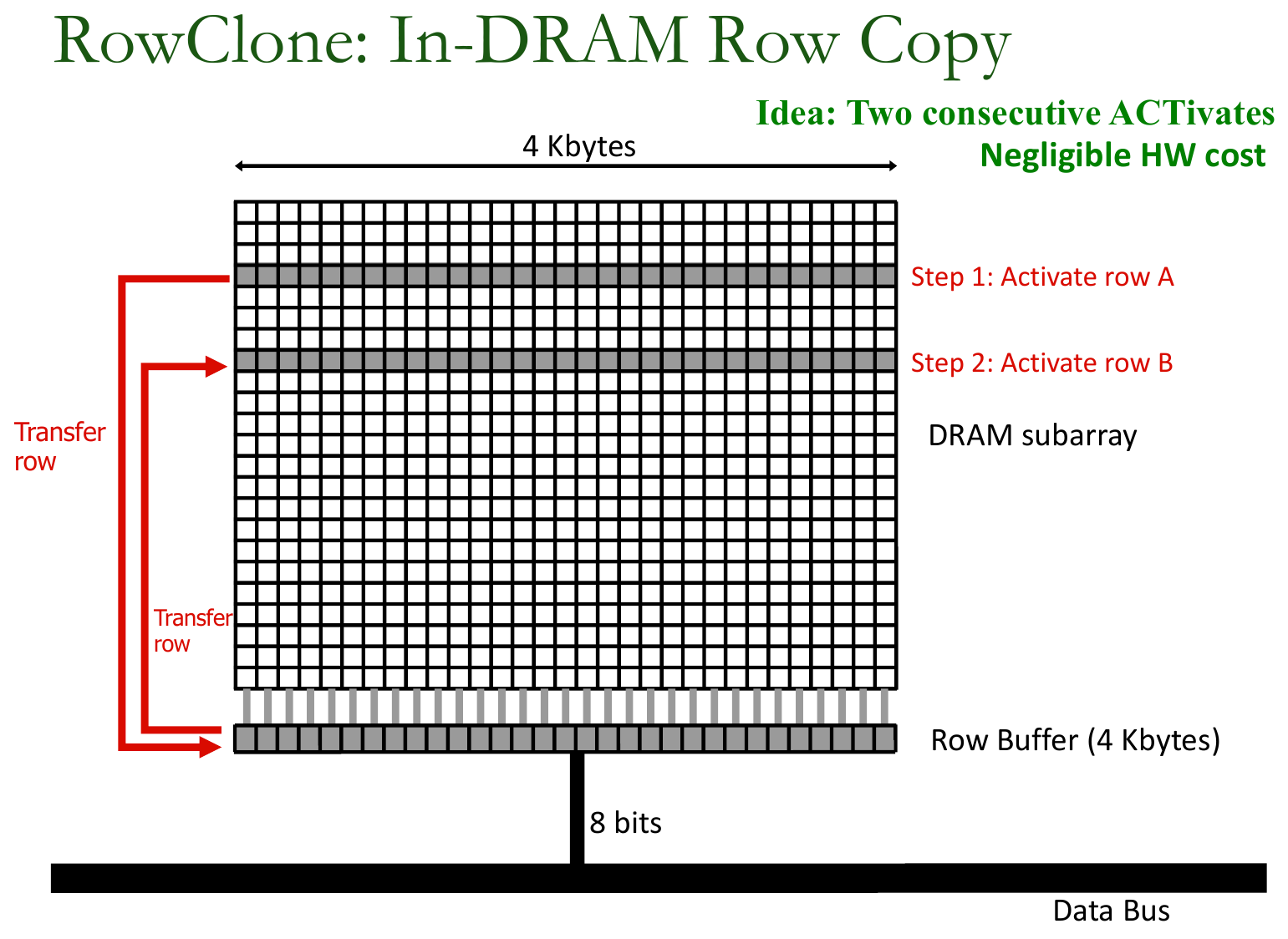}
\caption{RowClone Fast Parallel Mode. Reproduced from~\cite{mutlu.nsfpim20}.}
\label{fig:rowclone}
\end{figure}

RowClone consists of two mechanisms that take advantage of the
existing DRAM structure. The first mechanism, \juanrrr{\emph{Fast Parallel Mode}},
copies {the data of a row inside a subarray to another row}
inside the same DRAM subarray by issuing back-to-back activate (i.e.,
row open) commands to the source and the destination row.
Figure~\ref{fig:rowclone} illustrates the two steps of RowClone's Fast Parallel Mode. The first step activates source row $A$, which enables the \omv{capturing} of the entire row's data in the row buffer. The second step activates destination row $B$, which enables the copying of the contents of the row buffer into row $B$. Thus, the back-to-back activate in the same subarray enables the copying of source row $A$ to destination row $B$ by using the row buffer as a temporary buffer for row $A$'s contents.
The second mechanism \omv{(not shown)}, \juanrrr{\emph{Pipelined Serial Mode}}, can transfer an arbitrary number of bytes from a row in one bank to another row in another bank using the shared internal bus
among banks in a DRAM chip.

RowClone significantly reduces the raw latency and energy consumption
of bulk data copy and initialization, leading to $11.6\times$ latency
reduction and $74.4\times$ energy reduction for a {4kB bulk page
  copy (using the Fast Parallel Mode)}, at very low cost (only 0.01\%
DRAM chip area overhead)~\cite{seshadri2013rowclone}. This reduction
directly translates to improvement in performance and energy
efficiency of systems running copy or initialization-intensive
workloads. 
Our MICRO 2013 paper~\cite{seshadri2013rowclone} shows that the performance of six copy/\sgii{initialization}-\sg{intensive} benchmarks (including {the} fork system call, Memcached~\cite{memcached} and a MySQL \omv{database}~\cite{mysql}) improves between 4\% and 66\%. For the same six benchmarks, RowClone reduces the \omv{DRAM} energy consumption between 15\% and 69\%.
  
Recent works have improved upon the RowClone approach in various ways \juanrrr{(either performing RowClone in NVMs, proposing additional data movement mechanisms, or providing proofs-of-concept of RowClone in off-the-shelf DRAM chips)}. 
\juanrrr{First}, the \omv{Pinatubo} work~\cite{li.dac16} \juanrrr{shows} that RowClone can effectively be performed in emerging resistive memory chips, including Phase Change Memory (PCM)~\cite{lee-isca2009,qureshi.isca09}.
\juanrrr{Second}, Low-cost Interlinked Sub-Arrays (LISA)~\cite{chang.hpca16} 
provides mechanisms to enable the rapid transfer of data between one subarray to and adjacent subarray in the same bank, by enhancing the connectivity of subarrays using isolation transistors. LISA reduces inter-subarray copy latency by 9.2$\times$ and DRAM energy by 48$\times$, approaching the \textit{intra-subarray} latency and energy improvements of RowClone's Fast Parallel Mode. 
\juanrrr{Third}, FIGARO~\cite{wang2020figaro} improves upon LISA by enabling fine-grained (i.e., column granularity) data copy across subarrays within a bank using the shared global I/O structures of the bank as an intermediate location. This work~\omv{\cite{wang2020figaro}} shows significant benefit from FIGARO when its principles and techniques are used to build a highly-effective yet low-cost in-DRAM cache. 
\juanrrr{Fourth}, Network-on-Memory (NoM)~\cite{rezaei2020nom} improves the parallelism of bank-to-bank copy as well as bank read/write operations by providing more connectivity between different banks and chip I/O structures using the logic layer in 3D-stacked memory. 
\omv{Fifth, the CODIC work~\cite{orosa2021codic} shows that bulk data initialization can be sped up significantly in DRAM by enabling the memory controller to control internal DRAM timings in a fine-grained manner. By doing so, CODIC demonstrates that fast (in-DRAM) data initialization or destruction can \omvi{effectively} prevent cold boot attacks.}

\begin{figure}[!t]
    \centering
    \includegraphics[clip, trim=0.0cm 1.0cm 0.0cm 0.0cm, width=1.0\linewidth]{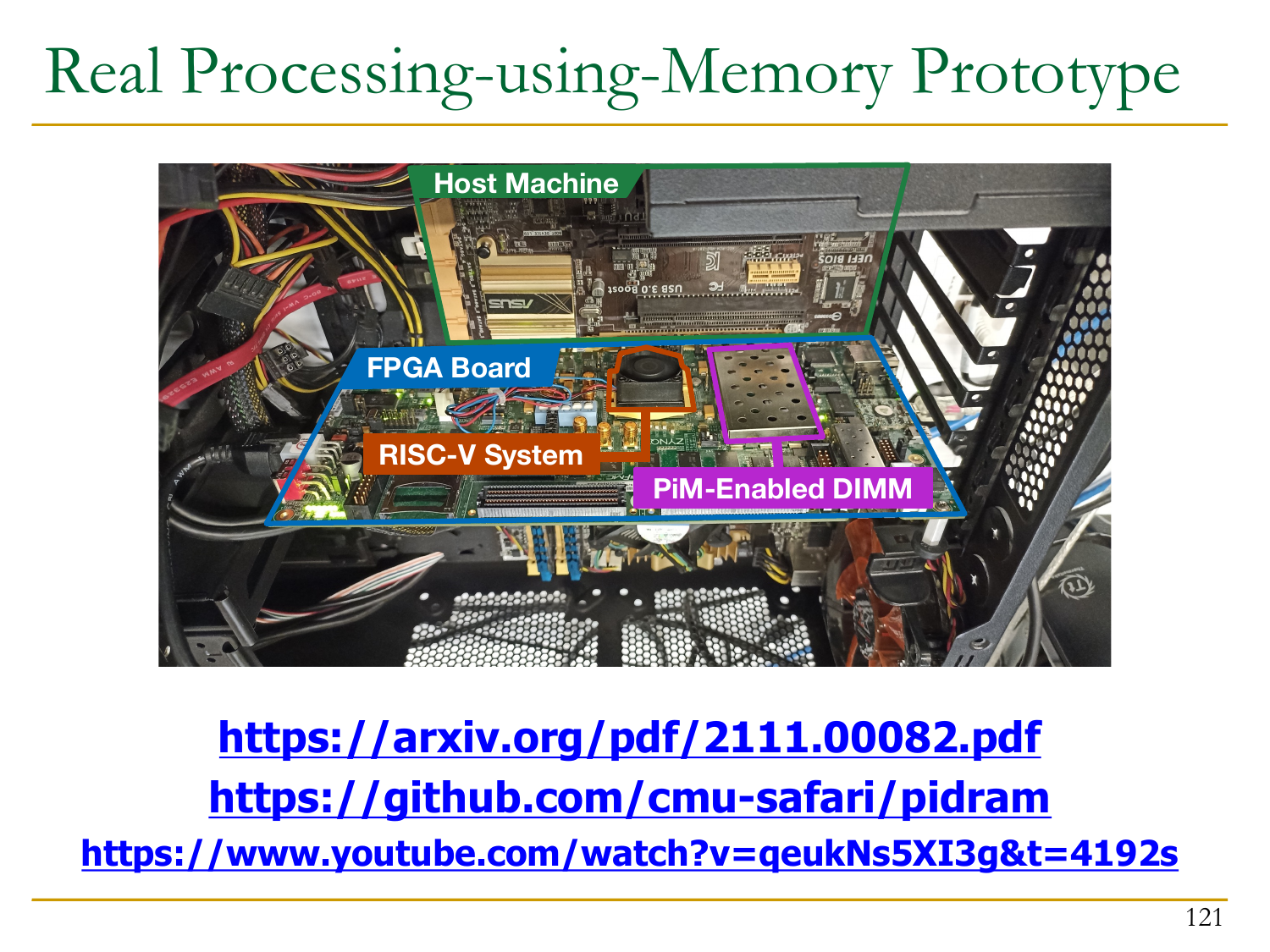}
    \caption{\gfix{PiDRAM's FPGA prototype. Figure reproduced from~\cite{mutlu.accml23.talk}.}}
    \label{fig:pidram}
\end{figure}

\gfv{Sixth}, the ComputeDRAM work~\cite{gao2020computedram} shows that one can mimic the effect of RowClone's back-to-back activation mechanism in off-the-shelf DRAM chips by violating the timing parameters such that two wordlines in a subarray are activated back-to-back as in \juanrrr{RowClone}. This work shows that such a version of RowClone can operate reliably in a variety of off-the-shelf DRAM chips tested using the SoftMC infrastructure~\cite{hassan2017softmc,softmc.github}. 
\juanrrr{By leveraging the findings of the ComputeDRAM work~\cite{gao2020computedram}, PiDRAM~\cite{olgun2021pidram, olgun2022pidram, olgun2021pidram_repo} prototypes the first flexible end-to-end framework that enables system integration studies and evaluation of real PUM techniques. 
PiDRAM provides software and hardware components to rapidly integrate PUM techniques across the whole system software and hardware stack (e.g., necessary modifications in the operating system, memory controller). 
PiDRAM provides an FPGA-based platform along with an open-source RISC-V system \omv{(as seen in Figure~\ref{fig:pidram})}. 
\omvii{We} demonstrate the flexibility and ease of use of PiDRAM by implementing and evaluating two state-of-the-art PUM techniques. The first use case implements in-memory copy and initialization using RowClone~\cite{seshadri2013rowclone}. 
\omvii{PiDRAM infrastructure provides} solutions to \omvii{system} integration challenges (e.g., memory coherence\omvii{, aligned data allocation into subarrays}) and \omvii{the PiDRAM work conducts} a detailed end-to-end implementation study. The second use case implements a true random number generator in DRAM based on D-RaNGe~\cite{kim.hpca19} (Section~\ref{sec:puf-trng}).} 
\gfv{In \omvi{more recent work}~\cite{yuksel2024simultaneous}, we show that \omvii{real unmodified} \gls{COTS} DRAM chips are capable of \omvii{executing} \emph{\omvi{M}ulti-RowCopy} operations as well, where one source row's content can \omvii{be} concurrently copied \omvii{into}
up to 31 destination rows \omvi{in the same subarray} with a \omvi{success rate greater} than 99.98\%. 
\omvii{To enable further research on in-DRAM data copy and initialization, we have open sourced our PiDRAM infrastructure as well as more recent work on \emph{Multi-RowCopy} at \url{https://github.com/CMU-SAFARI/PiDRAM} and \url{https://github.com/CMU-SAFARI/SiMRA-DRAM}. We \omviii{describe} this work and other PUD works on real COTS DRAM chips in Section~\ref{sec:simulation}.}}

We believe that RowClone provides very low-cost specialized support for a critical and often-used operation: data copy and initialization. In latency-critical systems, such as virtual machines, modern software is written to, as much as possible, avoid large amounts of data copy exactly because data copy is expensive in modern systems (because it goes through the processor over a bandwidth-bottlenecked memory bus). Eliminating copies as much as possible complicates software design, making it less maintainable and readable. If RowClone is implemented in real chips, perhaps the need for avoiding data copies will greatly diminish due to the more-than-an-order-of-magnitude latency reduction of page copy, leading to easier-to-write and easier-to-maintain software. 
\omv{Very fast data copy and initialization also \omvii{enables} the opportunity \omvi{to better protect the system} against various security attacks~\cite{orosa2021codic,yuksel2024simultaneous}.}
As such, we believe that an idea as simple as RowClone~\omvi{\cite{seshadri2013rowclone}} (and the \omvi{body of} work that builds on it~\omvi{\cite{orosa2021codic,yuksel2024simultaneous,li.dac16,chang.hpca16,wang2020figaro,rezaei2020nom,gao2020computedram,olgun2022pidram,kim.hpca19}}\omv{, such as \emph{\omvi{M}ulti-RowCopy} in real \gls{COTS} DRAM chips~\omvi{\cite{yuksel2024simultaneous}}}) can have exciting and forward-looking implications on making both systems and software much faster, more efficient\omvii{, more robust,} and overall better.

\subsubsection{Ambit}
\label{sec:ambit}

In addition to bulk data movement and initialization, many applications make use of
\emph{bulk bitwise operations}, i.e., bitwise operations on large bit
vectors~\cite{btt-knuth,hacker-delight}.  Examples of such
{applications} include bitmap
indices~\cite{bmide,bmidc,fastbit,bicompression} used in databases,  bitwise scan acceleration~\cite{bitweaving} in databases, accelerated document filtering for web search~\cite{bitfunnel}, DNA sequence
alignment~\omv{\cite{bitwise-alignment,xin.shd.bioinformatics15,alser.bioinformatics17,kim.bmc18,alser2020sneakysnake,cali2020genasm,alser2019shouji,myers1999fast,cali2022segram,lindegger2023scrooge}},
encryption algorithms~\cite{xor1,xor2,enc1}, graph
processing~\omv{\cite{li.dac16,besta2021sisa_micro}}, and
networking~\cite{hacker-delight}. Accelerating bulk bitwise operations
can {thus} significantly boost the performance and energy
efficiency of a wide range applications.

\begin{figure}[ht]
\centering
\includegraphics[width=1.0\linewidth]{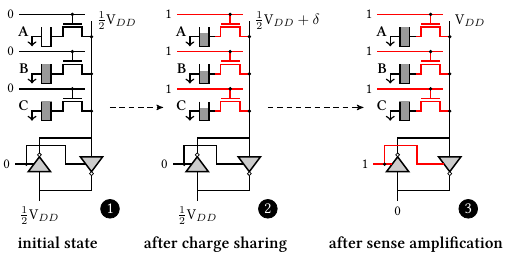}
\caption{Triple-row activation in Ambit. Reproduced from~\cite{seshadri.micro17}.}
\label{fig:ambit-tra}
\end{figure}

In order to avoid data movement bottlenecks when the system performs
these {bulk} bitwise operations, we have recently proposed a new
\textbf{A}ccelerator-in-\textbf{M}emory for bulk \textbf{Bit}wise
operations (Ambit)~\cite{Seshadri:2015:ANDOR,seshadri.arxiv16,seshadri.micro17,seshadri2020indram}.  
Unlike prior approaches \omvi{to \omvii{accelerating} such operations}, Ambit uses the analog
\omv{operational properties} of existing DRAM technology \omv{and circuitry} to perform bulk bitwise
operations.  Ambit consists of two components. 
The first component, Ambit--AND--OR, implements a new operation called \emph{triple-row activation}, where the memory controller simultaneously activates three rows. 
Triple-row activation, depicted in Figure~\ref{fig:ambit-tra}, performs a bitwise majority \juanrrr{(MAJ)} function across the cells in the three \omv{activated} rows, due to the charge sharing principles that govern the operation of the DRAM array. 
In the initial state, all three rows are closed \circled{1}. In the example of Figure~\ref{fig:ambit-tra}, two cells are in the charged state. When the three wordlines are raised simultaneously \circled{2}, charge sharing results in a positive deviation of the bitline. After sense amplification \circled{3}, the sense amplifier drives the bitline to $V_{DD}$, and as a result, fully charges the three cells. 
By controlling the initial value of one of the three rows (e.g., $C$), we can use triple-row activation to perform a bitwise AND or OR of the other two rows, since the bitwise majority function can be expressed as $C(A+B)+\bar{C}(AB)$. 
The second component, Ambit--NOT, takes advantage of the two
inverters that are part of each sense amplifier in a DRAM
subarray. 
Ambit--NOT exploits the fact that, at the end of the
sense amplification process, the voltage level of one of the
inverters \omv{in the sense amplifier} represents the negated logical value of the cell.  The
Ambit design adds a special row to the DRAM array, which is used to
capture the negated value that is present in the sense amplifiers.
One possible implementation of the special
row~\cite{seshadri.micro17} is a row of \emph{dual-contact cells} (a 2-transistor 1-capacitor cell~\cite{2t-1c-1,migration-cell}) that connect to both inverters inside the sense amplifier. 
With the ability to perform AND, OR, and NOT operations, Ambit is functionally complete: It can reliably perform \emph{any} bulk bitwise operation completely using DRAM technology, even in the presence of significant process variation (see \cite{seshadri.micro17} for details).
\omv{As such, any algorithm can likely be re-written to be \omvi{implemented} in the Ambit's functionally-complete execution substrate.}

Averaged across seven commonly-used bitwise operations ({\em not, and, or, nand, nor, xor, xnor}), Ambit with 8
DRAM banks improves bulk bitwise operation throughput by 44$\times$
compared to an Intel Skylake processor~\cite{intel-skylake}, and
32$\times$ compared to the NVIDIA GTX 745 GPU~\cite{gtx745}. Compared
to the DDR3 {standard}, Ambit reduces energy consumption of these
operations by 35$\times$ on average. Compared to HMC
2.0~\cite{hmc.spec.2.0}, Ambit improves bulk bitwise operation
throughput by 2.4$\times$. When integrated directly into the HMC 2.0
device, Ambit improves throughput by 9.7$\times$ compared to
processing in the logic layer of HMC 2.0.

The Ambit work also shows that porting bitmap-index based databases~\omv{\cite{bmide}} as well as the BitWeaving database~\omv{\cite{bitweaving}}  to execute Ambit can greatly improve query latencies. For example, Ambit reduces the end-to-end query latencies  by 5.4$\times$ to 6.6$\times$ for bitmap-based databases, with larger improvements coming from cases where more data needs to be scanned in the database. For the BitWeaving database, which is specifically designed to maximize bitwise operations so that the database can be relatively easily accelerated on modern GPUs, Ambit reduces the end-to-end query latencies by 4$\times$ to 12$\times$, again with larger improvements coming from cases where more data needs to be scanned in the database. These results are clearly very promising on two important data-intensive applications. 

A number of Ambit-like bitwise operation substrates have been proposed in recent years, making use of emerging resistive memory technologies, e.g., \sgii{phase-change memory}
(PCM)~\cite{lee-isca2009,lee.ieeemicro10,lee.cacm10,zhou.isca09,qureshi.isca09,yoon-taco2014},
SRAM~\cite{aga.hpca17,kang.icassp14,eckert2018neural,fujiki2019duality},
or specialized computational DRAM~\cite{li.micro17,angizi2019dna,xin2020elp2im,gao2020computedram, olgun2021pidram,angizi2019graphide, ferreira2021pluto}.
These substrates can perform bulk bitwise operations in a special DRAM array augmented
with computational circuitry~\cite{li.micro17,deng.dac2018} and in 
resistive memories~\cite{li.dac16} 
like \sgii{PCM}. 
Substrates similar to Ambit can perform simple arithmetic operations in SRAM~\cite{aga.hpca17, kang.icassp14} and arithmetic and logical operations in memristors~\cite{kvatinsky.tcasii14, kvatinsky.iccd11, kvatinsky.tvlsi14, shafiee2016isaac, levy.microelec14}. 
All of these works have shown significant benefits from performing bitwise operations using memory, for a wide variety of applications, including databases, machine learning, graph processing, genome analysis, and using a variety of different memory technologies, including DRAM, SRAM, PCM, memristors. 

Recently, the ComputeDRAM work~\cite{gao2020computedram} showed that carefully violating timing parameters between activation commands can mimic the triple-row-activation operation of Ambit in some existing off-the-shelf DRAM chips, using the SoftMC infrastructure~\cite{hassan2017softmc}. Thus, in-DRAM \omv{bulk bitwise} AND and OR operations \omv{(as envisioned by the Ambit work~\cite{seshadri.micro17,seshadri.arxiv16})} can be performed in some real off-the-shelf DRAM chips even though clearly such chips are \msvii{\emph{not}} designed to perform such Ambit operations. This proof-of-concept demonstration shows that the ideas presented in Ambit may not be far from reality: if some existing DRAM chips that are not even designed for in-DRAM bulk bitwise operations can perform such operations, then DRAM chips that are carefully designed for such operations will hopefully be even more capable!

\begin{figure}[ht]
    \centering
    \includegraphics[width=1\linewidth]{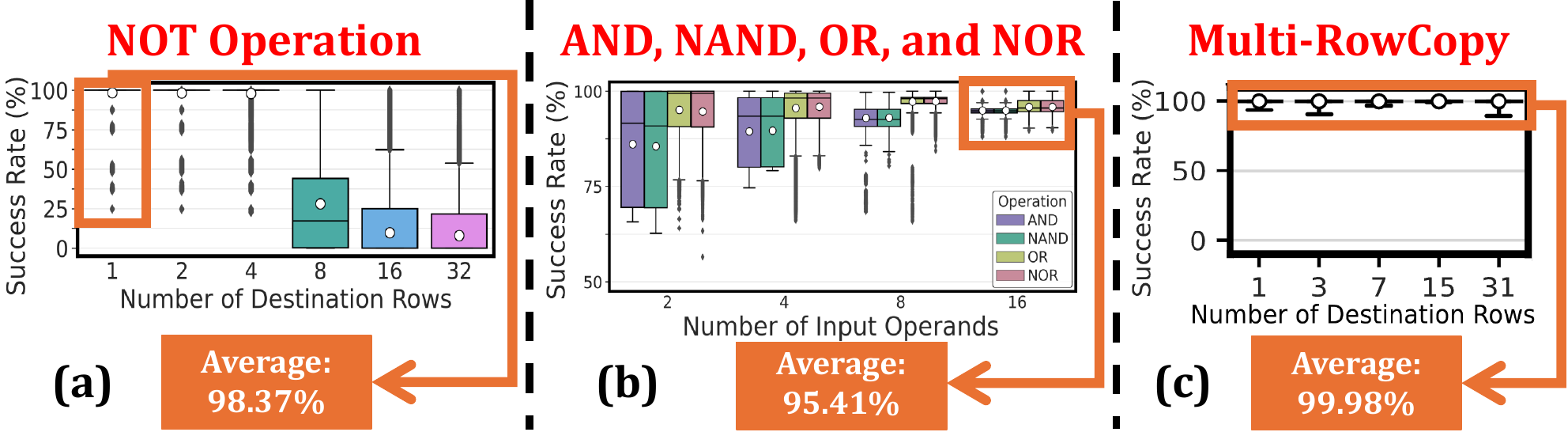}
    \caption{\gfvi{{Success rates of (a)~the \texttt{NOT} operation with varying numbers of destination rows, (b)~\texttt{AND}, \texttt{NAND}, \texttt{OR}, and \texttt{NOR} operations with varying numbers of input operands, (c)~the \emph{Multi-RowCopy} operation with varying numbers of destination rows, as measured in 224, 224, and 120 COTS DRAM chips, respectively.} 
    More results and experimental methodology are in~\cite{missingnot,yuksel2024simultaneous}. Reproduced from~\cite{mcciede2024}.}}
    \label{fig:fcdram_simra_results}
\end{figure}

\gfvi{\omvi{A recent} HPCA 2024 paper~\cite{missingnot} takes the realization of bulk bitwise Boolean PUM operations in real DRAM \omvii{chips} one step further by showing that, besides being able to perform \emph{Multi-RowCopy} operations~\omvii{\cite{yuksel2024simultaneous}} (as we describe in Section~\ref{sec:rowclone}), \gls{COTS} DRAM chips are capable of:
(1)~performing functionally-complete bulk-bitwise Boolean operations: \texttt{NOT}, \texttt{NAND}, and \texttt{NOR}; and
(2)~executing up to 16-input \texttt{AND}, \texttt{NAND}, \texttt{OR}, and \texttt{NOR} operations. 
The authors evaluate the robustness of these operations across data \omvii{various} patterns, \omvii{temperatures}, and voltage levels.
The evaluation results (\omvii{summarized in} Figure~\ref{fig:fcdram_simra_results}) show that COTS DRAM chips can perform these operations at high success rates ($>$94\%).
These fascinating findings demonstrate the fundamental computation capability of DRAM, even when DRAM chips are {\em not} designed for this purpose, and provide a solid foundation for building new and robust PUD mechanisms into future DRAM chips and standards.}
\omvii{This work~\cite{missingnot} is open sourced at \url{https://github.com/CMU-SAFARI/FCDRAM/} to enable future work in PUD systems.}


We believe it is extremely important to continue exploring such low-cost
Ambit-like \omvi{bulk bitwise execution} \sgii{substrates,} as well as more sophisticated computational substrates\sgii{,}
for all types of memory technologies, old and new. Resistive memory
technologies are fundamentally non-volatile and amenable to in-place
updates, and as such, can lead to even less data movement compared
to DRAM, which fundamentally requires some data movement to sense, amplify and restore
the data. \sgii{Thus}, we believe it is very promising to examine the
design of both charge-based conventional and emerging resistive memory chips that can incorporate
Ambit-like bitwise \sgii{operations} and other types of suitable computation
capability. 
\omvi{The fascinating fact that existing COTS DRAM chips, which are \omvii{\emph{not}} designed for computation purposes, and without \omvii{\emph{any}} changes to the chip at all, can \omvii{\emph{already}} perform many different bitwise operations demonstrates that the Ambit-like bulk bitwise approaches can have high practicality.}

\subsubsection{\juanrrr{SIMDRAM}}

\juanrrr{Going forward in the direction of Ambit-like PUM substrates, it is critical to research frameworks that can enable ease-of-programming of such substrates such that many algorithms can take advantage of the massive bit-level parallelism offered by Ambit-like substrates. 
In this direction, SIMDRAM~\cite{hajinazarsimdram, oliveira2022methodologies} provides a framework that (1)~enables the efficient PUM implementation of complex operations, and (2)~provides a flexible mechanism to support the implementation of arbitrary user-defined operations using DRAM.} 

\juanrrr{We build SIMDRAM on a DRAM substrate that enables two well-known techniques: (1)~vertical data layout, and (2)~majority-based computation. 
First, the vertical data layout (i.e., all bits of an operand in the same bitline) allows efficient bit-shift operations, which are necessary for complex computations (e.g., addition, multiplication), using RowClone~\cite{seshadri2013rowclone}. This eliminates the need for extra logic in DRAM for shifting~\cite{li.micro17, deng.dac2018}. 
Second, since MAJ and NOT operations represent a functionally complete set, and both operations are natively supported in Ambit~\cite{Seshadri:2015:ANDOR,seshadri.arxiv16,seshadri.micro17,seshadri2020indram}, a computation typically takes fewer DRAM commands using MAJ and NOT than using basic logical operations AND, OR, and NOT.} 

\begin{figure}[ht]
\centering
\begin{subfigure}[h]{0.46\textwidth}
    \centering
    \includegraphics[width=\textwidth]{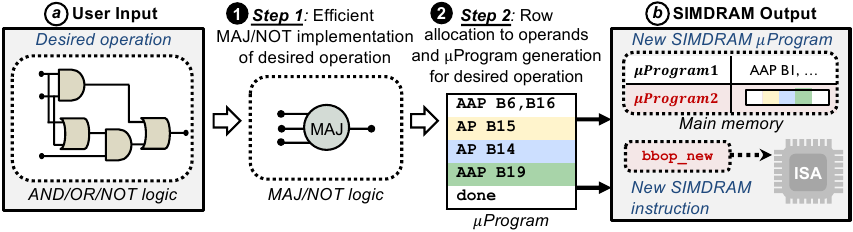}%
    \caption{SIMDRAM framework: Steps 1 and 2.}
    \label{fig_framework_1_2}
\end{subfigure}
\par\bigskip 
\begin{subfigure}[h]{0.46\textwidth}
    \centering
    \includegraphics[width=\textwidth]{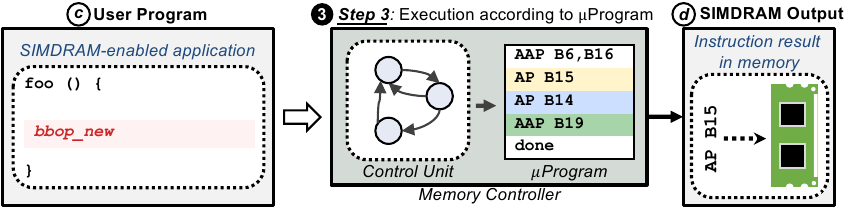}%
    \caption{SIMDRAM framework: Step 3.}
    \label{fig_framework_3}
\end{subfigure}
\caption{\omv{Overview of the SIMDRAM framework. Reproduced from~\cite{hajinazarsimdram,mcciede2024}.}}
\label{fig:simdram}
\end{figure}

\juanrrr{The SIMDRAM framework consists of three steps, shown in Figure~\ref{fig:simdram}.
The first step \gfv{(\circled{1} in Figure~\ref{fig:simdram})} obtains an optimized MAJ/NOT-based representation of the desired operation from its AND/OR/NOT-based representation \gfv{(\incircledd{a} in Figure~\ref{fig:simdram})}. 
This step employs logic optimization \omv{techniques} to minimize the number of logic primitives required for an operation (thus, minimizing the operation latency). 
The second step (\circled{2}) allocates DRAM rows to the input and output operands, and generates an optimized \gfv{$\mu$Program}, i.e., the optimized sequence of DRAM commands (ACTIVATE-ACTIVATE-PRECHARGE, AAP, and ACTIVATE-PRECHARGE, AP) that perform MAJ and NOT \omv{(and RowClone)} operations\gfv{, which is stored in main memory and will be used to execute the operation at runtime (\incircledd{b}).}
\gfv{The first and second steps are executed \emph{prior} to program execution, in a \emph{one-time \omvi{offline}} effort to map the user-desired operation to DRAM commands.}
The third step (\circled{3}) executes the \gfv{$\mu$Program} using a control unit in the memory controller, which 
issues the sequence of AAPs/APs to DRAM. 
\gfv{The third step happens \emph{during} program execution, when a \emph{bbop} instruction (\incircledd{c}), i.e., a SIMDRAM ISA extension that exposes SIMDRAM operations to the programmer, is executed.}
Once the \gfv{$\mu$Program} completes, the result of the operation is held in DRAM \gfv{(\incircledd{d})}.}

\juanrrr{SIMDRAM provides support for full system integration \gfv{of its PUM capabilities in current systems, in particular support for managing and storing the data required for in-DRAM computation in a vertical layout and ISA extensions for and programming interface of SIMDRAM operations.} 
First, we design a \emph{data transposition unit} that sits between the last-level cache and the memory controller of the host processor. It transposes horizontally laid-out data (used by traditional system software) into the vertical data layout (used by SIMDRAM). 
Second, we simplify program integration by providing ISA extensions that expose SIMDRAM operations to the programmer.} 
\gfv{We also briefly discuss other system integration challenges faced by SIMDRAM, including how SIMDRAM handles page faults, address translation, \omvi{interrupts, and} coherence (via flushing~\cite{guide2016intel, manual2010arm} \omvi{of cache lines that are needed by PUD execution}).}

\juanrrr{SIMDRAM \omv{provides} $2.0\times$ the throughput and $2.6\times$ the energy efficiency of Ambit for 16 complex operations (e.g., addition, multiplication, division, ReLU, predication, etc.). 
Compared to a modern CPU and a modern GPU, SIMDRAM provides (1) $88\times$ and $5.8\times$ the throughput, and $257\times$ and $31\times$ the energy efficiency of the CPU and the GPU, respectively, for 16 complex operations; (2) $21\times$ and $2.1\times$ the performance of the CPU and the GPU, over seven real-world applications (e.g., databases, CNNs, classification). 
Additional evaluations of reliability, area overhead, data movement overhead, and transposition overhead are detailed in~\cite{hajinazarsimdram}.}

\juanrrr{We believe SIMDRAM represents a big leap towards 
(1)~improving the performance and energy efficiency of PUM substrates, (2)~facilitating full system integration and \omv{easier} programmability, and 
(3)~inspiring future development \omvi{and \omvii{enhancement}} of real PUM substrates.}

\subsubsection{\gf{MIMDRAM}}
\label{sec:mimdram}

\gf{Processing-using-DRAM (PUD) is a PUM approach that {uses} {a} DRAM {array's} massive {internal}  parallelism to execute {very}-wide {(e.g., {16,384\omv{-bit to} 62,144-{bit-}wide}) } {data-parallel} operations{, in a \gfv{\gls{SIMD}} fashion}. However, DRAM rows' large and rigid granularity {limit} the effectiveness and applicability
of {PUD} in {three} ways. 
First, since {applications {have} varying degrees of SIMD parallelism} {(which is often smaller than the DRAM row granularity)}, {PUD}  execution {often} leads
to \omv{underutilization}, throughput loss, and energy waste. 
Second, due to {the} {high area cost of implementing} {interconnects} that connect {columns in a wide DRAM row}, 
most PUD architectures are limited to the execution of {parallel} \omv{\emph{map}} operations~\omv{\cite{mccool2012structured}}, where {a single {operation} is performed over equally-sized input and output arrays}.
{Third, the {need to feed the wide DRAM row with {tens of} thousands of data {elements} {combined with the} lack {of {adequate} compiler support for {PUD}} {systems create a {programmability barrier}}, since {programmers} need to {manually} extract SIMD parallelism from an application and map computation to the {PUD} {hardware}.}}}

\gf{To mitigate these issues, we propose MIMDRAM~\cite{mimdram}, a hardware/software co-designed PUD system that introduces new mechanisms to allocate and control {only} the {necessary} resources for {a given} PUD {operation}. The \emph{key idea} of MIMDRAM is to leverage fine-grained DRAM activation~\cite{cooper2010fine,Ani2010,zhang2014half,ha2016improving,lee2017partial,olgun2022sectored,o2021energy,oconnor2017fine,olgun2024sectored} for PUD.
\gfv{Prior works on fine-grained DRAM \gfvii{activation}~\cite{cooper2010fine,Ani2010,zhang2014half,ha2016improving,lee2017partial,olgun2022sectored,o2021energy,oconnor2017fine,olgun2024sectored} leverage the hierarchical design of a DRAM subarray to enable DRAM row accesses with \emph{flexible} granularity. 
A DRAM subarray is composed of multiple (e.g., 32--128) smaller 2D arrays of 512--1024 DRAM rows and 512--1024 columns, called \emph{DRAM mats}~\cite{standard2012jesd79,seshadri2020indram,seshadri.bookchapter17.arxiv,seshadri.bookchapter17,zhang2014half}.
During a row access, the DRAM access circuitry \emph{simultaneously} addresses \emph{all} columns across \emph{all} mats in a subarray, composing a \emph{large} DRAM row of size \texttt{[\#columns\_per\_mat $\times$ \#mats\_per\_subarray]}.
\emph{Fine-grained DRAM} modifies the DRAM access circuitry to enable reading/writing data from/to individual DRAM mats, \omvi{as opposed to \msvii{all} mats in a subarray}.
\omvi{Doing so allows fine-grained access to a much smaller portion of a DRAM subarray \omvii{(e.g., one or more mats)}, leading to efficiency benefits~\cite{olgun2024sectored}.}
Fine-grained DRAM \omvi{specifically} provides three main benefits for PUD execution.} 
First, it enables MIMDRAM to allocate \emph{only} the appropriate computation resources (based on the maximum vectorization factor of a loop) for a target loop, {thereby} reducing under-utilization and energy waste. 
Second, MIMDRAM can currently execute {multiple} independent operations inside a single DRAM subarray {\emph{independently} in \omv{\emph{separate}}} {DRAM mats}. This allows MIMDRAM to {operate} as a multiple-instruction multiple-data (MIMD)~\omv{\cite{smith1986pipelined,flynn1966very,thornton1964parallel}} PUD substrate, increasing overall throughput. 
Third, MIMDRAM implements low-cost interconnects that enable moving data across DRAM columns \emph{across} and \emph{within} DRAM mats by combining fine-grained DRAM activation with simple modifications to the DRAM I/O circuitry. This enables MIMDRAM to implement \omv{\emph{reduction}} operations~\omv{\cite{mccool2012structured}} in DRAM without any intervention of the host CPU cores \omv{or other logic outside DRAM}.}

\begin{figure}[ht]
    \centering
    \includegraphics[width=\linewidth]{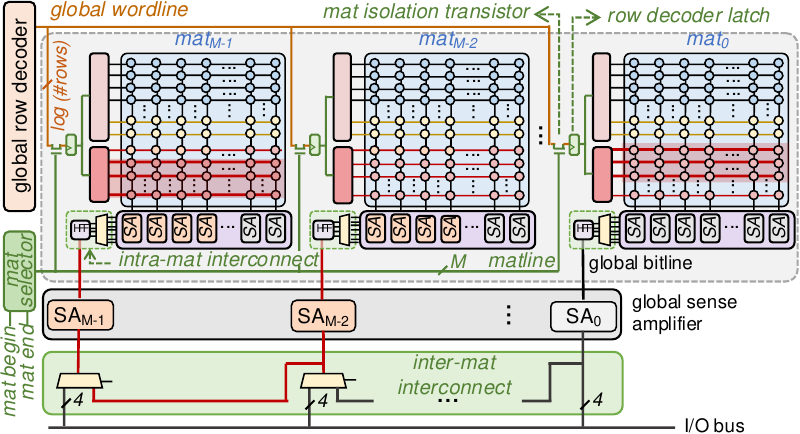}
    \caption{\gfv{MIMDRAM subarray and bank organization. {Green-colored boxes represent newly added hardware components.} Reproduced from~\cite{mimdram,mcciede2024}.}}
    \label{fig_subarray_matdram}
\end{figure}

\gf{Figure~\ref{fig_subarray_matdram} shows an overview of the DRAM organization of MIMDRAM. Compared to the baseline Ambit subarray organization, MIMDRAM adds four new components ({colored} in green) to a DRAM subarray and DRAM bank, which {enable} 
(1)~fine-grained PUD execution;
(2)~global I/O data movement; and
(3)~local  I/O data movement.} 

\gf{\paratitle{Fine-Grained PUD Execution} To enable fine-grained PUD execution, MIMDRAM modifies Ambit's subarray and the DRAM bank with three new hardware structures: the \emph{mat isolation transistor}, the \emph{row decoder latch}, and the \emph{mat selector}. 
At a high level, 
the \emph{mat isolation transistor} allows for the independent access and operation of \omv{different} DRAM mats within a subarray while the \emph{row decoder latch} enables the execution of a PUD operation in a range of DRAM mats that the \emph{mat selector} defines.}

\gf{\paratitle{Global I/O Data Movement} To enable data movement \omv{\emph{across}} different mats, MIMDRAM implements an  \emph{inter-mat {interconnect}} by slightly modifying the connection between the I/O bus and the global \gfv{sense amplifiers}.
The inter-mat {interconnect} relies on the observation that the \gfv{global} sense amplifiers  {have \emph{higher} drive} than the sense amplifiers in the local row buffer~\cite{keeth2007dram, wang2020figaro}, allowing to directly drive data from the \gfv{global sense amplifiers} into the local row buffer.
{To leverage this observation}, MIMDRAM adds a 2:1 multiplexer to the input/output port of each {set of \emph{four} 1-bit  \gfv{global} sense amplifiers}. The multiplexer selects whether the data that is written to the sense amplifier {set} $SA_i$  comes from the I/O bus or from the neighbor sense amplifier {set} $SA_{i-1}$.} 

\gf{\paratitle{Local I/O Data Movement} To enable data movement {across columns} \emph{within} a DRAM mat, MIMDRAM implements an  \emph{intra-mat {interconnect}}, which does \emph{not} require any hardware modifications. 
Instead, it modifies the sequence of steps DRAM executes during a column access operation. There are two \emph{key observations} that enable the intra-mat {interconnect}. 
First, we observe that the local bitlines {of a DRAM mat} \emph{already} share an interconnection path via the helper flip-flops (HFFs) and column select logic. 
Second, the HFFs in a DRAM mat can latch and \emph{amplify} the local row buffer's data~\cite{keeth2007dram,o2021energy}.} 

\omvi{\paratitle{Software Support}} \gfv{On the software side, MIMDRAM {implements} compiler passes to
(1)~automatically vectorize code regions that can benefit from PUD execution;
(2)~for such regions, generate PUD {operations} with the most appropriate SIMD granularity; and
(3)~schedule the concurrent execution of {independent} PUD operations {in} different DRAM mats. 
We discuss the concrete implementation of MIMDRAM compiler passes in Section~\ref{sec:mapping}.
In the MIMDRAM paper~\cite{mimdram, mimdramextended}, we further discuss how MIMDRAM expands on SIMDRAM's system integration support, including how MIMDRAM deals with 
(1)~data allocation {within} a DRAM subarray and 
(2)~mapping of a PUD's operands to guarantee high utilization of the \omvi{fine-grained} PUD substrate.
}

\begin{figure}[!t]
    \centering
    \includegraphics[width=\linewidth]{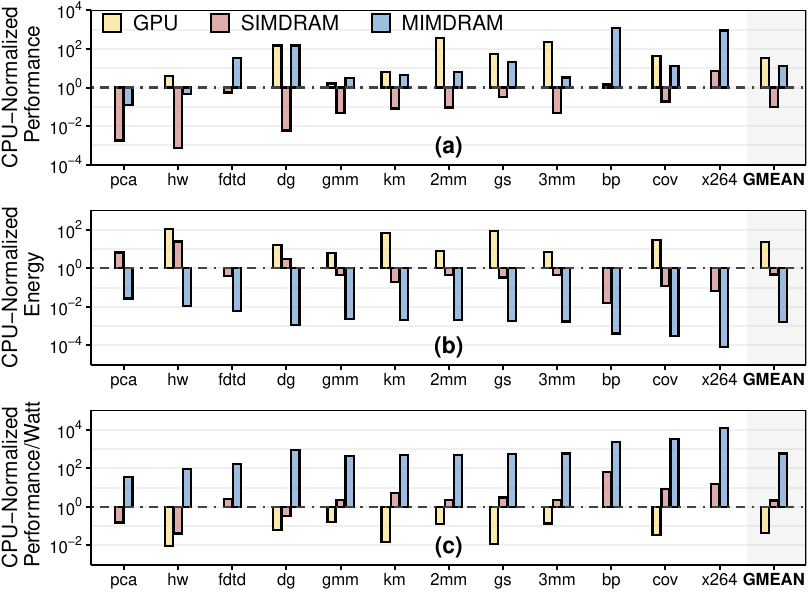}
    \caption{\gfv{CPU-normalized {performance}~(a), energy~(b), and energy efficiency~(performance/Watt)~(c) results for processor-centric (i.e., Intel Skylake CPU~\cite{intel-skylake} and NVIDIA A100 GPU~\cite{a100}) and memory-centric (i.e., SIMDRAM~\cite{hajinazarsimdram} and MIMDRAM) architectures executing 12 real-world applications. Reproduced from~\cite{mcciede2024}.}}
    \label{fig:mimdram_perf_energy}
\end{figure}

\gf{\gfv{We evaluate MIMDRAM using twelve real-world applications and 495 multi-programmed application mixes. 
When using 64 DRAM subarrays per bank and 16 banks for PUD computation in a DRAM chip, MIMDRAM provides (1)~13.2$\times$/0.22$\times$/173$\times$ the performance, 
(2)~0.0017$\times$/0.00007$\times$/0.004$\times$ the energy consumption, 
(3)~582.4$\times$/13612$\times$/272$\times$ the performance per Watt of the CPU~\cite{intel-skylake}/GPU~\cite{a100}/SIMDRAM~\cite{hajinazarsimdram} baseline (Figure~\ref{fig:mimdram_perf_energy}) and 
(4)~when using a single DRAM subarray, 15.6$\times$ the SIMD utilization, i.e., SIMD efficiency of the prior state-of-the-art PUD framework, SIMDRAM.} 
MIMDRAM {adds} {small} area cost {to} a DRAM chip (1.11\%) and CPU die (0.6\%).}
\omvi{More details on and evaluation of MIMDRAM are presented in~\cite{mimdram,mcciede2024}.}

\omvi{We believe MIMDRAM greatly advances the hardware and software needed to enable PUD systems.}
{We hope and believe that MIMDRAM's ideas and results will help {to enable} \omvi{much} more efficient and easy-to-program PUD systems. To this end, we open source
MIMDRAM at \url{https://github.com/CMU-SAFARI/MIMDRAM}.
\omvi{We hope that future work builds on the Ambit/SIMDRAM/MIMDRAM line of work to enable even more capable PUD systems.}}

\subsubsection{\gf{pLUTo}}
\label{sec:pluto}

\gf{Prior PUD works, such as Ambit~\cite{seshadri.micro17}, SIMDRAM~\cite{hajinazarsimdram}, and MIMDRAM~\cite{mimdram}, propose mechanisms for the execution of bulk bitwise operations (e.g., bitwise \texttt{MAJority,AND,OR,NOT}) and {bulk} arithmetic operations. 
{However, these proposals have two important limitations:
(1)~the execution of some complex operations (e.g., multiplication, division) incurs high latency and energy consumption~\cite{hajinazarsimdram}, and}
(2)~other complex operations (e.g., exponentiation, trigonometric functions) are \msvii{\emph{not}} \omv{(easily)} supported.}

\gf{To overcome these two limitations of prior PUD \omv{techniques}, we investigate the use of \emph{LUT-based computing}, i.e., the use of memory read operations \emph{(LUT queries)} to retrieve the results of complex operations from lookup tables that hold precomputed values, for PUM computing.
We \emph{extend the functionality of DRAM-based PUM {systems} to provide support for {general-purpose execution of} complex {operations.}}
\textbf{To this end}, we propose \textit{pLUTo~\cite{ferreira2021pluto,ferreira2022pluto}: \underline{p}rocessing-using-memory with lookup table (\underline{LUT}) \underline{o}perations,} a DRAM-based PUM architecture that leverages LUT-based computing {via bulk querying of LUTs} to perform complex operations beyond the scope of prior PUD proposals.
pLUTo introduces a novel LUT-querying mechanism, the \emph{pLUTo LUT Query}, which enables the simultaneous querying of multiple LUTs stored in a single DRAM subarray.
In pLUTo, the number of elements stored in each LUT may be as {large} as {the number of} rows in each {DRAM} subarray (e.g., 512{--}1024 rows~\cite{salp,kim2018solar,lee.sigmetrics17}).}

\begin{figure}[ht]
  \centering
  \includegraphics[width=\linewidth]{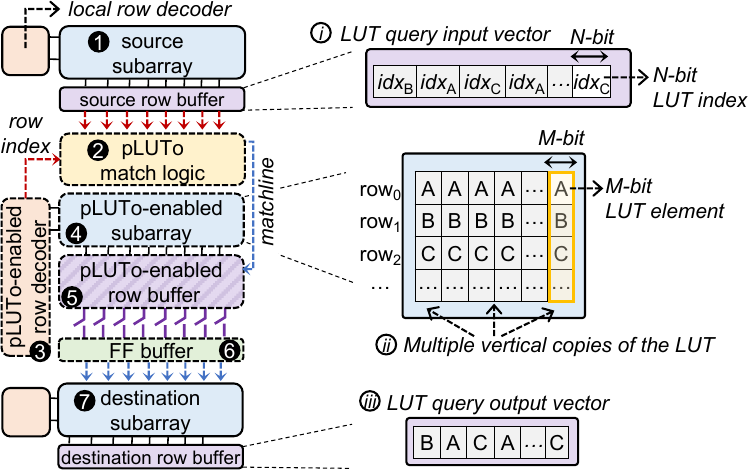}
  \caption{\gf{Main components of pLUTo. Reproduced from~\cite{ferreira2022pluto}.}}
  \label{fig:subarray_layout}
\end{figure}

\gf{{Figure~\ref{fig:subarray_layout} shows an overview of the DRAM structures required to {perform a} {\emph{pLUTo LUT Query}.%
}
First, the \emph{source subarray} (\circled{1} in Figure~\ref{fig:subarray_layout}) stores the \emph{LUT query input vector} (\incircledd{i} in Figure~\ref{fig:subarray_layout}), which consists of {a set of} $N$-bit LUT indices associated with LUT elements.
Second, the \emph{pLUTo Match Logic} (\circled{2}) comprises a set of comparators that identify matches between 
(1)~the row index of the currently activated row in the pLUTo-enabled subarray, and (2)~each LUT index in the {LUT query input vector (i.e., the source subarray's row buffer)}.
Third, the \emph{pLUTo-enabled row decoder} (\circled{3}) enables the {successive} activation {of} consecutive DRAM rows in the pLUTo-enabled subarray with a single DRAM command. It also outputs the row index of the currently activated row {as input} to the pLUTo Match Logic.
Fourth, the \emph{pLUTo-enabled subarray} (\circled{4}) stores multiple vertical copies of a given LUT (\incircledd{ii}), which consists of $M$-bit LUT elements.
Fifth, the \emph{pLUTo-enabled row buffer} (\circled{5}) allows {the reading of} individual LUT elements from the activated row in the \emph{pLUTo-enabled subarray}. This is possible by extending the DRAM sense amplifier design of the pLUTo-enabled row buffer with switches controlled by the pLUTo Match Logic (using the \emph{matchline} signal).
Sixth, the \emph{flip-\gfv{flop} (FF) buffer} (\circled{6}) enables {pLUTo to temporarily store select LUT elements by copying them from the pLUTo-enabled row buffer, conditioned on the output of the pLUTo Match Logic following each row activation.}
Seventh, {a LISA~\gfv{\cite{chang.hpca16}} operation \gfv{(described in Section~\ref{sec:rowclone})}  copies the entire contents of the FF buffer (i.e., the LUT query output vector, \incircledd{iii}) into the destination row buffer, i.e., the row buffer of the \emph{destination subarray} (\circled{7})}. {In contrast to the bit-serial paradigm employed by prior PUM architectures (e.g., SIMDRAM~\cite{hajinazarsimdram}), pLUTo operates in a \emph{bit-parallel} manner; in other words, the bits that make up {each LUT element (e.g., \texttt{A}) are stored \emph{horizontally} (i.e., in adjacent bitlines), and {all the} copies of each LUT element (i.e., \texttt{\{A,A,...,A\}}) take up \emph{one {whole} row} in the depicted pLUTo-enabled subarray {(\incircledd{ii})}.}}}
}

\gf{We implement three pLUTo designs{: 
(1)~pLUTo-BSA (\underline{B}uffered \underline{S}ense \underline{A}mplifier)  {incurs} moderate hardware overhead and provides intermediate performance {and energy efficiency gains;} 
(2)~pLUTo-GSA (\underline{G}ated \underline{S}ense \underline{A}mplifier)  {incurs} the lowest hardware overhead {but} provides the lowest performance {and energy efficiency;} and 
(3)~pLUTo-GMC (\underline{G}ated \underline{M}emory \underline{C}ell) {incurs} the highest hardware overhead {but} provides the highest performance {and energy efficiency}.}}
\omv{More details and evaluation of these designs are in the pLUTo paper~\cite{ferreira2022pluto}.}

\gf{To enable {the} {seamless} integration of pLUTo \omv{into} the system, we {methodically} describe {the changes} that {allow programmers} to offload their applications to {pLUTo.} {These changes comprise
(1)~pLUTo ISA {instructions} that enable support for each of the DRAM operations required for pLUTo's operation,
(2)~{the {pLUTo Library, an API library that includes} routines that programmers can use to conveniently express pLUTo operations at a high level of abstraction,}
(3)~the pLUTo Compiler, which analyzes {an} application's data dependency graph to {plan the in-memory placement {and alignment} of data}, and
(4)~the pLUTo Controller, a modified memory controller {that} supports the execution of pLUTo ISA {instructions}.
}}

\gf{{We evaluate pLUTo on a diverse range of real-world arithmetic, bitwise logic, cryptographic, image processing, and neural network workloads that {demonstrate} the limitations of existing PUD architectures and how pLUTo is able to overcome them. These workloads include
{(1)~}bitwise (\texttt{AND/OR/XOR}), arithmetic (addition, multiplication), and nonlinear operations (substitution tables, image binarization, and color grading); and
{(2)~}a quantized neural network.}
\gf{Our simulation infrastructure is available at~\cite{safariplutogithub} \omv{and \omvi{it is} fully artifact-evaluated and found to be functional and reproducible}.}
We compare pLUTo to state-of-the-art processor-centric architectures (CPU~\cite{intel_gold_cpu}, GPU~\cite{nvidia_3080ti}, FPGA~\cite{xilinxzcu102}) and PIM architectures (PNM~\cite{hmc.spec.1.1,hmc.spec.2.0}, PUM~\cite{seshadri.micro17,li.micro17,hajinazarsimdram}).
Our evaluations show that pLUTo outperforms optimized CPU and GPU baselines by an average {of 713$\times$ and 1.2$\times$, respectively, while simultaneously reducing energy consumption by an average of 1855$\times$ and 39.5$\times$. Across all workloads, pLUTo outperforms state-of-the-art PiM architectures by an average of 18.3$\times$.}
{We also show that different versions of pLUTo provide different levels of flexibility and performance at different {additional DRAM area overheads} (between 10.2\% and 23.1\%).}}

\omv{We believe pLUTo's three designs are important to greatly enhance the computation opportunities of \omvi{processing} using memory \omvi{(PUM)}. 
We call for more future work on different ways of performing LUT-based computation in memory with lower overheads and higher efficiency.}
\omvi{We believe there is significant opportunity for creating \msvii{novel} LUT-based computing designs and circuits in memory.}
  
\subsubsection{Gather-Scatter DRAM}
\label{sec:gs-dram}

Many applications access data structures with different access
patterns {at different points in time}. Depending on the layout
of the data structures in the physical memory, some access patterns
require non-unit strides. As current memory systems are optimized to
access {sequential} cache lines, non-unit strided accesses
exhibit low spatial locality, {leading to memory bandwidth
  waste} and cache space waste.

\begin{figure}[ht]
\centering
\includegraphics[width=1.0\linewidth]{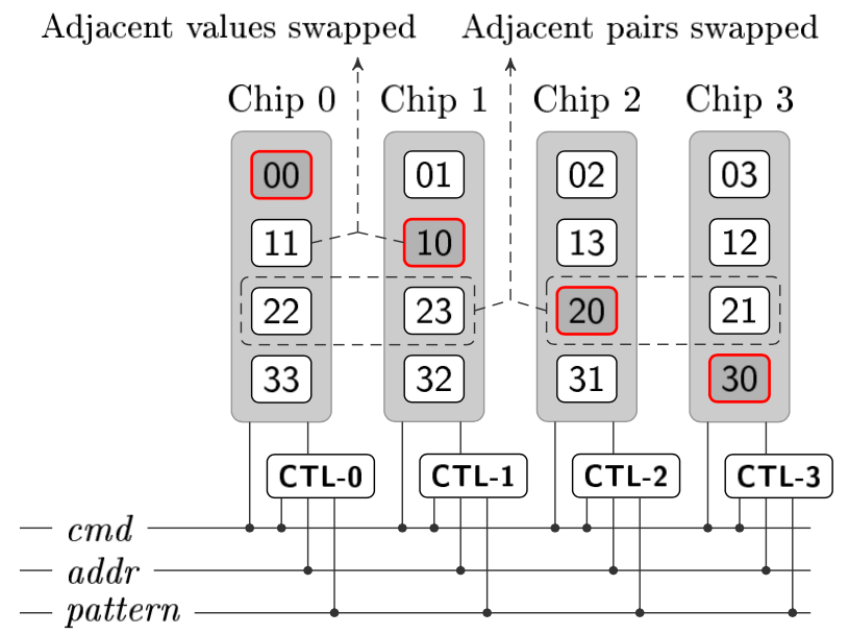}
\caption{GS-DRAM (Gather-Scatter DRAM) data mapping and chip control overview. CTL refers to Column Translation Logic hardware in the DRAM module. Reproduced from~\cite{GS-DRAM}.
}
\label{fig:gsdram}
\end{figure}

Gather-Scatter DRAM (GS-DRAM)~\cite{GS-DRAM} is a low-cost substrate
that addresses this problem. It {performs} in-DRAM data structure
reorganization by accessing multiple values that belong to a strided
access pattern using a single read/write command {in the memory
  controller}. GS-DRAM uses two key new mechanisms.  First, GS-DRAM
remaps the data of each cache line to different DRAM chips such that
multiple values of a strided access pattern are mapped to different
chips. {This enables the possibility of gathering different
  parts of the strided access pattern concurrently from different
  DRAM chips.}  Figure~\ref{fig:gsdram} show an example mapping on four DRAM chips. Adjacent values and/or adjacent pairs of values are swapped.
  Second, instead of sending separate requests to each chip,
the GS-DRAM memory controller communicates a pattern ID to the memory
module, as Figure~\ref{fig:gsdram} shows. With the pattern ID, each DRAM chip computes the address to be
accessed independently via a custom \emph{column translation logic (CTL)} hardware that is part of the DRAM module. This way, the returned cache line contains
different values of the strided pattern gathered from different DRAM chips.

GS-DRAM achieves near-ideal memory bandwidth and cache utilization in
real-world workloads, such as in-memory databases and matrix
multiplication.  {For in-memory databases, GS-DRAM outperforms a
  transactional workload with column-store layout by $3\times$
  and an analytics workload with row-store layout by $2\times$, thereby
  providing the best performance \sg{for} both transactional and analytical
  queries on databases (which in general benefit from different types
  of data layouts).  For matrix multiplication, GS-DRAM is 10\% faster
  than the best-performing tiled implementation of \sg{the matrix
  multiplication algorithm}.} We note that the idea of GS-DRAM is completely independent of memory technology, and thus GS-DRAM can be used in any type of memory module, including DRAM, SRAM, PCM, memristors, STT-MRAM, RRAM. 

\subsubsection{In-DRAM Security Primitives}
\label{sec:puf-trng}

Secure computation is of critical importance in modern computing systems. Therefore, it is important for a PIM system to support fundamental security primitives that enable secure computation and security functions. Doing so would enable PIM systems to execute a wider range of workloads and do so securely. To this end, recent work shows that \gf{PUM} can provide two basic security primitives: by carefully violating DRAM access timing parameters and taking advantage of the resulting characteristics of different DRAM cells (i.e., whether they always/never fail or fail randomly), it is possible to use DRAM to generate Physical Unclonable Functions (PUFs)~\cite{kim.hpca18, kim.hpca18talk, orosa2021codic} and true random numbers (TRNs)~\cite{kim.hpca19, olgun2021quactrng}.

\begin{figure}[ht]
\centering
\includegraphics[width=1.0\linewidth]{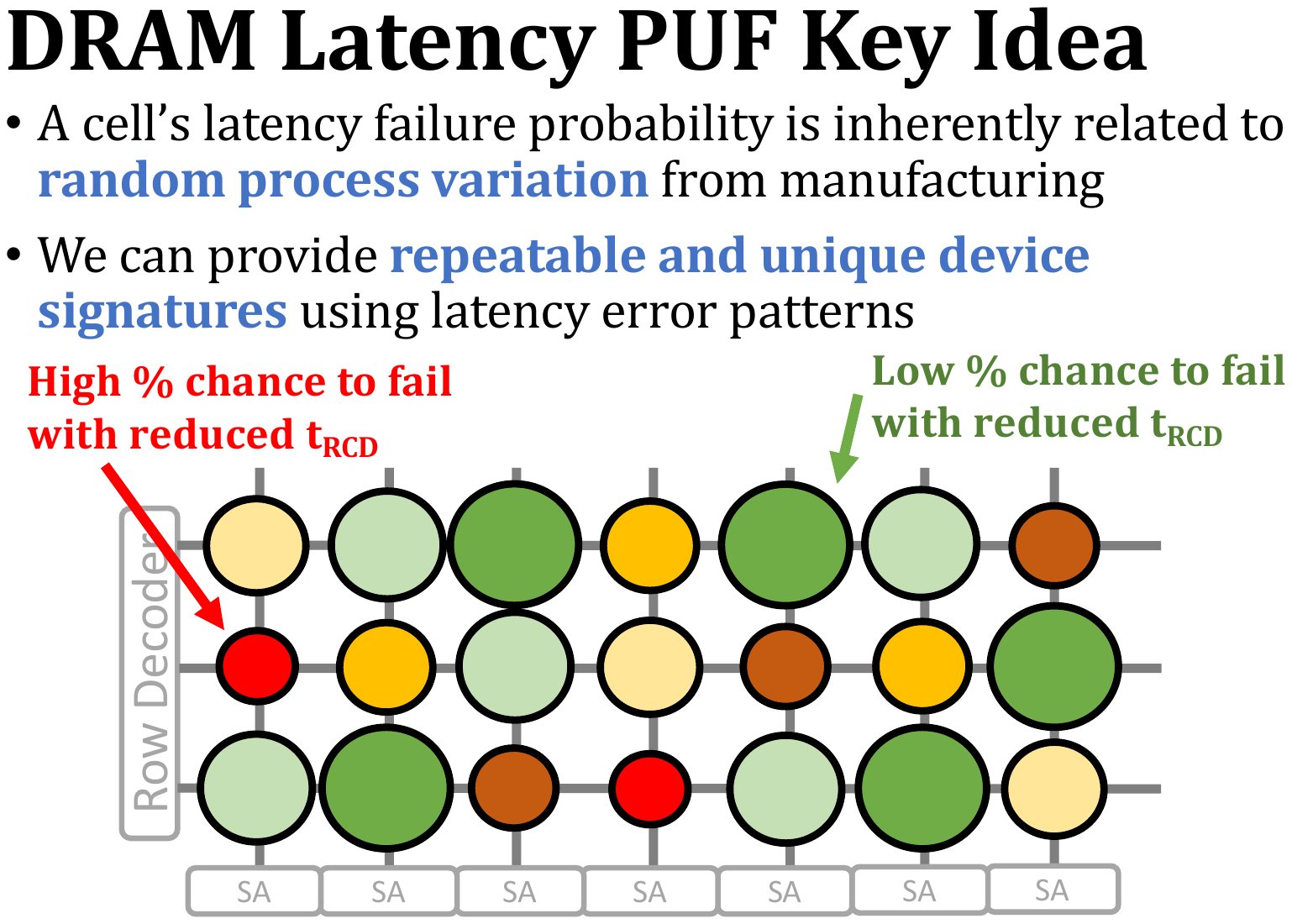}
\caption{Key idea of DRAM latency PUF. Reproduced from~\cite{kim.hpca18talk}.
}
\label{fig:dram-puf}
\end{figure}

\paratitle{\omvi{In-DRAM Physical Unclonable Functions}} PUFs are commonly used in cryptography to identify devices based on the uniqueness of their physical microstructures. 
DRAM-based PUFs have two key advantages: 
(1)~DRAM is present in many modern computing systems, and 
(2)~DRAM has high capacity and thus can provide many unique identifiers. However, traditional DRAM PUFs exhibit unacceptably high latencies and are not runtime-accessible. 
Our recent work, the DRAM Latency PUF~\cite{kim.hpca18},  proposes a new class of fast, reliable DRAM PUFs that are runtime-accessible, i.e., that can be used during online operation with low latency. 
The key idea is to reduce DRAM read access latency below the reliable datasheet specifications using software-only system calls. Doing so results in error patterns
that reflect the compound effects of manufacturing variations
in various DRAM structures (e.g., capacitors, wires, sense amplifiers). 
Some DRAM cells fail always \omv{after repeated accesses \omvi{with violated timing parameters} and some others never fail} at all.  
\omv{A} combination of a set of such \omv{consistently failing or never failing} cells can be used to generate a unique identifier for the device. 
Figure~\ref{fig:dram-puf} illustrates the key idea of using the pattern of predictable access latency failures in a DRAM subarray to generate a unique DRAM device identifier.
An experimental characterization of 223 LPDDR4 DRAM chips from all three major manufacturers shows that these error patterns (1) satisfy runtime-accesible PUF requirements, and (2) are quickly generated (i.e., at 88.2ms) irrespective of operating
temperature. The DRAM latency PUF does \omv{\emph{not}} require any modification to existing DRAM chips -- it only requires an intelligent memory controller that can change timing parameters and identify DRAM regions and cells that can be reliably used as PUFs.
\omv{Extensive characterization and analysis of the DRAM latency PUF on real \gls{COTS} DRAM chips is provided in~\cite{kim.hpca18}.}


\begin{figure}[ht]
\centering
\includegraphics[width=1.0\linewidth]{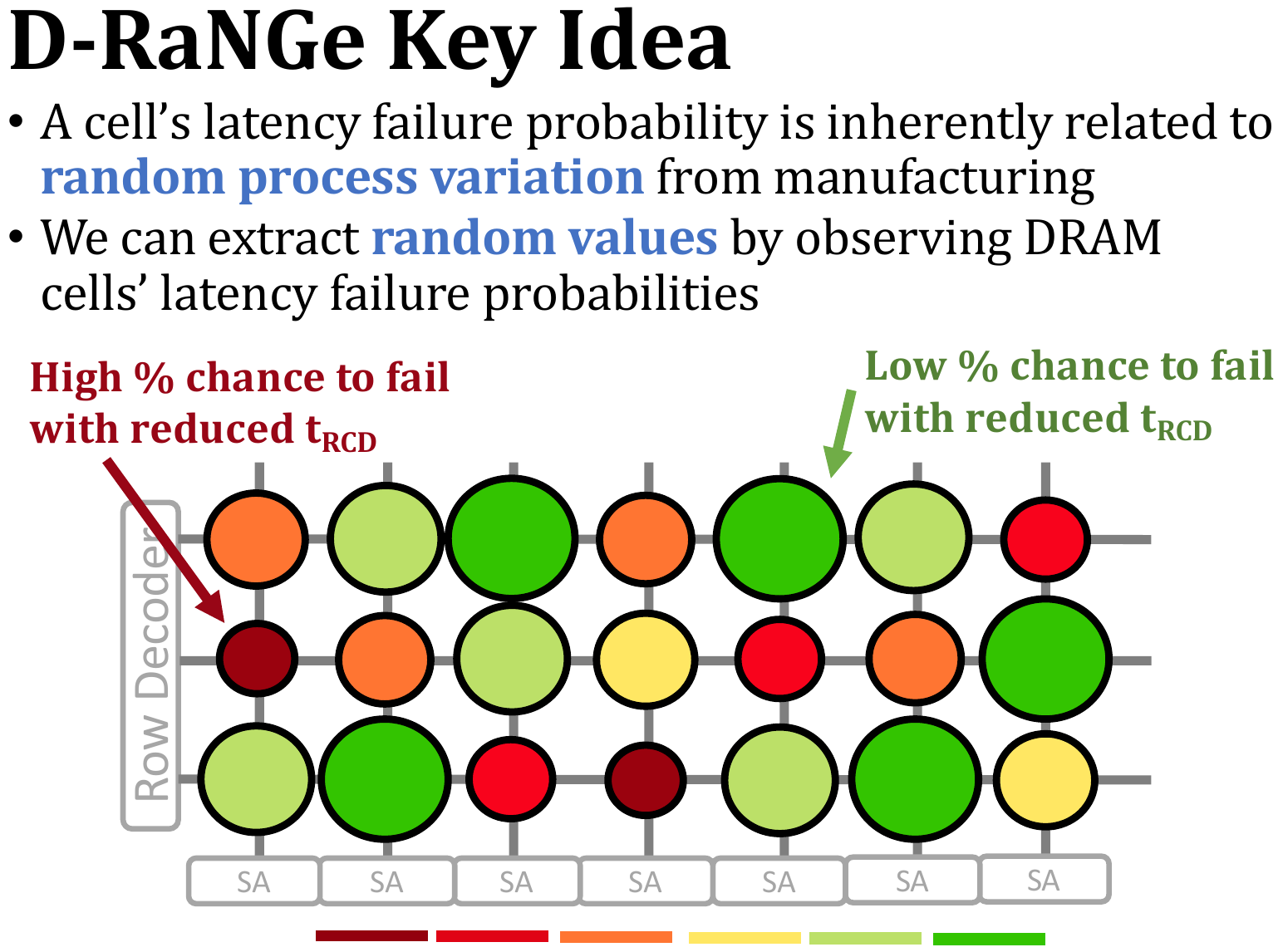}
\caption{Key idea of D-RaNGe. Reproduced from~\cite{kim.hpca19talk}.
}
\label{fig:d-range}
\end{figure}

\paratitle{\omvi{In-DRAM True Random Number Generation}} Intentionally violating DRAM access timing parameters can also be used to generate true random numbers inside DRAM. The technique we propose in~\cite{kim.hpca19} decreases the DRAM row activation latency (timing parameter tRCD) below the datasheet specifications to induce read errors, or activation failures. As a result, some DRAM cells, called TRNG (True Random Number Generator) cells, fail truly randomly. 
By aggregating the resulting data from multiple such TRNG cells, our technique, called D-RaNGe, provides a high-throughput and low-latency TRNG. Figure~\ref{fig:d-range} illustrates the key idea of D-RaNGe: finding and using the TRNG cells in a DRAM subarray to generate true random values.

We demonstrate the effectiveness of D-RaNGe in 282 LPDDR4 devices from the three major manufacturers, and observe that the produced random data remains robust over both time and temperature variation. 
D-RaNGe 
(1)~successfully passes all NIST statistical tests for randomness~\omv{\cite{rukhin2001statistical}}, and 
(2)~generates true random numbers with over two orders of magnitude higher throughput than the state-of-the-art DRAM-based TRNG. 
D-RaNGe does \omv{\emph{not}} require any modification to existing DRAM chips -- it only requires an intelligent memory controller that can change timing parameters and identify DRAM cells that can be reliably used as TRNG cells.
\gfv{Extensive characterization and analysis of \omvi{the DRAM latency TRNG} on real \gls{COTS} DRAM chips is provided in~\cite{kim.hpca19}.}


D-RaNGe~\omv{\cite{kim.hpca19}} and the DRAM Latency PUF~\omv{\cite{kim.hpca18}} show that commodity DRAM devices can be
reliably used to generate true random numbers and unique keys with
high throughput, low latency, and low power. As a result, PIM
systems can effectively generate true random numbers and unique keys  directly using DRAM itself. Doing so can improve the security and privacy
of the system: PIM applications can  directly
generate random numbers or unique keys within DRAM
and do not require off-DRAM devices to generate them and
transfer them over the CPU to DRAM bus. Thus, random numbers or unique keys are no longer transferred across buses, and security-critical computations can securely happen inside memory, which likely vastly improves the security guarantees of a PIM-enabled system. 

\begin{figure}[ht]
\centering
\includegraphics[width=1.0\linewidth]{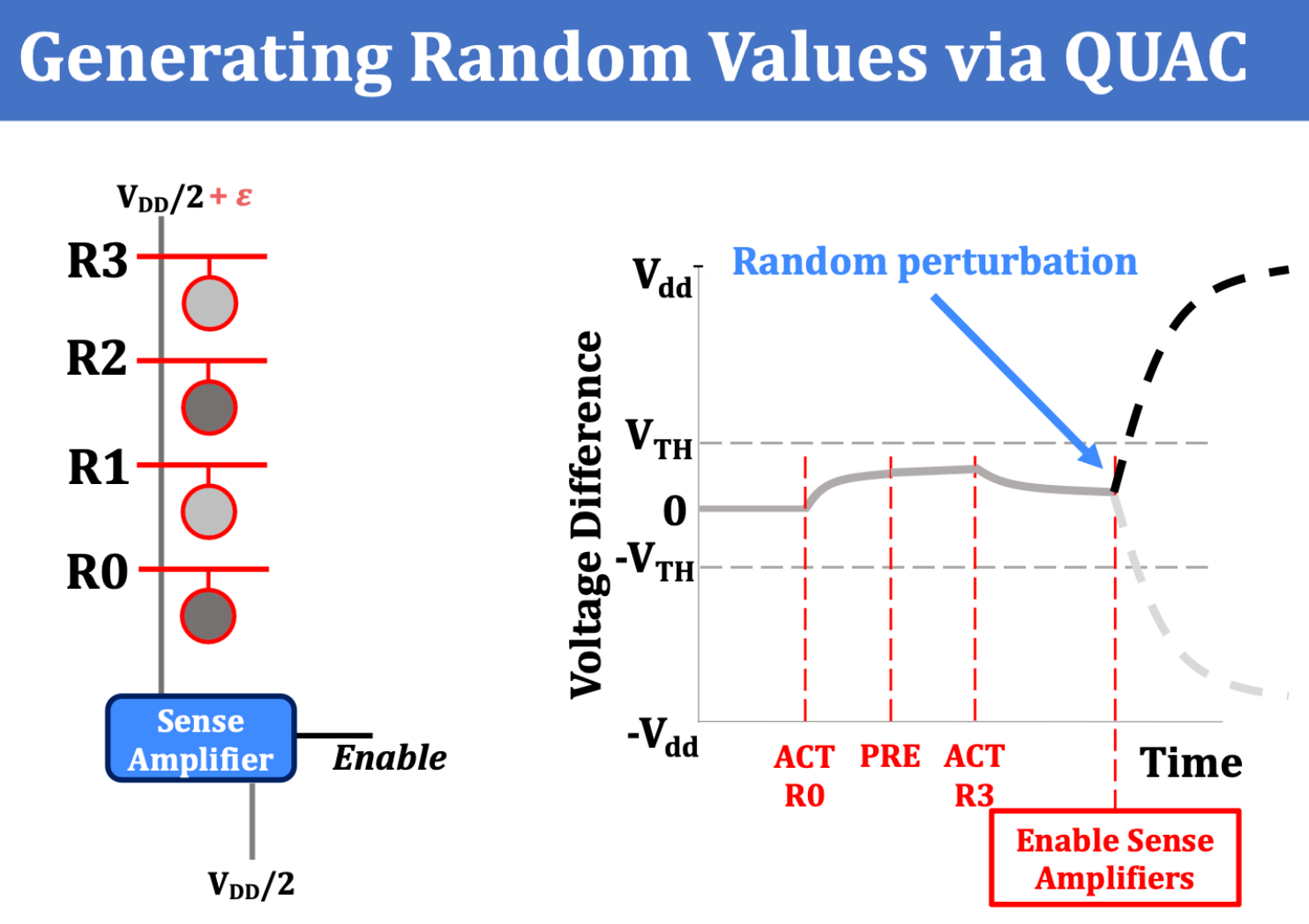}
\caption{\juanrrr{Key idea of QUAC\omix{-TRNG}. Reproduced from~\cite{olgun2021quactrng.talk}.}}
\label{fig:quac}
\end{figure}

\juanrr{The exploration of in-DRAM security primitives remains an active research direction. 
Recently, QUAC-TRNG~\cite{olgun2021quactrng} exploits the new observation that a carefully-engineered sequence of DRAM commands activates four consecutive DRAM rows in rapid succession \juanrrr{to generate random numbers}. 
This QUadruple ACtivation (QUAC) causes the bitline sense amplifiers to non-deterministically converge to random values when we activate four rows that store conflicting data because the net deviation in bitline voltage fails to meet reliable sensing margins.} 
\juanrrr{Figure~\ref{fig:quac} illustrates the QUAC operation. 
Initially, \omv{cells in} rows R0 and R2 are \emph{charged} ($V_{DD}$) and \omv{cells in} rows R1 and R3 are \emph{discharged} ($0$). 
As the timeline on the right of the figure shows, an ACT command to R0 is quickly interrupted by issuing a PRE command. 
Meanwhile, the cell on R0 shares its charge with the bitline, thus increasing slightly the voltage level of the bitline. 
Before the PRE command closes the row and precharges the bitline, we issue another ACT command to R3. 
This ACT command interrupts the PRE command and enables wordlines R1, R2, and R3 simultaneously, in addition to the already enabled R0. 
This simultaneous activation of rows is explained by the hierarchical design of wordlines in state-of-the-art DRAM chips and a hypothetical row decoder design (see~\cite{olgun2021quactrng} for more details \omv{and~\cite{yuksel2024simultaneous} for further information}). 
All four cells on rows R0, R1, R2, and R3 contribute to the bitline voltage. As a result, the bitline ends up with a voltage level below reliable sensing margins ($\pm V_{TH}$). When we enable the sense amplifier, the voltage level is sampled as a random value.}

\juanrrr{QUAC-TRNG reads the result of the QUAC operation from the sense amplifiers and performs the SHA-256 cryptographic hash function~\cite{1802.shs} to post-process the result and output random numbers. 
Our experimental evaluation using 136 real DDR4 DRAM chips shows that QUAC-TRNG generates an average of 7664 bits of random data per iteration and each iteration takes 1940 ns. 
Compared to prior work (e.g., \cite{kim.hpca19}), QUAC-TRNG enables (i) lower latency because it uses simultaneous activation of rows, which is very fast, and (ii) higher throughput because the QUAC operation induces metastability in many sense amplifiers in parallel.
\gf{QUAC-TRNG's source code is available at \url{https://github.com/CMU-SAFARI/QUAC-TRNG}.}}

\juanrr{Another recent work proposes DR-STRaNGe~\cite{bostanci2022dr}, an end-to-end system design for DRAM-based TRNGs that mitigates three key system integration challenges: 
(1) generating random numbers with DRAM-based TRNGs can degrade overall system performance by slowing down concurrently-running applications due to the interference between RNG and regular memory operations in the memory controller (i.e., RNG interference), (2) this RNG interference can degrade system fairness by causing unfair prioritization of applications that intensively use random numbers (i.e., RNG applications), and (3) RNG applications can experience significant slowdown due to the high latency of DRAM-based TRNGs. 
DR-STRaNGe proposes an \emph{RNG-aware scheduler} and a \emph{buffering mechanism} \juanrri{in the memory controller} to tackle these challenges.}

\begin{figure}[ht]
    \centering
    \includegraphics[width=1.0\linewidth]{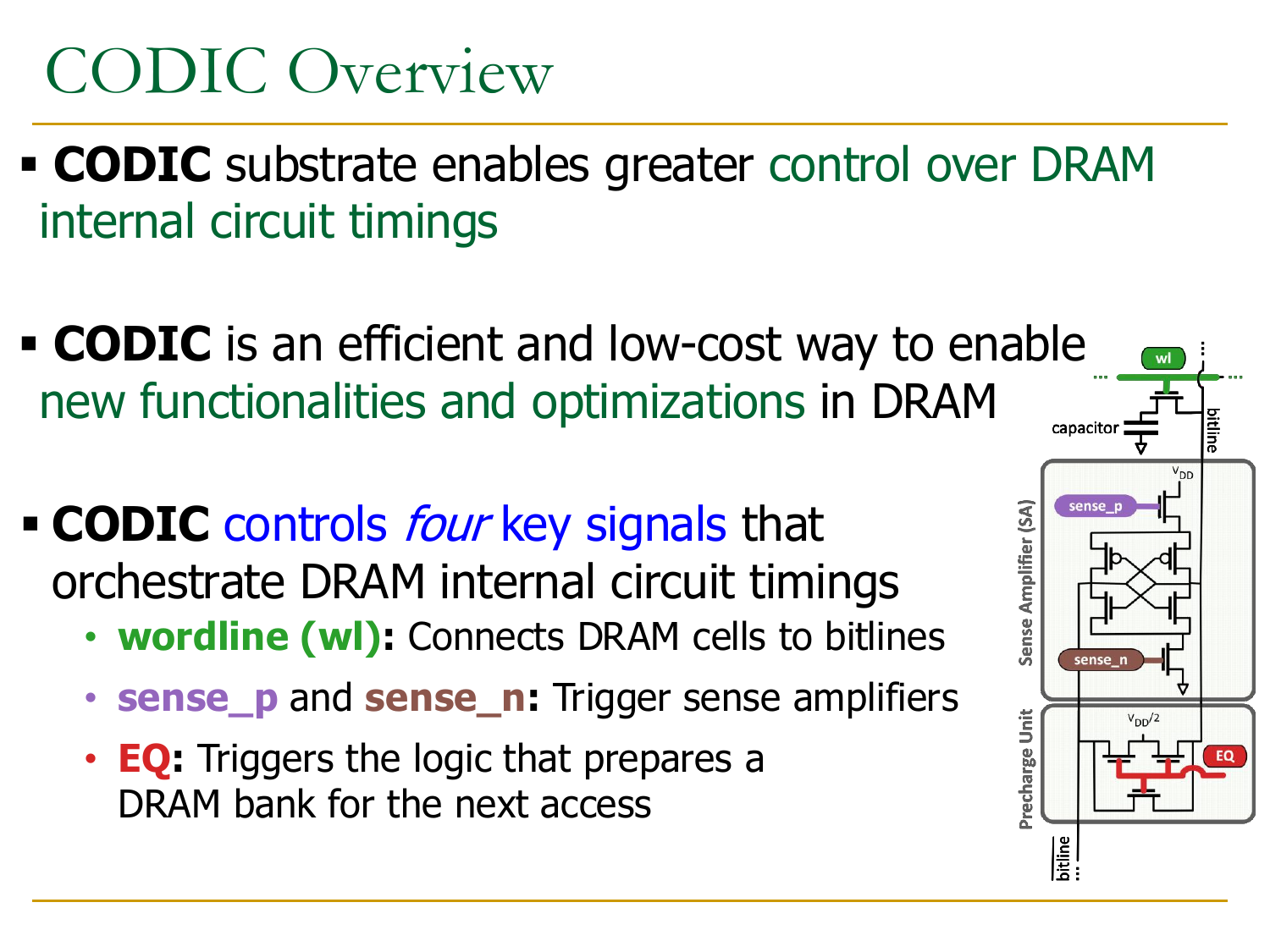}
    \vspace{-20pt}
    \caption{\gfv{Overview of CODIC. Reproduced from~\cite{orosa.isca21talk}.}}
    \label{fig:codic}
\end{figure}

\paratitle{\omvi{Changing the DRAM Interface to Enable Better In-DRAM Security Primitives and Other Functionalities}} \gfv{\omvi{In}~\cite{orosa2021codic}, we show how enabling programmable DRAM internal timings for controlling in-DRAM components can be used for improving system security at low-cost. 
\omvi{We} propose CODIC (fine-grained \underline{CO}ntrol over \underline{D}RAM \underline{I}nternal \underline{C}ircuit timings), a new low-cost DRAM substrate that enables fine-grained control over four previously
fixed internal DRAM signals (as Figure~\ref{fig:codic} illustrates) that are key to many DRAM operations, i.e., 
(1)~\emph{wl}, which connects DRAM cells to bitlines;
(2)~\emph{EQ}, which triggers the logic that prepares a DRAM bank for the next access; 
(3)~\emph{sense\_p} and 
(4)~\emph{sense\_n}, which trigger sense amplifiers.
By controlling the four internal DRAM signals, 
CODIC can control at least two existing commands (i.e., activate and precharge commands), and  two new CODIC variants, i.e., 
(1)~\texttt{CODIC-sig}, for generating digital signatures that depend on process variation, and 
(2)~\texttt{CODIC-det}, for generating deterministic values in memory.
}

\gfv{We use CODIC to develop two new applications for improving system security. 
First, we propose a new CODIC-based PUF, which leverages the \texttt{CODIC-sig} command to generate signature values that are used for PUF response.
We evaluate our CODIC-based DRAM PUF using 136 real commodity DRAM chips to validate the functionality of our mechanism.
Our evaluation shows that our CODIC-based PUF provides
1.8$\times$ higher throughput than the best state-of-the-art DRAM PUF~\cite{bahartalukder2019prelatpuf}, has similar resilience to temperature changes, and provides more repeatable PUF responses than the state-of-the-art DRAM PUF. 
Second, we propose a new CODIC-based mechanism to
prevent cold boot attacks~\cite{bauer2016lest,gruhn2013onthe,halderman2008lest,hilgers2014postmortem,lee2011correcting,lindenlauf2015cold,mcgregor2008braving,muller2010aesse,simmons2011security,villanuevapolanco2019cold,yitbarek2017cold}. 
In a cold boot attack, the attacker physically removes the DRAM module from the victim system and places it in a system under their control to extract secret information.
Because data in DRAM is stored in capacitors, the data can potentially remain in the cells long enough (during physical extraction) for data to be stolen. The \emph{key idea} of our CODIC-based mechanism is to automatically overwrite
the entire DRAM with values generated by CODIC when the DRAM chip is first powered-on using a pre-defined sequence of \texttt{CODIC-sig} and \texttt{CODIC-det} commands.
Our mechanism does \emph{not} incur any latency or power overhead at runtime, and it is 2.0$\times$ lower latency and 1.7$\times$ lower energy than the best state-of-the-art mechanisms (i.e., LISA~\cite{chang.hpca16} and RowClone~\cite{seshadri2013rowclone}) during DRAM power-on.
We conclude that CODIC can be used for implementing very efficient security applications at low cost. 
}

We hope and believe that CODIC will inspire and enable 
(1)~other new \omvi{in-}DRAM functionalities that enhance \omvi{system security, performance, and efficiency}, and 
(2)~new DRAM reliability, performance, and energy optimizations \omvi{that can be \omvii{effectively} leveraged by intelligent memory controllers}.

\subsection{\gf{Processing-Using-NVM}}
\label{sec:pum:punvm}

\gf{Processing-using-NVM (non-volatile memory) architectures leverage the analog \omv{operational} principles of different non-volatile memory technologies \omv{and devices}, such as emerging non-volatile \omv{memories} or {NAND} flash \omv{memories}, to implement different operations. 
In this section, we highlight the advances in processing-using-NVMs. 
First, we describe the implementation of bulk bitwise Boolean operations, current-based matrix-vector multiplication (MVM), and string matching operations using emerging NVMs (e.g., PCM, ReRAM, and memristors).
Second, we describe the implementation of bulk bitwise Boolean operations using NAND flash \omv{memory}.
}

\subsubsection{\gf{Using Emerging NVMs for Computation}}

\paratitle{\omv{Bitwise Operations in \gfv{NVMs}}} \gf{Several works~\cite{borghetti2010memristive, linn2012beyond, li.dac16,kvatinsky.tcasii14,kvatinsky.iccd11,kvatinsky.tvlsi14,lehtonen2009stateful,kim2011field,lehtonen2012applications,mahmoudi2013implication,kim2019single,xie2017scouting,gaillardon2016plim} investigate the implementation of \emph{stateful}~\cite{borghetti2010memristive,linn2012beyond,kvatinsky.tcasii14,kvatinsky.iccd11,kvatinsky.tvlsi14,lehtonen2009stateful,kim2011field,lehtonen2012applications,mahmoudi2013implication,kim2019single} and \emph{non-stateful}~\cite{li.dac16,xie2017scouting,gaillardon2016plim} bitwise \omvi{Boolean} operations in emerging NVM technologies. 
When implementing stateful Boolean bitwise operations, all input operands and the produced result are stored in terms of the resistance state variable. 
Hence, many Boolean bitwise operations can be executed back-to-back, at the expense of consuming the limited \omv{write} cycles \omv{(i.e., endurance)} of the emerging NVM device. 
In contrast, when implementing non-stateful Boolean bitwise operations, the input operands are stored as resistance state variables while the output value is represented as a voltage level. Such implementation \omv{avoids} \omv{write} cycles, since the produced output value is not directly written back into the NVM cells, but requires additional sensing circuitry when cascading \omv{consecutive} operations.
As \omv{examples}, we highlight the implementation of stateful (\omv{e.g.}, MAGIC~\cite{kvatinsky.tcasii14}) and non-stateful (\omv{e.g.}, Pinatubo~\cite{li.dac16}) NVM-based PUM architectures.} 

\gf{In memresistive devices, memory cells use resistance levels to represent logic-`1' (\omv{e.g.}, using a low resistance level $R_{low}$) and logic-`0' (\omv{e.g.}, using a high resistance level $R_{high}$). 
As in DRAM, NVM arrays organize memory cells in rows and columns, and a sense amplifier is used to convert resistance difference \gfv{of memory cells} into voltage levels or current signals based on a reference resistance level $R_{ref}$, determining the result between `\texttt{0}' and `\texttt{1}'. 
MAGIC~\cite{kvatinsky.tcasii14} implements a Boolean \texttt{NOR} operations as follows.
Initially, a \texttt{NOR} logic gate is constructed using two input memristors ($in_{1}$ and $in_{2}$) connected in parallel and an output memristor ($out$) connected in series. 
First, the output memristor $out$ is initialized with a logic-`1' value (i.e., $R_{out}$ = $R_{low}$).
Second, a voltage pulse $V_{0}$ is applied to the gateway of the \texttt{NOR} logic gate. 
Third, $V_{0}$ produces a current that passes through the
circuit. The logical state of the output memristor $out$ switches from the initial logic-`1' value to a logic-`0' value when \emph{either} $in_{1}$ or $in_{2}$ has a low resistance value (i.e., $in_{1}$\omv{=$R_{low}$} or \omv{$in_{2}$} = $R_{low}$), since the voltage/current will be greater than the memristor threshold voltage/current. 
MAGIC implements other Boolean operations (i.e., \texttt{OR}, \texttt{NAND}, and \texttt{AND}) by \gfv{changing} the connections between $in_{1}$, $in_{2}$, and $out$: 
\gfv{
(1)~a Boolean \texttt{OR} operation is implemented by connecting $in_{1}$, $in_{2}$, and $out$ as in a Boolean \texttt{NOR}, but initializing $out$ with a logic-`0' value, i.e., $R_{out}$ = $R_{high}$;
(2)~a Boolean \texttt{NAND} (\texttt{AND}) operation is implemented by connecting $in_{1}$, $in_{2}$, and $out$ in series, and initializing $out$ with a logic-`1' (logic-`0') value, i.e., $R_{out}$ = $R_{low}$ ($R_{out}$ = $R_{high}$).
A Boolean \texttt{NOT} operation is implemented by constructing an inverted logic gate using an input memristor ($in$) and an output memristor ($out$), which is initialized with a logic-`1' value, connected in series.
}
}

\gf{Pinatubo~\cite{li.dac16} implements \omv{bulk} bitwise Boolean operations by 
\gfv{
(1)~}performing \emph{multiple row activation}, similarly to Ambit~\cite{seshadri.micro17}, \gfv{and
(2)~modifying the array's sense amplifier by adding new reference resistance levels that are used during PUM execution to \omvi{enable} different Boolean operations.} 
For example, \omv{to compute bulk bitwise} \texttt{A} \texttt{OR} \texttt{B} \omv{of rows \texttt{A} and \texttt{B}}, Pinatubo simultaneously \omv{activates} the rows containing both input operands. 
\gfv{The newly-added} reference resistance level \gfv{for a Boolean \texttt{OR} operation (i.e., $R_{ref-or}$) then} allows the sense amplifier to output `\texttt{0}' \emph{only} when $R_{A} = R_{B} = R_{ref-or}$. 
Pinatubo follows \omv{a similar} approach to implement other Boolean operations, such as \texttt{OR}, \texttt{AND}, \texttt{XOR}, and \texttt{INV} operations.}

\gfv{As in DRAM-based PUM architectures (e.g., SIMDRAM~\cite{hajinazarsimdram} and MIMDRAM~\cite{mimdram}), NVM-based PUM architectures~\cite{kvatinsky.tcasii14,fernandez2024matsa,lin2022all,parveen2017lowpower,angizi2020exploring,angizi2019dna,angizi2019graphs,leitersdorf2023aritpim} can implement complex arithmetic operations by decomposing such operations into a series of
simple bitwise Boolean operations (e.g., \texttt{AND}, \texttt{NOR}, \texttt{XOR}) that are executed bit-serially over vertically-layout-out input operands.
We use such an approach to accelerate another important data-intensive application, i.e., time series analysis (TSA)~\cite{esling2012time} in our \omvi{recent} work \omvi{MATSA}~\cite{fernandez2024matsa}.}

\gfv{\omvi{MATSA}~\cite{fernandez2024matsa} \omvi{leverages} the computing capabilities of magnetoresistive
RAM (MRAM)~\cite{lin2022all,parveen2017lowpower,angizi2020exploring,angizi2019dna,angizi2019graphs,garello2014ultrafast,garzon20234} to enable high-performance and energy-efficient subsequence dynamic time warping (sDTW)~\cite{berndt1994using} execution (a key step during TSA~\cite{alaee2021time}).
MATSA \omvi{is} the first \underline{M}RAM-based \underline{A}ccelerator for \underline{TSA}.
MATSA derives its performance benefits from three key mechanisms.
First, \omvi{it} decomposes sDTW's computational kernel into simple bitwise Boolean operations and executes them in the MRAM array, significantly minimizing data movement overheads. 
MATSA implements key complex arithmetic operations employed during sDTW computation using MRAM-based PUM, including vertical and diagonal row copy, bit-serial addition/subtraction across columns, absolution calculation, and minimum-of-three value calculation. 
Second, we implement a novel data mapping that reduces the runtime memory footprint of sDTW from quadratic to linear based on four vectors, enabling computing the complete 2D dynamic programming matrix on-the-fly without storing it. 
Third, MATSA integrates an effective computation scheme that overcomes the inter-cell computation dependencies of the matrix by 
(1)~following an anti-diagonal approach and 
(2)~exploiting pipelining to increase parallelism.
We evaluate MATSA's performance based on state-of-the-art latency and energy characteristics of MRAM devices~\cite{yu2016emerging,gallagher2019recent}.
Our evaluation shows that MATSA improves performance by 7.35$\times$/6.15$\times$/6.31$\times$ and energy efficiency by 11.29$\times$/4.21$\times$/2.65$\times$ over server-class CPU, GPU, and processing-near-memory platforms, respectively.
}

\begin{figure}[ht]
    \centering
    \includegraphics[width=\linewidth]{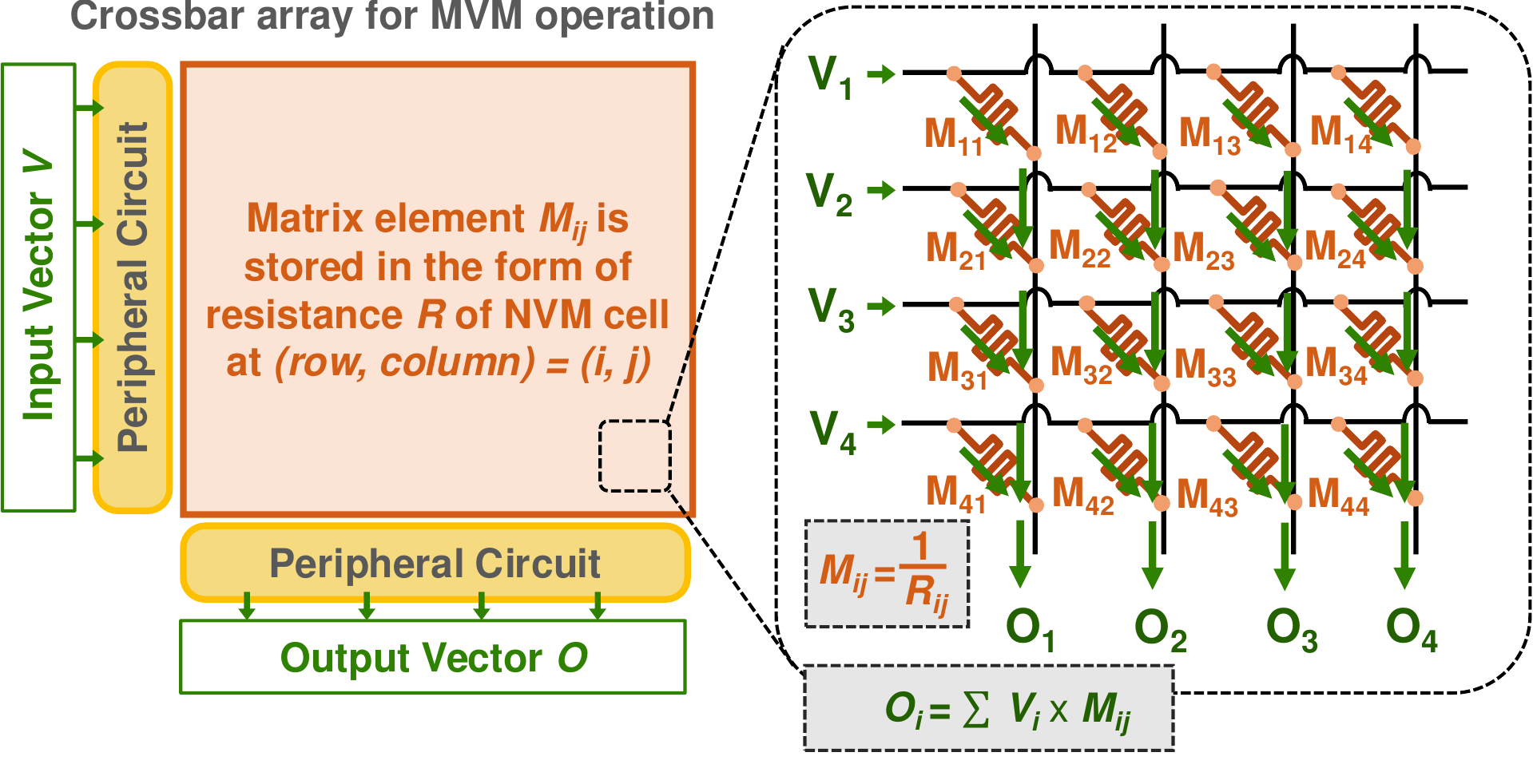}
    \caption{\gf{The basic structure of a \gfv{crossbar array based NVM architecture} {designed for computing an} MVM \omvi{(matrix-vector multiplication)} operation. Reproduced from~\cite{mao2022genpip}.}}
    \label{fig:mvmarray}
\end{figure}

\paratitle{\gf{In-Memory Crossbar Array \omv{Matrix-Vector Multiplication (MVM) Operations}}} 
Crossbar array based NVM architectures can natively execute MVM operations based on \omv{analog operational} principles~\gfv{\cite{song2018graphr,imani2020dual,challapalle2020gaas,bojnordi2016memristive,feinberg2018enabling,shafiee2016isaac,chi2016prime,ielmini2018memory,wan2022compute,jung2022crossbar,sebastian2020memory,le202364,ankit2019puma,joshi2020accurate,song2017pipelayer,li2019long,valavi201964,imani2019floatpim,le2018mixed,yang2019sparse,jia2020programmable,tang2017binary,marinella2018multiscale,yuan2021forms,wen2020ckfo,ankit2020panther,wen2019memristor,long2018reram,chou2019cascade,feinberg2018making,li2020timely,angizi2019mrima,wang2018snrram,zhu2019configurable,yang2020retransformer,tang2017aepe,xia2016switched,xia2017fault,huang2017highly,cheng2017time,bojnordi2016memristive,chen2018regan,cai2018training,mao2018lergan,nag2018newton}}\omv{, specifically Kirchhoff's Law~\cite{kirchhoff1859uber,kirchhoff1978verhaltnis,kirchhoff1860relation}}. 
Figure~\ref{fig:mvmarray} shows the basic structure of crossbar array based NVM architectures designed for the MVM operation~\cite{chi2016prime} \gfv{between the input vector $V$ and the input matrix $M$, producing an output vector $O$ (i.e., $O = V \times M$).
Before execution, the input matrix $M$ is stored in the crossbar array as resistance levels, where each matrix element $M_{i,j}$ is represented as \omvi{a} resistance level corresponding \omvi{to the value of the element} i.e., $M_{i,j} = \frac{1}{R_{i,j}}$.
Then, the crossbar array} performs \omvi{a} \emph{in-situ} MVM operation by applying 
(1)~a voltage level $V_i$, which represents the input vector, on the wordlines of the array that stores the matrix $M$ and 
(2)~sensing the output vector $O$ on the bitlines.
Based on Kirchhoff's Law~\omvi{\cite{kirchhoff1859uber,kirchhoff1978verhaltnis,kirchhoff1860relation}}, the current level sensed on the bitlines will be equal to \gfvi{$O_i = \sum\limits_{j} V_i \times M_{i,j} = \sum\limits_{j} V_i \times \frac{1}{R_{i,j}}$}. 
Using a crossbar array based NVM, an MVM operation can be performed in nearly a single NVM read cycle (as long as the matrix fits in the crossbar array).

\gf{A large number of works \omv{leverage crossbar array based} NVM's ability to natively perform MVM operations to accelerate different applications~\gfvii{\cite{alibart2012high,angizi2019mrima,ankit2019puma,ankit2020panther,bojnordi2016memristive,cai2018training,challapalle2020gaas,chen2018regan,cheng2017time,chi2016prime,chou2019cascade,feinberg2018enabling,feinberg2018making,holmes1993use,hu2012hardware,huang2017highly,ielmini2018memory,imani2019floatpim,imani2020dual,jia2020programmable,joshi2020accurate,jung2022crossbar,le2018mixed,le202364,li2019long,li2020timely,long2018reram,mao2018lergan,marinella2018multiscale,nag2018newton,sebastian2020memory,shafiee2016isaac,song2017pipelayer,song2018graphr,tang2017aepe,tang2017binary,valavi201964,wan2022compute,wang2014energy,wang2018snrram,wen2019memristor,wen2020ckfo,xia2017fault,yang2019sparse,yang2020retransformer,yuan2021forms,zhu2019configurable,li2013memristor,prezioso2015training,kim2015reconfigurable,chen2015optimized,burr2015large,xia2016switched}}, in particular, neural network (NN) inference and training~\gfvi{\cite{shafiee2016isaac,chi2016prime,ielmini2018memory,wan2022compute,jung2022crossbar,sebastian2020memory,le202364,ankit2019puma,joshi2020accurate,song2017pipelayer,li2019long,valavi201964,imani2019floatpim,le2018mixed,yang2019sparse,jia2020programmable,tang2017binary,marinella2018multiscale,yuan2021forms,wen2020ckfo,ankit2020panther,wen2019memristor,long2018reram,chou2019cascade,feinberg2018making,li2020timely,angizi2019mrima,wang2018snrram,zhu2019configurable,yang2020retransformer,tang2017aepe,xia2016switched,xia2017fault,huang2017highly,cheng2017time,bojnordi2016memristive,chen2018regan,cai2018training,mao2018lergan,nag2018newton,prezioso2015training,burr2015large}}. 
Even though such proposals are \omvi{intellectually} promising, NVM-based acceleration of MVM operations needs to take into account inherent device non-idealities (e.g., non-ideal \omvi{analog-to-digital and} digital-to-analog converters, write resistance due to imperfect writes, non-ideal sensing \omvi{circuits})\omvi{,} architectural limitations (e.g., limited write endurance)\omvi{, and cost \& latency considerations (e.g., due to analog-to-digital and digital-to-analog conversion and sensing circuitry)} that might impact the accuracy \omv{and robustness} of the target application~\cite{shahroodi2023swordfish,feinberg2018making}\omvi{ as well as the cost and scalability of the system}.   }

\begin{figure}[ht]
    \centering
    \includegraphics[width=\linewidth]{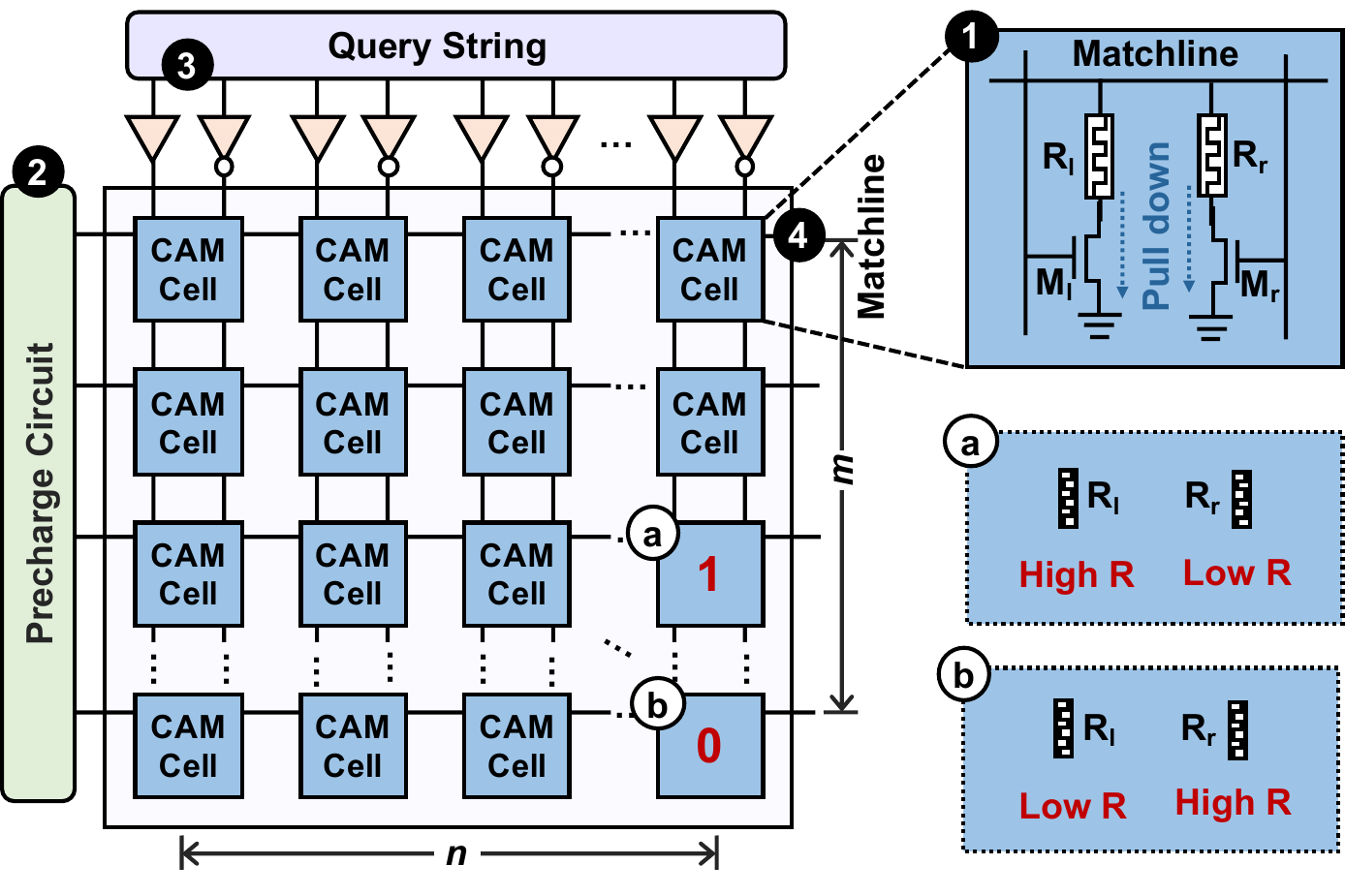}
    \caption{\gf{An example NVM-based CAM array for string matching. Reproduced from~\cite{mao2022genpip}.}} \label{fig:cam}
\end{figure}

\paratitle{\gf{In-Memory Crossbar Array String Matching Operations}} \gf{\gfv{Crossbar array based NVM} architectures \omv{can also} perform in-situ string matching operations. 
\omv{To enable these operations}, the \gfv{crossbar array} \omv{is operated} as a content addressable memory (CAM). Figure~\ref{fig:cam} {shows an example NVM-based CAM used for string matching. The CAM array consists of $m \times n$ CAM cells that house \emph{m} reference strings, each of which is \emph{n-bit} long. Each CAM cell stores \emph{one} bit and has two programmable resistors, $R_{l}$ and $R_{r}$, and two transistors, $M_{l}$ and $M_{r}$ (\circled{1} in Figure~\ref{fig:cam}).
To store `\texttt{1}' (or `\texttt{0}') in a CAM cell, $R_{l}$ and $R_{r}$ are programmed to high and low (or low and high) resistance \omv{levels}, respectively (\incircledd{a}--\incircledd{b} in Figure~\ref{fig:cam}).}} 

\gf{The NVM-based CAM array is able to query the existence of an \emph{n-bit} string in parallel across all \emph{m} rows in four steps. First, the CAM array precharges the matchline signals to \emph{high} voltage (\circled{2}). 
Second, each bit in the input string and its {complement} drive the gate voltages of $M_{l}$ and $M_{r}$ transistors of the CAM cells in the corresponding column, respectively (\circled{3}). 
Third, each CAM cell compares its stored bit to the corresponding bit in the input string. If these two bits are different, the pull down network in the CAM cell is turned on and the matchline becomes `\texttt{0}'. Otherwise, the matchline keeps its precharged \emph{high} voltage. 
Fourth, if all bits of the input string match with all corresponding CAM cells in a row, the matchline remains high, indicating an {\emph{exact match}} between the input string and the reference string stored in the CAM array (\circled{4}).}

\gf{In our work \omv{called} GenPIP~\cite{mao2022genpip}, we leverage NVM's ability to perform in-situ \omv{MVM} and string matching operations to accelerate genome analysis. 
In the genome analysis pipeline, basecalling and read mapping are two of the most time-consuming steps because they rely on computationally-intensive algorithms, i.e.,  basecalling commonly uses deep neural network\omv{s} (DNN\omv{s})~\omv{\cite{singh2024rubicon,cavlak2024targetcall,cali2018nano}} while read mapping depends on dynamic programming (DP)-based algorithms~\omv{\cite{alser2021technology,alser2020accelerating,alser2022molecules}}. 
In GenPIP, we tightly integrate these two key steps of genome analysis inside a single \omv{NVM-based} in-memory processing architecture, which 
(1)~minimizes data movement by eliminating the need to store intermediate results and 
\msvii{(2)}~minimizes computation in the genome analysis pipeline that leads to unused outputs.
GenPIP incorporates a state-of-the-art NVM-based PIM array~\cite{lou2020helix} to implement basecalling (by performing \omvi{deep neural network} inference using the NVM array in-situ MVM operations) and chaining~\cite{chen2020parc} (by performing the \omvi{dynamic programming} algorithm using the NVM array as a CAM for string matching operations).}
\omvi{For more detail on the GenPIP design and its architecture, we refer the reader to our MICRO 2022 paper~\cite{mao2022genpip}, which provides detailed information and results.}

\gfv{In our MICRO 2023 \omvi{work} called Swordfish~\cite{shahroodi2023swordfish}, we develop a hardware/software framework to accelerate DNN-based \omvi{genomic} basecallers while taking into account the possible adverse effects of inherent NVM device non-idealities, which can greatly degrade basecalling accuracy.
Such device non-idealities include:
(1)~non-ideal digital-to-analog converter (DAC)~\cite{jain2020rxnn};
(2)~variation of synaptic conductance, which includes both imperfect
programming \omvi{operations} and process variation that exists in memristors~\cite{alibart2012high,chen2014rram,lee2019exploring,zhang2011stt};
(3)~wire resistance \omvi{variation} and sneak paths, due to imperfect wires (i.e., wires
with different resistances) and \omvi{fluctuations} in the voltages of the
internal nodes while performing an MVM operation~\cite{jeong2017parasitic,zhang2011stt}; and 
(4)~non-ideal sensing circuits or analog-to-digital converters (ADCs)~\cite{jain2020rxnn,zahedi2022system}.
The Swordfish framework is composed of four main modules.
First, the \emph{Partition \& Map} module partitions and maps the vector-matrix-multiplication (VMM) operations of the target DNN-based basecaller to the underlying crossbar array architecture.
Second, the \emph{VMM Model Generator} module generates an end-to-end
model for possible non-idealities and errors of a VMM operation
considering the underlying technology.
Third, the \emph{Accuracy Enhancer} module implements online (i.e., random sparse adaptation online retraining~\cite{charan2020accurate}) and offline (i.e., analytical variation-aware offline training~\cite{chen2017accelerator,klachko2019improving,liu2015vortex,long2019design}, knowledge distillation-based variation-aware training~\cite{hinton2015distilling,charan2020accurate}, and read-verify-write training~\cite{liu2014reduction}) mitigation techniques to counter accuracy loss.
Fourth, the \emph{System Evaluator} module analyzes the accuracy and throughput of \omvi{the genomic} basecaller while also providing area overhead \omvi{estimates}.
We leverage Swordfish to comprehensively investigate the potential for \omvi{\emph{accurately}} accelerating a state-of-the-art \omvi{genomic} basecaller (Bonito~\cite{bonito}) on a crossbar array-based NVM architecture
(PUMA~\cite{ankit2019puma}) while accounting for the non-idealities of the underlying
devices and technologies. 
Our evaluations using Swordfish show that, on a wide range of real
\omvi{genomic} datasets, PUMA realistically (i.e., when we consider essential mitigation techniques to prevent huge accuracy loss) accelerates Bonito by an
average of 25.7$\times$, while suffering from a 6.01\% accuracy loss, compared to \omvi{the execution of Bonito on} an NVIDIA V100 GPU~\cite{nvidia2017nvidia}. 
We conclude that the Swordfish framework effectively facilitates the development and adoption of memristor-based PUM designs for \omvi{genomic} basecalling \omvi{and DNN acceleration}, which we hope will be leveraged by future work. We also believe that our framework is applicable to other DNN-based applications and hope future work takes advantage of this.}

\gfv{\omvi{We conclude that} PUM architectures with emerging non-volatile memories offer a promising approach for accelerating different classes of data-intensive workloads by minimizing data movement overhead \omvi{and enabling \emph{in-situ} matrix-vector multiplication and bitwise operations}. 
This approach demonstrates significant potential for performance and energy efficiency improvements, yet further work is needed to address the NVM devices' non-idealities, improve overall robustness \omvi{\& cost efficiency}, and investigate applicability to \omvi{a wider variety of} workloads. We believe that future research can build on these advancements \omvi{we described} by investigating scalable \omvi{\& robust} designs and novel programming models for broader applicability.}

\subsubsection{\gf{Flash-Cosmos: In-Flash Bulk Bitwise Operations}}


\gf{Processing data \emph{inside} NAND flash chips, i.e., \emph{in-flash processing (IFP)}, can fundamentally reduce the data movement that bottlenecks the execution of bulk bitwise operations. When processing large amounts of data that do not fit in main memory, IFP significantly reduces data movement across the entire memory hierarchy by performing computation within the underlying storage media (i.e., NAND flash chips) and transferring only the result (when needed, to main memory and CPUs/GPUs).}

\gf{A recent work, ParaBit~\cite{gao2021parabit}, proposes an in-flash processing technique for bulk bitwise operations. We identify that ParaBit has two major limitations. First, it falls short of maximally exploiting the bit-level parallelism of bulk bitwise operations that could be enabled by leveraging the unique cell-array architecture and operating principles of NAND flash memory\omv{.}
Second,~it is unreliable because it is not designed to take into account the
highly error-prone nature of NAND flash memory.}

\gf{We propose Flash-Cosmos (\underline{Flash} \underline{C}omputation with \underline{O}ne-\underline{S}hot \underline{M}ulti-\underline{O}perand \underline{S}ensing)~\cite{flashcosmos}, a new in-flash processing technique that significantly increases the performance and energy efficiency of bulk bitwise operations while providing high reliability. Flash-Cosmos introduces two key mechanisms that can be easily supported in modern NAND flash chips: (1)~\underline{M}ulti-\underline{W}ordline \underline{S}ensing (MWS), which enables bulk bitwise operations on a large number of operands (tens of operands) with a \emph{single} sensing operation, and (2)~\underline{E}nhanced \underline{S}LC-mode \underline{P}rogramming (ESP), which enables reliable computation inside NAND flash memory.}

\begin{figure}[ht]
    \centering
    \includegraphics[width=\linewidth]{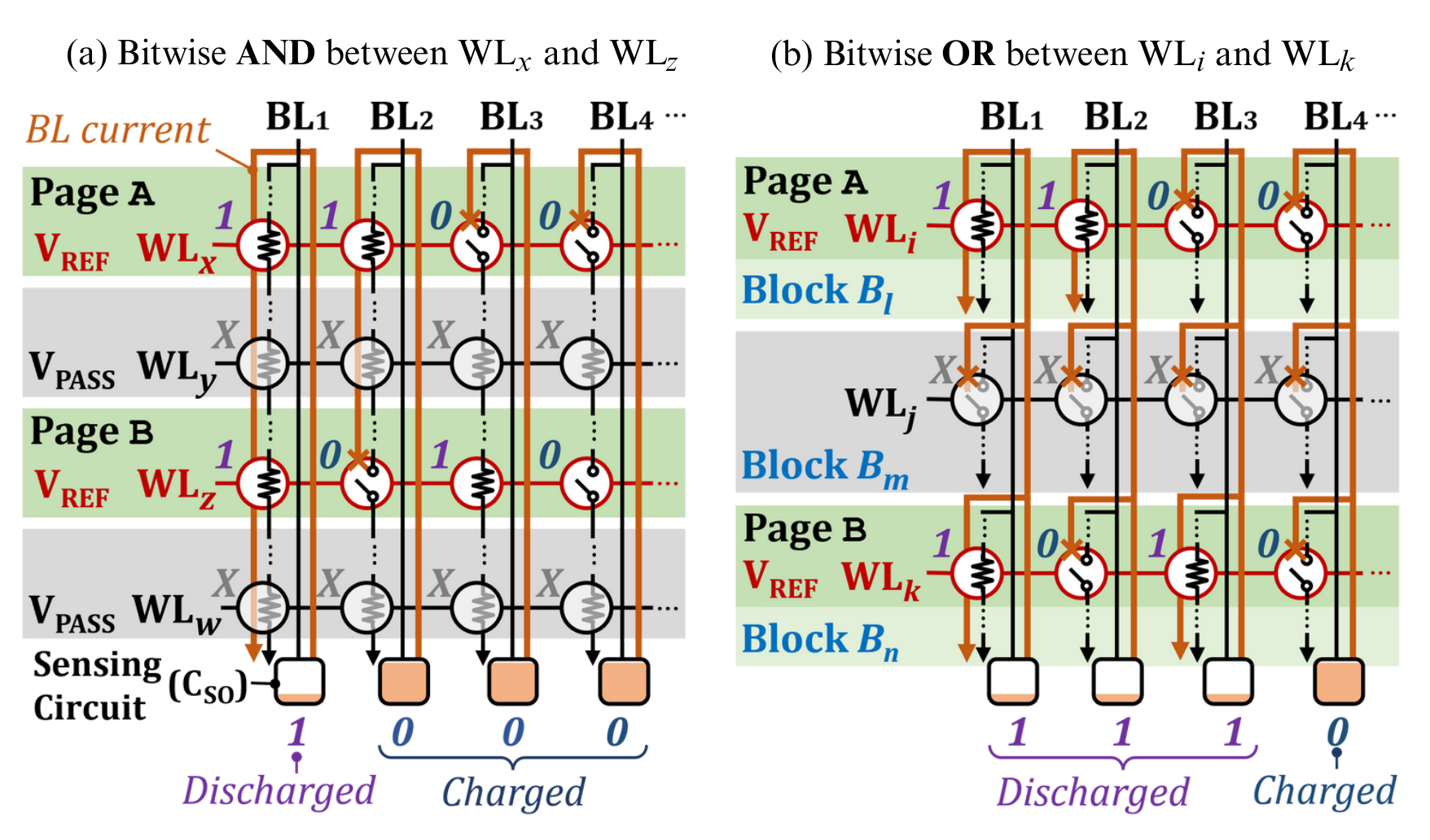}
    \caption{\gfvi{Overview of (a)~intra-block MWS (leading to bitwise \texttt{AND}) and (b)~inter-block MWS (leading to bitwise \texttt{OR}). Reproduced from~\cite{flashcosmos}.}} \label{fig:flashcosmos}
\end{figure}

\gf{MWS, \gfv{which Figure~\ref{fig:flashcosmos} illustrates,} leverages the two fundamental cell-array structures of NAND flash memory to perform in-flash bulk bitwise operations on a large number of operands with a \emph{single} sensing operation:
(1)~a number of flash cells (e.g., 24--176 cells) are serially connected to form a NAND string (similar to digital \texttt{NAND} logic); 
(2)~thousands of NAND strings are connected to the same bitline (similar to digital \texttt{NOR} logic).
Under these cell-array structures, simultaneously sensing \emph{multiple} wordlines automatically results in 
(i)~bitwise \texttt{AND} of \emph{all} the sensed wordlines if they are in the same NAND string \gfv{(Figure~\ref{fig:flashcosmos}a)} or 
(ii)~bitwise \texttt{OR} of \emph{all} the wordlines if they are in different NAND strings \gfv{(Figure~\ref{fig:flashcosmos}b)}.
\omv{Our paper demonstrates that we can perform 48-input \texttt{AND} and 4-input \texttt{NOR} operations reliably in existing NAND flash chips with only $<$10\% increase in sensing latency.}}

\gf{ESP effectively avoids raw bit errors in stored data via more precise programming-voltage control \omv{using the NAND flash controller}.
A flash cell stores bit data as a function of the level of its threshold-voltage (V$_{\text{TH}}$). Reading a cell incurs an error if the cell's V$_{\text{TH}}$ level moves to another V$_{\text{TH}}$ range that corresponds to a different bit value than the stored value, due to various reasons, such as program interference~\gfv{\cite{cai2013program,cai2017vulnerabilities, park-dac-2016,cai.procieee17}}, data retention loss~\gfv{\cite{cai2015data, cai2012flash, cai2013error, luo2018improving,cai.procieee17}}, read disturbance~\gfv{\cite{cai2015read, ha-ieeetcad-2015,cai.procieee17}}, and cell-to-cell interference~\gfv{\cite{cai2017vulnerabilities,cai.procieee17}}.
ESP maximizes the margin between different V$_{\text{TH}}$ ranges by carefully leveraging two existing approaches.
First, to store data for in-flash processing, it uses the single-level cell (SLC)-mode programming scheme~\cite{lee-atc-2009, kim2020evanesco}. 
Doing so guarantees a large V$_{\text{TH}}$ margin by forming only two V$_{\text{TH}}$ ranges (for encoding `\texttt{1}' and `\texttt{0}') within the fixed V$_{\text{TH}}$ window. 
Second, ESP enhances the SLC-mode programming scheme by using (i)~a higher programming voltage to increase the distance between the two V$_{\text{TH}}$ ranges and (ii)~more programming steps to narrow the high V$_{\text{TH}}$ range.}  
\omv{Using ESP makes bulk bitwise computation \omvi{operations} using flash \omvi{memory} completely reliable \omvi{(i.e., no errors)} on the NAND flash chips we tested in our work~\cite{flashcosmos}.}

\gf{In Flash-Cosmos, we enhance our basic MWS mechanism in two ways to make it more general purpose. 
First, we support bitwise \texttt{NAND}/\texttt{NOR}/\texttt{XOR}/\texttt{XNOR} by using MWS along with 
(i)~the \emph{inverse sensing} mechanism~\cite{lee-ieeejssc-2002} and 
(ii)~internal \texttt{XOR} logic~\cite{kim-ieeejssc-2012}, both of which are already supported in most NAND flash chips.
Second, we relax the data location constraints of the basic MWS mechanism (e.g., bitwise OR/NOR operations are possible only for wordlines in different NAND strings) 
by 
(i)~storing each operand's inverse data and 
(ii)~leveraging De Morgan's laws.}

\gf{We demonstrate the feasibility \omv{and robustness} of performing bulk bitwise operations with high reliability in Flash-Cosmos by testing 160 real 3D NAND flash chips \omv{ and demonstrating that they are capable of all operations described without \omvi{\emph{any}} errors.} 
Our \omv{system-level} evaluation shows that Flash-Cosmos improves average performance and energy efficiency by 3.5$\times$/32$\times$ and 3.3$\times$/95$\times$, respectively, over ParaBit~\omvi{\cite{gao2021parabit}}/outside-storage processing techniques across three real-world applications.}

\omv{We believe that \omvi{enabling} computation using flash \omvi{memory} to become more capable and programmable is a promising direction future research can take.
Compared to processing-using-DRAM, we believe processing-using-flash is still in its infancy. 
More work is needed to demonstrate the possibility of more complex computations, improve processing-using-flash programmability, and address system integration challenges.} 
\omvi{The good news is that in flash memory, the \omvii{memory} controller can have much more sophisticated capabilities to \omix{make} computation reliable, as existing flash controllers \omvii{already} use many mechanisms to greatly enhance robustness and avoid errors~\omviii{\cite{cai.procieee17,cai.bookchapter18.arxiv,yucai.bookchapter18}}\omix{. For example, existing flash memory controllers perform} \msvii{data randomization, wear leveling, \omix{adaptive refresh}, \omix{block remapping,} garbage collection, \omix{careful} tuning \omix{of} read reference voltage, and \omix{sophisticated} error correction}~\cite{cai.procieee17,do2013query,lee2020smartssd,kang2013enabling,Cho_2013,liang2019ins,ghiasimegis2024,mansouri2022genstore,pei2019registor,jun2018grafboost, seshadri2014willow,kim2016storage, gu.isca16, wang2019project,jun2015bluedbm, jun2016bluedbm, torabzadehkashi2019catalina, ajdari2019cidr, koo2017summarizer,cho2013xsd,jeong2019react, jun2016storage,cai2013error,cai2012error,cai2012flash,cai2013program,cai2013threshold,cai2014neighbor,cai2015data,cai2015read,cai2017vulnerabilities,yucai.bookchapter18,kim2020evanesco,park-dac-2016,nadig2023venice}}. \msvii{The ESP (enhanced SLC-mode programming) technique proposed in Flash-Cosmos~\cite{flashcosmos} is a great example of using already  supported mechanisms in flash memory controllers to improve the reliability of processing-using-flash \omix{techniques}. 
We believe there is significant opportunity for creating novel \omviii{and} robust processing-using-flash techniques by \omviii{exploiting} and enhancing \omix{the sophisticated capabilities of} flash memory controllers.}

\subsection{\gf{Processing-Using-SRAM}}
\label{sec:pum:pusram}

\gf{SRAM arrays can also be used to implement bitwise Boolean operations~\gfv{\cite{aga.hpca17,kang2015energy,jeloka201628,eckert2018neural,fujiki2019duality,simon2020blade,nag2019gencache,1802.shs,si2019dual,wang2019bit,al2020towards,kang.icassp14,kim2021colonnade,jiang2020c3sram,wang2023infinity,agrawal2018x}}.
\omv{Similarly to processing-using-DRAM\omvi{,} processing-using-NVM}, \omvi{and processing-using-NAND-flash,} a Boolean operation is realized by simultaneously activating \omv{multiple} rows\omv{, specifically two rows} containing the input operands and exploiting the parasitic bitline capacitance produced when sensing the produced bitline voltage level.
If \emph{both} the activated rows store a logic-`1' value, the bitline voltage stays high and thus the sense amplifier senses `1'. 
However, if either one of the \omv{activated} rows store a logic-`0' value, the bitline voltage \omv{goes} below the reference sensing voltage $V_{ref}$, and the sense amplifier will sense a `0'. 
A similar process happens when sensing the complementary bitline, i.e., the sense amplifier connecting the complementary bitline outputs `1' \emph{only} when both activated rows store a logic-`0' value. 
Therefore, a multiple row activation operation in SRAM array \omv{leads} to the \omv{computation} of logic \texttt{AND} and \texttt{NOR} operations.}

\gfv{One promising direction for processing-using-SRAM is to leverage the SRAM arrays already present in the cache hierarchy of modern computers for \emph{in-situ} computation. We illustrate such an approach to processing-using-SRAM with three examples \omvi{from recent scientific literature}: 
Compute Caches~\cite{aga.hpca17}, Neural Cache~\cite{eckert2018neural}, and Duality Cache~\cite{fujiki2019duality}. } 
\gf{Compute Caches~\cite{aga.hpca17} exploits the ability \omv{to} \omv{implement} bitwise Boolean operations in SRAM by turning the last-level cache (LLC) of a processor into an in-memory SIMD engine, where each bitline of SRAM subarray becomes a SIMD lane. 
It also extends the functionatilies of SRAM-based PUM to other operations, including \texttt{NOR}, compare, search, copy, and carryless multiplication operations.
Neural Cache~\cite{eckert2018neural} and Duality Cache~\cite{fujiki2019duality} build on top of Compute Caches by implementing bit-serial arithmetic inside LLC slices.}
\gfv{The Neural Cache architecture~\cite{eckert2018neural} targets accelerating neural network (NN) inference by re-purposing LLC slices into a SIMD engine capable of performing bit-serial arithmetic operations over integer or fixed-point data. 
Similar to \omvi{Ambit~\cite{seshadri.micro17}/}SIMDRAM~\cite{hajinazarsimdram}\omvi{/MIMDRAM~\cite{mimdram}}, Neural Cache stores and processes data vertically across the cache rows, with bits from multiple data elements distributed across different wordlines, allowing massive parallelism to be exploited: for example, a 35~MB LLC re-purposed as a bit-serial SIMD engine can execute an arithmetic operation over up to 1,146,880 1-bit input elements \emph{simultaneously} while incurring only 2\% area overhead when in computation mode~\omvi{\cite{eckert2018neural}}.}

\gfv{Duality Cache~\cite{fujiki2019duality} further enhances Neural Cache by targeting the execution of general-purpose data-parallel program in a \gls{SIMT} execution model.
In Duality Cache's execution model, an LLC slice is divided into (1)~control blocks, each capable of executing 1,024 threads simultaneously;
(2)~each \omvi{control block} is further subdivided into thread blocks;
(3)~each \omvi{thread block} consists of 256 threads; and 
(4)~each thread maps to a bitline in the cache slice.
Each bitline in the cache slice represents a thread lane, with multiple threads executing the same bit-serial instruction \emph{simultaneously}.
Duality Cache further extends the computing capabilities of Compute Caches and Neural Cache by supporting floating-point and transcendental operations using bit-serial computation and CORDIC algorithms~\cite{volder1959cordic,walther1971unified} for functions such as sine and cosine. 
To ease programmability, Duality Cache also proposes a specialized compiler that translates CUDA~\cite{cheng2014professional} \omvi{and} OpenACC~\cite{OpenACCA1:online} code to the Duality Cache ISA. 
}

\gfv{Processing-using-SRAM (processing-using-cache) architectures are quite attractive solutions for PUM, since \omvi{differently from} processing-using-DRAM and processing-using-NVM, they can be manufactured and tested using standardized and commercially-available CMOS-based \omvi{electronic design automation} tools and library cells. 
However, the \omvi{much} lower density and \omvi{much higher} cost-per-bit of SRAM compared to DRAM and NVM are the major drawbacks of processing-using-SRAM architectures for two main reasons.
First, they limit the applicability of processing-using-SRAM architectures to workloads with relatively small datasets that can fit within megabyte-sized SRAM arrays. Otherwise, the processing-using-SRAM architecture needs to bring data in/out the memory hierarchy, requiring extra data movement compared to processing-using-DRAM and processing-using-NVM architectures.
Second, they lead to a design that provides lower \omvi{computation} density. 
For example, while a 35~MB LLC can be deployed as a processing-using-cache SIMD engine with 1,146,880 SIMD lanes, an 8~GB DRAM \omvi{chip} deployed as a processing-using-DRAM SIMD engine has up to 67,108,864 SIMD lanes.\footnote{\gfv{Assuming that a DRAM row is 8~KB, and the DRAM module has 16 DRAM banks, each of which with 64 DRAM subarrays.}} 
In our SIMDRAM paper~\cite{hajinazarsimdram}, we show that when we consider DRAM-to-cache data movement, a processing-using-DRAM architecture (SIMDRAM) can outperform a processing-using-SRAM architecture (Duality Cache) by up to 52.9$\times$ for bit-serial addition operations when data movement is realistically taken into account. 
}

\gfv{We believe that the key to fully unlocking the potential of PUM architectures lies in investigating the holistic integration of processing-using-SRAM, processing-using-DRAM, processing-using-NVM\omvi{, and processing-using-NAND-flash} architectures into a memory-centric system to accelerate memory-intensive workloads with various degrees of \omvi{locality,} memory parallelism and computation demands. 
As we discussed in this section, several of such architectures already share a common execution model (i.e., \omvi{bulk-bitwise computation and} bit-serial SIMD \omvi{execution}). Thus, we believe that future solutions can likely map and schedule different PUM operations across different technology-specific PUM architectures in a collaborative fashion, minimizing overall data movement \omvi{across different types of memories used for different purposes (e.g., caches, main memory, storage)} and \omvii{thereby} significantly improving performance\omvii{,} energy efficiency\omvii{, and system scalability}.}

%% file: sections/07-PNM.tex
\section{Processing-Near-Memory (PNM)}
\label{sec:PNM}
\label{sec:3dstacked}


{\em \Gls{PNM}} involves adding or integrating PIM logic (e.g., accelerators, simple processing cores, reconfigurable logic) close to or inside the memory (e.g.,~\gfv{\cite{fernandez2020natsa, cali2020genasm, kim.bmc18, ahn.pei.isca15, ahn.tesseract.isca15, boroumand.asplos18, boroumand2021google, boroumand2021google_arxiv, boroumand2019conda, boroumand2016pim, boroumand.arxiv17, singh2019napel, asghari-moghaddam.micro16, JAFAR, chi2016prime, farmahini2015nda, gao.pact15, DBLP:conf/hpca/GaoK16, gu.isca16, guo2014wondp, hashemi.isca16, cont-runahead, hassan.memsys15, hsieh.isca16, kim.isca16, kim.sc17, DBLP:conf/IEEEpact/LeeSK15, liu-spaa17, morad.taco15, nai2017graphpim, pattnaik.pact16, pugsley2014ndc, zhang.hpdc14, zhu2013accelerating, DBLP:conf/isca/AkinFH15, gao2017tetris, drumond2017mondrian, dai2018graphh, zhang2018graphp, huang2020heterogeneous, zhuo2019graphq, herruzo2021enabling, boroumand2021polynesia, boroumand2022icde, syncron, besta2021sisa_micro, besta2021sisa, asgarifafnir, upmem2018, devaux2019, shin2018mcdram, cho2020mcdram, denzler2021casper, gomez2022machine, giannoula2022towards, fernandez2022exploiting, oliveira2022heterogeneous, balasubramonian2014near, jacob2016compiling, nair2015active, lloyd2018dse, gokhale2015rearr, lloyd2015memory, rodrigues2016scattergather, lloyd2017keyvalue, landgraf2021combining, nair2015evolution, kwon202125, lee2021hardware, kim2021aquabolt, ke2021near, lee2022improving, loh2013processing, DBLP:conf/sigmod/BabarinsaI15, impica, kim.arxiv17, sura.cf15, singh2020nero, singh2021fpga, RVU, NIM, singh2021accelerating, gu2016leveraging, amiraliphd, niu2022isscc, azarkhish2016logic, azarkhish2018neurostream, de2018design, akin2014hamlet, liu2018processing, tsai:micro:2018:ams, gu2020ipim, DRAMA_CAL_2014, Asghari-Moghaddam_2016, huang2019active, kersey2017lightweight, li2019pims, lim2017triple, smc_sim, HIVE, jang2019charon, hadidi2017cairo, santos2018processing, hadidi2017demystifying, gu2020dlux, asgari2020mahasim, baskaran2020decentralized, ahmed2019compiler, picorel2017near, dai2022dimmining, min2019neuralhmc, zhou2022flexidram, hall1999mapping, he2020newton, pugsley2014comparing, devic2022pim, subramaniyan2017parallel, IRAM_WML_1997, NMP_2005, ke2019recnmp, kim2017heterogeneous, sun2021abc, gomez2023evaluating, boroumand2021mitigating, gupta2023evaluating, oliveira2023dappa, park2024attacc, seo2024ianus, li2024pim, lopes2024pim, leepresto2024, baekpsyncpim2024, wangndsearch2024, liuisca2024, yueisca2024, tianndpbridge2024, zhaoumpim2024, ghiasimegis2024,li2024stream,schwedock2024leviathan,lee2024pim,huo2024pifs,ham2024low,zhao2024pim,mahapatra2024storage,heo2024neupims}).}
Many of these works place PIM logic inside the logic layer of 3D-stacked memories~\omv{\cite{lee.taco16,HBM,jedec.hbm.spec,hmc.spec.1.1,hmc.spec.2.0,HMC2,loh2008stacked,jeddeloh2012hybrid}}.  This
\emph{PIM processing logic}, which we also refer to as \emph{PIM
  cores} or \emph{PIM engines}, interchangeably, can execute portions
of applications (from individual instructions to functions) or entire
{threads and applications,} depending on the design of the
architecture. Such PIM engines have high-bandwidth and low-latency
access to the memory stacks that \sg{are on} top of them, since the logic
layer and the memory layers are connected via high-bandwidth vertical
connections~\omv{\cite{lee.taco16,HBM,jedec.hbm.spec,hmc.spec.1.1,hmc.spec.2.0,HMC2,loh2008stacked,jeddeloh2012hybrid}}, e.g., through-silicon vias \omv{(as described in Section~\ref{sec:tech-enablers})}. 

In this section, we discuss several examples of how systems can make use of relatively simple PIM engines \gf{near memory} to avoid data movement and {thus obtain} significant performance and energy {improvements} on a
wide variety of application domains.
\gf{We first focus our discussion on PNM architectures that \omv{add} PIM engines near the main memory of a system (often consisting of a 3D-stacked DRAM device).
Then, we briefly discuss how PNM architectures can be implemented at other levels of the memory hierarchy (i.e., near caches and storage devices).}

\subsection{\gf{PNM \omx{in} DRAM-Based Main Memory}}
\label{sec:pnm:pnd}

\gf{We discuss and highlight PNM architectures implemented \omv{in} the main memory of a system. 
A common feature across \omv{the} presented PNM architectures is the use of high bandwidth 3D-stacked DRAM devices, which allows for the implementation of custom logic at its logic layer. 
However, each architecture allows for different offloading granularities (i.e.,  the granularity in which computation is offloaded to the PNM architecture), including 
(1)~coarse-grained application-level offloading (where the entire application is offloaded to PIM engines),
(2)~function-level offloading (where functions of an application are offloaded to PIM engines), and
(3)~fine-grained instruction-level offloading (where a single instruction is offloaded to PIM engines). Such varying offloading granularities \gfv{lead} to trade-offs in terms of performance gains, complexity, and cost.}

\subsubsection{Tesseract: Coarse-Grained Application-Level PNM Acceleration \omv{(}of Graph Processing\omv{)}}
\label{sec:tesseract}


A promising approach to using PNM is to enable coarse-grained acceleration of {\em entire applications} that are heavily memory bound. In such a fundamentally coarse-grained (i.e., application-granularity) approach, an entire application is re-written to completely execute on the PNM substrate, potentially using a specialized programming model and specialized architecture/hardware. This approach is especially promising because it can provide the maximum performance and energy benefits achievable from PNM acceleration of a given application, since it enables the customization of the entire PNM system for the application. We believe this approach can be especially promising for widely-used data-intensive applications, such as \gfvi{machine learning \omvi{(including neural networks and large language models)}, databases, graph analytics, genome analysis, high-performance computing, security, data manipulation, and a wide variety of mobile and server-class workloads.}

A {popular modern} application is large-scale graph
processing~\omv{\cite{salihoglu.ssdbm13, tian.vldb13, low.vldb12,
  hong.asplos12, malewicz.sigmod10, harshvardhan.pact14,
  gonzalez.osdi12, ligra, Seraph, graphlab, nai2017graphpim,ahn.tesseract.isca15,besta2021sisa_micro,kyrola2012graphchi,nguyen2013lightweight}}.
{Graph processing has} broad applicability and use in many
domains, from social {networks} to machine learning, from
{data analytics} to bioinformatics.  Graph analysis workloads
are known to put {significant} pressure on memory bandwidth due
to 
(1)~large amounts of random memory accesses across large memory
regions (leading to very limited cache efficiency and very large
amounts of unnecessary data transfer on the memory bus) and 
(2)~small amounts of computation per each data item {fetched from
  memory} (leading to \gfv{a} limited ability to hide long memory
latencies {and exacerbating the energy bottleneck by exercising
  the huge energy disparity between memory access and computation}).
These two characteristics make it very challenging to scale up such
workloads despite their inherent parallelism, especially with
conventional architectures based on large on-chip caches and
{relatively} scarce off-chip memory bandwidth {for random
  access}.

\begin{figure}[ht]
\centering
\includegraphics[width=1.0\linewidth]{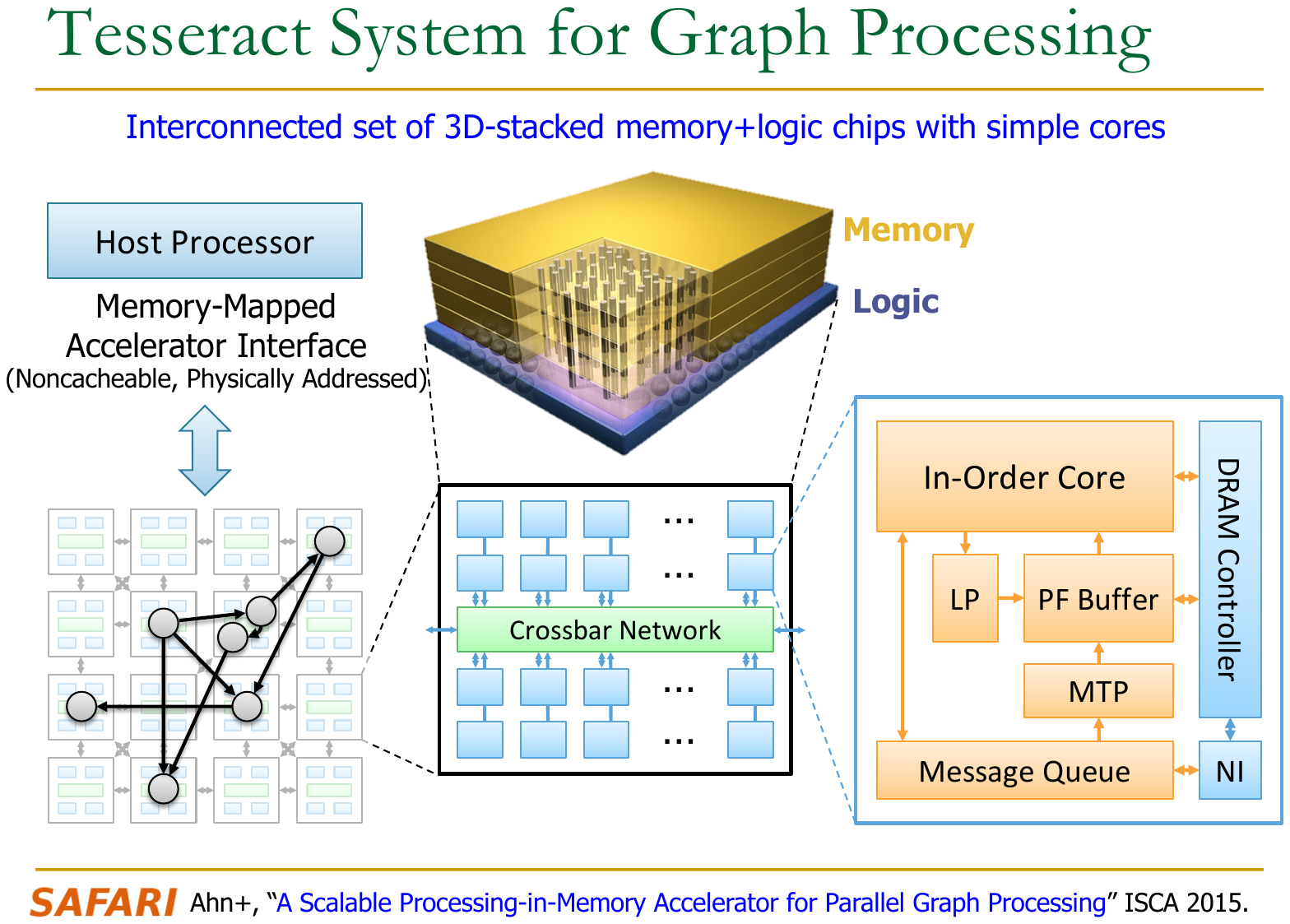}
\caption{Overview of the Tesseract system for graph processing. Reproduced from~\cite{mutlu.nsfpim20}. Originally presented in~\cite{ahn.tesseract.isca15,ahn.tesseract.isca15talk}.
}
\label{fig:tesseract}
\end{figure}

We can exploit the high bandwidth as well as \sg{the} potential computation
capability available within the logic layer of 3D-stacked memory to
overcome the limitations of conventional architectures for graph
processing.  To this end, we design a programmable PNM accelerator for
large-scale graph processing, called
Tesseract~\omv{\cite{ahn.tesseract.isca15,mutlu2023tesseractretrospective}}, depicted at a high level in Figure~\ref{fig:tesseract}.  

Tesseract consists of (1) a
new hardware architecture that {effectively} utilizes the
available memory bandwidth in 3D-stacked memory by placing simple
in-order processing cores in the logic layer and enabling each core
\sg{to manipulate} data only on the memory partition it is assigned to
control, (2) an efficient method of communication between different
in-order cores within a 3D-stacked memory to enable each core to
request computation on data elements that reside in the memory
partition controlled by another core, and (3) a {message-passing
  based} programming interface, {similar to how modern
  distributed systems are programmed}, which enables remote function
calls on data that resides in each memory partition. The Tesseract
design moves functions (i.e., computations and temporary values) to data that is to be updated rather than moving data elements across
different memory partitions and cores. It also includes two hardware
prefetchers specialized for memory access patterns of graph
processing, which operate based on the hints provided by our
programming model. 

Our comprehensive evaluations using five
state-of-the-art graph processing workloads with large real-world
graphs show that the proposed Tesseract PIM architecture improves
average system performance by {$13.8\times$} and achieves 87\%
average energy reduction over conventional systems.

A significant amount of research has built upon Tesseract to enable the graph processing PNM system to be much more powerful~\cite{dai2018graphh,zhang2018graphp,huang2020heterogeneous,zhuo2019graphq}. 
Among these, GraphP~\cite{zhang2018graphp} proposes a new graph partitioning scheme that greatly reduces the costly communication across 3D-stacked memory chips. 
Better partitioning is also proposed in GraphH~\cite{dai2018graphh}, together with a reconfigurable double mesh network that provides higher bandwidth across 3D-stacked memory chips. 
GraphQ~\cite{zhuo2019graphq} employs static and structured communication patterns to eliminate irregular communication, which is one of the key bottlenecks of Tesseract. 
Hetraph~\cite{huang2020heterogeneous} combines memristor-based analog computation units and CMOS-based digital compute cores on the logic layer of 3D-stacked memory chips, in order to use the most suitable one for each phase of computation.
Overall, combining the multiple proposals reported by these works, using the Tesseract-based PNM approach to accelerate graph processing can lead to more than two orders of magnitude \omv{improvements} both in performance as well as energy efficiency compared to a conventional processor-centric system with high-bandwidth memory. This demonstrates the potential promise of designing an entire PNM system from the ground up completely for important data-intensive \omv{applications}.

\omv{A recent invited retrospective paper describes the \omvi{influence} of \omvi{and lessons learned from} the Tesseract paper~\cite{mutlu2023tesseractretrospective}.
This retrospective demonstrates that our ISCA 2015 paper's principled rethinking of system design to accelerate an important class of data-intensive workloads provided significant value and
enabled/influenced a large body of follow-on works and ideas.
We expect that such rethinking of system design for key
workloads, especially with a focus on ``\emph{maximal acceleration
capability},'' (where we set ourselves free to change things as much as needed in the system in order to explore the maximal performance and energy efficiency benefits we can gain from a PNM accelerator) will continue to be critical as pressing technology
and application scaling challenges increasingly require us to
think differently to substantially improve performance and
energy (as well as other metrics).}


\subsubsection{\gf{\gfv{Polynesia: Coarse-Grained} Application-Level PNM Acceleration \gfv{(}of Hybrid Transactional/Analytical Processing\gfv{)}}}

\gf{Many application domains, such as business intelligence~\cite{sql-htap,snappy-data,sahay2008real}, healthcare~\cite{chisholm2014adopting,ta2016big}, personalized recommendation~\cite{wiser,zhou2017kunpeng}, \omvii{fraud detection~\cite{cao2019titant,qiu2018real,quah2008real},} and IoT~\cite{wiser}, have a critical need to perform 
\emph{real-time data analysis}, where data analysis needs to be performed using the most recent version of data~\cite{sql-htap,huang2020tidb}. To enable real-time data analysis, state-of-the-art database management systems (DBMSs) leverage \emph{hybrid transactional and analytical processing} (HTAP)~\cite{htap-gartner,sap-hana-evolution,htap}. An HTAP DBMS is a single-DBMS
solution that supports both transactional and analytical
workloads~\gfv{\cite{htap-gartner,peloton,batchdb,htap-survey,sql-htap,huang2020tidb,lahiri2015oracle,lee2017parallel,wiser,htap}}.} 

\gf{Ideally, an HTAP system should have three properties~\cite{batchdb} to guarantee efficient execution of transactional and analytical workloads:
(1)~allow for workload-specific optimizations (e.g., algorithms, data structures\omv{, hardware}); 
(2)~guarantee access to the most recent version of data for analytical workloads; and
(3)~ensure that the latency and throughput of both the transactional workload and
the analytical workload are the same as if each of them were run in isolation. 
We extensively study \omv{in~\cite{boroumand2021polynesia}} state-of-the-art HTAP systems~\gfv{\cite{htap-gartner,peloton,batchdb,htap-survey,sql-htap,huang2020tidb,lahiri2015oracle,lee2017parallel,wiser,htap}} and observe that
(1)~the mechanisms used to provide data freshness and consistency
induce a large amount of data movement between the CPU cores and
main memory and
(2)~HTAP systems suffer from severe performance interference because of the high resource contention, which prevent them from achieving all three properties of an ideal HTAP system.}

\begin{figure}[ht]
    \centering
        \centering
        \includegraphics[width=\linewidth]{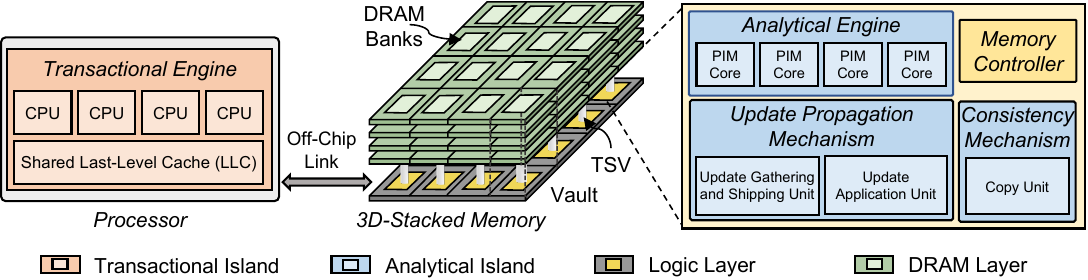}%
    \caption{\gf{High-level organization of Polynesia hardware \omix{for accelerating hybrid transactional/analytical processing (HTAP) databases}. Reproduced from~\cite{boroumand2021polynesia}.}}
    \label{fig:high-level-hw}
\end{figure}

\gf{To solve \omv{these} issues, we propose a novel system for HTAP databases called 
\emph{Polynesia}~\omv{\cite{boroumand2021polynesia}}. The key insight behind Polynesia is to partition the computing resources into two isolated new custom processing \emph{islands}: \emph{transactional islands} and \emph{analytical islands}. 
Each island consists of 
(1)~a replica of data for a specific workload, 
(2)~an optimized execution engine, and
(3)~a set of hardware resources
that cater to the execution engine and its memory access patterns. {Figure~\ref{fig:high-level-hw} shows the high-level organization of Polynesia, which includes one transactional island and one analytical island.  
Polynesia meets all desired properties from an HTAP system in three ways. 
First, by employing processing islands, Polynesia enables workload-specific optimizations for both transactional and analytical workloads. 
Second, we design new  accelerators and modified algorithms to 
propagate transactional updates to analytical islands and to maintain a consistent view of data across the system.
Third, we tailor the design of transactional and analytical islands to fit the characteristics of transactional and analytical workloads. 
The transactional islands use dedicated CPU hardware resources (i.e., multicore CPUs and multi-level caches) to execute transactional workloads since transactional queries have cache-friendly access patterns~\cite{boroumand2019conda,boroumand2016pim,amiraliphd}. The analytical islands leverage PNM techniques due to the large data traffic analytical workloads produce. 
We equip the analytical islands with a new PIM-based analytical engine that includes simple in-order PIM cores added to the logic layer of a 3D-stacked memory, software to handle data placement, and runtime task scheduling heuristics. 
Our new design enables the execution of transactional and analytical workloads at low latency and high throughput.}}

\gf{In our evaluations, we show the benefits of each component of Polynesia,
and compare its end-to-end performance and energy usage to three
state-of-the-art HTAP systems (modeled after Hyper~\cite{hyper}, AnkerDB~\cite{ankerdb}, and Batch-DB~\cite{batchdb}).
Polynesia outperforms all three, with higher 
transactional throughput (2.20$\times$/1.15$\times$/1.94$\times$; mean of 1.70$\times$) and 
analytical throughput (3.78$\times$/5.04$\times$/2.76$\times$; mean of 3.74$\times$), while consuming 48\% lower energy than the prior lowest-energy HTAP system. 
We open source Polynesia at \url{https://github.com/CMU-SAFARI/Polynesia}.} 

\omv{Going forward, we believe that such heterogeneous hardware, customized for different parts of a workload in a hardware/software co-designed manner, can be enabled for a wide variety of important workloads (e.g., ML workloads~\cite{boroumand2021google}\omvi{, large language models~\cite{heo2024neupims,zhou2022transpim,park2024attacc,yang2020retransformer,park2024lpddr,samsunghc23}, graph analytics~\cite{salihoglu.ssdbm13, tian.vldb13, low.vldb12,
  hong.asplos12, malewicz.sigmod10, harshvardhan.pact14,
  gonzalez.osdi12, ligra, Seraph, graphlab, nai2017graphpim,ahn.tesseract.isca15,besta2021sisa_micro,kyrola2012graphchi,nguyen2013lightweight}}) to achieve the highest performance, efficiency, and robustness \omvi{above and beyond} both processor-centric \omvi{systems} and memory-centric systems.}

\subsubsection{\gfv{NATSA: Coarse-Grained} Application-Level PNM Acceleration \gfv{(}of Time Series Analysis\gfv{)}}
\label{sec:natsa}

\begin{figure}[ht]
\centering
\includegraphics[width=1.0\linewidth]{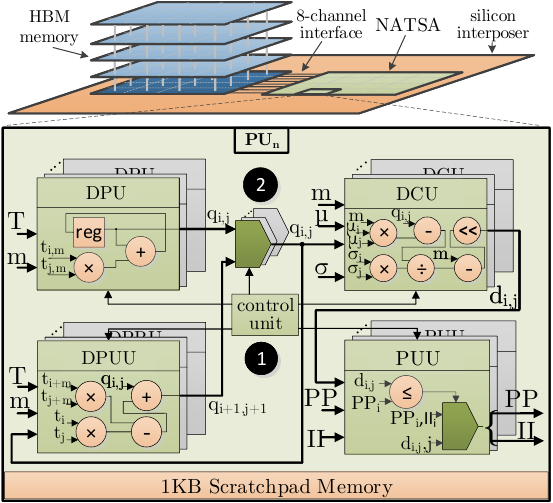}
\caption{NATSA design and integration near HBM memory. A PU is a NATSA Processing Unit that can \omv{perform} energy-efficient floating point arithmetic for time series analysis. Its components are described in~\cite{fernandez2020natsa}. Figure reproduced from~\cite{fernandez2020natsa}.}
\label{fig:natsa}
\end{figure}

NATSA~\cite{fernandez2020natsa, fernandez2022exploiting} is a near-memory processing accelerator for time series analysis. Time series analysis~\omv{~\cite{esling2012time}} is a powerful technique for extracting and predicting events with applications in epidemiology, genomics, neuroscience, astronomy, environmental sciences, economics, \omv{political science and more}. 
NATSA implements \emph{matrix profile}~\cite{yeh2016matrix}, the state-of-the-art algorithm for time series analysis fully via PNM. 
\emph{Matrix profile} operates on large amounts of time series data, but it has low arithmetic intensity. 
As a result, data movement represents a major performance bottleneck and energy waste, which NATSA alleviates by performing the complete time series analysis processing near memory using specialized accelerators. 
NATSA places energy-efficient floating point arithmetic processing units (PUs in Figure~\ref{fig:natsa}) close to 3D-stacked HBM memory~\cite{jedec.hbm.spec,lee.taco16}, 
connected via silicon interposers, 
as Figure~\ref{fig:natsa} shows.
NATSA improves performance by up to 14.2$\times$ (9.9$\times$ on average) and reduces energy by up to 27.2$\times$ (19.4$\times$ on average) over the state-of-the-art multi-core implementation \omv{of \omvi{the} matrix profile algorithm}. 
NATSA also improves performance by 6.3$\times$ and reduces energy by 10.2$\times$ over a general-purpose PNM platform with 64 in-order cores. \gf{The source code of NATSA is available at \url{https://github.com/CMU-SAFARI/NATSA}.}

\subsubsection{Function-Level PNM Acceleration \gfv{(}of Mobile Consumer Workloads \omv{and Edge Neural Networks}\gfv{)}}
\label{sec:google}

Another promising approach to using PNM, {\em function-level offloading}, is less intrusive than Tesseract's application-granularity approach described in Section~\ref{sec:tesseract}. This approach can still be coarse-grained since the function that is offloaded to the PNM logic can potentially \omv{be} \gfv{arbitrarily} long. However, the entire application does not need to be re-written. This approach is promising because it can enable easier adoption of PNM while still providing significant benefits. The key question in this approach is which functions in an application should be offloaded for PNM acceleration. Several works tackle this question for various applications, e.g., mobile consumer workloads~\cite{boroumand.asplos18}, GPGPU workloads~\cite{pattnaik.pact16,hsieh.isca16},
graph processing and in-memory database workloads~\cite{boroumand2019conda,boroumand2016pim},
and a wide variety of workloads from many domains~\omv{\cite{oliveira2021pimbench, oliveira2021pimbench_arxiv, oliveira2021.SLS,ghiasi2022alp}}.
We will discuss function-level PNM acceleration of mobile consumer workloads \omv{as well as edge neural network (NN) inference models} in this section, focusing on our recent \gfv{works} on the topic~\gfv{\cite{boroumand.asplos18, boroumand2021google}}.  

\paratitle{\gf{\gfv{PNM} Offloading for Consumer Workloads}} A very popular domain of computing today consists of consumer devices, which include smartphones, tablets, web-based computers such as Chromebooks, and wearable devices.
In consumer devices, energy efficiency is a first-class concern due to the limited battery capacity and {the stringent} thermal power budget.  We find that \emph{data movement} is a major contributor to the total system energy {and execution time} in modern consumer devices.  
Across all of the popular modern mobile consumer applications we study (described in the next paragraph), we find that 62.7\% of the total system energy, on average, is spent on data movement {across the memory hierarchy~\cite{boroumand.asplos18}.} As described before, this large fraction consumed on data movement is directly the result of the processor-centric design paradigm of modern computing systems. 

We comprehensively analyze the energy and performance impact of data
movement for several widely-used Google consumer
workloads~\cite{boroumand.asplos18}, which account for a significant
portion of the applications executed on consumer devices.  These
workloads include (1)~{the Chrome web browser}~\cite{chrome}, which is
a very popular browser used in mobile devices and laptops;
(2)~{TensorFlow Mobile}~\cite{mobile-tensorflow}, Google's machine
  learning framework, which is used in services such as Google
  Translate, Google Now, and Google Photos; (3)~{the VP9 video
    playback engine}~\cite{vp9-specification}, and (4)~{the VP9
    video capture engine}~\cite{vp9-specification}, both of which are
  used in many video services such as YouTube~\omvi{\cite{youtube}} and Google Hangouts~\omvi{\cite{google-hangouts}}.  We
  find that {offloading key functions to the logic layer} \msvii{of 3D-stacked DRAM} can
  greatly reduce data movement in all of these workloads.  However,
  there are challenges to introducing PIM in consumer devices, as
  consumer devices are extremely stringent in terms of the area and
  energy budget {they can accommodate for any new hardware
    enhancement}.  As a result, we need to identify what kind of
  in-memory logic can both (1)~\emph{maximize energy efficiency} and
  (2)~be implemented at \emph{minimum possible cost, in terms of both
    area overhead and complexity}.

We find that many of the target functions for PIM in consumer workloads are comprised of simple operations such as \emph{memcopy}, \emph{memset},
basic arithmetic and bitwise operations, {and simple data
  shuffling and reorganization routines}.  Therefore, we can
{relatively easily} implement these PIM target functions in
{the logic layer of 3D-stacked} memory using either (1)~a small
low-power general-purpose embedded core or (2)~a group of small
fixed-function accelerators.  
\omv{The functions we accelerate in memory are 
(1)~texture tiling, color blitting, and compression/decompression used by Google Chrome web browser~\cite{chrome};
(2)~packing/unpacking and quantization used by Google TensorFlow machine learning framework~\cite{mobile-tensorflow}; and 
(3)~video decoding/encoding used by Google's video playback and capture codecs~\cite{vp9-specification}.}
Our analysis shows that the area of a
PIM core and a PIM accelerator take up no more than 9.4\% and 35.4\%,
respectively, of the area available for PIM logic in a 3D-stacked \omv{DRAM} architecture.  Both the PIM
core and PIM accelerator eliminate a large amount of data movement,
and {thereby} significantly reduce total system energy (by an
average of 55.4\% across {all} the workloads) and execution time
(by an average of 54.2\%). 

\paratitle{\gf{\gfv{PNM} Offloading for Edge \omvi{Machine Learning} Workloads}} \gf{Modern consumer devices make widespread use of machine learning (ML), including neural network (NN) inference. 
Due to their resource-constrained nature, edge computing platforms now employ specialized energy-efficient accelerators for on-device inference (e.g., Google Edge Tensor Processing Unit, TPU~\cite{edge-tpu}; NVIDIA Jetson~\cite{jetson};
Intel Movidius~\cite{movidius}). 
At the same time, neural network (NN) algorithms are evolving rapidly, which has led to many types of NN models. Despite the wide variety of NN model types, \omv{many ML accelerators, including} Google's state-of-the-art Edge TPU~\cite{edge-tpu}\omv{,} \omv{provide a} \emph{one-size-fits-all} design (i.e., a monolithic accelerator with a fixed, large number of processing elements (PEs) and a fixed \emph{dataflow},
which determines how data moves within the accelerator) that caters to edge device area and energy constraints. Unfortunately,  it is very challenging to achieve
high energy efficiency, \omv{high} computational throughput, and \omv{high} area efficiency for \omvi{every} NN model with this one-size-fits-all design.} 

\gf{We conduct an in-depth analysis of ML inference execution on a commercial Edge TPU, across 24 state-of-the-art Google edge NN models spanning four popular NN model types:
(1)~convolutional neural networks (CNNs), 
(2)~long short-term memories (LSTMs)~\cite{lstm-google}, 
(3)~Transducers~\cite{he.icassp2019, transducer3, transducer4}, and 
(4)~recurrent CNNs (RCNNs)~\cite{rcnn-google,lrcn}. 
\omv{We \omvi{find} that DRAM \omvi{access} and off-chip \omvi{interconnects} are responsible for more than 90\% of the total system energy \omvi{when executing} large edge NN models.}
\omv{A} \emph{key takeaway} from our extensive analysis of Google edge NN models on the Edge TPU is that \emph{all key components} of an edge accelerator (i.e., PE array, dataflow, memory system)
must be co-designed and co-customized based on specific layer characteristics to achieve high utilization and energy efficiency.} 

\begin{figure}[ht]
    \centering
    \includegraphics[width=\linewidth]{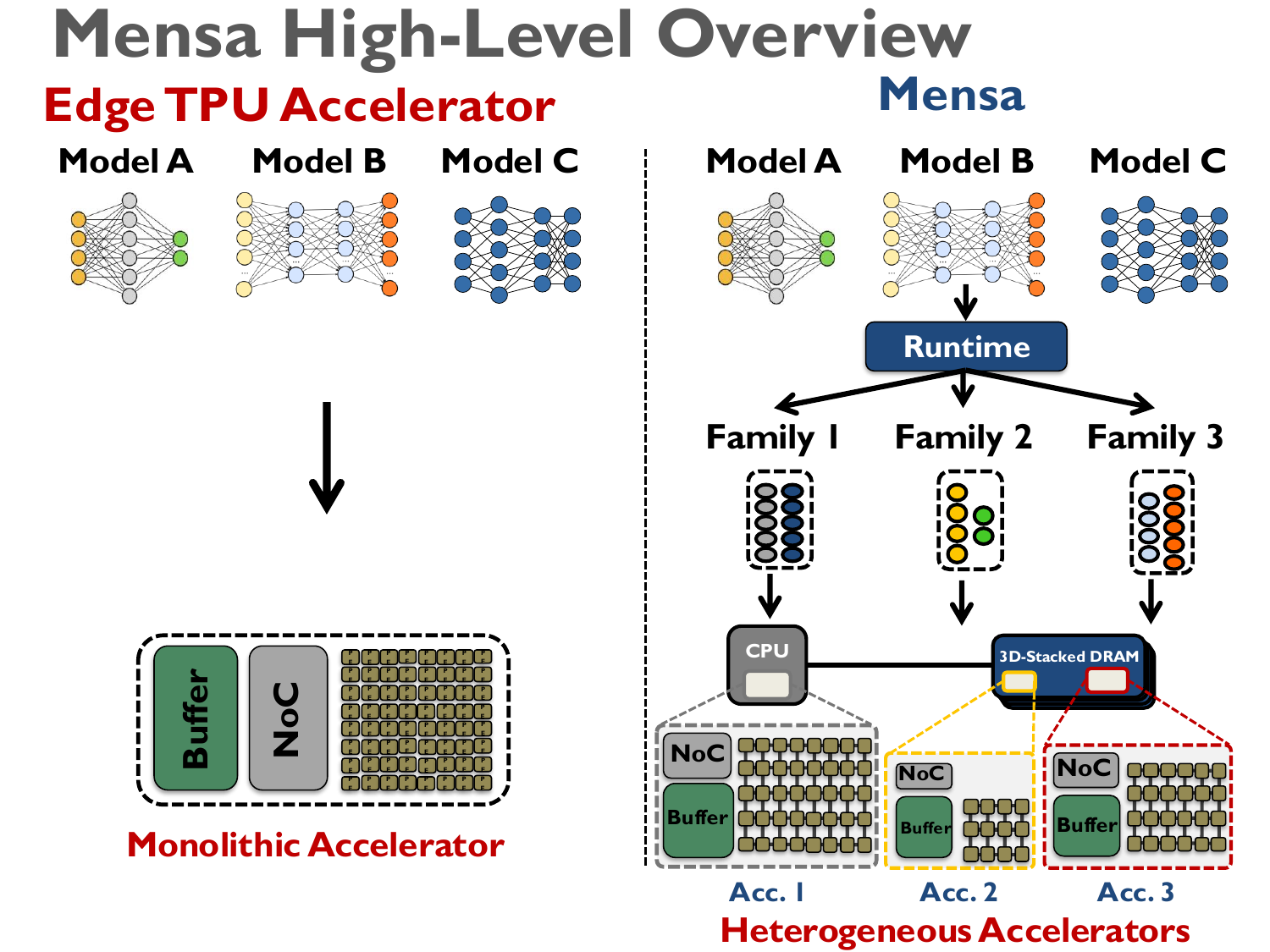}
    \caption{\gfix{\omvi{Conventional monolithic accelerator (e.g., Google Edge TPU) versus heterogeneous accelerator (e.g., Mensa) design principles.} 
    \gfvi{Mensa implements \omvii{different} hardware accelerators\omvii{, each} tailored for a particular set (i.e., family) of NN layers' characteristics. 
    \omvii{The figure depicts \omx{an} example with three accelerators, two of which are memory-centric (PNM) and one processor-centric.} To determine which hardware accelerator should execute which NN layer, Mensa implements a \emph{software runtime scheduler}.
    More details on \omvii{the} Mensa software runtime scheduler and hardware design can be found in~\cite{boroumand2021google}.} Reproduced from~\cite{mutlu.IMACAW23.talk}.}}
    \label{fig:mensa_overview}
\end{figure}

\gf{To this end, we propose Mensa~\cite{boroumand2021mitigating,boroumand2021google}, the first general hardware/software \omv{co-designed and} composable framework for ML acceleration in edge devices. The key idea of Mensa is to perform NN layer execution \omv{heterogeneously} across \emph{multiple} \omv{processor-centric} and \omv{memory-centric} accelerators, each of which is small and tailored to the characteristics of a particular subset (i.e., family) of layers\gfv{, as Figure~\ref{fig:mensa_overview} illustrates}. 
Our experimental study of the characteristics of \omv{all} layers in the \omv{studied} Google edge NN models reveals that the layers naturally group into a small number of clusters that are based on a subset of these characteristics.
This new insight allows us to significantly limit the number of different accelerators required in a Mensa design. 
We design a runtime scheduler for \omv{Mensa} to determine which of these accelerators should execute which NN layer,
using information about
(1)~which accelerator is best suited to the layer's characteristics, 
(2)~inter-layer dependencies\gfv{, and
(3)~\omvi{dataflow characteristics}}.}

\begin{figure}[ht]
    \centering
    \includegraphics[width=\linewidth]{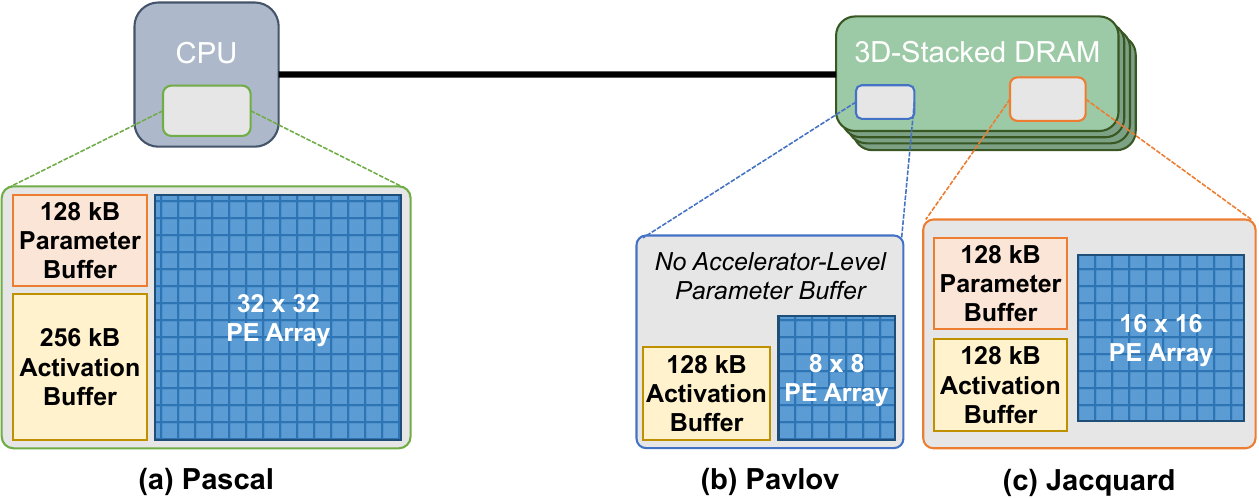}
    \caption{\gf{Mensa-G accelerator design. \omvi{The figure demonstrates three accelerators, one \omvii{processor}-centric, \emph{Pascal}~(a), and two  \omvii{memory}-centric, \emph{Pavlov}~(b) and \emph{Jacquard}~(c).
    Different accelerators are described in detail in~\cite{boroumand2021google}. Reproduced from~\cite{oliveira2022accelerating}.}}}
    \label{fig:accel_all}
\end{figure}

\gf{Using our new insight about layer clustering, we develop \emph{Mensa-G}, an example design for Mensa optimized for Google edge NN models.
We find that the design of Mensa-G's underlying accelerators should center around two
key layer characteristics (memory boundedness, and  activation/parameter reuse opportunities). This allows us to provide efficient inference execution
for \emph{all} of the Google edge NN models using \emph{only three} accelerators in Mensa-G. 
Figure~\ref{fig:accel_all} shows the design of the three Mensa-G accelerators, called \emph{Pascal}, \emph{Pavlov}, and \emph{Jacquard}.}

\gf{Pascal (Figure~\ref{fig:accel_all}a), for compute-centric layers, maintains the high PE utilization that these layers achieve in the Edge TPU, but does so using an optimized dataflow that both reduces the size of the on-chip buffer (16$\times$ smaller than in the Edge TPU)
and the amount of on-chip network traffic.
Pavlov (Figure~\ref{fig:accel_all}b), for LSTM-like data-centric layers, employs a dataflow that enables the temporal reduction of output activations, and enables the parallel execution of layer operations in a way that increases parameter reuse, greatly reducing off-chip memory traffic.
Jacquard (Figure~\ref{fig:accel_all}c), for other data-centric layers, significantly reduces the size of the on-chip parameter buffer (by 32$\times$) using a dataflow that exposes reuse opportunities for parameters. 
\omv{Since} both Pavlov and Jacquard are optimized for data-centric layers, which require significant memory bandwidth and are unable to utilize a significant fraction of PEs in the Edge TPU, we place the accelerators in the logic layer of 3D-stacked \omv{DRAM}~\gfv{\cite{lee.taco16,HBM,jedec.hbm.spec,hmc.spec.1.1,hmc.spec.2.0,HMC2,loh2008stacked,jeddeloh2012hybrid}} and use significantly smaller PE arrays compared to the PE array in Pascal, unleashing significant performance \emph{and} energy benefits.}

\gf{Our evaluation shows that compared to the baseline \omv{monolithic} Edge TPU, Mensa-G reduces total inference energy by 66.0\%, improves energy efficiency (TFLOP/J) by 3.0$\times$ and increases computational throughput (TFLOP/s) by 3.1$\times$, averaged across all 24 Google edge NN models. Mensa-G improves inference energy 
efficiency and throughput by 2.4$\times$ and 4.3$\times$ over Eyeriss~v2~\gfv{\cite{eyerissv2}}, a state-of-the-art accelerator.}
\gfvi{Mensa-G also reduces cost and improves area efficiency compared to the baseline monolithic Edge TPU accelerator.}
\gfvi{Overall, by leveraging a PNM design, Mensa-G \omvi{can improve performance, energy, and area efficiency, all at the same time}.}

\paratitle{\omv{Summary of Function-Based PNM Acceleration}} As evident from these results, function-level \omv{PNM} acceleration provides significant performance and energy benefits\omv{. As expected,} the benefits are not as high as full application-level offloading and customization of the PNM system, as we have shown for Tesseract~\omv{\cite{ahn.tesseract.isca15}} in Section~\ref{sec:tesseract}. This is expected since function-level offloading makes much fewer changes to the system and the programming model than application-level offloading, customization and rethinking of the system.
\omvi{As such, function-level offloading is likely easier to adopt in the short-term for a wide variety of workloads and systems.}

\subsubsection{Programmer-Transparent Function-Level PNM Acceleration of GPU Applications}
\label{sec:gpu}

In the last decade, Graphics Processing Units (GPUs)~\gfv{\cite{lindholm2008nvidia}} have become the
accelerator of choice for {a wide variety of} data-parallel
applications. They deploy thousands of in-order, {SIMT (Single
  Instruction Multiple Thread)}~\gfv{\cite{lindholm2008nvidia,mutlu2018recent,bingol2024gatekeeper,lindegger2023scrooge,darabi2022morpheus,pan2021exploring,lindegger2022algorithmic,olabi2022compiler,sadrosadati2021highly,alser2020sneakysnake,de2019automatic,sadrosadati2019itap,vijaykumar2018locality,mask,sadrosadati2018ltrf,mosaic,veynu.2011,nandita.2016,nandita.2015,jog.2016}} cores that run lightweight
threads. The heavily-multithreaded GPU architecture is devised to hide the long
latency of memory accesses by interleaving threads that execute
arithmetic and logic operations. Despite that, many GPU applications
are {still} very
\rachata{memory-bound~\gfv{\cite{veynu.2011,jog.asplos2013,jog.isca2013,medic,nandita.2015,nandita.2016,jog.2016,mask,mosaic,rachata-thesis,li-asplos19,wahib2014scalable,davis2012spmv}},}
because the limited off-chip pin bandwidth cannot supply enough data
to the running threads.

\begin{figure}[ht]
\centering
\includegraphics[width=\linewidth]{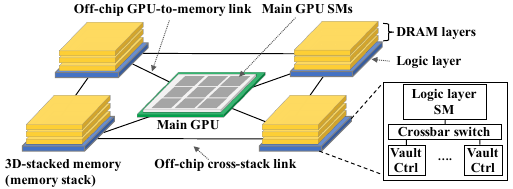}
\caption{Overview of a PNM GPU system with a powerful main GPU and less powerful logic-layer GPUs distributed across four 3D-stacked memories. Reproduced from~\cite{hsieh.isca16}.}
\label{fig:tom}
\end{figure}

\gf{\Gls{PNM}} in 3D-stacked memory architectures presents a promising opportunity to alleviate the memory bottleneck in GPU systems. GPU cores placed {in the logic layer of a 3D-stacked memory} can be directly
connected to the DRAM layers with high-bandwidth (and low-latency)
connections. 
Figure~\ref{fig:tom} presents an example configuration with a main GPU system connected to four 3D-stacked memories. In the logic layer of each 3D-stacked memory, there are GPU cores (also known as streaming multiprocessors, SMs) connected to memory vault controllers via a crossbar switch.
In order to leverage the potential performance benefits
of such systems, it is necessary to enable computation offloading and
data mapping to multiple {such compute-capable} 3D-stacked
memories, {such} that GPU applications can benefit from processing-in-memory capabilities in the logic layers {of such memories}.

TOM (Transparent Offloading and Mapping)~\cite{hsieh.isca16} proposes
two mechanisms to address computation offloading and data mapping
{in such a system} in a programmer-transparent manner.  First,
it introduces {new} compiler analysis {techniques} to
identify code sections in GPU kernels that can benefit from 
offloading to PIM engines. The compiler estimates the potential memory bandwidth
savings for each code block. To this end, {the compiler} compares
the bandwidth consumption of the code block, when executed on the
regular GPU cores, to the bandwidth cost of transmitting/receiving
input/output registers, when offloading to the GPU cores in the logic
layers. {At runtime, a final} offloading decision is made based
on dynamic system conditions, such as contention for processing resources in
the logic layer.  Second, a software/hardware {cooperative}
mechanism predicts the memory pages that will be accessed by offloaded
code, and places {such pages} in the same 3D-stacked memory 
where the code will be executed.  The goal is to make PIM effective by
ensuring that the data needed by the PIM cores is in the same memory
stack as the code that needs it.  {Both mechanisms are completely transparent to the
  programmer, who only needs to write regular GPU code without any
  explicit PIM instructions or any other modification to the code.}
We find that TOM improves the average performance {of a variety of GPGPU
  workloads} by 30\% and reduces the average energy consumption by
11\% with respect to a baseline GPU system without PIM offloading
capabilities.

A related work~\cite{pattnaik.pact16} identifies GPU kernels that are
suitable for PIM offloading by using a regression-based affinity
prediction model. A concurrent kernel management mechanism uses the
affinity prediction model and determines which kernels {should}
be scheduled concurrently {to maximize performance}. This way,
{the proposed} mechanism enables the simultaneous exploitation
of the regular GPU cores and the in-memory GPU cores. {This
  scheduling technique improves} performance and energy efficiency by
an average of 42\% and 27\%, respectively.


\subsubsection{Function-Level PNM Acceleration of Genome Analysis Workloads}
\label{sec:grim}

Genome analysis is a critical data-intensive domain that can greatly benefit from acceleration~\omv{\cite{alser.bioinformatics17,alser2019shouji,alser2020sneakysnake,xin.shd.bioinformatics15,cali2020genasm,cali2018nano,turakhia2018darwin,fujiki2018genax, cali2022segram, mansouri2022genstore.arxiv, mansouri2022genstore, diab2022high, diab2022framework,mutlu.aacbbtalk19,singh2021fpga,alser2022molecules,mutlu2023accelerating,alser2020accelerating,ghiasimegis2024,singh2024rubicon,firtina2024aphmm,cavlak2024targetcall,lindegger2023scrooge,xin.bmcgenomics13,liu20173d,doblas2023gmx,kaplan2017aresistive,angizi2020pimaligner,nag2019gencache,mao2022genpip,shahroodi2023swordfish}},
specifically processing-in-memory acceleration. 
We find that function-level PNM acceleration via algorithm-architecture co-design is especially beneficial for data-intensive genome analysis workloads, as demonstrated in 
 our \omv{various} recent works~\gfvi{\cite{kim.bmc18,cali2020genasm, cali2022segram,mansouri2022genstore,ghiasimegis2024,singh2021fpga,diab2023framework,alonso2024bimsa,singh2024rubicon}}.

    GRIM-Filter~\cite{kim.bmc18} is an in-memory accelerator for \juanrrri{\emph{genome seed filtering}}. \gfv{T}o read the genome (i.e., DNA sequence) of an organism, geneticists often need to reconstruct the genome from small segments of DNA known as reads, as current DNA extraction techniques are unable to extract the entire DNA sequence. A genome read mapper can perform the reconstruction by matching the reads against a reference genome, and a core part of read mapping is a computationally-expensive dynamic programming algorithm that aligns the reads to the reference genome. One technique to significantly improve the performance and efficiency of read mapping is seed filtering~\omv{\cite{xin.bmcgenomics13,xin.shd.bioinformatics15,alser2019shouji,alser.bioinformatics17,alser2020sneakysnake,alser2022molecules}}, which reduces the number of reference genome seeds (i.e., segments) that a read must be checked against for alignment by quickly eliminating seeds with no probability of matching. GRIM-Filter proposes a state-of-the-art filtering algorithm, and places the algorithm inside memory~\cite{kim.bmc18}. 

\begin{figure}[ht]
\centering
\includegraphics[width=1.0\linewidth]{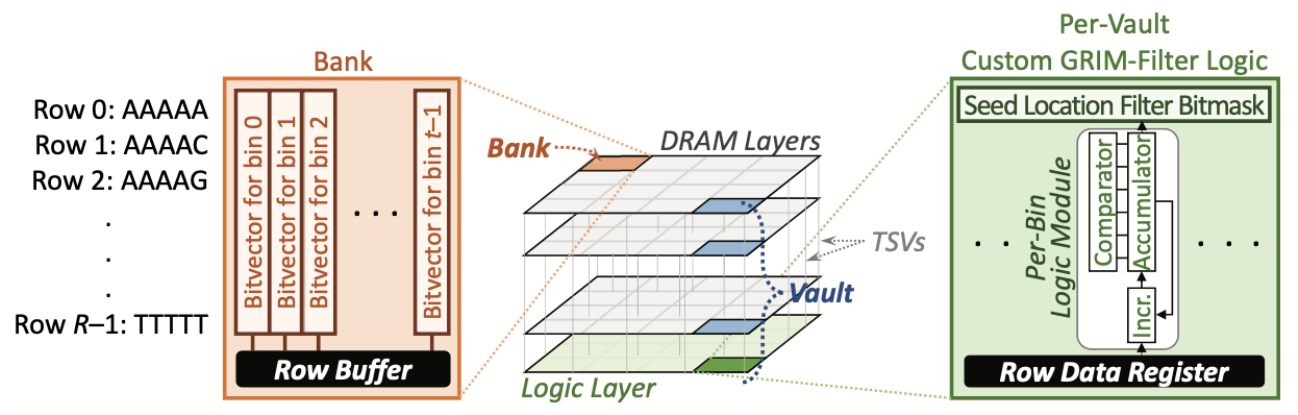}
\caption{Left block: GRIM-Filter bitvector layout within a DRAM bank. Center block: 3D-stacked DRAM with tightly integrated logic layer stacked underneath with TSVs for high inter-layer data transfer bandwidth. Right block: Custom GRIM-Filter logic placed in the logic layer, for each vault. Reproduced from~\cite{kim.bmc18}.}
\label{fig:grim}
\end{figure}

GRIM-Filter represents the entire reference genome by dividing it into short continuous segments, called \emph{bins}, and performs analyses on metadata associated to each bin. This metadata, represented as a \emph{bitvector}, stores whether or not a particular \emph{token} (a short DNA sequence) is present in the associated bin. 
Bitvectors are stored in DRAM in column order, such that a DRAM access to a row fetches the bits of the same token across many bitvectors, as the left block of Figure~\ref{fig:grim} shows. GRIM-Filter places custom logic for each vault in the logic layer of 3D-stacked memory (center block of Figure~\ref{fig:grim}). In each vault, there are multiple \emph{per-bin logic modules} which operate on the bitvector of a single bin. Each logic module consists of an incrementer, accumulator, and comparator, as the right block of Figure~\ref{fig:grim} shows.

GRIM-Filter introduces a communication protocol between the read mapper and the filter. The communication protocol allows GRIM-Filter to be integrated into a full genome read mapper (e.g., 
mrFAST~\cite{alkan.naturegenetics09}, 
BWA-MEM~\cite{li2009fast}, Minimap2~\cite{li2018minimap2}\gf{)},
by allowing (1) the read mapper to notify GRIM-Filter about the DRAM addresses on which to execute customized in-memory filtering operations, (2) GRIM-Filter to notify the read mapper once the filter generates a list of seeds for alignment. Across 10 real genome read sets, GRIM-Filter improves the performance of a full state-of-the-art read mapper by 3.65$\times$ over a conventional CPU-only system~\cite{kim.bmc18}. \gf{GRIM-Filter's simulation infrastructure is available at \url{https://github.com/CMU-SAFARI/GRIM}.}

\begin{figure}[ht]
\centering
\includegraphics[width=1.0\linewidth]{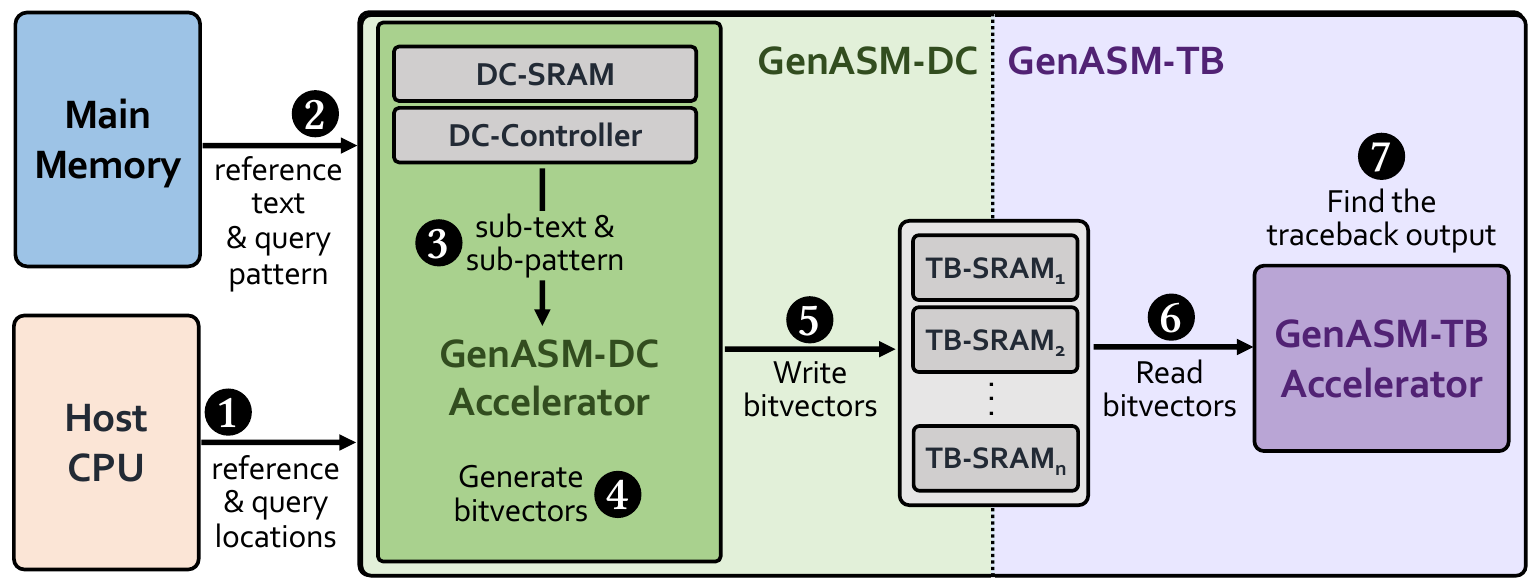}
\caption{Overview of GenASM. Different components are described in detail in~\cite{cali2020genasm}. Figure reproduced from~\cite{cali2020genasm}.}
\label{fig:genasm}
\end{figure}

In a more recent work~\cite{cali2020genasm}, we develop an algorithm-architecture co-design to accelerate {\em approximate string matching (ASM)}, which is used at multiple points during the mapping process of genome analysis. ASM enables read mapping to account for sequencing errors and genetic variations in the reads. Our work, GenASM, is the first ASM acceleration framework for genome sequence analysis. GenASM performs bitvector-based ASM, which can efficiently accelerate multiple steps of genome sequence analysis. 
We modify the underlying ASM algorithm (Bitap~\cite{baeza1992new,wu1992fast}) 
to significantly increase its parallelism and
reduce its memory footprint. We accelerate this modified ASM algorithm, called \emph{GenASM-DC} for Distance Calculation, using an accelerator that performs very efficient Distance Calculation between two input strings. 
We also develop a novel Bitap-compatible algorithm for \emph{traceback} (i.e., a method to collect information about the different types of alignment errors, or differences, between two input strings), called \emph{GenASM-TB}. 
Using our modified \juanrrr{GenASM-DC} algorithm and the new GenASM-TB algorithm, we design the first hardware accelerator for Bitap. 
Figure~\ref{fig:genasm} illustrates a high-level overview of GenASM, depicting the flow of input and intermediate data in the system as well as  the communication paths of the two accelerators for GenASM-DC and GenASM-TB.
Our hardware accelerator, which is placed in the logic layer of 3D-stacked memory to minimize data movement overheads, consists of specialized systolic-array-based~\omv{\cite{kung1982systolic}} compute units and on-chip SRAMs that are designed to match the rate of computation with memory capacity and bandwidth, resulting
in an efficient design whose performance scales linearly as we
increase the number of compute units working in parallel. 
Our detailed performance and energy evaluations demonstrate that GenASM provides significant performance
and power benefits for three different use cases in genome
sequence analysis, outperforming the best prior hardware accelerators as well as \omv{best} software baselines by one or more orders of magnitude. We believe these results are quite promising and point to the need for further exploration of \omv{hardware-software co-designed} PIM accelerators in genome analysis. \gf{To this end, we open source GenASM at \url{https://github.com/CMU-SAFARI/GenASM}.}

\begin{figure}[ht]
\centering
\includegraphics[width=1.0\linewidth]{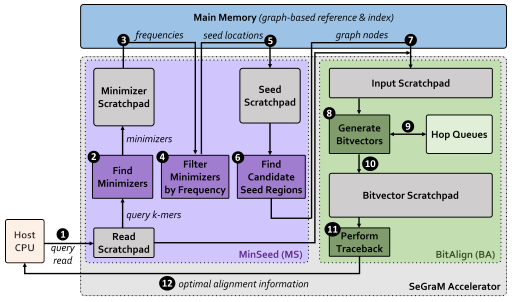}
\caption{\juanrrr{Overview of SeGraM. Different components are described in detail in~\cite{cali2022segram}. Figure reproduced from~\cite{cali2022segram}.}}
\label{fig:segram}
\end{figure}

\juanrrri{In another recent work, we present} 
SeGraM~\cite{cali2022segram}, an accelerator for \emph{genomic sequence-to-graph and sequence-to-sequence mapping}.
Sequence-to-graph mapping is a recent trend~\cite{paten2017genome, pevzner2001eulerian, ameur2019goodbye, rakocevic2019fast, kaye2021genome, eizenga2020pangenome} that replaces the linear reference sequence with a graph-based representation of the reference genome, which captures genetic variations and diversity across individuals. 
Sequence-to-graph mapping results in significant quality improvements in genome analysis. 
Our work~\cite{cali2022segram} identifies key bottlenecks of the seeding and alignment steps of sequence-to-graph mapping. Specifically, seeding suffers from long DRAM latency while alignment shows high cache miss rate. 
To alleviate these bottlenecks, SeGraM follows an algorithm/hardware co-designed approach to create the first hardware accelerator for sequence-to-graph mapping and sequence-to-sequence mapping (in practice, a special case of sequence-to-graph mapping). 
As depicted in Figure~\ref{fig:segram}, SeGraM consists of two main components. 

First, \emph{MinSeed} (\emph{MS}) finds the minimizers (i.e., representative substrings of a sequence) for a given query read, fetches the candidate seed locations for the selected minimizers, and for each candidate seed, fetches the subgraph surrounding the seed. 
Second, \emph{BitAlign} (\emph{BA}) aligns the query read to the subgraphs identified by MinSeed, and finds the optimal alignment. BitAlign adapts GenASM's bitvector-based algorithm~\cite{cali2020genasm} to perform sequence-to-graph alignment. 
SeGraM couples arrays of SeGraM accelerators (each one with MS and BA) with stacks of High Bandwidth Memory (HBM2E)~\cite{jedec.hbm.spec}, which ensures low-latency and high-bandwidth memory access for each SeGraM accelerator. 
We demonstrate that SeGraM provides significant improvements in performance and reduction of power consumption with respect to state-of-the-art sequence-to-graph mapping software. 
\gf{SeGraM's source code is available at \url{https://github.com/CMU-SAFARI/SeGraM}.}
In summary, SeGraM is a promising framework that shows the great potential of algorithm/hardware co-design research for graph-based genome analysis.

\subsubsection{\gfv{PEI: Fine-Grained} Instruction-Level PNM Acceleration}
\label{sec:pei}


A finer-grained approach to using PNM is {\em instruction-level offloading}. With this approach, individual instructions can be offloaded to the PNM engine and accelerated. As we describe below, this fine-grained approach can have significant benefits in terms of potential adoption since existing processor-centric execution models already operate (i.e., perform computation) at the granularity of individual instructions and all such machinery can be reused to aid offloading to be as seamless as possible with existing programming models and system mechanisms. 
PIM-Enabled Instructions (PEI)~\cite{ahn.pei.isca15} aims to provide
the minimal processing-in-memory support to take advantage of PIM
using 3D-stacked memory, in a way that can achieve significant
performance and energy benefits without changing the computing system
significantly.  To this {end}, PEI proposes a collection of
simple instructions, which introduce small changes to the
computing system and no changes to the programming model or the
virtual memory system, in a system with 3D-stacked memory. These
instructions, generated by the compiler or programmer to indicate potentially PIM-offloadable operations in the program, are operations that can be executed either in a traditional host CPU (that fetches and decodes them) or the PIM engine in 3D-stacked memory.

PIM-Enabled Instructions are based on two key ideas. First, a PEI is a
cache-coherent, virtually-addressed host processor instruction that
operates on only a single cache block. It requires no changes to the
sequential execution and programming model, no changes to virtual
memory, minimal changes to cache coherence, and no need for special
data mapping to take advantage of PIM (because each PEI is restricted
to a single memory module due to the single cache block restriction it
has). Second, a Locality-Aware Execution runtime mechanism decides
dynamically where to execute a PEI (i.e., {either} the host
processor or the PIM logic) based on simple locality characteristics
and simple hardware predictors. {This runtime mechanism executes 
the PEI at the location that maximizes performance.}  In summary,
PIM-Enabled Instructions provide the illusion that PIM operations are
executed as if they were host instructions: the programmer may not even be aware that the code is executing on a PIM-capable system and the exact same program containing PEIs can be executed on conventional systems that do not implement PIM.

\begin{figure}[ht]
\centering
\includegraphics[width=1.0\linewidth]{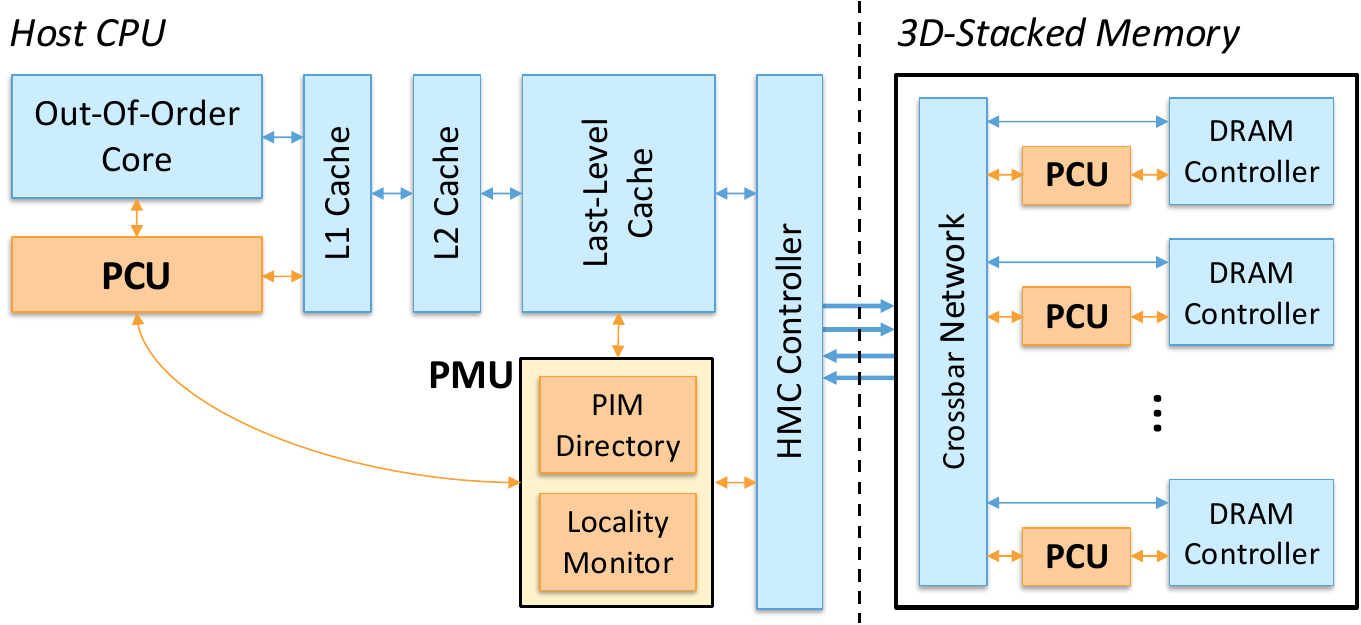}
\caption{\gfv{Example architecture for PIM-enabled instructions \omvi{(\textbf{PEI}s)}. \omvi{\textbf{PCU}: PEI Computation Unit. \textbf{PMU}: PEI Management Unit.} Reproduced from~\cite{ghose2019arxiv}. Originally presented in~\cite{ahn.pei.isca15,ahn.pei.isca15talk}.}
}
\label{fig:pei}
\end{figure}

Figure~\ref{fig:pei} shows an example architecture that can be used to enable PEIs. In this architecture, a PEI is executed on a PEI Computation Unit (PCU). To enable PEI execution in either the host CPU or in memory, a PCU is added to each host CPU and to each vault in an HMC-like \omv{(or HBM-like)} 3D-stacked memory. While the work done in a PCU for a PEI might have required multiple CPU instructions in the baseline CPU-only architecture, the CPU only needs to execute a single PEI instruction, which is sent to a central PEI Management Unit (PMU in Figure~\ref{fig:pei}). The PMU, which is in charge of the Locality-Aware Execution, launches the appropriate PEI operation on one of the PCUs, either on the CPU or in 3D-stacked memory. 

Examples of PEIs are integer increment, integer minimum,
floating-point addition, hash table probing, histogram bin index,
Euclidean distance, and dot product~\cite{ahn.pei.isca15}.
Data-intensive workloads such as graph processing, in-memory data
analytics, machine learning, and data mining can {significantly}
benefit from these PEIs.  Across 10 {key data-intensive}
workloads, we observe that the use of PEIs in these workloads, 
in combination with the Locality-Aware Execution runtime mechanism, 
leads to an average performance improvement of 47\% and an average
energy reduction of 25\% over a baseline \sg{CPU}, on reasonably large data set sizes. 

As such, the benefits provided by the fine-grained PEI approach are quite promising: with minimal changes to the system, performance and energy improve significantly. We therefore believe that the PEI mechanism can ease the adoption of PIM systems going into the future, a key issue we discuss in detail next.

\subsection{\gfix{PNM \omv{in} Caches \& Storage}}
\label{sec:pnm:pnnvm}
\label{sec:pnm:pnsram}

\gf{Many prior works implement PNM architectures at different levels of the memory hierarchy other than main memory, including SRAM caches (i.e., \emph{near-cache computing)}~\gfvi{\cite{wang2023affinity,wang2021stream,wang2022near,pattnaik2019opportunistic,denzler2021casper,lockerman2020livia,schwedock2022tako,kumar2015dasx,orenes2023dalorex,yuan2019halo,vieira2021compute,eggermann202316,schwedock2024leviathan,nori2021reduct,schwedock2023optimizing}} and storage (i.e., \emph{near-storage})~\gfvi{\cite{riedel2001active,acharya.1998,keeton.1998,riedel.1998,mansouri2022genstore,pei2019registor,jun2018grafboost, do2013query, seshadri2014willow,kim2016storage, gu.isca16, kang2013enabling, wang2019project,jun2015bluedbm, jun2016bluedbm, torabzadehkashi2019catalina, lee2020smartssd, ajdari2019cidr, koo2017summarizer,cho2013xsd,jeong2019react, jun2016storage,De_2013,liang2019ins,Kim2018HowMC,ghiasimegis2024}}, using both disk~\cite{riedel2001active,acharya.1998,keeton.1998,riedel.1998} and NAND flash memory~\gfvi{\cite{mansouri2022genstore,pei2019registor,jun2018grafboost, do2013query, seshadri2014willow,kim2016storage, gu.isca16, kang2013enabling, wang2019project,jun2015bluedbm, jun2016bluedbm, torabzadehkashi2019catalina, lee2020smartssd, ajdari2019cidr, koo2017summarizer,cho2013xsd,jeong2019react, jun2016storage,ghiasimegis2024}}. 
The primary benefit of near-cache computing is to reduce the energy and latency associated with moving data back-and-fourth L1/L2/L3 caches. 
Compared to main-memory-based PNM architectures, implementing near-\omv{cache} PNM systems can be a simpler task since:
(1)~logic and memory can be manufactured using the same CMOS process,
(2)~the PNM cores can leverage the same mechanisms as the CPU cores to maintain data coherence\omv{,
(3)~PNM cores can use the on-processor-chip mechanisms for virtual memory address translation and access control.}
Near-storage computing systems often 
(1)~make use of the already present computing resources at SSD controllers (i.e., simple in-order cores that are used to execute the SSD firmware code and the flash translation layer) to perform computation or
(2)~add simple logic units to the flash controller.
To illustrate the design of non-main-memory-based PNM architectures, we highlight \omv{three major} prior works: 
(1)~Casper~\cite{denzler2021casper}, a near-cache PNM architecture for stencil computation,
(2)~GenStore~\cite{mansouri2022genstore}, a near-storage PNM architecture for genome sequence analysis\omv{,
(3)~MegIS~\cite{ghiasimegis2024}, a near-storage PNM \omvi{system} for \omvi{accelerating} metagenomic analysis.} }

\begin{figure}[ht]
    \centering
\includegraphics[width=\columnwidth]{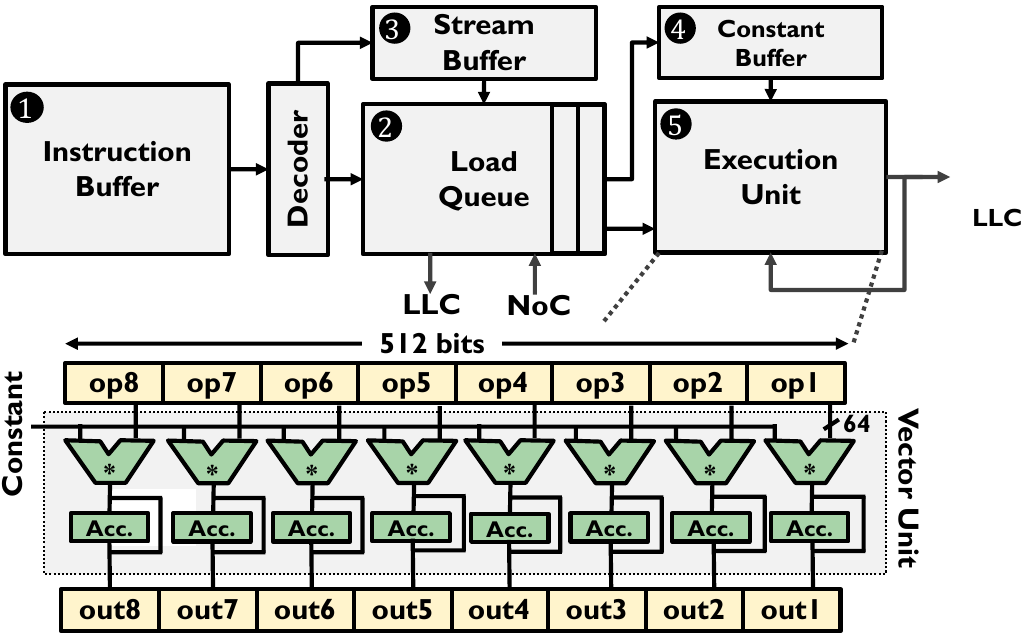}
    \caption{\gf{Main components \omvi{of} Casper's stencil processing unit (SPU). \omvi{Details of each component can be found in~\cite{denzler2021casper}.} Reproduced from~\cite{denzler2021casper}.}}
    \label{fig:spu}
\end{figure}

\gf{In~\cite{denzler2021casper}, we propose Casper, a near-cache accelerator consisting of specialized stencil {computation} units connected to the last-level cache (LLC) of a traditional CPU. Casper is based on two key ideas: 
(1)~avoiding the cost of moving rarely reused data {throughout} the cache hierarchy, and 
(2)~exploiting the regularity of the data accesses and the inherent parallelism of stencil {computations} \omv{across many SRAM arrays} to increase overall performance. 
With {small} changes in LLC address decoding logic and data placement, Casper performs stencil computations at the peak {LLC bandwidth}. 
Casper's hardware is composed of a set of stencil processing units (SPUs), placed in each cache slice of the LLC. Similar to a regular LLC, SPUs can load data from other levels of the cache hierarchy into the LLC. 
Figure~\ref{fig:spu} shows the architecture of \omv{an} SPU. The main building blocks {of an {SPU}} are the instruction buffer (\circled{1} in Figure~\ref{fig:spu}), which holds instructions to compute a stencil;
the load queue (\circled{2}), which is responsible for issuing data
requests to the memory subsystem;  
the stream buffer (\circled{3}), which holds
the current state of the stencil streams (i.e., a sequence of data elements of the same type located in consecutive physical memory addresses); 
the constant buffer (\circled{4}), which holds the constant values needed for the
multiply-accumulate (MAC) operation; and 
the execution unit (\circled{5}), which comprises of a 512-bit vector unit that performs double-precision floating-point MAC operations. 
The complete SPU is pipelined to maintain single-cycle instruction throughput and controls the complete computation, from instruction fetch until \omv{instruction retirement}. }  

\gf{Casper improves performance of stencil kernels by $1.65\times$ on average ({up to $4.16\times$}) compared to a commercial high-performance multi-core processor, while reducing {system} energy consumption by $35$\% on average ({up to $65\%$}). Casper provides $37\times$ ({up to $190\times$}) improvement in performance-per-area compared to a state-of-the-art GPU.}

\begin{figure}[ht]
    \centering
\includegraphics[width=\columnwidth]{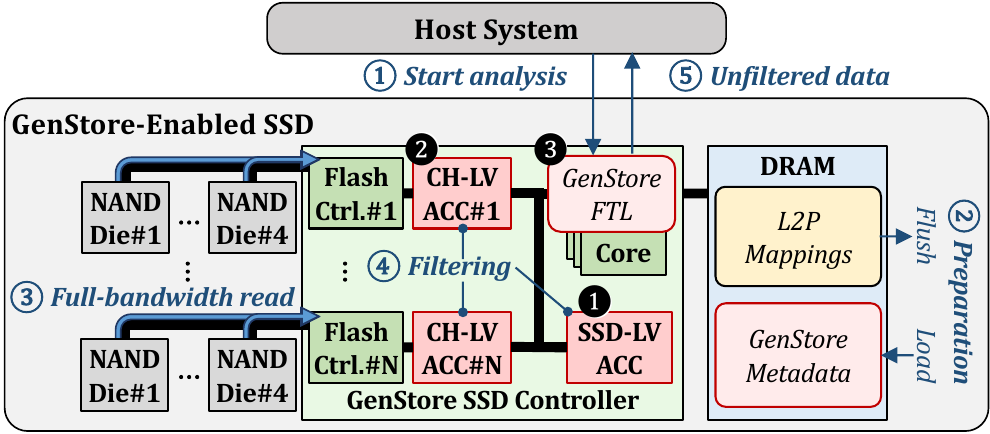}
    \caption{\gf{Overview of GenStore. Reproduced from~\cite{mansouri2022genstore}.}}
    \label{fig:genstore}
\end{figure}

\gf{In~\cite{mansouri2022genstore}, we aim to improve the performance of genome sequence analysis by effectively reducing unnecessary data movement from the storage system. To this end, we propose \emph{GenStore}, the
first near-storage (i.e., \omv{near} NAND flash \omv{chips}) processing system designed for genome sequence analysis. 
The key idea of GenStore is to exploit low-cost in-storage accelerators to accurately filter out the reads that do \omv{\emph{not}} require the expensive alignment step in read mapping and thus significantly reduce unnecessary data movement from the storage system to main memory and processors \omv{or accelerators}. 
Figure~\ref{fig:genstore} shows the overall architecture of GenStore and how it interacts with the host system. GenStore employs two types of hardware accelerators: 
\circled{1}~a single SSD-level accelerator and 
\circled{2}~channel-level accelerators, each of which is dedicated to a channel. The GenStore-FTL \circled{3} communicates with the host system and manages the metadata and data flow over the SSD hardware components (i.e., NAND flash chips, internal DRAM, and in-storage accelerators).
}

\gf{{Once} the host system {indicates that the SSD should start analysis as required by} a read mapping application {(\incircledd{1} in Figure~\ref{fig:genstore}}),  GenStore prepares {for operation} as an accelerator {(\incircledd{2})}. It flushes the conventional FTL metadata necessary to operate as a regular SSD (e.g., L2P mappings~\cite{kim2020evanesco}), while loading the GenStore metadata necessary for each use case.
After finishing the preparation, GenStore starts the filtering process.
{It}  keeps concurrently reading the data to process from \emph{all} {NAND flash} chips {(\incircledd{3})} via multi-plane operations (i.e., it exploits the SSD's full internal bandwidth{, which} is much higher than the I/O bandwidth between the SSD and host system~\omv{\cite{koo2017summarizer, mailthody2019deepstore,nadig2023venice}}),  while  filtering out reads {(\incircledd{4})} that do not have to undergo {further analysis ({{e.g.}, \omv{approximate string matching} computation})}.
Doing so is possible due to multiple channel-level accelerators that provide {computational} throughput matching the SSD's internal bandwidth even for the most complicated computation required for filtering. 
The host system {performs} further computation as soon as GenStore sends unfiltered {reads} {(\incircledd{5})}, which removes GenStore's filtering process almost completely from the critical path of the application.}  
\gf{We design GenStore to support two in-storage filtering mechanisms in a single SSD: 
(1)~GenStore-EM, which filters exactly-matching reads, i.e., reads that exactly match subsequences of a reference genome;  and (2)~GenStore-NM, which filters most of the non-matching reads, i.e., reads that would not align to any subsequence in the reference genome.
}

\gf{Our evaluation using a wide range of real genomic datasets shows that GenStore, {when implemented in three {modern} NAND flash-based SSDs,} significantly improves the read mapping performance of state-of-the-art software (hardware) baselines by {2.07--6.05}$\times$ ({1.52--3.32}$\times$) for {read sets with high similarity to the reference genome} and {1.45--33.63}$\times$ (2.70-19.2$\times$) for {read sets with low similarity to the reference genome}.
\gfv{By filtering exact matches in a short read set with 80\% exactly-matching read rate,
GenStore-EM reduces the energy consumption by (up
to) 3.92$\times$ (3.97$\times$), on average across all storage configurations. 
By filtering non-matching reads in a long read set with a read alignment rate of 0.35\%, 
GenStore-NM
reduces the energy consumption by (up to) 27.17$\times$
(29.25$\times$), on average across all storage configurations.
The total hardware area needed for GenStore is very small
(0.20~$mm^2$ at 65~nm and 0.02~$mm^2$  at 14~nm, only 0.006\% of a 14~nm Intel Processor~\cite{arafa2019cascade}).}
\gf{GenStore's source code is available at \url{https://github.com/CMU-SAFARI/GenStore}.}}

\gf{\omvi{Building on insights and experience gained from designing} GenStore, in~\cite{ghiasimegis2024}, we design a \omvi{more comprehensive} storage-based \gls{PNM} solution targeting metagenomic analysis~\cite{hhrlich2011metahit,sunagawa2015structure,fierer2017embracing} (i.e., the key tasks of determining the species present in a sample and their relative abundances). \msvii{Metagenomic analysis~\cite{hhrlich2011metahit,sunagawa2015structure,fierer2017embracing} suffers from significant data movement overhead because it requires accessing \omix{very} large amounts of
low-reuse data. Since we do not know the species present in a
metagenomic sample, metagenomic analysis requires searching large databases \omix{(whose sizes could be hundreds of TBs or larger)} that contain information on different organisms’ genomes. Our motivational
analysis of state-of-the-art metagenomic analysis tools~\omix{\cite{ghiasimegis2024}} executing on processor-centric systems 
shows that data movement overhead from the storage system
significantly impacts their end-to-end performance. Due to its low reuse, the data needs to move all the way from the storage
system to the main memory\omix{, caches,} and processing units for its first use, and it will likely not be used again or reused very little during analysis. This unnecessary data movement, combined with the low computation intensity of metagenomic analysis and the limited I/O (input/output) bandwidth, leads to large storage I/O overheads for metagenomic analysis\omix{, as shown in our ISCA 2024 paper~\cite{ghiasimegis2024}}.} 

\msvii{Our goal in~\cite{ghiasimegis2024} is to improve metagenomic analysis performance by reducing the large data movement overhead
from the storage system in a cost-effective manner. To this end, we propose MegIS~\omv{\cite{ghiasimegis2024}}, the first storage-based PNM system designed to reduce the data movement overheads inside the end-to-end metagenomic analysis pipeline. The \emph{key idea} of MegIS is to enable cooperative storage-based PNM system for metagenomics, where we do not solely focus on processing inside the storage system but, instead, we capitalize on the strengths of processing both inside and outside the storage system. We enable cooperative storage-based PNM system via a synergistic
hardware/software co-design between the storage system and the host system.}

\msvii{The hardware/software co-designed accelerator framework in MegIS~\cite{ghiasimegis2024} consists of five aspects. 
First, we \omix{\emph{partition and map}}
different parts of the metagenomic analysis pipeline to the host
and the storage-based PNM system such that each part is executed on the most
suitable architecture. 
Second, we \omix{\emph{coordinate the data \omx{\&} computation flow}} between the host and the SSD such that MegIS
(i)~completely overlaps the data transfer time between them
with computation time to reduce the communication overhead between different parts, 
(ii)~leverages SSD bandwidth
efficiently, and 
(iii)~does not require large DRAM inside the
SSD or a large number of writes to the flash chips. 
Third, we devise \omix{\emph{storage technology-aware metagenomics algorithm
optimizations}} to enable efficient access patterns to the SSD.
Fourth, we design \omix{\emph{lightweight in-storage \omix{hardware} accelerators}} to perform MegIS' storage-based PNM functionalities while minimizing the required
SRAM \omix{and} DRAM buffer spaces inside the SSD. 
Fifth, we design an \omix{\emph{efficient data mapping scheme and Flash Translation Layer
(FTL)}} specialized to the characteristics of metagenomic analysis to leverage the SSD’s full internal bandwidth. For more detail on the MegIS design and architecture, we refer the reader to
our ISCA 2024 paper~\cite{ghiasimegis2024}, which provides detailed information and results.}


Our evaluation shows that MegIS~\cite{ghiasimegis2024} 
(i)~outperforms the state-of-the-art performance-optimized (P-Opt)~\cite{wood2019improved} and accuracy-optimized (A-Opt)~\cite{lapierre2020metalign} software metagenomic tools by 2.7$\times$--37.2$\times$ and 6.9$\times$--100.2$\times$ (while matching the accuracy of the accuracy-optimized tool), respectively; and
(ii)~achieves 1.5$\times$--5.1$\times$ speedup compared to the state-of-the-art metagenomic hardware-accelerated (using PIM) tool~\cite{wu2021sieve}, while achieving significantly higher accuracy.  
\gfv{MegIS significantly improves energy efficiency compared to state-of-the-art software and hardware metagenomic tools. 
Across our
evaluated SSDs and datasets, MegIS leads to 5.4$\times$ (9.8$\times$), 15.2$\times$ (25.7$\times$), and 1.9$\times$ (3.5$\times$) average (maximum) energy reduction
compared to P-Opt~\cite{wood2019improved}, A-Opt~\cite{lapierre2020metalign}, and the PIM-accelerated P-Opt~\cite{wu2021sieve}. \msvii{MegIS's benefits come at a low area cost of 1.7\% over the area of the three cores~\cite{cortexr4} in an SSD controller~\cite{samsung860pro}.} 

We also show that MegIS \omix{\emph{increases system cost-efficiency}}. 
MegIS \omix{outperforms} a \msvii{costly} performance-optimized \msvii{processor-centric} system (a system with \omix{an expensive high-bandwidth} PCIe Gen4 SSD~\omx{\cite{specification2002pci}} and 1~TB DRAM) by up to 7.2$\times$, even when running on a cost-optimized system (a system with a \omix{cheaper low-bandwidth} SATA3 SSD~\omx{\cite{sata}} and \omix{only} 64~GB DRAM). \msvii{This experiment underscores that PNM can improve \omix{all three major metrics, i.e.,} performance, energy efficiency, and cost efficiency\omix{,} of \omix{a state-of-the-art computing} system by enabling computation to be done in memory or in storage.} \msvii{MegIS's source code is available at \url{https://github.com/CMU-SAFARI/MegIS}.}
}
}

\paratitle{\omix{Summary and Future Outlook}} \gfv{Going forward, we believe that enabling the storage system as a first-class \omix{citizen} in performing application and system computation tasks, including \msvii{executing entire} workloads, is central to building higher performance, lower energy, \msvii{as well as} low-cost sustainable systems.} \omix{Modern storage systems consist of multiple technologies where processing can be performed, such as processing-using-flash/DRAM and processing-near-flash/DRAM. Our vision of storage-centric computing sees the entire storage system as a specialized-enough accelerator that can execute many types of computation efficiently to accelerate important workloads. As such, storage-centric computing can make use of special-purpose accelerators as well as general-purpose computation mechanisms along with multiple different memory types inside the storage system. Most importantly, the storage system becomes a first-class citizen where computation takes place when it makes sense to do so from a user/application- and system-level perspective, which can 
 greatly improve different  metrics, such as performance, energy efficiency, system cost, and sustainability, at the same time, as demonstrated by recent works, e.g., GenStore~\cite{mansouri2022genstore} and MegIS~\cite{ghiasimegis2024}. We believe similar principles followed in GenStore~\cite{mansouri2022genstore} and MegIS~\cite{ghiasimegis2024} can be applied to many other data-intensive workloads, such as 
 \gfix{machine learning (including generative AI and LLMs), databases, graph analytics, high-performance computing, and a wide variety of mobile and server-class workloads,}
 to fundamentally and sustainably accelerate different data-intensive applications. However, storage-centric designs also face several key limitations, such as reliability of computation, limited endurance (i.e., limited number of program and erase cycles) \omx{and higher latencies} of flash memories, small internal DRAMs, and limited power and area budgets, as we have faced and addressed in our recent storage-centric designs~\cite{mansouri2022genstore,ghiasimegis2024,flashcosmos}. We believe there is significant opportunity for creating novel and robust storage-centric \omx{computing system} designs by exploiting rich computational possibilities inside the modern storage systems, while addressing the key design limitations.}

%% file: sections/08-enabling-adoption.tex
\section{Enabling the Adoption of PIM}
\label{sec:adoption}

\begin{figure}[ht]
\centering
\includegraphics[width=\linewidth]{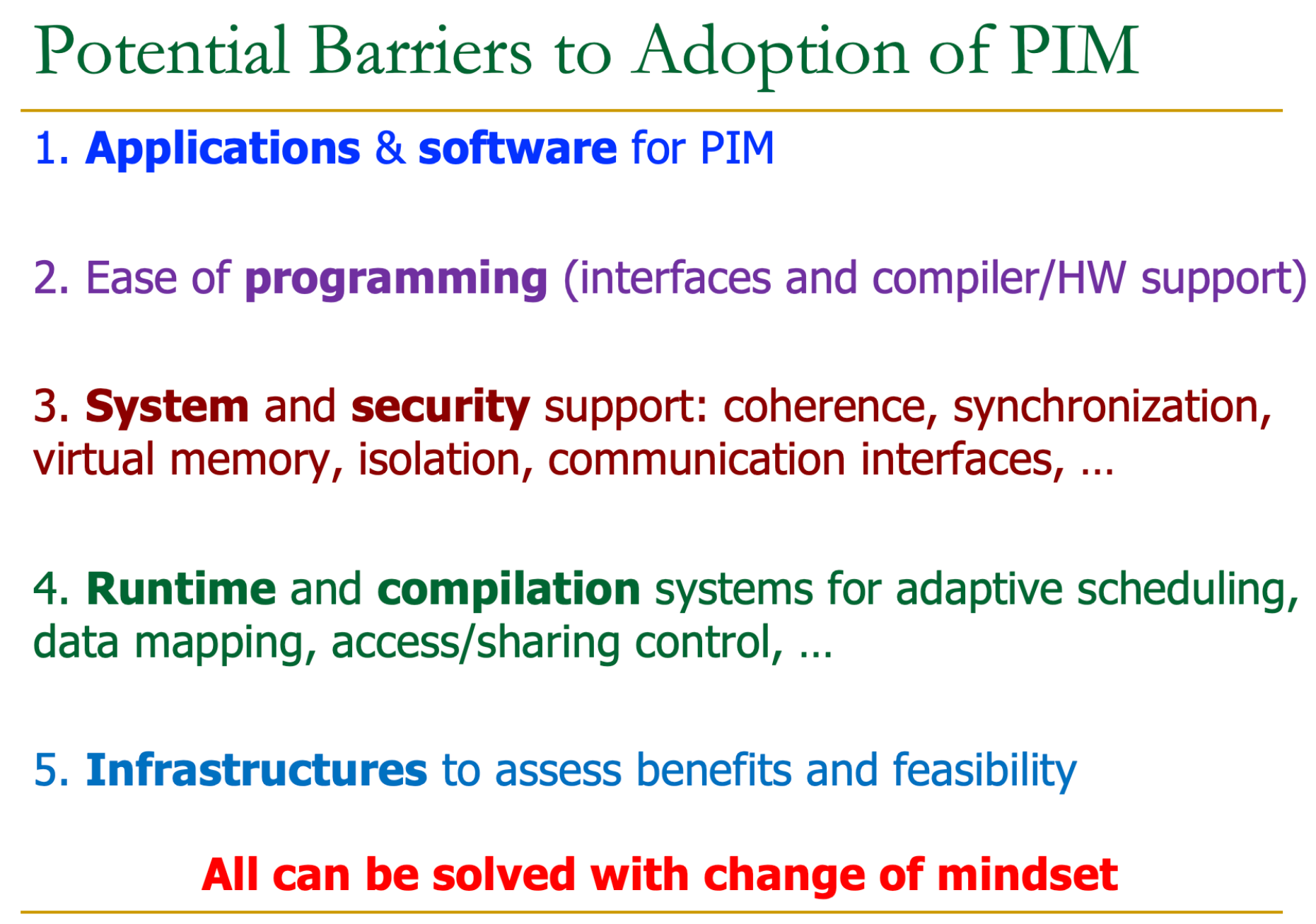}
\caption{Potential barriers to adoption of PIM. Reproduced from~\cite{mutlu.nsfpim20,mutlu.iccdtalk19,mutlu.meerut22}.}
\label{fig:pim-adoption}
\end{figure}

Pushing some or all of the computation for a program from the CPU to
{memory} introduces new challenges for system architects
{and programmers} to overcome.  
Figure~\ref{fig:pim-adoption} lists some of these key challenges. 
These challenges must be
addressed carefully and systematically in order for PIM to be adopted as a mainstream architecture
in a wide variety of systems and workloads, and in a seamless manner
that does not place \omv{a} heavy burden on the vast majority of programmers.
In this section, we discuss several of these system-level {and programming-level} challenges, and highlight a number of works
that \omv{address} these challenges for a wide range of PIM
architectures. 
We believe future research should \omv{continue to} examine solutions to these challenges with an open mindset that is keen on enabling adoption, since the widespread success of the PIM paradigm critically depends on effective solutions to these challenges. 

\subsection{Programming Models and Code Generation for PIM}
\label{sec:mapping}




Two open research questions to enable the adoption of PIM are 
(1)~what should the programming {model\omv{(s)}} be \omv{to enable ease of use by a wide range of programmers}, and 
(2)~how can compilers and libraries \omv{(as well as hardware support for them)} alleviate the programming burden?

While PIM-Enabled Instructions~\cite{ahn.pei.isca15} \omv{(see Section~\ref{sec:pei})} work well for
offloading fine-grained and small amounts of computation to memory, they can
potentially \chii{introduce overheads \omv{when one wants to take}} advantage of PIM
for large tasks, due to the need to frequently exchange information
between \chiii{the PIM processing logic} and the CPU.  Hence, there is
a need for researchers to investigate how to integrate PIM\omv{-enabled} instructions \omv{(or more broadly, PIM-enabled tasks)} with other compiler-based methods or library calls that
can support PIM integration, and how these approaches can ease the
burden on the programmer, by enabling seamless offloading of \omv{tasks of various granularities (e.g.,} instructions\omv{,} function/library calls\omv{, or even larger code patterns)}.

Such solutions can often be platform-dependent. 
\gf{We illustrate how different hardware platforms (i.e., GPUs, CPUs, and PIM architectures) can \omv{exploit} different mechanisms for PIM offloading.}
\omv{We \omvi{provide examples of some} models that reduce \omvi{programmer} burden in exploiting \gfv{PNM and PUM systems}.}

\subsubsection{\gfv{Programming Support for PNM Systems}}

First, one of our recent works~\cite{hsieh.isca16} \omv{(see Section~\ref{sec:gpu})} examines compiler-based mechanisms to decide what portions of code should be offloaded to PIM processing logic in a GPU-based system in a manner that is transparent to the GPU programmer.  Another recent work~\cite{pattnaik.pact16} examines system-level techniques that decide which GPU application kernels are suitable for PIM execution.

\gf{Second, in our work ALP~\cite{ghiasi2022alp}, we consider the case where the execution of an application is divided \gfv{into \emph{segments} (a sequence of instructions), and different segments can execute either in} CPU cores \gfv{or} PIM cores. 
Such an execution model introduces inter-segment data movement overhead~\omvi{\cite{suleman2010data,suleman2011data}}, i.e., \omv{performance and energy overheads due to} the cost of moving data between segments of an application that executes in different \omv{types of hardware} (i.e., CPU \omv{cores} and PIM cores). 
ALP alleviates the inter-segment data movement overhead by \emph{proactively and accurately} transferring the required data between the segments mapped \omv{onto CPU cores} and PIM cores. ALP uses a compiler pass to identify these instructions and uses specialized hardware support to transfer data between the CPU \omv{cores} and PIM cores at runtime. Using both the compiler and runtime information, ALP  efficiently maps application segments to either CPU or PIM cores, considering 
(1)~the properties of each segment \gfv{(e.g., L1/L3 cache misses)}, 
(2)~the inter-segment data movement overhead between different segments and 
(3)~whether this inter-segment data movement overhead can be alleviated proactively and in a timely manner.}
\omvi{The ALP paper~\cite{ghiasi2022alp} describes in detail the corresponding mechanisms and results obtained on real workloads.}

\gfvi{Third, we can develop high-level APIs that abstract away underlying hardware characteristics from the programmer. To facilitate the adoption of PIM system, we propose \omv{two} high-level programming frameworks, SimplePIM~\cite{chen2023simplepim} and DaPPA~\cite{oliveira2023dappa}.}

\gfv{SimplePIM \omvi{is a high-level programming framework \gfvi{\omvii{whose} \emph{key idea is to}} make use of software abstractions widely used in distributed system \omvii{programming} frameworks (e.g., Spark~\cite{RDD}, MapReduce~\cite{dean2004mapreduce}) to aid PIM programmability. SimplePIM} offers three key interfaces to support PIM systems. 
The {\emph{management interface}} stores {metadata} for the PIM-resident arrays, which can be accessed by the {programmer} and other parts of SimplePIM as needed. 
The {\emph{communication interface}} provides abstractions for both Host-PIM and PIM-PIM communication patterns. 
These patterns are similar to communication patterns in other distributed frameworks, such as MPI~\cite{mpi}, {which makes it easier for SimplePIM to be adopted}. 
{The \emph{processing interface}} leverages PIM's high memory bandwidth and parallelism to execute \texttt{map}, \texttt{reduce}, and \texttt{zip} iterators on PIM arrays. 
{Programmers} can combine these iterators to implement many widely-used workloads ranging from simple vector addition to complex machine learning model training.} 
\gfv{We implement and evaluate SimplePIM on a real PIM system, UPMEM~\cite{upmem} (which we describe in Section~\ref{sec:upmem}), with six different applications: reduction, vector addition, histogram, linear regression, logistic regression, and K-means.
These applications have previously been implemented on UPMEM~\cite{gomezluna2021repo, gomez2022machine, gomez2023evaluating, gomez2022experimental,gomezluna2022ieeeaccess} \omvi{by hand}, providing a baseline for comparing performance, correctness, and code complexity.
SimplePIM offers a {programmer}-friendly interface and requires $4.4\times$ fewer lines of code, on average, compared to {the best} existing open-source {hand-optimized} implementations.
In addition, we apply several code optimizations to tailor our SimplePIM implementation to the underlying hardware, making it \omvi{more} suitable \omvii{for} \omvi{and higher \omvii{performance} on the} UPMEM \omvi{system}.
Our evaluation results show that SimplePIM performs similarly to hand-optimized implementations in three applications, and outperforms them in the remaining three, despite its general-purpose \omvi{and easy-to-use} design. Specifically, for vector addition, logistic regression, and K-means, SimplePIM performs 1.10$\times$, 1.17$\times$, and 1.37$\times$ faster than {the best prior hand-optimized implementations} in weak scaling tests and 1.15$\times$, 1.22$\times$, and 1.43$\times$ faster in strong scaling tests.} {\gf{The source code of SimplePIM is available at \url{https://github.com/CMU-SAFARI/SimplePIM}.} }
\omvi{We hope and believe such a high-level framework will enable PIM hardware to be more easily exploited by a wide variety of workloads.}

\begin{figure}[ht]
    \centering
    \includegraphics[width=1.0\linewidth]{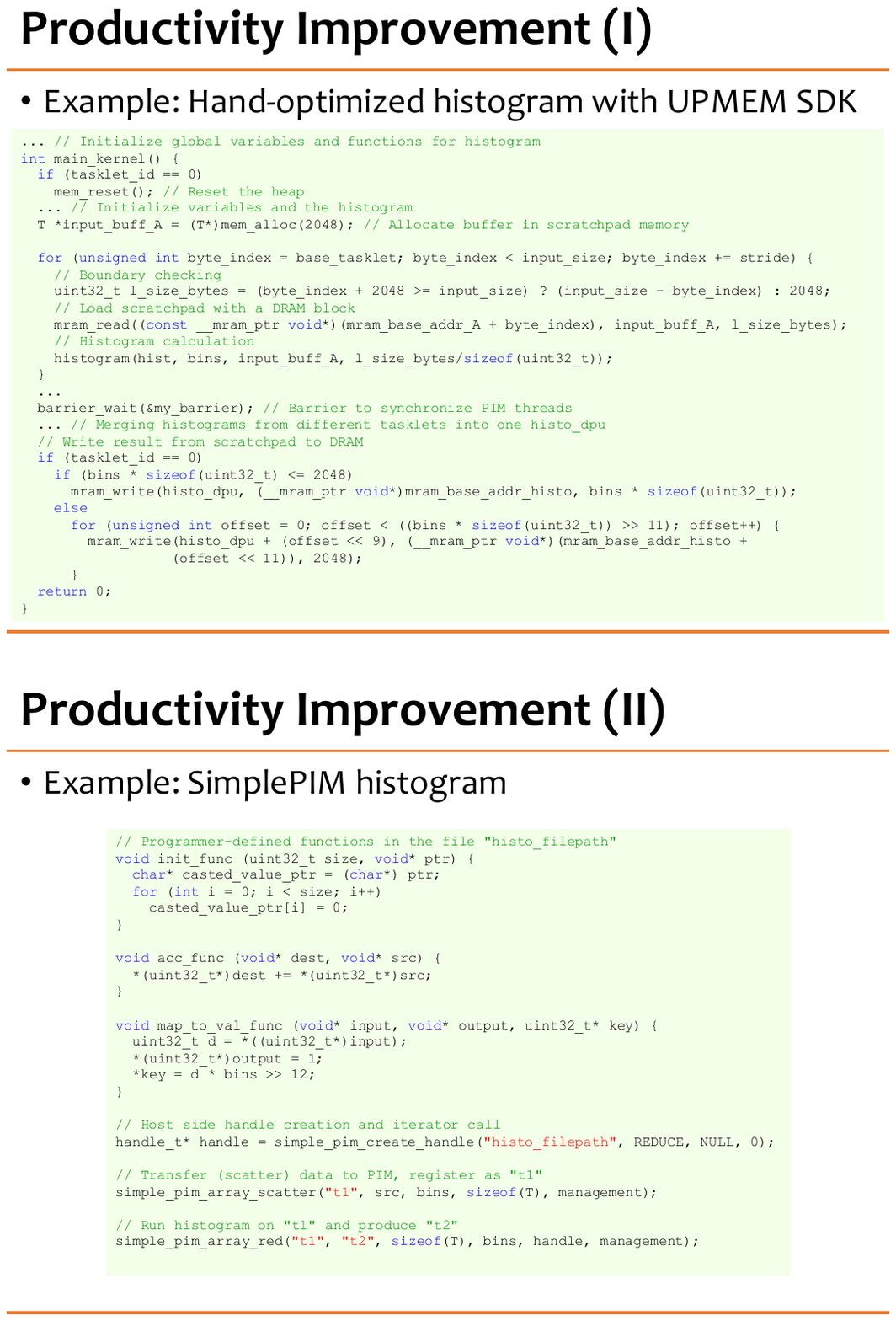}
    \caption{\gfix{Comparison between hand-optimized \texttt{histogram} code using the UPMEM SDK~\cite{upmem} (top) and SimplePIM \texttt{histogram} code (bottom). 
    \omx{The SimplePIM version (bottom) leads to 5.43$\times$ \omxi{fewer} lines of code  and is also easier to read and understand.} Reproduced from~\cite{simplepim.talk}.}}
    \label{fig:simplepimcode}
\end{figure}

\begin{figure}[!t]
    \centering
    \includegraphics[width=\linewidth]{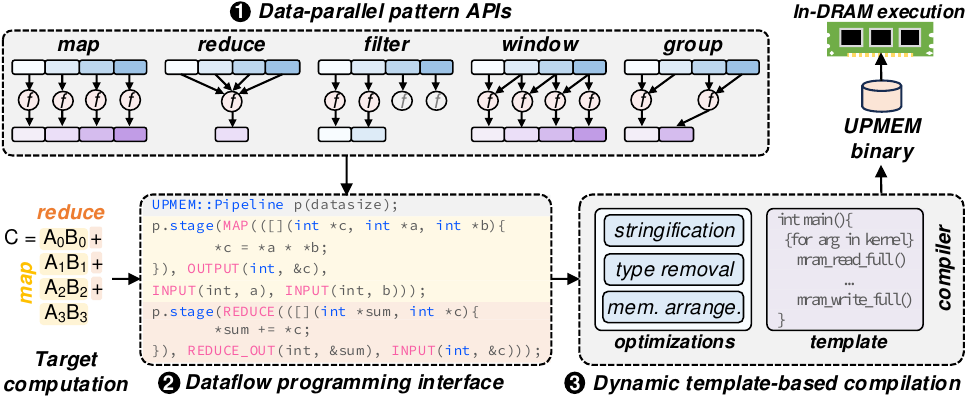}
    \caption{\gfvii{Overview of the DaPPA framework. Reproduced from~\cite{oliveira2023dappa}.}}
    \label{fig:dataframework_overview}
\end{figure}

\gfv{DaPPA~\cite{oliveira2023dappa} (\underline{\omvi{Da}}ta-\underline{\omvi{P}}arallel \underline{\omvi{P}}rocessing-in-memory \underline{\omvi{A}}rchitecture) is a framework that can, for a given application, \emph{automatically} distribute input \omvi{data} and gather output data, handle memory management, and parallelize work across the DPUs. 
The \emph{key idea} behind DaPPA is to remove the responsibility of managing hardware resources from the programmer by providing an intuitive data-parallel pattern-based programming interface~\cite{cole1989algorithmic,cole2004bringing} that abstracts the hardware components of the PIM system.
\gfvi{Using this key idea, DaPPA transforms a data-parallel pattern-based application code into the appropriate PIM-target code, including the required APIs for data management and code \omvi{partitioning} (which are the programmer’s responsibility in SimplePIM~\omvi{\cite{chen2023simplepim}}). Then, DaPPA compiles the data-parallel pattern-based application into a PIM binary \emph{transparently} from the programmer.}
While generating PIM-target code, DaPPA implements several code optimizations to improve end-to-end performance.
Figure \ref{fig:dataframework_overview} shows an overview of \omvii{the} DaPPA framework, which consists of three main components:
(1)~data-parallel pattern APIs,
(2)~dataflow programming interface, and
(3)~dynamic template-based compilation.
DaPPA's data-parallel pattern APIs (\circled{1} in Figure~\ref{fig:dataframework_overview}) are a collection of pre-defined functions that implement high-level data-parallel pattern primitives\omvi{, i.e.,} \texttt{map}, \texttt{filter}, \texttt{reduce}, \texttt{window}, and \texttt{group}.
DaPPA's dataflow programming interface (\circled{2}) allows the programmer to represent an application as a \texttt{Pipeline}, which represents a sequence of data-parallel \omvi{pattern} \texttt{stage}s that will be executed in order on the PIM system.
DaPPA's dynamic template-based compilation (\circled{3}) generates PIM binaries using \gfvi{\emph{code templates}}, which are dynamically filled to represent each implemented  \texttt{Pipeline} in the application. 
\gfvi{\emph{\omvii{C}ode templates} contain hardware-specific code routines required for PIM execution, including  input/output data \& code partitioning, Host--PIM \& PIM--PIM data orchestration and synchronization.}
Our evaluation results show that DaPPA improves end-to-end performance by 2.1$\times$, on average across six hand-optimized PrIM \omvi{benchmarks}~\cite{gomezluna2021repo, gomez2022machine, gomez2023evaluating, gomez2022experimental,gomezluna2022ieeeaccess} running on a real UPMEM PIM system.
DaPPA's performance improvement is due to code optimizations, such as data transfer \omvi{parallelization} and \omvi{better} workload partition between CPU and PIM cores. 
DaPPA also \emph{significantly} reduces programming complexity (measured using \omvi{lines}-of-code) on average by 94.4\% (min. 92.3\%, max. 96.1\%).
\omvi{More information about DaPPA can be found in~\cite{oliveira2023dappa}.}} 

As described in Section~\ref{sec:PNM} with multiple promising examples, different granularities of code offloading in \gf{\gls{PNM}} architectures have different implications for performance and energy as well as system complexity. These different granularities also have implications on programming and code generation complexity. Adoption-minded solutions should clearly take into account the granularity of code offloading and how a PNM system supports code execution. 

\omvi{We believe that high-level programming frameworks like SimplePIM and DaPPA are central to develop and expand into the future to greatly reduce the burden of programming and automate the task of code \omvii{generation and} optimization for PIM hardware.}

\subsubsection{\gfv{Programming Support for PUM Systems}}

Similarly, programming and code generation frameworks \omv{that are specialized} for \gf{PUM} approaches like Ambit~\omv{\cite{seshadri.micro17} (Section~\ref{sec:ambit})} are also critical for such approaches to become widely adopted. Programming model, compiler and library support for expressing, extracting and generating \omv{PIM operations (e.g.,} bulk bitwise operations\omv{, matrix-vector multiplication, or LUT-based execution)} in a program can greatly help the adoption of \omv{PUM} execution models like Ambit. We believe there is exciting research to do in these directions. \omv{We provide an example of how to seamlessly compile code into bulk bitwise PUM operations, via our new hardware/software co-designed MIMDRAM framework~\cite{mimdram}.}

\begin{figure}[ht]
    \centering
    \includegraphics[width=\linewidth]{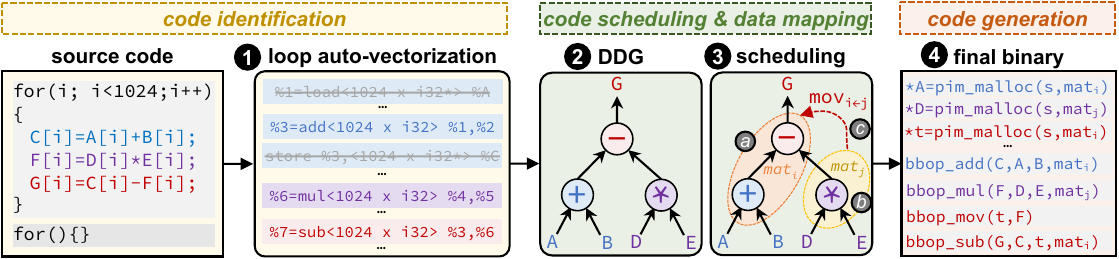}
    \caption{\gf{MIMDRAM's compilation flow. Reproduced from~\cite{mimdram}.}}
    \label{fig_compilation_flow}
\end{figure}

\gf{MIMDRAM~\cite{mimdram} \omv{provides} a step towards automatic code generation for Ambit-like \omv{bulk bitwise PUM} architectures.}
\gf{Figure~\ref{fig_compilation_flow} illustrates MIMDRAM compilation flow, which is implemented using LLVM~\cite{lattner2008llvm}{: it takes} a C/C++ application's source code as input, {performs} \gf{three} transformation passes, and {outputs} a binary with a mix of CPU and {{PUD} instructions.} The first pass is responsible for \emph{code identification}. Its goal is to identify 
(1)~loops that can be successfully auto-vectorized and 
(2)~the appropriate vectorization factor of a given vectorized loop. The code identification pass takes as input the application's LLVM intermediate representation (IR) generated by the compiler's front-end. It produces as output an optimized IR containing SIMD instructions that will be translated to Ambit's bulk bitwise instructions \omv{(see Section~\ref{sec:ambit})}.  MIMDRAM leverages the native LLVM's loop auto-vectorization pass~\cite{AutoVect65:online} to identify and transform loops into their vectorized form {(\circled{1} in \gf{Figure}~\ref{fig_compilation_flow})}. The \gf{second} pass is responsible for \emph{code scheduling and data mapping} (\circled{2}--\circled{3}). Its goal is to improve overall SIMD utilization by allowing the distribution of independent PUD instructions across DRAM mats. 
The code scheduling pass operates on the following premises. 
Since PUD instructions operate directly on the data stored in DRAM, the DRAM mat where the data is allocated determines the efficiency and utilization of the PUD SIMD engine. If operands of {independent} instructions are distributed across different DRAM mats, such instructions can be executed concurrently. Likewise, operands of dependent instructions are mapped to the same DRAM mat.
The {third} pass is responsible for (1)~\emph{data allocation} and (2)~\emph{code generation} (\circled{4}). It takes as input the LLVM IR containing both CPU and \emph{bbop} instructions {(with metadata)} and produces a binary to the target ISA. 
}
\gfvi{As we show in Section~\ref{sec:mimdram}, MIMDRAM \emph{significantly} improves energy efficiency and system throughput of a wide range of general-purpose applications. The MIMDRAM compiler \omvii{serves} an \emph{integral} part in obtaining these results.}

\subsubsection{\gfv{Summary \& Future Work}}

\gf{Despite the significant effort and \omvi{especially recent} progress on tools and frameworks that can ease PIM programmability~\gfvi{\cite{farzaneh2024c4cam,qu2024cim,mimdramextended,ahn.pei.isca15,hsieh.isca16,pattnaik.pact16,ghiasi2022alp,chen2023simplepim,oliveira2023dappa,wang2023infinity,schwedock2024leviathan,ahmed2019compiler,hadidi2017cairo,giannoula2024pygim,peng2023chopper,leitersdorf2024pypim}},} \omvi{finding} {effective \omvi{and easy-to-adopt} programming interfaces} and the necessary (as well as useful) compiler/library support to effectively \omvi{exploit PIM hardware} \omvii{continue to be central} open research\omvi{, design and development issues}, which are important for \omvi{ongoing and} future works to tackle.

\gfv{In summary,  there are several key research questions that should be investigated {in programming models and code generation for PIM}. \omvi{These include, but are not limited\omvii{,} to the following}:}
\begin{itemize}
[leftmargin=5mm,itemsep=0mm,parsep=0mm,topsep=0mm]
    \item \gfv{How can compilers effectively map high-level abstractions onto diverse PIM hardware \omvi{(e.g., different types of PNM and PUM \omvii{at different parts of the system})} while minimizing data movement \omvi{and maximizing parallelism}?}
    
    \item \gfv{How can compiler frameworks optimize data placement and movement to minimize memory access latency in PIM?  How can code generation strategies balance computation offloading and synchronization overhead in PIM systems?}

    \item \gfv{What strategies can be used to automate the parallelization of loops and data partitioning for PIM execution? \omvii{For example}, what techniques can be employed to efficiently map irregular loops (and workloads) onto heterogeneous PIM cores \omvi{or heterogeneous CPU--PIM systems}?}  

    \item \gfv{How can compilers ensure efficient utilization of both in-memory and host processing units in hybrid PIM architectures? In particular, how can code generation approaches handle varying memory bandwidth and compute capabilities in a system with different PIM devices (e.g., a PIM system with both PNM and PUM capabilities \omvi{or a system with different types of PIM capabilities \omvii{at different parts of the system}})?}   
\end{itemize}


\subsection{PIM Runtime: Scheduling and Data Mapping}
\label{sec:scheduling}
We identify \omv{at least} four key runtime issues in PIM: 
(1)~what {code} to
execute near data, 
(2)~when to schedule execution on PIM (i.e., when
is it worth offloading computation to the PIM cores), 
(3)~how to map
data to memory \omv{arrays (at different granularities, including module, bank, subarray, mat)} such that PIM execution is viable and
effective, and (4)~how to effectively share/partition PIM
mechanisms/accelerators at runtime across multiple threads/cores to
maximize performance and energy efficiency.  We have proposed
several approaches to solve these four issues, yet much research remains to be done to enable a robust\omvi{,} effective\omvi{, and efficient} PIM runtime system that can be effective under many conditions.

\begin{figure}[ht]
\centering
\includegraphics[width=1.0\linewidth]{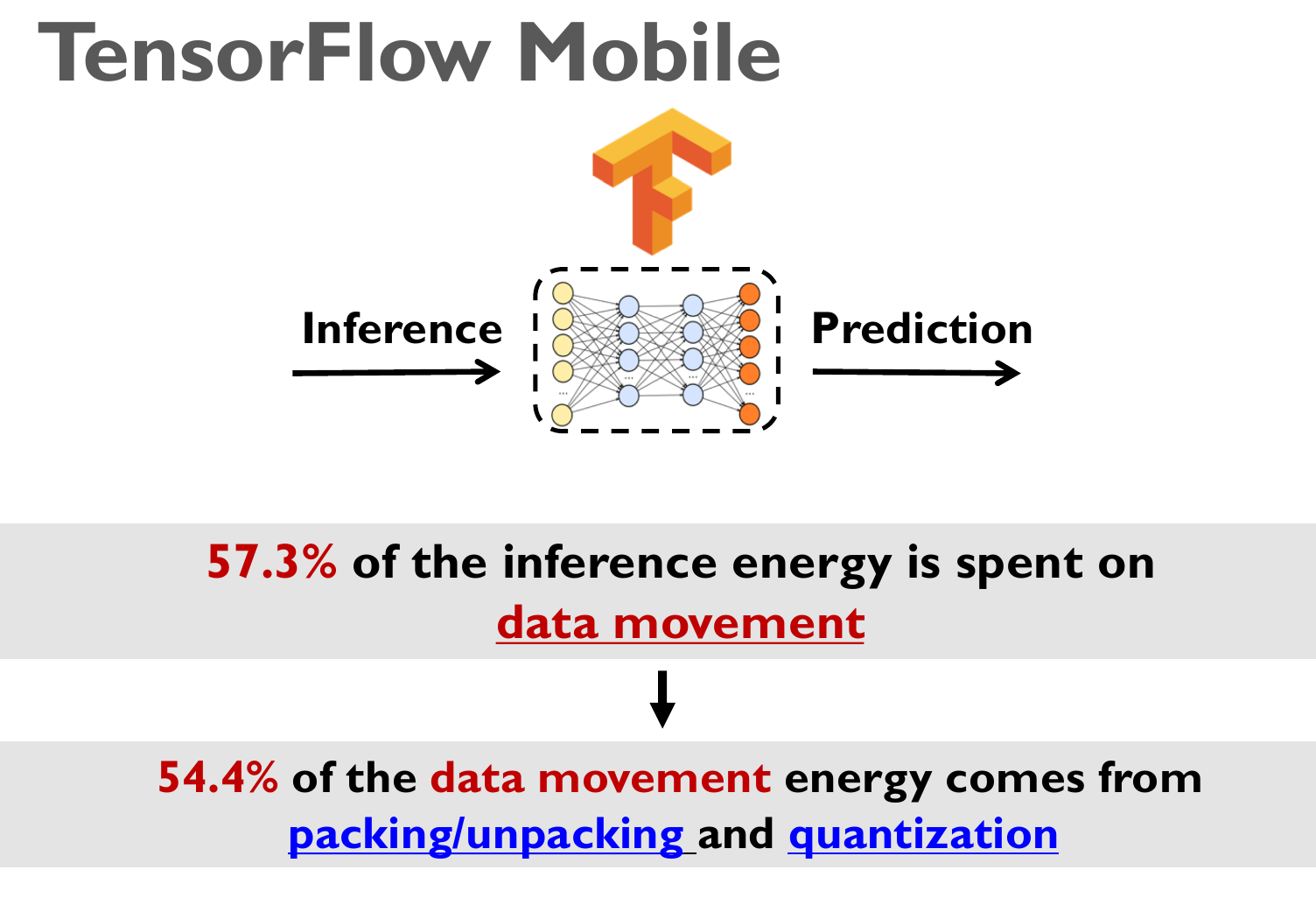}
\caption{A majority of the data movement energy in TensorFlow Mobile machine learning inference framework~\cite{abadi2016tensorflow,mobile-tensorflow} 
is caused by two key functions. Reproduced from~\cite{mutlu.nsfpim20}. 
Originally presented in~\cite{boroumand.asplos18,boroumand.asplos18talk}.
}
\label{fig:google-methodology}
\end{figure}

The first key issue is to identify which portions of an application are suitable for PIM. We call such portions \emph{PIM offloading candidates}. While PIM offloading candidates can be identified manually by a programmer, the identification would require significant programmer effort along with a detailed understanding of the hardware tradeoffs between CPU cores and PIM cores. For architects who are adding custom PIM logic (e.g., fixed-function accelerators, which we call PIM accelerators) to memory, the tradeoffs between CPU cores and PIM accelerators may not be known before determining which portions of the application are PIM offloading candidates, since the PIM accelerators are tailored for the PIM offloading candidates.
To alleviate the burden of manually identifying PIM offloading candidates, we develop a systematic toolflow for identifying PIM offloading candidates in an application~\gfv{\cite{boroumand.asplos18,boroumand.arxiv17,boroumand2016pim,boroumand2019conda,amiraliphd}}. This toolflow uses a system that executes the entire application on the CPU to evaluate whether each PIM offloading candidate meets the constraints of the system under consideration. For example, when we evaluate workloads for mobile consumer devices (e.g., Chrome web browser, TensorFlow Mobile, video playback, and video capture)~\cite{boroumand.asplos18}, we use hardware performance counters and our energy model to identify candidate functions that could be PIM offloading candidates. A function is a PIM offloading candidate in a mobile consumer device if it meets the following conditions:
\begin{enumerate}
[leftmargin=5mm,itemsep=0mm,parsep=0mm,topsep=0mm]
\item It consumes a significant fraction (e.g., more than 30\%) of the overall workload energy consumption, since energy reduction is a primary objective in mobile systems and workloads.
\item Its data movement consumes a significant fraction (e.g., more than 30\%) of the total workload energy to maximize the potential energy benefits of offloading to PIM. 
\item It is memory-intensive (e.g., its last-level cache misses per kilo instruction, or MPKI, is greater than 10~\cite{chou2015reducing,kim2010atlas,muralidhara2011reducing,kim2010thread}), as the energy savings of PIM is higher when more data movement is eliminated.
\item Data movement is the single largest component of the function’s energy consumption.
\end{enumerate}
Figure~\ref{fig:google-methodology} shows two example functions in Google's Mobile TensorFlow machine learning inference framework~\cite{mobile-tensorflow,abadi2016tensorflow} 
that are identified to be PIM offloading candidates using the afore-described methodology:  {\em packing/unpacking} and {\em quantization}~\cite{boroumand.asplos18}. Note that these functions are together responsible for more than 54\% of the data movement energy in the examined neural networks for this workload, which spend more than 57\% of their execution energy on data movement.  

Some of our other recent works in PIM identify suitable 
PIM offloading candidates 
with different granularities. 
\omvi{The} PIM-Enabled
Instructions \omv{work}~\cite{ahn.pei.isca15} {\omv{proposes} various} operations
that can benefit from execution near or inside memory, such as integer
increment, integer minimum, floating-point addition, hash table
probing, histogram bin index, Euclidean distance, and dot product.
\gfv{In ALP~\cite{ghiasi2022alp}, we perform \emph{online profiling} of code segments in an application running on the host CPU to identify their PIM suitability. During the online profiling, we measure the number of L1 cache misses and LLC misses, and if the ratio of the L1 cache misses to the ratio of the LLC misses (a metric called \emph{last-to-first miss-ratio}, introduced by~\cite{oliveira2021pimbench}) is close to one \omvi{(i.e., caching effectiveness beyond \omvii{the} Level 1 cache is very low)}, we
offload the execution of such code segment to a PIM core.}
GPU applications also contain several parts \omv{(e.g., functions, \omvi{warps})} that are suitable
{for offloading to PIM
  engines}~\cite{hsieh.isca16,pattnaik.pact16}.
Bulk memory operations (copy,
\sg{initialization}) and {bulk} bitwise operations are good candidates
for Ambit-like processing-using-DRAM approaches~\cite{seshadri2013rowclone,seshadri.micro17, hajinazarsimdram,Seshadri:2015:ANDOR, seshadri.thesis16}, as we discussed earlier \omv{(in Section~\ref{sec:pum:pud})}. 
For PUM approaches that can execute more complex operations (e.g., addition, multiplication) using memory, the \emph{operation complexity} (i.e., the latency of an operation for a certain data type) can determine how beneficial offloading to PUM can be compared to CPU execution. 
A recent analytical model~\cite{korgaonkar2019bitlet} helps to evaluate such offloading trade-offs in memristor-based PUM \omv{systems}~\cite{kvatinsky.tcasii14,kvatinsky.iccd11,kvatinsky.tvlsi14}.

In several of our research works, we propose runtime mechanisms for
\emph{dynamic scheduling} of PIM offloading candidates, i.e.,
mechanisms that decide whether or not to actually offload code that is
marked to be potentially offloaded to PIM engines.
In~\cite{ahn.pei.isca15}, we develop a locality-aware scheduling
mechanism {for PIM-enabled instructions}.  
\gfv{In~\cite{ghiasi2022alp}, we add 
(1)~an \emph{Offload Management Unit} to the host chip to handle the
offload of code segments to PIM cores and
(2)~a \emph{Monitor Unit} in both the PIM and
CPU cores to collect the necessary runtime information (e.g., L1 cache misses, LLC misses) to feed the  \emph{Offload Management Unit}.}
For GPU-based
systems~\cite{hsieh.isca16,pattnaik.pact16}, we explore the
combination of compile-time and runtime mechanisms for identification
and dynamic scheduling of PIM offloading candidates.

The best \emph{mapping of data and code} that enables the maximal
benefits from PIM depends on the applications and the computing system
configuration as well as the type of PIM employed in the system. 
For instance, in order to be able operate on two source arrays inside DRAM with PUM approaches~\cite{chang.hpca16,seshadri.micro17, hajinazarsimdram,seshadri2013rowclone,seshadri2020indram,seshadri.bookchapter17.arxiv,seshadri.bookchapter17,seshadri.arxiv16,Seshadri:2015:ANDOR,angizi2019graphide, ferreira2021pluto}, one key issue is how to guarantee the alignment of the two arrays inside the same DRAM subarray. Practical solutions for this issue need to involve both the memory controller and the operating system to enable that arrays aligned in virtual memory can also be physically aligned in DRAM. \omv{T}he programmer and/or the compiler also likely need to carefully annotate and communicate computation patterns on large data blocks so that the system software and the memory controller can cooperatively map the data blocks in an appropriate manner that is amenable to bulk bitwise computation via PUM.
\gfv{We address such data mapping and alignment for PUM in both our PiDRAM~\cite{olgun2022pidram} and MIMDRAM~\cite{mimdram} works. 
In both works, we implement a new memory allocation API that can guarantee that the source and destination operands of a PUM operation are mapped to the same DRAM subarray. 
At a high level, the new memory allocation API in PiDRAM (i)~splits
the source and destination operands into page-sized virtually-addressed memory blocks, 
(ii)~allocates two physical pages in
different DRAM rows in the same DRAM subarray, and (iii)~assigns
these physical pages to virtual pages that correspond to the
source and destination memory blocks at the same index. 
MIMDRAM's memory allocation API further extends PiDRAM's memory allocation API to guarantee that a PUM operation's source and destination operands are mapped to the same DRAM mat within a DRAM subarray.}
Another key issue is how to move partial results generated in one DRAM subarray to other DRAM subarrays to continue the execution with other input operands residing in those subarrays. Several of our recent works~\omvi{\cite{seshadri2013rowclone,chang.hpca16,rezaei2020nom,wang2020figaro, yuksel2024simultaneous}} propose mechanisms for in-DRAM internal data movement that can facilitate gathering of data in appropriate rows/subarrays/banks in a DRAM chip.

Programmer-transparent data and code mapping mechanisms are especially desirable for PIM adoption. {In~\cite{hsieh.isca16}, we present}
a software/hardware cooperative mechanism to map data and code to several
3D-stacked memory chips in regular GPU applications with relatively
regular memory access patterns. This work also deals with effectively
\emph{sharing PIM {engines} across multiple threads}, as GPU
code sections can be offloaded from different GPU cores to the PNM GPU cores in 3D-stacked memory chips.  Developing
{new approaches} to data/code mapping and scheduling for a wide
variety of applications and possible core and memory configurations is
still necessary.

In summary,  there are still several key research questions that should be investigated {in runtime systems for PIM, which perform scheduling and data/code mapping}:
  
\begin{itemize}
[leftmargin=5mm,itemsep=0mm,parsep=0mm,topsep=0mm]
\item What are simple mechanisms to enable and disable PIM execution? How can PIM execution be throttled for highest performance gains? How should data locations and access patterns affect whether \omv{and where} PIM execution should occur?
\item Which parts of {a given application's} code should be executed on PIM? What are simple mechanisms to identify when those parts of the application code can benefit from PIM?
\item What are scheduling mechanisms to share PIM {engines} between multiple requesting cores to maximize \chvii{benefits obtained from PIM}? 
\item What are simple mechanisms to {manage access to a memory that serves both CPU requests and PIM requests?}
\item \omv{What are the communication mechanisms that seamlessly enable fast and efficient data and code movement between PIM cores?}
\end{itemize}

\subsection{Memory Coherence}
\label{sec:coherence}


In a traditional multithreaded execution model that makes use of
shared memory, writes to memory must be coordinated between multiple
CPU cores, to ensure that threads do not operate on stale data values.
Since CPUs include per-core private caches, when one core writes data
to a memory address, cached copies of the data held within the caches
of other cores must be updated or invalidated, using a
{mechanism} known as \emph{cache coherence}.  Within a modern
chip multiprocessor, the per-core caches perform coherence actions
over a shared interconnect, with hardware coherence protocols.

Cache coherence is a major system challenge for enabling PIM
architectures as general-purpose execution engines, \omv{since} \chiii{PIM
  processing logic} can modify the data \chiii{it processes}, and this
data may also be needed by \omv{other} cores \omv{(e.g., CPU, GPU, accelerators)}.  If \chiii{PIM processing logic
  is} coherent with the processor, the PIM programming model is
relatively simple, as it remains similar to conventional \ch{shared
  memory} multithreaded programming\ch{, which makes PIM architectures
  easier to adopt in general-purpose systems}.  Thus, allowing
\chiii{PIM processing logic} to maintain such a simple and traditional
shared memory programming model can facilitate the \ch{widespread}
adoption of PIM.  However, employing traditional fine-grained cache
coherence {(e.g., a cache-block based MESI
  protocol~\cite{mesi1984})} for PIM forces a large number of
coherence messages to traverse {the narrow processor-memory
  bus}, potentially undoing the benefits of high-bandwidth \chvi{and
  low-latency} PIM \chv{execution}.  \chv{Unfortunately,} solutions
for coherence proposed by prior PIM
works~\gfv{\cite{ahn.pei.isca15,hsieh.isca16,ahn.tesseract.isca15,drumond2017mondrian,farmahini2015nda,gao2017tetris,nai2017graphpim,pugsley2014ndc,gao.pact15,lee.hpca01,tsai:micro:2018:ams,zhang.hpdc14,JAFAR,boroumand.asplos18}} either place some
restrictions on the programming model (by eliminating coherence and
requiring message passing based programming) or limit the performance
\ch{and energy} gains achievable by a PIM architecture.

\begin{figure}[ht]
\centering
\includegraphics[width=1.0\linewidth]{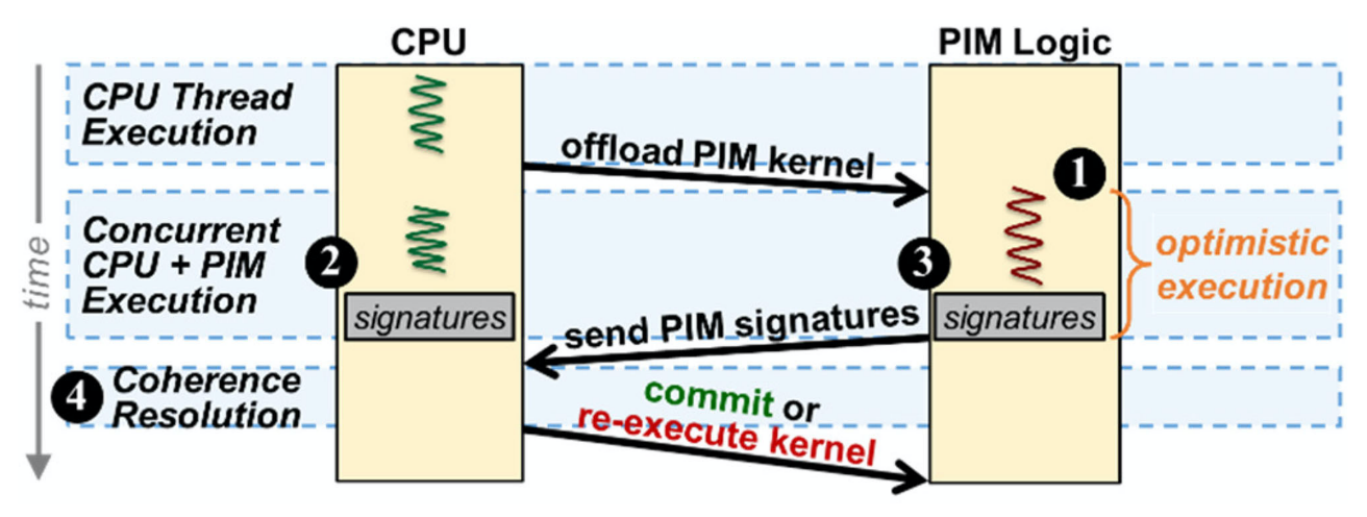}
\caption{High-level operation of CoNDA, a new coherence mechanism for near-data accelerators, including PNM and PUM. Reproduced from~\cite{ghose2019arxiv}. Originally presented in~\cite{boroumand2019conda}.}
\label{fig:conda}
\end{figure}

We have developed a new coherence protocol,
\emph{CoNDA}~\cite{boroumand2019conda,boroumand2016pim,boroumand.arxiv17}, that maintains
cache coherence between PIM processing logic and CPU cores
\emph{without} sending coherence requests for every memory access.
Instead, as shown in Figure~\ref{fig:conda}, CoNDA enables efficient coherence by having the
PIM logic:
\begin{enumerate}
[leftmargin=5mm,itemsep=0mm,parsep=0mm,topsep=0mm]
\item \emph{speculatively} acquire coherence permissions for multiple memory operations over a given period of time (which we call \emph{optimistic execution}; \circled{1} in the figure);
\item \emph{batch} the coherence requests from the multiple memory operations into a set of compressed coherence \emph{signatures} (\circled{2} and 
\circled{3});
\item send the signatures to the CPU to determine whether the speculation violated any coherence semantics.
\end{enumerate}
Whenever the CPU receives compressed signatures from the PIM core (e.g., when the PIM kernel finishes), the CPU performs \emph{coherence resolution} (\circled{4}), where it checks if any coherence conflicts occurred. If a conflict  \omv{is found to exist}, any dirty cache line in the CPU that caused the conflict is flushed, and the PIM core rolls back and re-executes the code that was optimistically executed.

As a result {of this ``lazy'' checking of coherence violations}, CoNDA approaches near-ideal coherence behavior: {the performance and energy consumption of a PIM architecture with CoNDA are, respectively, within 10.4\% and 4.4\% the performance and energy consumption of} a system where coherence is performed at zero latency and energy cost.
\omv{The CoNDA paper~\cite{boroumand2019conda} also reports speedups of 38.3\% over the highest-performance prior coherence mechanism~\cite{goodman.isca83,mesi1984} (and 8.4$\times$/7.7$\times$ over CPU-only/PIM-only execution), and comes
within 10.2\% of an ideal no-cost coherence mechanism in four workloads on large datasets \omvii{(see~\cite{boroumand2019conda})}.}
\gfv{For modest size datasets \omvii{(see~\cite{boroumand2019conda})}, we observe that CoNDA 
(1)~reduces energy consumption by 18.0\% 
over the best prior coherence mechanism \omvi{that provides the lowest} memory system energy~\cite{farmahini2015nda,JAFAR}, 
(2)~achieves nearly all of the energy reduction potential of a no-cost coherence mechanism (coming within 4.4\%), and 
(3)~reduces energy consumption by 43.7\% over
a CPU-only execution.}

Despite the leap that CoNDA~\cite{boroumand2019conda,boroumand2016pim,boroumand.arxiv17} represents for memory coherence in computing systems with PIM support, we believe
that it is still necessary to explore other solutions for memory
coherence that can efficiently deal with all types of workloads and
PIM offloading granularities as well as different approaches to PIM \omv{(e.g., bulk bitwise execution, as we have seen in Section~\ref{sec:PUM})}. 
\juanr{In this direction, the design of new interfaces featuring memory coherence support across devices and memory (e.g., CXL~\cite{van2019hoti}, OpenCAPI~\cite{openCAPI, singh2020nero, singh2021fpga, singh2021accelerating}, OMI~\cite{coughlin2021higher}) can enable faster adoption of PIM by providing} \juanrr{a communication substrate on top of which efficient coherence and programming support can be built.}
\omv{Simpler \omvi{and even more efficient} coherence mechanisms are also \omvi{critical} to investigate.}

\subsection{Virtual Memory Support}
\label{sec:virtualmemory}


When an application needs to access its data inside the main memory,
the CPU core must first perform an \emph{address translation}, which
converts the data's virtual address into a \emph{physical} address
within main memory.  If the translation {metadata} is not
available in the CPU's translation lookaside buffer (TLB), the CPU
must invoke the page table walker in order to perform a long-latency
page table walk that involves multiple \emph{sequential} reads to the
main memory and lowers the application's performance~\omvi{\cite{kanellopoulos2023victima,kanellopoulos2023utopia,basu2013efficient,karakostas2014performance,kanellopoulos2024virtuoso,li-asplos19,mosaic-osr,mask,mosaic}}. In modern
systems, the virtual memory system also provides access protection
mechanisms.

\gfv{Enabling a unified virtual address space between PIM \omvi{cores} and host \omvi{cores}, where the memory address space is shared across all computing elements in the system, is key for enabling a more flexible programming model (i.e., ``\emph{a point is a pointer'' semantics}~\cite{che2016challenges}), with increased utilization of memory capacity and throughput. 
However, developing a high-performance and scalable unified virtual memory address space \omvi{for both} conventional host systems and PIM architectures can be challenging, since it requires a distributed memory management approach where different components of the system are able to perform address translation while guaranteeing \omvi{access} protection \omvi{mechanisms} for every memory access.}

%


\gfv{One naive solution to the unified Host--PIM virtual memory address space issue is to make \chiii{PIM cores} reliant on
 existing CPU-side address translation mechanisms.
However, in this approach, any performance
gains from performing in-/near-memory operations could easily be nullified, as the PIM cores need to send a long-latency translation request to the CPU via the off-chip channel for each memory access.
The translation can sometimes require a page table walk that issues multiple memory requests back to the memory, thereby leading to increased memory traffic on the main memory channels.}
  
\gfv{To improve \omvi{over} such \omvi{a} naive solution and reduce the} overhead of page walks\gfv{, we could} utilize
PIM engines to perform page table walks. This 
 can be done by
duplicating the content of the TLB and {moving} the page walker
{to} the PIM processing logic in main memory.  Unfortunately,
this is either difficult or expensive for three reasons. First,
coherence {has} to be maintained between the CPU's TLBs and the
memory-side TLBs. This introduces extra complexity and off-chip
requests.  Second, duplicating the TLBs increases the storage and
complexity overheads {on the memory side, which should be
  carefully contained}.  Third, if main memory is shared across
{CPUs with} different types of architectures, page table
structures and the implementation of address translations can be
different across {the different} architectures. Ensuring
compatibility between the in-memory TLB/page walker and all possible
types of {virtual memory} architecture designs can be
complicated and often not even practically feasible.

To address these concerns and reduce the overhead of virtual memory,
we explore a tractable solution for PIM address translation as part of
our in-memory pointer chasing accelerator, IMPICA~\cite{impica}.
IMPICA exploits the high bandwidth available within 3D-stacked memory
to traverse a chain of virtual memory pointers within DRAM,
\emph{without} having to look up virtual-to-physical address
translations in the CPU translation lookaside buffer (TLB) and without
using the page walkers within the CPU.  {IMPICA's key ideas are
  1) to use a region-based page table, which is optimized for PIM
  acceleration, and 2) to decouple address calculation and memory
  access with two specialized engines.  IMPICA improves the
  performance of pointer chasing operations in three commonly-used
  linked data structures (linked lists, hash tables, and B-trees) by
  92\%, 29\%, and 18\%, respectively. On a real database application,
  DBx1000, IMPICA improves transaction throughput and response time by
  16\% and 13\%, respectively. IMPICA also reduces overall system
  energy consumption (by 41\%, 23\%, and 10\% for the three
  commonly-used data structures, and by 6\% for DBx1000).
  \gf{IMPICA's source code is available at \url{https://github.com/CMU-SAFARI/IMPICA}.}}

Beyond pointer chasing operations that are tackled by
IMPICA~\cite{impica}, providing efficient mechanisms for PIM-based
virtual-to-physical address translation (as well as access protection)
remains a challenge for the generality of applications,
especially those that access large amounts of virtual
memory~\omv{\cite{mask,mosaic,mosaic-osr,kanellopoulos2023utopia,kanellopoulos2023victima,basu2013efficient}}.  

\begin{figure}[ht]
\centering
\includegraphics[width=1.0\linewidth]{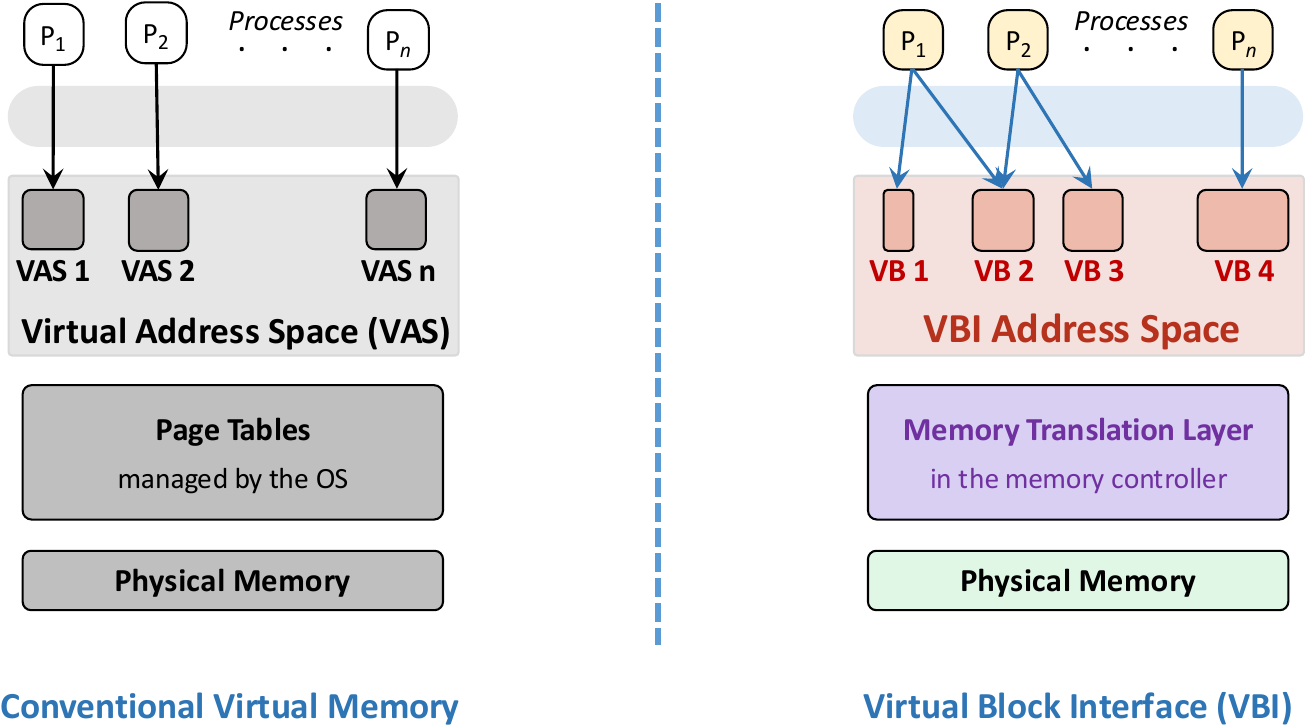}
\caption{The Virtual Block Interface \omxi{(right)} versus conventional virtual memory \omxi{(left)}. Reproduced from~\cite{hajinazar2020virtual}.}
\label{fig:vbi}
\end{figure}

\omv{We believe virtual memory, which \omvii{was} introduced \omvi{more than six decades ago~\cite{fotheringham1961dynamic,kilburn1962one,daley1968virtual,bensoussan1972multics}} and which did not change much since then (for a myriad of reasons, including backward compatibility with legacy systems and software ecosystem stability), needs to be fundamentally rethought in both processor-centric and memory-centric systems. To this end}, we introduced a fundamentally-new virtual memory framework, the Virtual Block Interface (VBI)~\cite{hajinazar2020virtual}, which proposes to delegate physical memory management duties completely to the memory controller hardware as well as other specialized hardware. 
Figure~\ref{fig:vbi} compares VBI to conventional virtual memory at a very high level.
Designing VBI-based PIM units that manage memory allocation and address translation can help fundamentally overcome this important virtual memory challenge of PIM systems. We refer the reader to our VBI work~\cite{hajinazar2020virtual} for details.

\subsection{Data Structures for PIM}
\label{sec:datastructures}

Current systems with many cores run applications with concurrent data
structures to achieve high performance and scalability, with
significant benefits over sequential data structures. Such concurrent
data structures \sg{are often used in} 
heavily-optimized server systems today, where high performance is
critical. To enable the adoption of PIM in such many-core systems, it
is necessary to develop concurrent data structures that are
specifically tailored to take advantage of PIM.

\emph{Pointer chasing data structures} and \emph{contended data
  structures} require careful analysis and design to leverage the high
bandwidth and low latency of 3D-stacked memories~\cite{liu-spaa17}.
First, pointer chasing data structures, such as linked-lists and
skip-lists, have a high degree of inherent parallelism and low
contention, but a naive implementation in PIM cores is burdened by
hard-to-predict memory access patterns. By combining and partitioning
the data across 3D-stacked memory vaults, it is possible to fully
exploit the inherent parallelism of these data structures.  Second,
contended data structures, such as FIFO queues, are a good fit for CPU
caches because they expose high locality. However, they suffer from
high contention when many threads access them concurrently.
{Their performance on traditional CPU systems can be improved
  using a new FIFO queue \omv{designed for PIM systems}~\cite{liu-spaa17}.  The proposed
  PIM-based FIFO queue uses a PIM core to perform enqueue and dequeue
  operations requested by CPU cores. The PIM core can pipeline
  requests from different CPU cores for improved performance.}
{As \sg{recent work~\cite{liu-spaa17}} shows, PIM-managed concurrent data
  structures can outperform state-of-the-art concurrent data
  structures that \sg{are designed} for and executed on multiple cores.}
\omv{Another recent work \gfv{(which we \omvi{describe} in Section~\ref{sec:upmem})} tackles how to optimize sparse matrix operations and data structures for near-bank PIM systems~\cite{giannoula2022sparsep}.}


\msvii{\omvi{I}n our SISA paper~\cite{besta2021sisa_micro}, we design set-based data structures for set operations in \emph{graph mining} algorithms~\cite{besta2021graphminesuite,graph_mining1,graph_mining2}, including \emph{graph pattern matching}~\cite{graph_mining2}, which focuses on finding certain specific subgraphs (also called motifs or graphlets) and \emph{graph learning}~\cite{graph_mining1} with
problems such as unsupervised learning or clustering~\cite{graph_clustering}. These algorithms are memory-bound and thus could be fundamentally  accelerated by \gfix{PIM} systems. However, graph mining algorithms also come with non-straightforward parallelism and complicated memory access patterns, \omix{which makes their implementation on} PIM \omix{systems} challenging. 
In~\cite{besta2021sisa}, we address this problem with a simple yet
surprisingly powerful observation: operations on \omix{\emph{sets of vertices}}, such as intersection or union, form a large part of many complex graph mining algorithms, and can offer rich and simple parallelism at multiple levels. This observation drives our cross-layer \omx{programming-algorithm-architecture-hardware co-}design,
in which we 
(1)~expose \omix{\emph{set operations}} using a novel programming paradigm, 
(2)~express and execute these operations efficiently with carefully designed \omix{\emph{\omx{S}et-centric \gls{ISA} extensions}} called SISA, and 
(3)~use PIM \omix{(both PUM and PNM)} to accelerate SISA instructions. 
SISA benefits from two types of PIM: in-DRAM bulk bitwise computing (\omix{PUM, \emph{à la} Ambit~\cite{seshadri.micro17}})
for bitvectors representing high-degree vertices, and near-memory logic layers (\omix{PNM, similar to Tesseract~\cite{ahn.tesseract.isca15}}) for integer arrays representing low-degree vertices. Set-centric SISA-enhanced algorithms are efficient and outperform
hand-tuned baselines, e.g., offering more than 10$\times$ speedup over the established Bron-Kerbosch algorithm~\cite{bron-kerbosch} for listing maximal cliques. 
We \omix{provide} more than 10 set-centric \omx{SISA} algorithm formulations, illustrating SISA's wide applicability \omix{to many graph mining problems}. SISA~\cite{besta2021sisa} is a great example of how \gfix{(i)~the co-design of algorithms, hardware, and ISA alongside 
(ii)~the \omx{synergistic} combination \omx{and exploitation} of PUM and PNM techniques can \emph{significantly} accelerate complex data-intensive applications, such as graph mining.} 
}

  
We believe and hope that future work will enable other types of data structures (e.g., hash tables, search trees, priority queues\omvii{, irregular and sparse data structures}) to \omv{largely} benefit from PIM designs.
\omv{Data structure designs to benefit from bulk bitwise PIM execution are also promising to research.}
\omvii{Such designs can also enable new formulations of algorithms for PIM systems (as our SISA paper~\cite{besta2021sisa_micro} demonstrates).}

\subsection{Benchmarks and Simulation Infrastructures}
\label{sec:simulation}



To ease the adoption of PIM, it is critical that we {accurately}
assess the benefits and shortcomings of PIM. Accurate assessment of
PIM requires (1)~a {preferably large} set of real-world
memory-intensive applications that have the potential to benefit
significantly when executed near memory, (2)~a rigorous methodology to
(automatically) identify PIM offloading candidates, and
(3)~simulation/evaluation infrastructures that allow architects and
system designers to {accurately} analyze the benefits and
overheads of adding PIM processing logic to memory and executing code
on this processing logic.

In order to explore what processing logic should be introduced near
memory, and to know what properties are ideal for PIM kernels, we
believe it is important to begin by developing a \emph{real-world
  benchmark suite} of a wide variety of applications that can
potentially benefit from PIM. While many data-intensive applications \omv{and kernels},
such as \omv{database operations, graph processing}\omvi{,} pointer chasing\omvi{,} and bulk memory copy, can potentially benefit
from PIM, it is crucial to examine important candidate applications
for PIM execution, and for researchers to agree on a common set of
these candidate applications to focus the efforts of the community
{as well as to enable reproducibility of results, which is
  important to assess the relative benefits of different ideas
  developed by different researchers}. We believe that these
applications should come from a number of popular and emerging
domains. Examples of potential domains~\omv{\cite{oliveira2021pimbench}} include data-parallel
applications, \gfv{databases,} neural networks, machine learning, graph processing,
data analytics, search/filtering, mobile workloads, bioinformatics, Hadoop/Spark programs, security/cryptography, \gfv{\omvi{physics modeling} \& simulation, video/image processing,} and in-memory data
stores. Many of these applications have large data sets and can
benefit from high memory bandwidth\omv{,} low memory latency\omv{, and high array-level parallelism} benefits
provided by computation near memory. 
In our prior work, we have
started identifying several applications that can benefit from PIM in
graph processing
frameworks~\omv{\cite{ahn.pei.isca15,ahn.tesseract.isca15,besta2021sisa_micro}}, 
pointer chasing~\cite{cont-runahead,impica}, databases~\gfv{\cite{boroumand2016pim,boroumand2019conda,boroumand.arxiv17, impica, GS-DRAM, boroumand2021polynesia}}, 
consumer workloads~\cite{boroumand.asplos18}, 
time series analysis~\gfv{\cite{fernandez2020natsa,fernandez2024matsa}}, 
genome analysis~\gfv{\cite{kim.bmc18,cali2020genasm,mansouri2022genstore,ghiasimegis2024,diab2023framework}}, 
machine learning~\gfv{\cite{boroumand.asplos18, boroumand2021google, boroumand2021google_arxiv,rhyner2024pimopt,gomez2022machine,gogineni2024swiftrl},
security primitives~\cite{gupta2023evaluating},
weather prediction~\cite{singh2021accelerating,singh2020nero}}, and GPGPU
workloads~\cite{hsieh.isca16,pattnaik.pact16}. 
However, there is significant room for methodical development of a large-scale PIM benchmark suite, which our 
recent \juanrrr{DAMOV} work~\cite{oliveira2021pimbench, oliveira2021pimbench_arxiv, oliveira2021.SLS} 
takes the first steps for, \juanrrr{as we explain below}.

\begin{figure}[ht]
\centering
\includegraphics[width=1.0\linewidth]{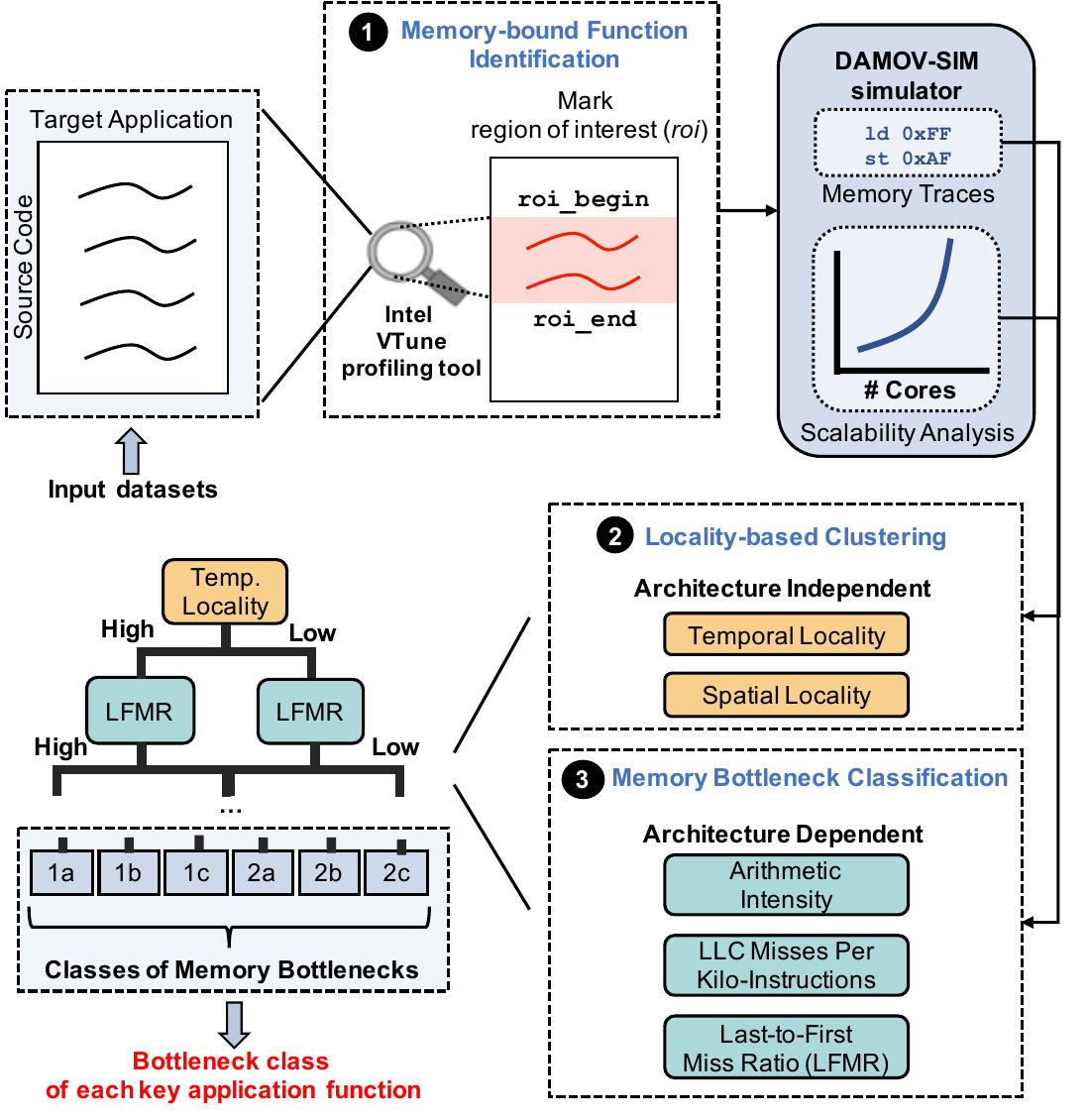}
\caption{\juanrrr{Overview of DAMOV three-step workload characterization methodology. Components are explained in detail in~\cite{oliveira2021pimbench}. Figure adapted from~\cite{oliveira2021pimbench}.}}
\label{fig:damov}
\end{figure}

A systematic \emph{methodology} for (automatically) identifying
potential PIM kernels (i.e., code portions that can benefit from PIM)
within an application can, among many other benefits, (1) ease the burden of programming PIM architectures by aiding the programmer to identify what should be offloaded, (2) ease the burden of and improve the reproducibility of PIM research, (3) drive the design and implementation of PIM functional units that many types of
applications can leverage, (4) inspire the development of tools that programmers and compilers can use to automate the process of offloading portions of existing applications to PIM processing logic, and (5) lead the community towards convergence on PIM designs and offloading candidates. 
In 
\juanrrr{our DAMOV work}~\cite{oliveira2021pimbench, oliveira2021pimbench_arxiv, oliveira2021.SLS, oliveira2022methodologies},
we take the first steps in developing such a methodology and the first benchmark suite for PIM. 
\juanrrr{DAMOV's workload characterization methodology consists of three main steps, depicted in Figure~\ref{fig:damov}.}

\juanrrr{The first step, \emph{memory-bound function identification}, aims to identify the functions of an application that suffer from data movement bottlenecks (this step is optional if these functions are known \emph{a priori}).
There are various potential sources of memory boundedness, e.g., cache misses, cache coherence traffic, and long queueing latencies. 
In this step, the user of DAMOV's methodology can use hardware profiling tools that characterize the application \omv{behavior} on a computing system. In particular, we use the Intel VTune Profiler~\cite{vtune}. VTune implements \emph{top-down analysis}~\cite{yasin_ispass2014}, which uses available CPU hardware counters to identify different sources of CPU system bottlenecks. 
A relevant metric that VTune provides is \emph{Memory Bound}~\cite{MemoryBo20}, which measures the percentage of pipeline slots that are \emph{not} utilized due to any issue related to data access.}

\juanrrr{The second step, \emph{locality-based clustering}, analyzes \emph{spatial} and \emph{temporal locality} of an application (or the function/s identified in the first step) in an architecture-independent manner. 
These two properties are related to how well an application can exploit the memory hierarchy in computing systems and how accurate prefetchers can be. 
We analyze these two properties in an architecture-independent manner to isolate the application's \omv{behavior} from possible effects derived from limitations of the memory subsystem (e.g., limited cache size, inaccurate prefetching policies). 
We use the definitions of spatial and temporal locality presented in~\cite{weinberg2005quantifying, shao2013isa}, and integrate them into our fast, scalable, and cycle-accurate open-source simulator DAMOV-SIM~\cite{damov2021_repo}.}

\juanrrr{The third step, \emph{memory bottleneck classification}, allows us to understand how hardware architectural features can also result in memory bottlenecks. 
This step performs a scalability analysis of the functions selected in the first step. 
The scalability analysis uses three different system configurations, which are simulated with DAMOV-SIM: 
(1)~a host CPU with a deep cache hierarchy, 
(2)~a host CPU with a deep cache hierarchy and a stream prefetcher, and 
(3)~a \gfv{processing-near-memory (PNM)} \omvi{core} with a single level of cache and no prefetchers. 
For the three configurations, we sweep the number of cores from 1 to 256. 
The analysis provides measurements of three key \emph{architecture-dependent} metrics: (1) \emph{Arithmetic Intensity} (\emph{AI}), (2) \emph{Misses per Kilo-Instruction} (\emph{MPKI}), and (3) \emph{Last-to-First Miss-Ratio} (\emph{LFMR}), a new metric that accurately quantifies how efficient the cache hierarchy is in reducing data movement.}

\juanrrr{By combining the data obtained in the three steps, we can systematically classify the leading causes of data movement bottlenecks in an application or function into different bottleneck classes. 
In~\cite{oliveira2021pimbench, oliveira2021pimbench_arxiv}, we analyze 345 applications (with a total of 77K functions) from 37 different workload suites. 
We \omv{analyze} in detail 144 functions from 74 different \omv{applications}, which are memory-bound according to the first step of our methodology. 
We found six main classes of \omv{bottlenecks that are present in applications, such as} DRAM bandwidth, DRAM latency, cache capacity, and cache contention (see~\cite{oliveira2021pimbench, oliveira2021pimbench_arxiv} for the detailed analysis of these \omv{bottleneck} classes).}

\juanrrr{The 144 representative functions identified in our study constitute the first open-source benchmark suite for data movement, called DAMOV Benchmark Suite~\cite{damov2021_repo}. 
This benchmark suite can aid the study of open research problems for PIM architectures. 
For example, in~\cite{oliveira2021pimbench, oliveira2021pimbench_arxiv}, we evaluate four case studies that use DAMOV benchmarks: (i) study of the impact of load balance and inter-vault communication in 3D-stacked PIM systems, (ii) study of PIM accelerators compared to general-purpose PIM cores, (iii) study of different core models for PIM, (iv) identification of simple PIM instructions (\emph{à la} PEI~\cite{ahn.pei.isca15, ahn.pei.isca15talk}).}

We believe 
\juanrrr{our DAMOV} work opens up many more steps to extend the methodology and develop other new methodologies for identifying PIM kernels as well as automatic tools (e.g., profilers, compilers, runtime systems) that implement these methodologies, generate optimized code for PIM (potentially with help from programmer annotations), coordinate offloading to PIM cores, etc.

\begin{figure}[ht]
\centering
\includegraphics[width=1.0\linewidth]{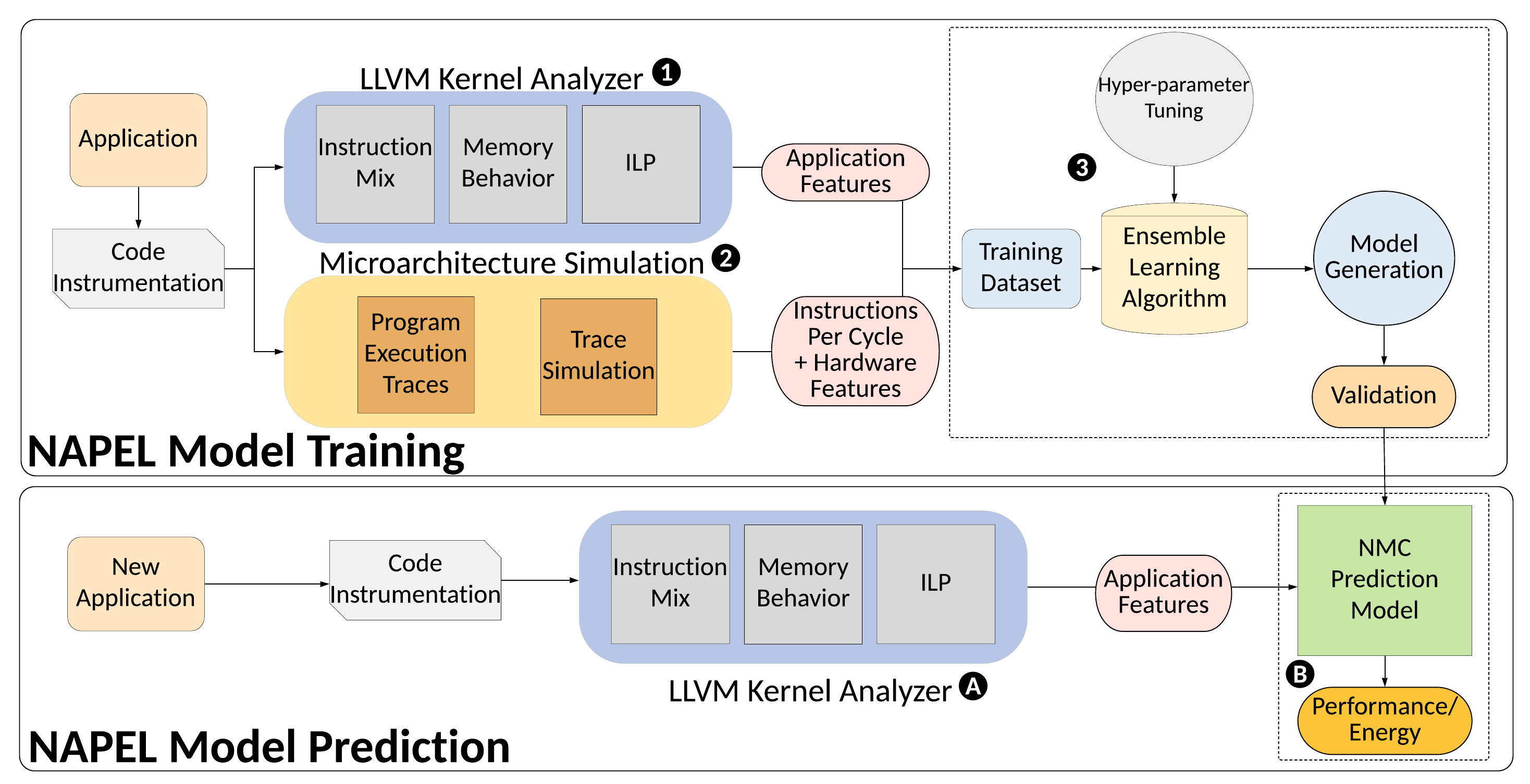}
\caption{Overview of NAPEL training and prediction. Components are explained in detail in~\cite{singh2019napel}. Figure reproduced from~\cite{singh2019napel}.}
\label{fig:napel}
\end{figure}

Along these lines, NAPEL~\cite{singh2019napel} is an early example of an ML-based performance and energy estimation framework for PNM. NAPEL leverages ensemble learning techniques to generate PNM performance and energy prediction models that are based on microarchitecture parameters and application characteristics. Figure~\ref{fig:napel} shows the high-level overview of NAPEL training and prediction, the components of which are explained in detail in~\cite{singh2019napel}. 
Our evaluations show that NAPEL can make fast yet accurate predictions of PIM offloading suitability for previously-unseen applications on general-purpose PNM architectures.

{We also} need \emph{simulation infrastructures} to accurately
model the performance and energy of PIM hardware structures, available
memory bandwidth, and communication overheads when we execute code
near or inside memory. Highly-flexible and commonly-used memory
simulators (e.g., Ramulator~\cite{ramulator, ramulator.github}, \omv{Ramulator 2.0~\cite{luo2023ramulator,safari2023ramulator2}) and \omvi{FPGA-based memory modeling} prototypes (e.g.,} SoftMC~\cite{hassan2017softmc, softmc.github}\omv{, DRAM Bender~\cite{olgun2023drambender,safari-drambender}}) can be combined with system \omv{and processor} simulators (e.g., gem5~\cite{GEM5}, zsim~\cite{zsim},
gem5-gpu~\cite{gem5-gpu}, GPGPUSim~\cite{gpgpusim}\omv{, Sniper~\cite{carlson2011sniper}, \gfv{Accel-Sim~\cite{khairy2020accel}, MARSS~\cite{marss}}}) to provide a
robust environment that can evaluate how various PIM architectures
affect the \emph{entire compute stack}, and can allow designers to
identify memory, workload, and system characteristics that affect the
efficiency of PIM execution. 
A powerful open-source simulation infrastructure that provides such environment is Ramulator-PIM~\cite{ramulator-pim}, first introduced by our NAPEL framework~\cite{singh2019napel}, which combines Ramulator~\cite{ramulator, ramulator.github} and zsim~\cite{zsim}. 
Ramulator-PIM can simulate a wide range of configurations of PIM in-order and out-of-order cores and accelerators with different memory technologies. 
\juanrr{DAMOV-SIM~\cite{damov2021_repo} augments Ramulator-PIM with additional configurations and a more user-friendly interface.}
\omv{Ramulator 2.0~\cite{luo2023ramulator,safari2023ramulator2}, a successor to Ramulator 1.0~\cite{ramulator, ramulator.github}, can also enable further research by being developed to model PIM further \omvi{and more easily, due to its more modular structure and easier-to-handle software engineering}.}

\subsection{Real PIM Hardware Systems and Prototypes}
\label{sec:realpim}

As industry and academia push toward enabling the PIM paradigm, it will be important to also provide real PIM hardware or prototypes. Such hardware can greatly enable and accelerate evaluations of both adoption and research issues in PIM, leading to learnings from real workloads executed on real systems and thus better PIM systems over time. 
\omv{Especially software\omvi{-level} and algorithm-level research can be catalyzed and accelerated with the existence of real PIM hardware and prototypes.}
Such real hardware for PIM is very much useful for both PUM and PNM approaches. 

We are aware of 
\juanrr{a handful of} such real \omv{PIM} hardware systems. 
\juanrrr{First, there are several successful attempts in academia to perform PUM operations in off-the-shelf DRAM chips~\gfv{\cite{gao2020computedram, kim.hpca18, kim.hpca19, olgun2021pidram, olgun2021quactrng, olgun2022pidram, missingnot,yuksel2024simultaneous}} (Section~\ref{sec:PUMprot}). 
Second, the UPMEM company~\cite{upmem} commercializes a PIM architecture that integrates simple processors into DDR4 DRAM chips (Section~\ref{sec:upmem}).} 
Third, there have been many prototypes of real PNM DRAM chips developed by major vendors in industry, including Samsung~\gf{\cite{kwon202125, lee2021hardware, kim2021aquabolt, ke2021near, lee2022improving,samsunghc23}}, SK Hynix~\cite{lee2022isscc} and Alibaba~\cite{niu2022isscc}, in 2021-202\gf{4}. 
We briefly explain \omvii{the} different \omvii{DRAM-based PNM} prototypes we are aware of \omvii{(Sections~\ref{sec:fimdram} to~\ref{sec:hbpnm})}. 
\gfvi{Other industry players and startups are currently investigating the implementation of PIM system using memory technologies other than DRAM, such as Axelera AI's SRAM-based~\cite{axeleraai} and Mythic's NVM-based~\cite{mythic} PIM architectures targeting artificial intelligence (AI) applications.}

\subsubsection{\juanrrr{PUM Prototypes}}
\label{sec:PUMprot}

ComputeDRAM~\cite{gao2020computedram},
which is based on the SoftMC memory controller infrastructure~\omv{\cite{hassan2017softmc,softmc.github}}  
can potentially provide the opportunity to test \omv{the} \gf{RowClone}~\omv{\cite{seshadri2013rowclone}} (Section~\ref{sec:rowclone})  
and Ambit~\omv{\cite{seshadri.micro17}} (Section~\ref{sec:ambit}) 
PUM approaches on real workloads, albeit likely at reduced reliability since it exploits off-the-shelf DRAM chips, as we discussed in Section~\ref{sec:ambit}.

\juanr{PiDRAM~\cite{olgun2021pidram, olgun2021pidram_repo, olgun2022pidram} is a flexible end-to-end \juanrr{FPGA-based} experimental framework that leverages ComputeDRAM's idea of implementing PUM approaches by violating timing parameters.} 
\juanrr{PiDRAM enables the study of end-to-end benefits of PUM techniques such as RowClone~\cite{seshadri2013rowclone} and D-RaNGe~\cite{kim.hpca19}.} 
\juanrrr{PiDRAM can potentially enable end-to-end studies of other PUM techniques (e.g., QUAC-TRNG~\cite{olgun2021quactrng}) and frameworks (e.g., SIMDRAM~\cite{hajinazarsimdram}). \gf{The source code of PiDRAM is available at \url{https://github.com/CMU-SAFARI/PiDRAM}.}}

\begin{figure}[ht]
    \centering
    \includegraphics[width=1.0\linewidth]{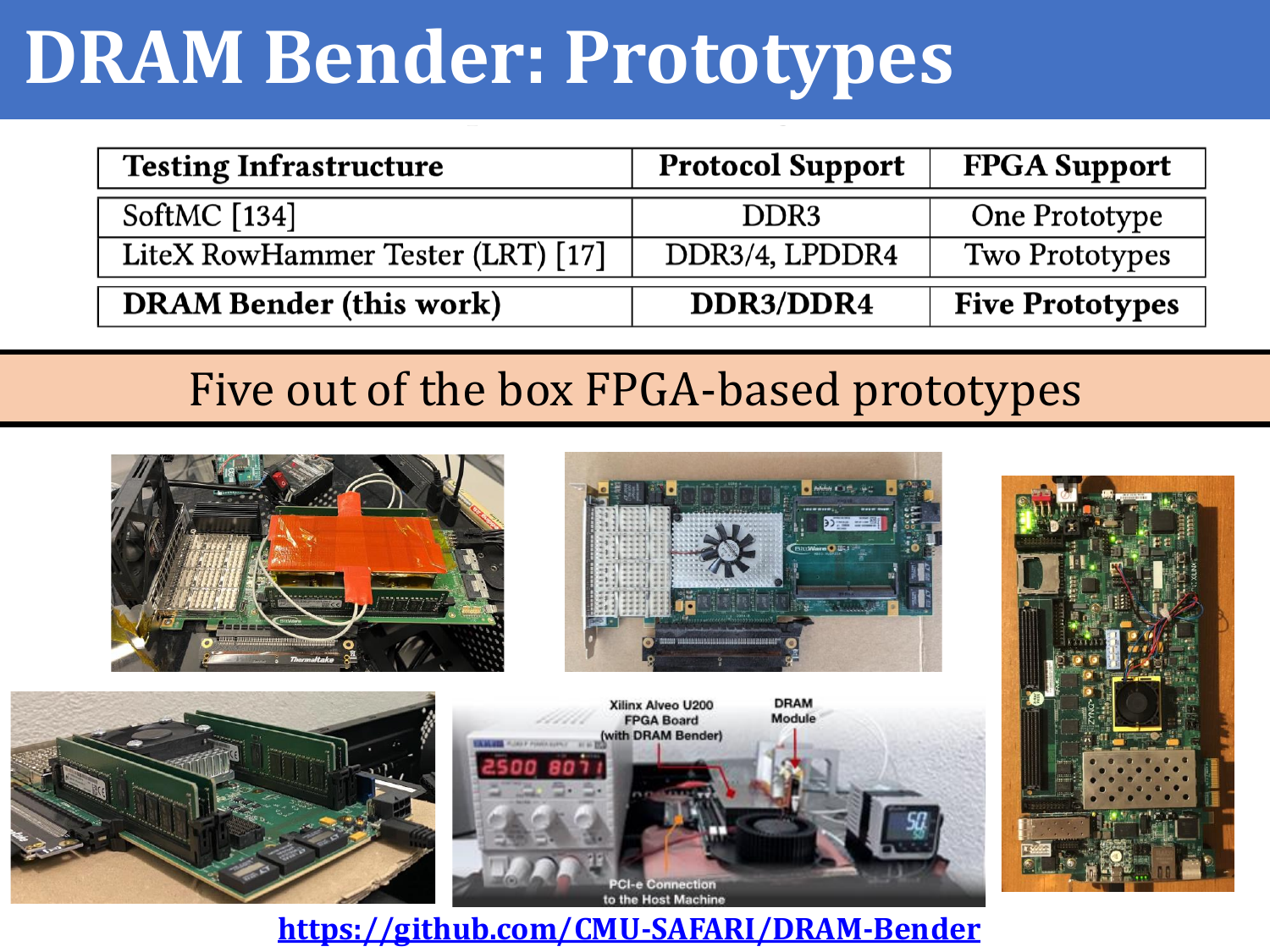}
    \caption{\gfix{Overview of DRAM Bender~\omvii{\cite{safari-drambender,olgun2023drambender}}\omx{, the state-of-the-art open-source FPGA-based infrastructure that enables testing and characterization of DDR3/DDR4 and HBM2~\cite{olgun2024read} DRAM chips}. Figure reproduced from~\cite{lecturedrambender}.}}
    \label{fig:drambander}
\end{figure}

\gfv{In our recent works~\cite{missingnot,yuksel2024simultaneous}, we experimentally demonstrate the PUM capabilities of hundred \gls{COTS} DRAM chips using DRAM Bender~\cite{olgun2023drambender,safari-drambender}, an FPGA-based DDR4 testing infrastructure that provides precise control of DDR4 commands issued to a DRAM module (which Figure~\ref{fig:drambander} illustrates).
We demonstrate that COTS DRAM chips are capable of 
(1)~performing functionally-complete bulk-bitwise Boolean operations: NOT, NAND, and NOR,
(2)~executing up to 16-input AND, NAND, OR, and NOR operations, and (3)~copying the contents of a DRAM row (concurrently) into up to 31 other DRAM rows wit $>$99.98\% reliability.
We evaluate the robustness of these operations across data patterns, temperature, and voltage levels. Our results 
show that COTS DRAM chips can perform these operations at high success rates ($>$94\%).
These fascinating findings demonstrate the fundamental computation capability of DRAM, even when DRAM chips are {\em not} designed for this purpose, and provide a solid foundation for building new and robust PUD mechanisms into future DRAM chips and standards.
To aid future
research and development, we open-source our infrastructures at \url{https://github.com/CMU-SAFARI/SiMRA-DRAM} and 
 \url{https://github.com/CMU-SAFARI/FCDRAM}.
}

\subsubsection{\juanrrr{UPMEM PIM Architecture}}
\label{sec:upmem}

\begin{figure}[h]
\centering
\includegraphics[width=1.0\linewidth]{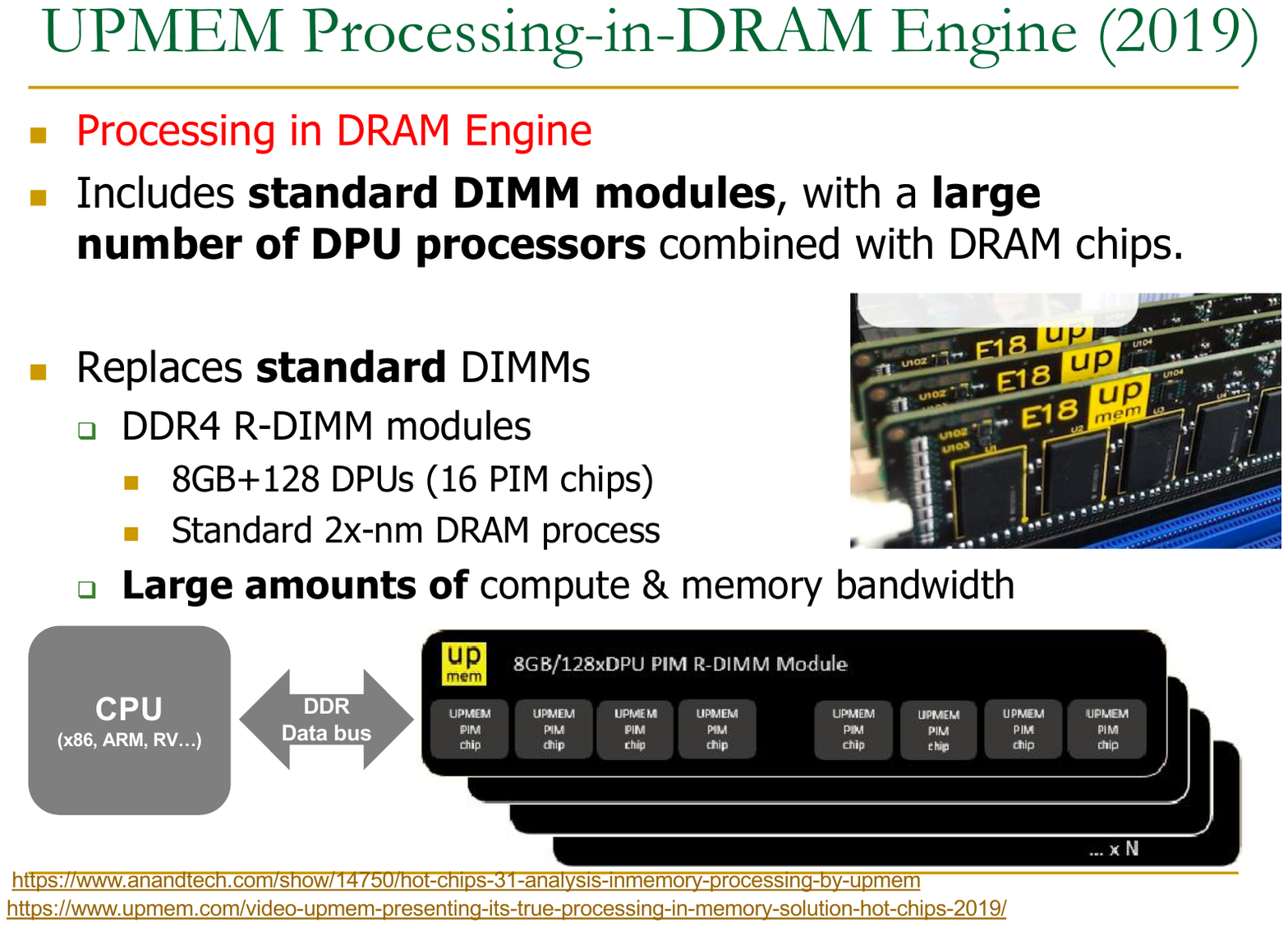}
\caption{UPMEM PIM architecture and hardware. Reproduced from~\cite{mutlu.nsfpim20}.}
\label{fig:upmem}
\end{figure}

The UPMEM PIM architecture~\cite{upmem2018, devaux2019}, shown in Figure~\ref{fig:upmem}, is the first real-world publicly-available PIM architecture. This PNM system consists of one simple processor (called \emph{DRAM Processing Unit}, \emph{DPU}) implemented next to each bank in a DRAM chip. A DPU has high-bandwidth, low-latency, low-energy access to all the data in its corresponding bank.

\begin{figure*}[t]
    \centering
    \includegraphics[width=\linewidth]{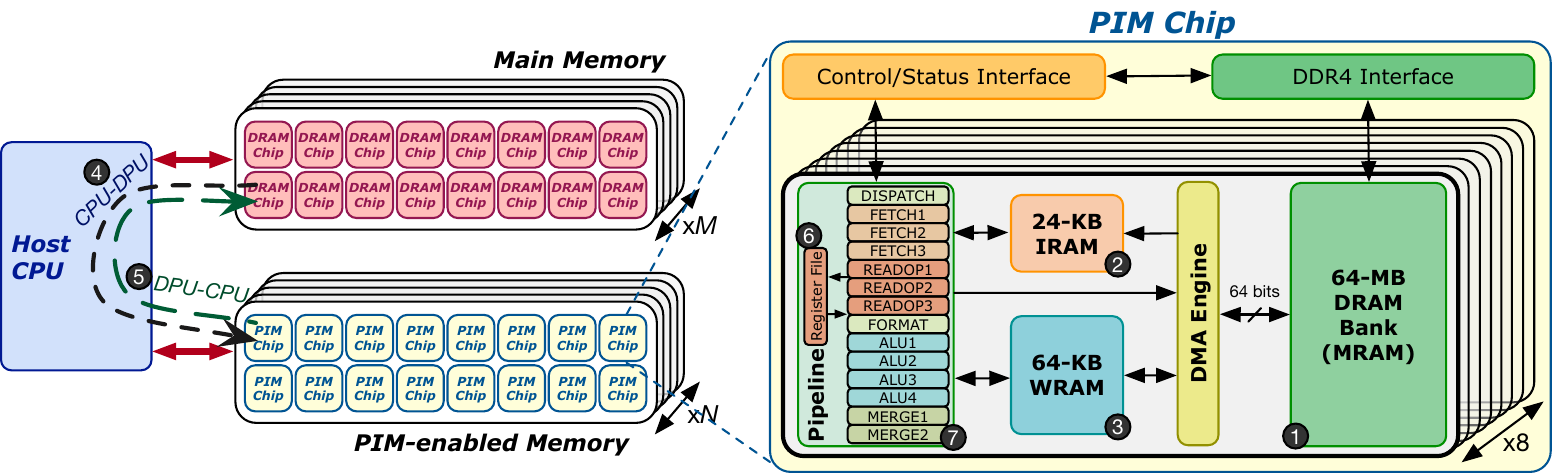}
    \caption{\juanrrr{UPMEM-based PIM system with a host CPU, standard main memory, and PIM-enabled memory (left), and internal components of a UPMEM PIM chip (right)~\cite{upmem2018, devaux2019}. Reproduced from~\cite{gomezluna2022ieeeaccess}.}}
    \label{fig:upmem-scheme}
\end{figure*}

UPMEM has produced real DRAM modules \juanrrr{(standard DDR4-2400 DIMMs~\cite{jedec2012ddr4})} that contain 16 PNM-capable DRAM chips each. Each DRAM chip includes eight 64-MB DRAM banks, each of which has a DPU attached running at a few hundred MHz. A full-blown UPMEM system configuration 
\juanr{contains} 2,560 DPUs capable of operating on 160 GB of DRAM memory.

\juanrrr{Figure~\ref{fig:upmem-scheme} (left) shows an UPMEM PIM system with (1) a host CPU, (2) standard main memory (DRAM memory modules), and (3) PIM-enabled memory (UPMEM modules). 
Each PIM chip (Figure~\ref{fig:upmem-scheme} (right)) contains 8 DPUs. Each DPU has exclusive access to (1) a 64-MB DRAM bank, called \emph{MRAM} \circled{1}, (2) a 24-KB instruction memory, called \emph{IRAM} \circled{2}, and (3) a 64-KB scratchpad memory, called \emph{WRAM} \circled{3}. 
The DPU pipeline \circled{6}, \circled{7} supports natively 32-bit addition/subtraction while 32-bit multiplication/division and floating-point operations are emulated by the runtime library~\cite{upmem-guide}. 
The host CPU can access MRAM (Figure~\ref{fig:upmem-scheme} (left)) to copy input data (from main memory to MRAM) \circled{4}~ and to retrieve results (from MRAM to main memory) \circled{5}. 
There is no support for direct communication between DPUs. All inter-DPU communication takes place through the host CPU.}

\juanr{Thorough \emph{architecture characterization and benchmarking}, \juanrr{as we have performed in our recent works}~\omv{\cite{gomezluna2021benchmarking, gomezluna2022ieeeaccess, gomez2021cut, gomezluna2021.SLS, gomezluna2021repo, gomezluna2021.CUT, giannoula2022sparsep, giannoula2022sigmetrics, gomez2022machine, gomez2022experimental,gomez2023evaluating}}, is necessary to understand the potential of the UPMEM PIM system and propose programming recommendations and architecture \& hardware improvements for future PIM systems.} 
\juanrr{To this end, we have carried out the first comprehensive analysis~\gfvi{\cite{gomezluna2021benchmarking, gomezluna2022ieeeaccess, gomez2021cut, gomezluna2021.SLS, gomezluna2021repo, gomezluna2021.CUT}} of the UPMEM PIM system. Our work makes two key contributions.} 

{First, we conduct an experimental characterization of the UPMEM-based PIM system using microbenchmarks to assess various architecture limits such as compute throughput and memory bandwidth, yielding new insights. Second, we present \emph{PrIM} \omv{benchmarks} (\emph{\underline{Pr}ocessing-\underline{I}n-\underline{M}emory benchmarks})~\gf{\cite{gomezluna2021repo}}, a benchmark suite of 16 workloads from different application domains (e.g., dense/sparse linear algebra, databases, data analytics, graph processing, neural networks, bioinformatics, image processing), which we identify as memory-bound.} 

\begin{figure}[ht]
\centering
\includegraphics[width=1.0\linewidth]{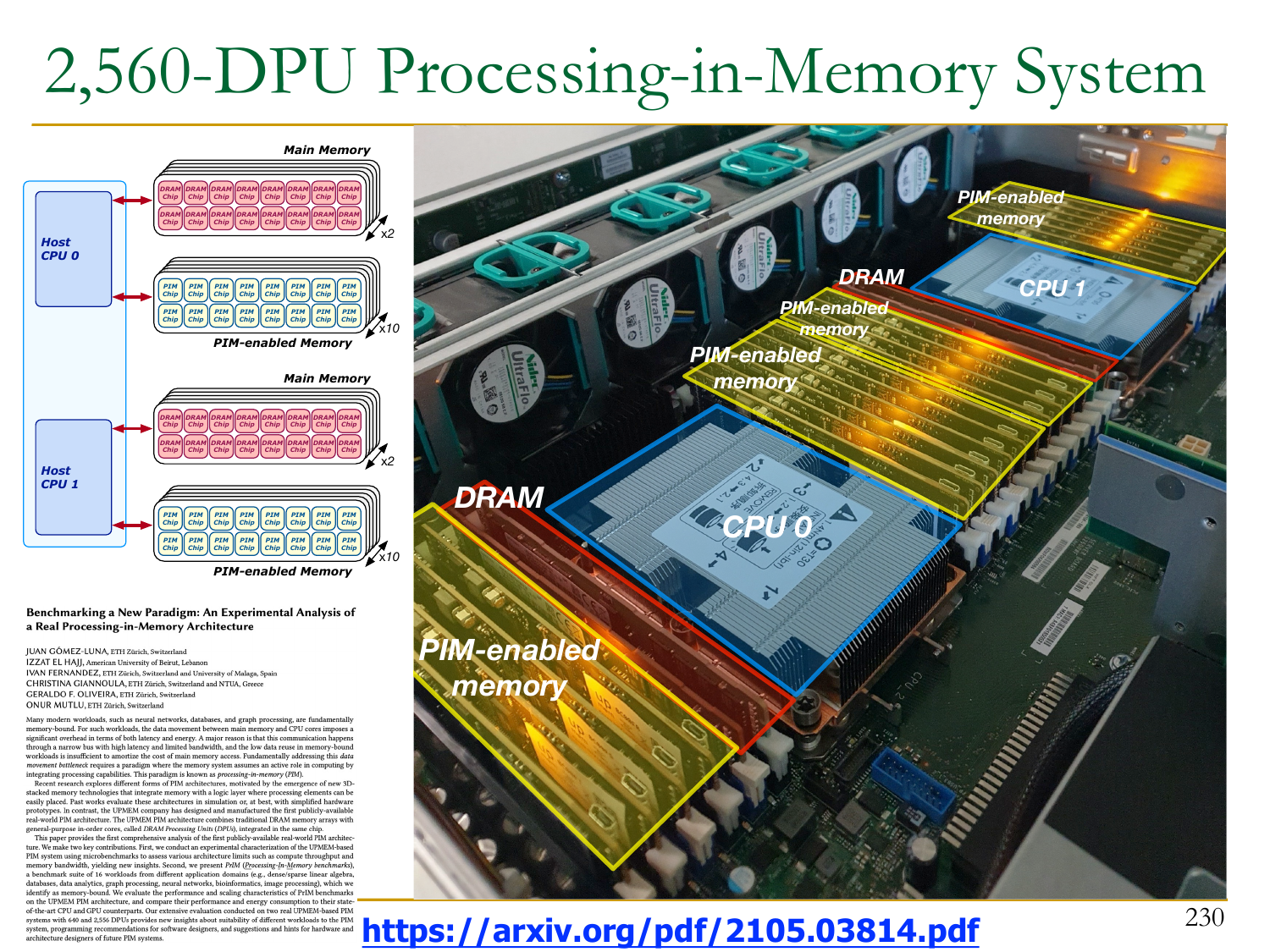}
\caption{\gfvi{\omvii{Photograph of a real} UPMEM PIM system with 2,560 DRAM Processing Units (DPUs). Reproduced from~\cite{mutlu.accml23.talk}}.}
\label{fig:upmemsystem}
\end{figure}

{We evaluate the performance and scaling characteristics of PrIM benchmarks on the UPMEM PIM architecture, and compare their performance and energy consumption to their state-of-the-art CPU and GPU counterparts. Our extensive evaluation conducted on two real UPMEM-based PIM systems with 640 and 2556 DPUs \gfvi{(which Figure~\ref{fig:upmemsystem} illustrates)} provides new insights about suitability of different workloads to the PIM system, programming recommendations for software designers, and suggestions and hints for hardware and architecture designers of future PIM systems.} 

\juanrrr{We highlight here \emph{four key takeaways} that come out of our work~\cite{gomezluna2021benchmarking, gomezluna2022ieeeaccess, gomez2021cut, gomezluna2021.SLS, gomezluna2021repo, gomezluna2021.CUT}:
\begin{enumerate}
[leftmargin=5mm,itemsep=0mm,parsep=0mm,topsep=0mm]
\item The UPMEM PIM architecture is \emph{fundamentally compute bound}. As a result, the most suitable workloads are memory-bound in processor-centric systems (i.e., CPU, GPU).
\item The most well-suited workloads for the UPMEM PIM architecture use \emph{no arithmetic operations} or use \emph{only simple operations} (e.g., bitwise operations and integer addition/subtraction). 
\item The most well-suited workloads for the UPMEM PIM architecture require \emph{little or no communication across DPUs}. 
\item UPMEM PIM systems outperform modern CPUs in terms of performance and energy efficiency on most of PrIM benchmarks and outperform modern GPUs on a majority (10 out of 16) of PrIM benchmarks, and the outlook is even more positive for future PIM systems. UPMEM-based PIM systems are more energy-efficient than modern CPUs and GPUs on workloads that they provide performance improvements over the CPUs and the GPUs. 
\end{enumerate}}

\juanrr{We are also exploring in depth important \emph{application domains and their suitability to the UPMEM PIM system}.} 
\juanrrr{We introduce next \omv{several} recent studies of the suitability of the UPMEM PIM architecture to sparse linear algebra, bioinformatics, machine learning\gf{, and homomorphic \omv{encryption}} applications\gfv{, as well as software support for transcendental functions.}}

\begin{figure}[ht]
    \centering
    \includegraphics[width=1.0\linewidth]{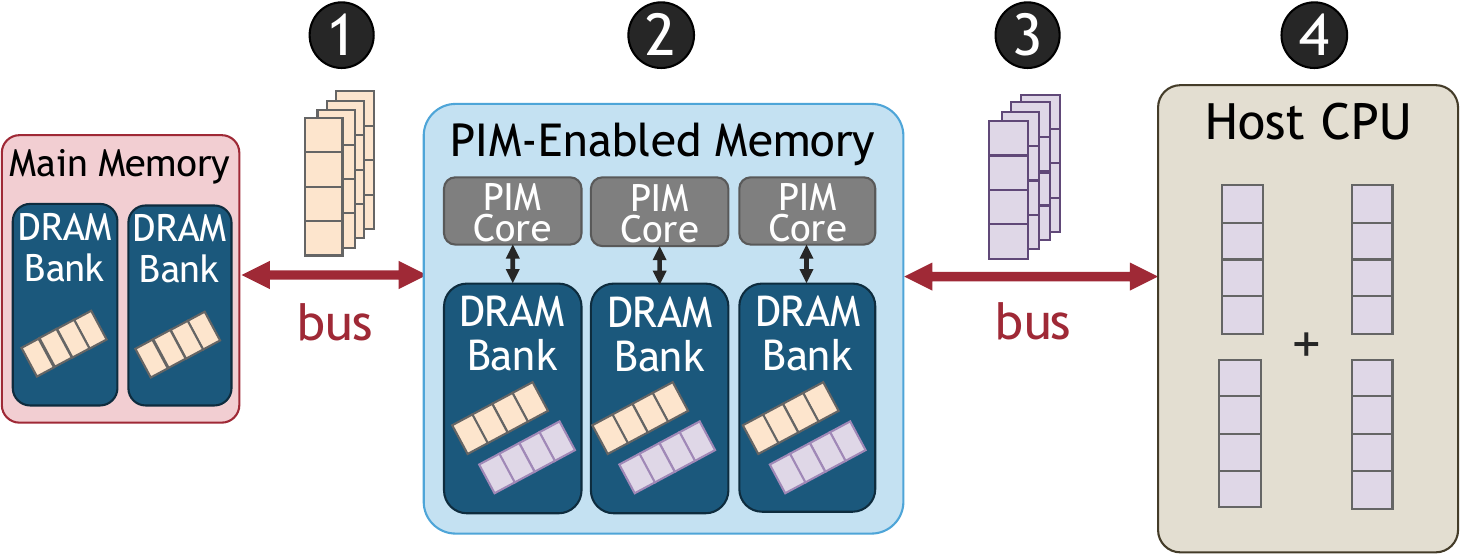}
    \caption{\gfvi{SparseP's execution of an SpMV kernel on a real PIM system in four main steps: \ding{202}~load the input vector into the PIM-enabled memory; \ding{203}~execute the SpMV kernel on PIM cores; \ding{204}~retrieve the output vectors from the PIM-enabled memory; \ding{205}~merge partial results and assemble the input output vector on the host CPU. Reproduced from~\cite{giannoula2022sparsep}.}}
    \label{fig:sparsep}
\end{figure}

\juanrr{First, we have performed an extensive analysis of \juanrri{\emph{Sparse Matrix-Vector multiplication} (\emph{SpMV})}, an important memory-bound kernel, on the UPMEM PIM system. We introduce \emph{SparseP}~\cite{giannoula2022sparsep, giannoula2022sigmetrics, giannoula2022towards, giannoula2022repo}, the first SpMV library for real PIM architectures. 
\gfv{Figure~\ref{fig:sparsep} shows a high-level overview of SparseP's execution of an SpMV kernel on a real PIM system.}
We make two key contributions. 
(1)~We design efficient SpMV algorithms to accelerate the SpMV kernel in current and future PIM systems, while covering a wide variety of sparse matrices with diverse sparsity patterns. 
(2)~We provide the first comprehensive analysis of SpMV on a real PIM architecture\omv{, examining various matrix compression, \omvi{data types,} data partitioning, load balancing, and synchronization approaches.}} 
\omv{Our results demonstrate that while the UPMEM PIM system can efficiently eliminate data movement overheads, it suffers from the lack of a good multiplier in hardware \omvi{(based on the design choices made for the first generation UPMEM PIM system)}.}
\gfv{Concretely, in our results, we compare the performance and energy of SpMV on the state-of-the-art UPMEM PIM system with 2528 PIM cores to state-of-the-art CPU and GPU systems. 
SpMV execution
achieves less than 1\% of the peak performance on processor-centric CPU and GPU systems, while it achieves on average 51.7\% of the peak performance on the UPMEM PIM system, thus better leveraging the computation capabilities of \omvi{the} underlying hardware. 
The UPMEM
PIM system also provides high energy efficiency on the SpMV kernel.
The SparseP software library, which is publicly and freely available at \url{https://github.com/CMU-SAFARI/SparseP}, provides 25 SpMV kernels for real PIM systems supporting the four most widely used compressed matrix formats, i.e., CSR~\cite{bjorck2024numerical,pooch1973survey}, COO~\cite{pooch1973survey,sengupta2007scan}, BCSR~\cite{im1999optimizing} and BCOO~\cite{pooch1973survey}, and a wide range of data types, i.e., 8-/16-/32-/64-bit integer and 32-/64-bit float data types.}
\omvi{Much more detailed analyses are provided in the SparseP paper~\cite{giannoula2022sparsep}.}

\juanrrr{Second, we have evaluated the suitability of the UPMEM PIM system for accelerating sequence alignment algorithms, such as \emph{Needleman-Wunsch} (\emph{NW})~\cite{needleman1970general}, \emph{Smith-Waterman-Gotoh} (\emph{SWG})~\cite{gotoh1982improved}, GenASM~\cite{cali2020genasm}, and the \emph{wavefront algorithm} (\emph{WFA})~\cite{marco2021fast}, which is currently the state-of-the-art gap-affine pairwise alignment algorithm. 
Our recent work~\gfv{\cite{diab2023framework,diab2022high,diab2023repo}} introduces a framework for PIM-based sequence alignment, where the host CPU dispatches a large number of sequence pairs across the DPUs available in the PIM system. 
In each DPU, we assign different sequence pairs to different PIM threads, which perform the alignments \omv{in parallel}. 
Our framework\gf{, whose source code is available at \url{https://github.com/CMU-SAFARI/alignment-in-memory},} supports NW, SWG, GenASM, WFA, and \emph{WFA-adaptive} (a heuristic variant of WFA)~\cite{marco2021fast}. 
Each of these algorithms has alternate implementations that manage the UPMEM memory hierarchy (i.e., MRAM and WRAM) differently and are suitable for different read lengths.}
\omv{Our extensive experimental evaluations show that the UPMEM PIM system can accelerate these sequence alignment algorithms by 1.8$\times$--28$\times$ (over CPUs) and 1.2$\times$--2.7$\times$ (over GPUs)~\cite{diab2023framework}.
Our very recent work, BIMSA~\omvi{\cite{10.1093/bioinformatics/btae631,alonso2024bimsa}}, demonstrates that further algorithmic implementation optimizations of the WFA~\cite{marco2021fast} and BiWFA algorithms~\cite{marco2023optimal} lead to even larger performance improvements with the UPMEM system over state-of-the-art processor-centric implementations~\cite{10.1093/bioinformatics/btae631,alonso2024bimsa}.}

\begin{figure}[ht]
    \centering
    \includegraphics[clip, trim=0.0cm 1.0cm 0.0cm 0.0cm, width=1.0\linewidth]{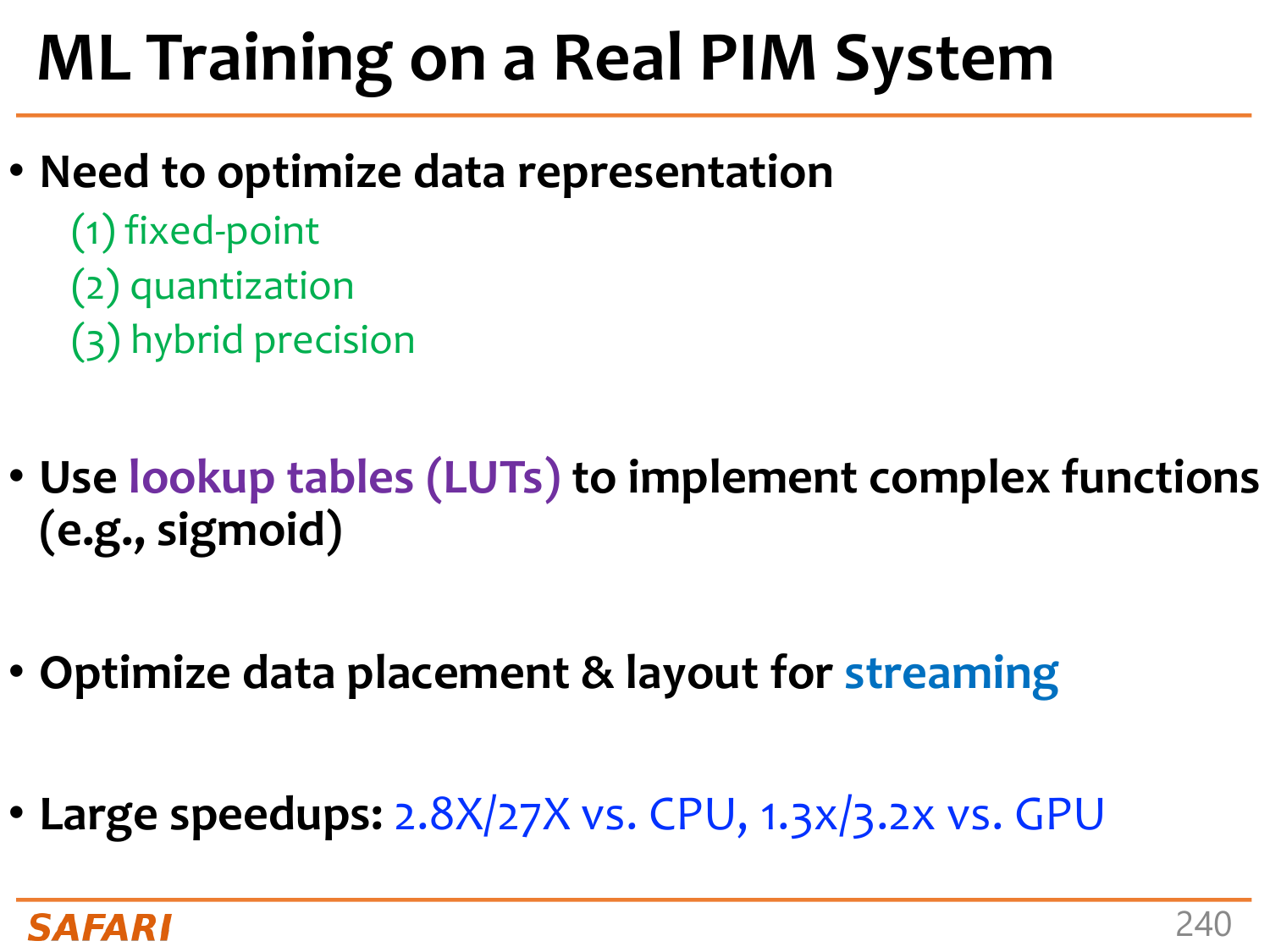}
    \caption{\gfix{Summary of the key \omvi{findings} of our ISPASS 2023 paper~\cite{gomez2023evaluating} on accelerating \omx{classical} machine learning training using the UPMEM PIM system. \omvii{A longer version of the paper with more detailed analyses is available on arXiv~\cite{gomez2022experimental}.} Reproduced from~\cite{mutlu.IMACAW23.talk}.}}
    \label{fig:mlupmem}
\end{figure}

\juanrrr{Third, we have performed a comprehensive analysis~\omv{\cite{rhyner2024pimopt,gomez2023evaluating,gomezluna2022isvlsi}} of the potential of the UPMEM PIM architecture to accelerate machine learning training \gfv{(Figure~\ref{fig:mlupmem} summarizes \omvi{our key findings})}. 
To do so, we 
(1)~implement several representative \omv{classical} machine learning algorithms (namely, linear regression, logistic regression, decision tree, K-Means clustering) on an UPMEM PIM system, 
(2)~apply several optimizations \omvi{(see Figure~\ref{fig:mlupmem})} to overcome the limitations of the current UPMEM PIM architecture (e.g., limited instruction set \omvii{and computation capability}, no direct inter-DPU communication), 
(3)~rigorously evaluate and characterize \omvi{our new UPMEM PIM implementations} in terms of accuracy, performance and scaling, and 
(4)~compare to their counterpart implementations on CPU and GPU. 
\omvi{Our results demonstrate that, after careful optimization, we can accelerate classical ML training  on UPMEM significantly over CPU and GPU implementations (e.g., 2.8$\times$/27$\times$ vs. CPU and 1.3$\times$/3.2$\times$ vs. GPU).}
Our \gfv{work}~\omv{\cite{gomez2023evaluating,gomezluna2022isvlsi}}\gf{, whose source code is available at \url{https://github.com/CMU-SAFARI/pim-ml},} provides several key observations, takeaways, and recommendations that can inspire users of machine learning workloads, programmers of PIM architectures, and hardware designers and architects of future PIM systems.}

\gfv{We \omvi{have greatly} extended our analysis of \omvi{different types of} machine learning \omvi{algorithm} execution on the UPMEM PIM systems in our recent PACT 2024 paper, called PIM-Opt~\cite{rhyner2024pimopt}, and SIGMETRICS 2025 paper, called PyGim~\cite{giannoula2024pygim}.
In PIM-Opt~\cite{rhyner2024pimopt}, we aim to understand the capabilities and characteristics of popular distributed stochastic gradient descent (SGD) algorithms~\cite{robbins1951stochastic}, which are used in data-intensive ML training workloads, on a real-world PIM architectures.
To do so, we implement\omvi{, optimize,} and rigorously evaluate 12 representative
ML training workloads, commonly used in the ML community, on
the real-world UPMEM PIM architecture.
Our results demonstrate three major findings: 
(1)~The UPMEM PIM system can be a viable alternative to state-of-the-art CPUs and GPUs for many memory-bound ML training
workloads, especially when operations and datatypes are natively
supported by PIM hardware; 
(2)~it is important to carefully choose the optimization algorithms that best fit \omvi{the} PIM \omvi{system}; and 
(3)~the UPMEM PIM system does \emph{not} scale approximately linearly with the number of PIM cores for many data-intensive ML training workloads. 
We open
source all our code to facilitate future research at \url{https://github.com/CMU-SAFARI/PIM-Opt}.}

\gfv{In PyGim~\cite{giannoula2024pygim}, we aim to efficiently map graph neural networks (GNNs)~\cite{hamilton2017inductive,kipf2016semi,xu2018powerful,zheng2020distdgl} to the UPMEM PIM system. 
PyGim provides high system performance in real
Host-PIM \omvi{cooperative execution} of GNNs, and bridges the gap between ML engineers, who prefer high-level
programming interfaces (e.g., Python), and real PIM systems, \omvi{which} typically provide complex and
low-level APIs and may \omvi{require} deep knowledge of PIM hardware.
PyGim improves system efficiency by running compute-bound \omvi{kernels on processor-centric hardware (i.e., CPUs/GPUs) and} memory-bound kernels on memory-centric hardware \omvii{(e.g., UPMEM PIM system)}\omvi{.}
\omvi{PyGim provides} highly\omvi{-}efficient parallelization strategies in GNN aggregation (a key step in GNN execution, which aggregates the input feature vectors of the neighboring vertices for each vertex in a graph via a permutation-invariant operator, such as average) tailored for
real PIM systems. 
We extensively evaluate PyGim on a real-world PIM system that has 16 PIM DIMMs with 1992 PIM cores connected to a host CPU. In GNN inference, we demonstrate that \omvi{PyGim} outperforms prior state-of-the-art PIM works by on average 4.38$\times$ (up to 7.20$\times$), and state-of-the-art PyTorch running on host by on average 3.04$\times$
(up to 3.44$\times$). 
PyGim improves energy efficiency by 2.86$\times$ (up to 3.68$\times$) and 1.55$\times$ (up to 1.75$\times$) over prior PIM and PyTorch host schemes, respectively.
PyGim is publicly and freely available at \url{https://github.com/CMU-SAFARI/PyGim}.
}

\gf{Fourth, in~\cite{gupta2023evaluating}, we accelerate the
Brakerski-Fan-Vercauteren (BFV) scheme~\cite{halevi2019improved,wibawa2022bfv} for homomorphic addition and multiplication using the UPMEM PIM system. Our evaluation shows
that the UPMEM PIM system accelerates the homomorphic addition operation by 50--100$\times$ over a state-of-the-art CPU and by
2--15$\times$ over a state-of-the-art GPU. 
For the homomorphic multiplication operation, the real PIM system provides a speedup of 30--50$\times$ over the CPU, but lags 10--15$\times$ behind the GPU due to the lack of  sufficiently wide \omvii{and powerful} \omvi{native} multiplication support on the evaluated first-generation UPMEM PIM \omv{hardware}.
We also evaluate our implementation of three statistical workloads (mean, variance, linear regression) using \omv{homomorphic} addition and homomorphic multiplication. In our evaluation, the real PIM system achieves up to 300$\times$ speedup over the CPU for all workloads and up to 30$\times$ over the GPU for \gfvi{a kernel that performs} arithmetic mean \gfvi{computation}. 
}

\omx{Fifth, in \gfx{SwiftRL}~\cite{gogineni2024swiftrl}, we accelerate reinforcement learning (RL)~\cite{sutton2018reinforcement} algorithms and their training phases on the \gfx{UPMEM PIM system}. 
Concretely, we implement two RL algorithms, Tabular Q-learning~\cite{sutton2018reinforcement,watkins1992q,li2020sample} and SARSA~\cite{sutton2018reinforcement}, on the UPMEM PIM system alongside performance optimization strategies to improve PIM performance.
SwiftRL performance optimization strategies include
(1)~approximating the Q-value update function, which avoids high performance costs due to runtime instruction emulation used by the UPMEM runtime library; and 
(2)~incorporating PIM-specific routines needed by the underlying algorithms. 
We experimentally evaluate RL workloads on \omxi{the} OpenAI GYM~\cite{1606.01540}
\omxi{environment} using UPMEM hardware. 
\info{GF: Correct.}Our experimental results show that SwiftRL's PIM-based implementations of RL algorithms outperform implementations of RL algorithms on an Intel
Xeon Silver 4110 CPU~\cite{IntelXe74:online} and an NVIDIA RTX 3090 GPU~\cite{ampere} by \emph{at least} 1.62$\times$ and 4.84$\times$, respectively.
SwiftRL is publicly and freely available at \url{https://github.com/kailashg26/SwiftRL}.
}

\gf{\omv{W}e \omv{also} address \omv{an important} limitation of the UPMEM PIM system caused by its limited instruction set. In UPMEM, some complex operations should be emulated by the runtime \omv{software} library (e.g., integer multiplication/division and floating-point arithmetic) or are not even supported. This is usually true for transcendental functions (e.g., trigonometric, hyperbolic, exponentiation, logarithm) and other hard-to-calculate functions (e.g., square root). To solve this limitation, we develop \emph{TransPimLib}~\cite{oliveira2023transpimlib}, an open-source library of transcendental and other hard-to-calculate functions for PIM. TransPimLib implements CORDIC-based methods, LUT-based methods, and combinations and variations of them (with and without interpolation). In total, TransPimLib uses eight different methods for trigonometric functions (sine, cosine, tangent), hyperbolic functions (sinh, cosh, tanh), exponentiation, logarithm, square root, and
GELU (Gaussian Error Linear Unit). We open source TransPimLib to facilitate reproducibility and future research~\cite{transpimlib2023repo}.}

\omv{Many other recent works have demonstrated applications and optimizations~\cite{chen2023uppipe,lavenier2020variant,lavenier2016blast,kang2023pim,baumstark2023accelerating,baumstark2023adaptive,lim2023design,bernhardt2023pimdb,lee2024spid,nider2022bulk,das2022implementation,zarif2023offloading} on real UPMEM PIM systems, including 
other genomics \& bioinformatics workloads (e.g., RNA-seq qualification~\cite{chen2023uppipe}, variant calling~\cite{lavenier2020variant}, BLAST~\cite{lavenier2016blast}), analytics \& databases~\cite{kang2023pim,baumstark2023accelerating,baumstark2023adaptive,lim2023design,bernhardt2023pimdb,lee2024spid}, imagine processing~\cite{nider2022bulk}, and other machine learning workloads (e.g., deep neural networks~\cite{das2022implementation} and recommendation models~\cite{zarif2023offloading}).} 
\omvi{Several other works~\cite{lee2024pim,hyun2024pathfinding,noh2024pid,kim2023virtual} propose improvements over the UPMEM \omvii{PIM} system, focusing \omvii{especially on benchmarking and} communication mechanisms between processing units}.

\subsubsection{\juanrrr{Samsung Function-In-Memory DRAM (FIMDRAM)}}
\label{sec:fimdram}

\juanrrr{Samsung has recently introduced FIMDRAM, also known as HBM-PIM~\cite{kwon202125, lee2021hardware, kim2021aquabolt}, a PIM architecture targeted to accelerate machine learning inference. 
FIMDRAM embeds one 16-bit floating-point SIMD unit with 16 lanes, called \emph{Programmable Compute Unit} (\emph{PCU}), next to two DRAM banks in HBM2 layers~\cite{jedec.hbm.spec}. 
PCUs support a \omv{small} instruction set (FP16 add, multiply, multiply-accumulate, multiply-and-add).} 

\begin{figure}[ht]
\centering
\includegraphics[width=1.0\linewidth]{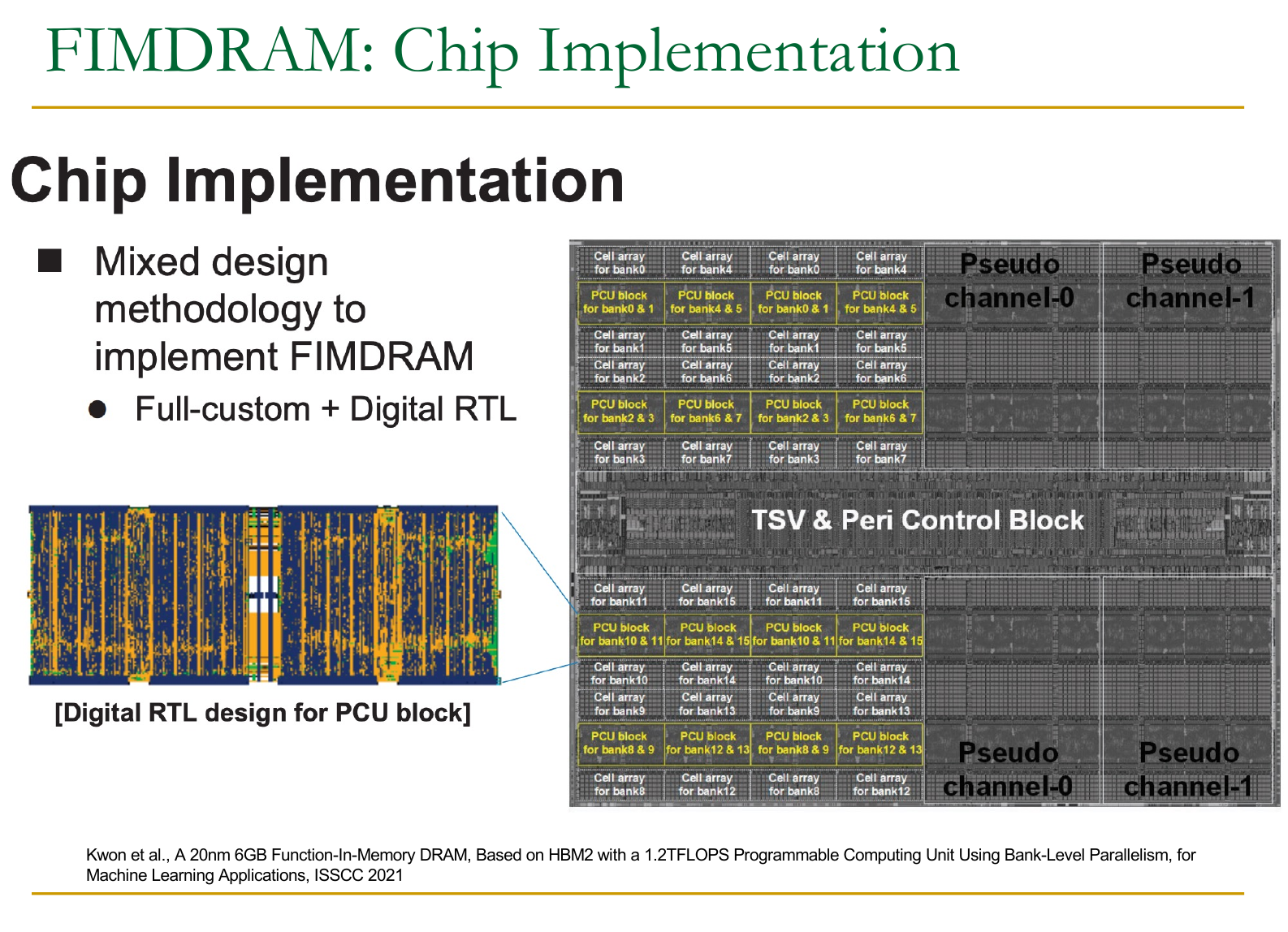}
\caption{\juanrrr{\omx{Samsung's Function-in-Memory DRAM (FIMDRAM)} chip implementation. Reproduced from~\cite{PIM.Spring2022.fimdram.lecture}.}}
\label{fig:fimdram}
\end{figure}

\juanrrr{Figure~\ref{fig:fimdram} shows a view of FIMDRAM chip implementation~\cite{kwon202125}, where HBM2 layers are modified to place one PCU block between two DRAM banks.} 

\begin{figure}[ht]
\centering
\includegraphics[width=1.0\linewidth]{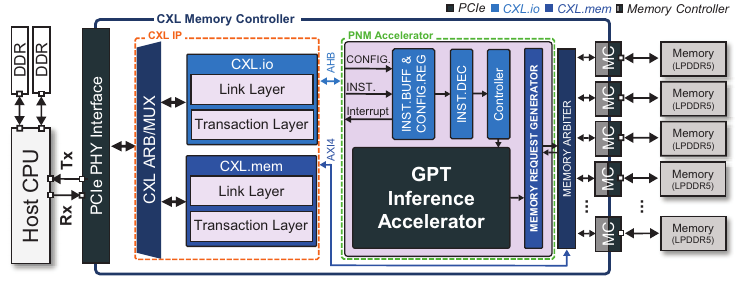}
\caption{\gfvi{Overview of Samsung's CXL-PNM architecture presented in~\cite{park2024lpddr}. The CXL-PNM architecture integrates an LLM inference accelerator (called \omvii{the} CXL-PNM controller) within LPDDR5X-based CXL memory modules~\cite{van2019hoti,datasheetlpdddr5x}.  The main components of the CXL-PNM controller are: (1)~a control unit, (2)~a register file manager, (3)~a matrix processing unit, and (4)~a vector processing unit. Reproduced from~\cite{park2024lpddr}.}}
\label{fig:samsungcxl}
\end{figure}

\gf{In the HotChips 2023 conference~\cite{samsunghc23}, Samsung has introduced two main extensions of the FIMDRAM architecture targeting generative artificial \omvi{intelligence} (AI)~\gfv{\cite{goodfellow2014generative,kingma2013auto,vaswani2017attention}} applications\gfv{, e.g., large language models (LLMs) such as GPT-4~\cite{achiam2023gpt}, GPT-J~\cite{mesh-transformer-jax}, T5~\cite{raffel2020exploring}, BERT~\cite{devlin2019bert,devlin2018bert}, and Llama-2~\cite{touvron2023llama}.} 
The new architectures include:
(1)~\omv{LPDDR}-PIM, which adds FIMDRAM's PCU to LPDDRx~\omvi{\cite{lpddr4}} memories and targets on-device generative AI \omv{inference}, and
(2)~CXL-PNM, which adds PNM engines to \omv{the} compute express link (CXL)~\cite{van2019hoti} memory expander. 
Samsung presented two different implementations of their CXL-PNM architecture. The first architecture adds PNM engines to the CXL controller, while the second architecture adds PNM engines to the memory chip itself. The PNM engines are composed of a PE array (targeting the acceleration of \omv{general matrix \omvi{multiplication}} operations) and an \omv{adder tree} (targeting the acceleration of \omv{general matrix vector \omvi{multiplication}} operations). 
The CXL-PNM architecture can provide a capacity of 512~GB and a memory bandwidth of 1.1~TB/s.
\gfv{Samsung provides further details on their CXL-PNM architecture in their HPCA 2024 paper~\cite{park2024lpddr}, where they describe their current realization of the CXL-PNM architecture using LPDDR5X-based CXL memory \gfvi{(which Figure~\ref{fig:samsungcxl} illustrates)}. 
The paper shows that the proposed architecture (equipped with \omvii{eight} CXL-PNM devices)
achieves 23\% lower latency, 31\% higher throughput, and 2.8$\times$ higher energy efficiency at 30\% lower hardware cost compared to a GPU-based architecture (with \omvii{eight} A100 NVIDIA GPUs~\cite{a100}) during LLM inference.}
}

\subsubsection{\juanrrr{Samsung Acceleration DIMM (AxDIMM)}}
\label{sec:axdimm}

\juanrrr{AxDIMM~\cite{ke2021near, kim2021aquabolt, lee2022improving}, also from Samsung, is a DIMM-based solution which places an FPGA fabric in the buffer chip of the DIMM. 
AxDIMM has been tested for DLRM recommendation inference~\cite{naumov2019deep, ke2021near} and for database operations~\cite{lee2022improving}.} 

\begin{figure}[ht]
\centering
\includegraphics[clip, trim=0.0cm 0.9cm 0.0cm 0.0cm, width=1.0\linewidth]{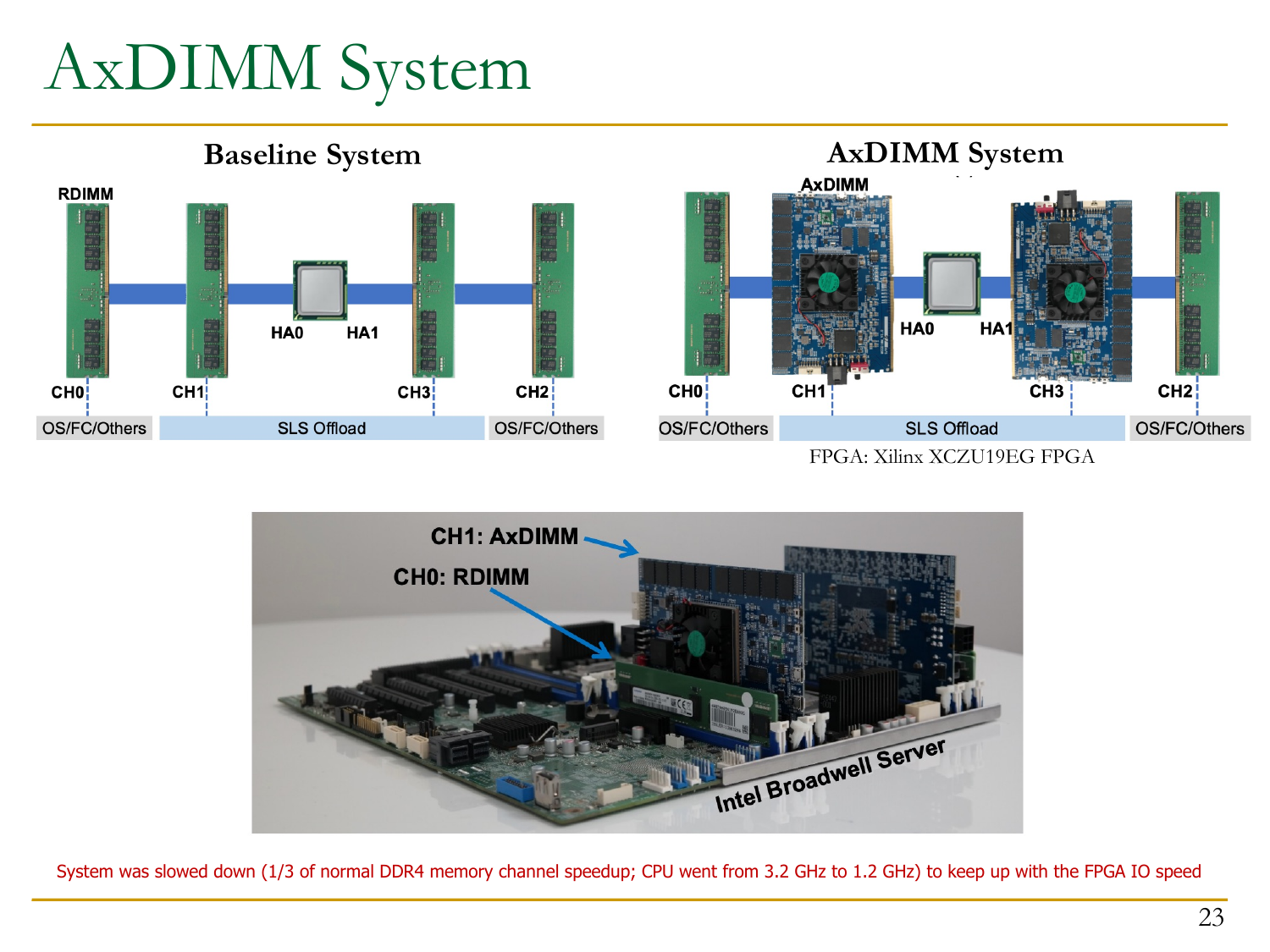}
\caption{\gfix{\omx{Samsung's Acceleration DIMM (AxDIMM)} hardware architecture. Reproduced from~\cite{PIM.Spring2022.axdimm.lecture}.}}
\label{fig:axdimm}
\end{figure}

\juanrrr{Figure~\ref{fig:axdimm} shows the AxDIMM module with two ranks of DRAM, an FPGA, and standard DRAM interface. For DLRM~\cite{ke2021near}, the FPGA implements two near-memory accelerators (one per rank) that execute element-wise summation of embedding table entries, which represent sparse features learned by the recommendation system~\cite{naumov2019deep}.}

\subsubsection{\juanrrr{SK Hynix Accelerator-in-Memory (AiM)} \gf{\& CXL-Memory Solution (CMS)}}
\label{sec:aim}

\juanrrr{Another major DRAM vendor, SK Hynix, has recently introduced \gf{two PNM solutions:
(1)~}Accelerator-in-Memory~\cite{lee2022isscc}, a GDDR6-based PIM architecture with specialized units for multiply-and-accumulate and lookup-table-based activation functions for deep learning applications\gf{; and
(2)~CXL-memory solution (CMS)~\cite{sim2022computational}, a CXL-based PIM architecture.}} 

\begin{figure}[ht]
\centering
\includegraphics[width=1.0\linewidth]{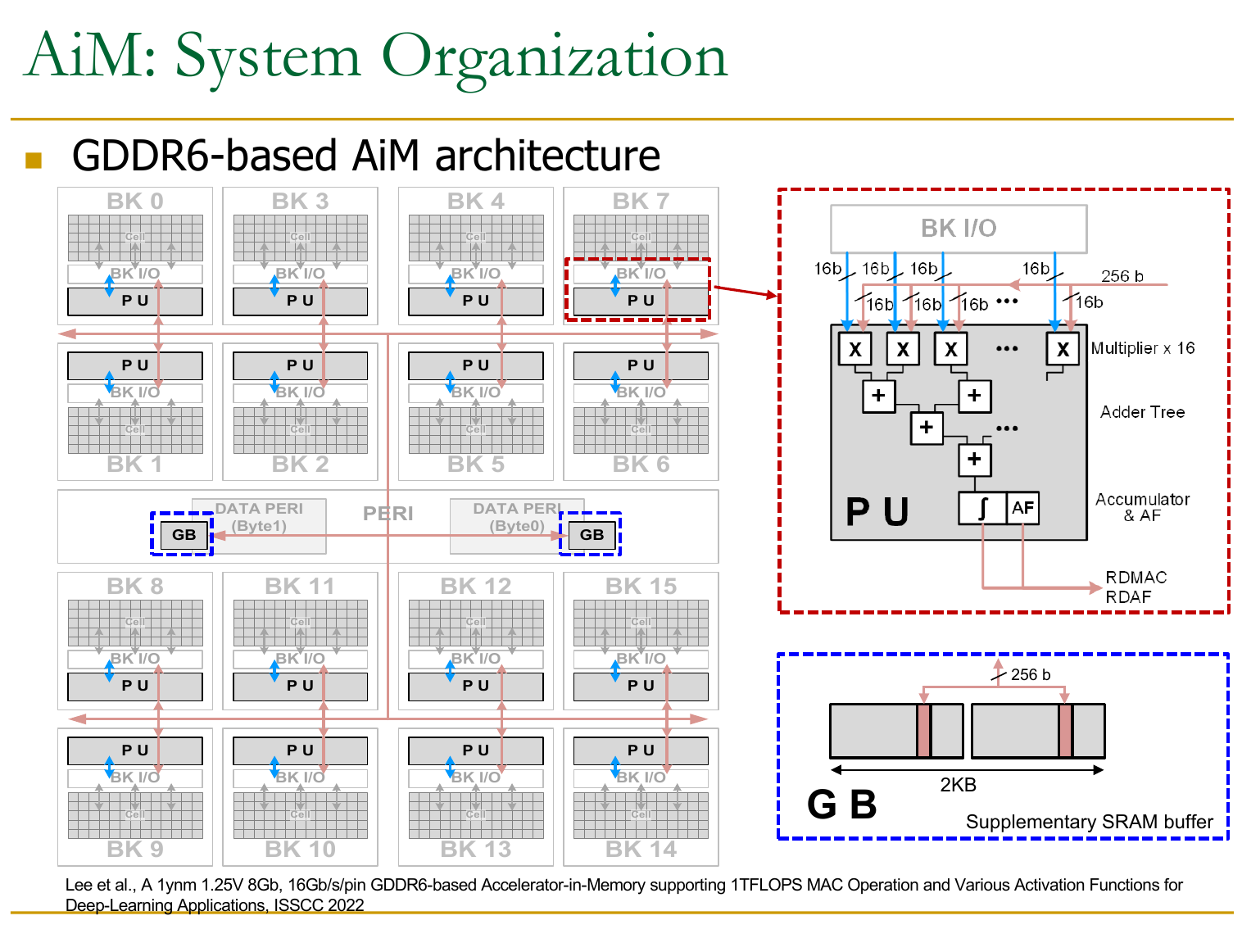}
\caption{\juanrrr{\omx{SK Hynix's Accelerator-in-Memory (AiM)} system organization. Reproduced from~\cite{PIM.Spring2022.aim.lecture}.}}
\label{fig:aim}
\end{figure}

\juanrrr{Figure~\ref{fig:aim} shows AiM system organization~\cite{lee2022isscc}. 
Near each DRAM bank, there is a \emph{processing unit} (\emph{PU}) that is composed of an array of 16 16-bit floating-point multipliers, an adder tree, an accumulator, and the necessary logic for activation functions. 
The chip also contains a supplementary 2-KB SRAM buffer, called \emph{global buffer} (\emph{GB}), which can store input vectors or serve as an intermediate buffer for copy operations between DRAM banks.} 
\juanrrri{This inter-bank copy operation is a limited form of a RowClone-like operation~\cite{seshadri2013rowclone} (Section~\ref{sec:rowclone}). In particular, it resembles RowClone's Pipelined Serial Mode (\omvii{which transfers data} from bank to bank, \omvii{see Section~3.2 of the RowClone paper~\cite{seshadri2013rowclone}}).}

\gf{The CMS architecture~\cite{sim2022computational} comprises of a CXL controller, internal DDR memory, a custom load balancer, and a PIM engine. CMS has been prototyped using a Xilinx Alveo U250 FPGA board~\cite{AlveoU2590}.
The custom load balancer is designed to, transparently from the host CPU, maximize memory bandwidth utilization by interleaving data across each channel in units of DDR access granularity (i.e., 64~bytes).
The PIM engine comprises of three custom logic \omv{units} (i.e., dot product, kNN calculator, and top-K unit) targeting \omv{the acceleration of} kNN application \omv{kernels}.  }

\subsubsection{\juanrrr{Alibaba Logic-to-DRAM Hybrid Bonding with PNM (HB-PNM)}}
\label{sec:hbpnm}

\juanrrr{Alibaba has recently presented HB-PNM~\cite{niu2022isscc}, a PNM system with specialized engines for recommendation systems, which is composed of a DRAM die and a logic die vertically integrated via hybrid bonding (HB)~\cite{fujun2020stacked}.}

\begin{figure}[ht]
\centering
\includegraphics[width=1.0\linewidth]{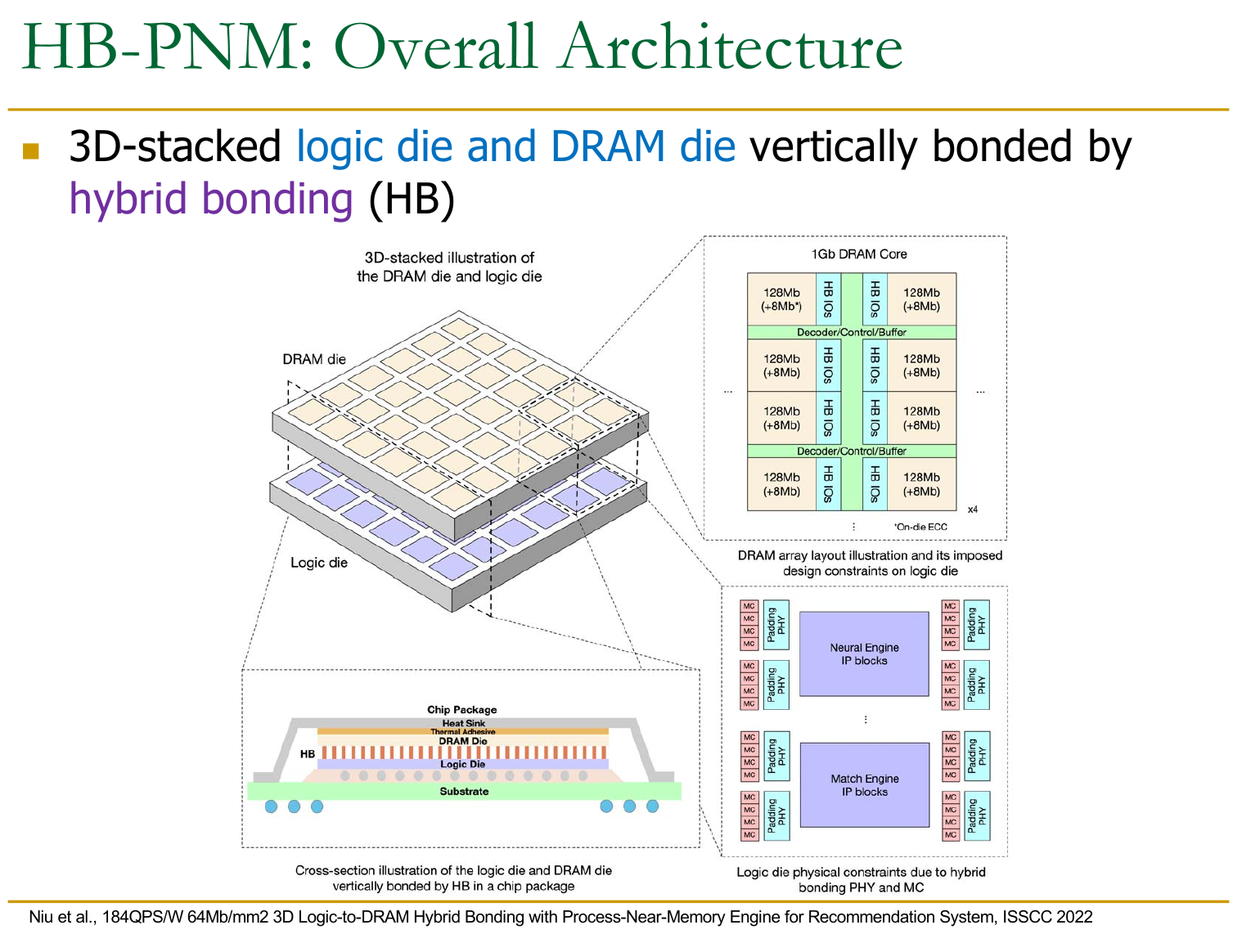}
\caption{\juanrrr{\omx{Alibaba's Logic-to-DRAM Hybrid Bonding with PNM (HB-PNM) architecture, including \gfx{a} logic die and \gfx{a} DRAM die and their structures.} Reproduced from~\cite{PIM.Spring2022.hbpnm.lecture}.}}
\label{fig:hb-pnm}
\end{figure}

\juanrrr{Figure~\ref{fig:hb-pnm} shows the logic die and the DRAM die (top left), and the cross-section of a chip package with the logic die and the DRAM die vertically bonded by HB (bottom left)~\cite{niu2022isscc}. 
The DRAM die contains \gfv{36} 1~Gb DRAM cores \gfv{(organized in a $6 \times 6$ 2D-array of DRAM cores)} with 8 banks each (top right of Figure~\ref{fig:hb-pnm}). 
The logic die contains multiple processing elements called \emph{match} and \emph{neural engines} (bottom right of Figure~\ref{fig:hb-pnm}) that perform, respectively, matching and ranking in a recommendation system.} 

\subsubsection{\juanrrr{Summary}}
\label{sed:realpim_summary}
\juanrrr{As we discuss in this section, PIM systems are finally becoming a reality in the form of PUM \omv{and PNM} prototypes and PNM solutions. 
It is exciting that the industry is taking up PNM \omv{and both researchers and designers are building prototype PIM systems}. We foreshadow that there will be more to come. 
We believe the existence of such real PIM hardware can greatly enable and accelerate software and adoption-related research for PIM, and can set a promising and useful baseline for future research in \omv{both PUM and} PNM systems.}

\subsection{Security Considerations}
\label{sec:security}

As a new processing paradigm, PIM introduces new security considerations related to its integration in real-world computing systems. 
First, there is a need to provide security guarantees in systems with PIM capabilities so that applications that offload code can execute securely in PIM computation units. Naively providing access to PIM computation units for all concurrently-executing applications may lead to potentially unforeseen \omv{information} leakage \omv{(e.g., see our recent work that provides an example of amplified \omvi{information} leakage in PIM~\cite{kanellopoulos2024amplifying})} and other \omv{confidentiality and integrity} issues. Second, the ability to perform computation inside or near memory using PIM can enable the opportunity to specialize such computation mechanisms to enhance system security (as briefly discussed in Section~\ref{sec:puf-trng}). 
We cover each of these topics briefly but envision many future ideas related to them in future PIM research and designs. 

First, PIM computation units should provide at least as good security primitives as processor-centric computation units of today. This means that there should at least be isolation between concurrently-executing processes on PIM computation units and access control to PIM resources (both data storage and computation units) should be securely managed. Partitioning of \omv{memory across computation units}, as done \omv{for example} in ~\omvi{\cite{hajinazar2020virtual,muralidhara2011reducing}}, can enable isolation. We believe new approaches to virtualization and cross-layer design that provide extensive hardware management capabilities in the memory controllers, such as the Virtual Block Interface  (VBI)~\cite{hajinazar2020virtual} or Expressive Memory~\cite{vijaykumar2018xmem,vijaykumar2018locality} 
can not only make PIM security mechanisms easier and more effective to implement but can provide much more enhanced PIM security mechanisms than existing systems. 

As in existing systems, \emph{reliability} and \emph{data integrity} are important in PIM systems, especially in PUM approaches, where memory rows can be frequently activated and deactivated. The RowHammer vulnerability~\gfv{\cite{kim-isca2014,mutlu2017rowhammer,mutlu2020retrospective,kim2020revisiting, yauglikcci2021blockhammer, hassan2021uncovering, orosa2021deeper,cojocar2020susceptible,frigo2020trr,mutlu2023fundamentally,mutlu2023retrospective,giray-thesis}} (Section~\ref{sec:majortrends}) 
can potentially become exacerbated in PIM systems but it can also be more easily preventable using an intelligent memory controller\juanr{~\cite{yauglikcci2021blockhammer}}\omv{, which is the core idea of PIM.} The cell wearout problem due to endurance limitations in some modern NVM technologies~\omv{\cite{luo2018heatwatch,luo2018improving,luo2015warm,cai.procieee17,fukami2017improving,luo2016enabling,cai2013error,cai2012error,cai2012flash,cai2013program,cai2013threshold,cai2014neighbor,cai2015data,cai2015read,cai2017vulnerabilities,yucai.bookchapter18}} can limit the reliability and thus effectiveness of NVM-based PUM approaches~\omv{\cite{kvatinsky.iccd11,kvatinsky.tcasii14,kvatinsky.tvlsi14,li.dac16,bhattacharjee2017revamp,flashcosmos,gao2021parabit,borghetti2010memristive,linn2012beyond,lehtonen2009stateful,kim2011field,lehtonen2012applications,mahmoudi2013implication,kim2019single,xie2017scouting,gaillardon2016plim,song2018graphr,imani2020dual,challapalle2020gaas,bojnordi2016memristive,feinberg2018enabling,shafiee2016isaac,chi2016prime,ielmini2018memory,wan2022compute,jung2022crossbar,sebastian2020memory,le202364,ankit2019puma,joshi2020accurate,song2017pipelayer,li2019long,valavi201964,imani2019floatpim,le2018mixed,yang2019sparse,jia2020programmable,tang2017binary,marinella2018multiscale,yuan2021forms,wen2020ckfo,ankit2020panther,wen2019memristor,long2018reram,chou2019cascade,feinberg2018making,li2020timely,angizi2019mrima,wang2018snrram,zhu2019configurable,yang2020retransformer,tang2017aepe,xia2016switched,xia2017fault,huang2017highly,cheng2017time,chen2018regan,cai2018training,mao2018lergan,nag2018newton}} \omv{(Section~\ref{sec:pum:punvm})} and thus needs to be addressed. Employing in-memory error correcting code (ECC) \rachata{techniques~\omv{\cite{binarystar,patel2019understanding,patel2020bit, patel2021harp,cai.bookchapter18.arxiv,yucai.bookchapter18,cai.procieee17,kang.memoryforum14,cilasunpim2024,leitersdorf2021efficient}}} 
is likely necessary in future PIM approaches\omv{.} 
PIM systems should likely be designed to support ECC \omv{and other robustness} techniques to maintain data reliability in the presence of \omv{both} computation mechanisms using/near memory and increasing noise and reliability problems due to technology scaling.

\msvii{\omix{The} adoption of PIM solutions provides a new way to directly access main memory, which malicious user applications can potentially exploit. In~\cite{kanellopoulos2024amplifying}, we show that
this new way to access main memory opens opportunities for high-throughput timing attack vectors that are hard-to-mitigate without significant performance overhead. We introduce IMPACT~\cite{kanellopoulos2024amplifying}, a set of high-throughput main
memory-based timing attacks that leverage characteristics of PIM architectures to establish covert and side channels. IMPACT enables high-throughput communication and private information leakage by exploiting the shared DRAM row buffer.
To achieve high throughput, IMPACT (1)~eliminates expensive cache bypassing steps required by processor-centric main memory and cache-based timing covert and side-channel attacks
and (2)~leverages the intrinsic parallelism of PIM operations. We showcase two \omix{examples} of IMPACT. First, we build two covert-channel attack variants that run on the host CPU and leverage different \omix{PIM} approaches (i.e., processing-near-memory and processing-using-memory) to gain direct and fast access to main memory and establish high-throughput communication covert channels. Second, we showcase a side-channel
attack that leaks private information of concurrently running victim applications that are accelerated with PIM. Our results demonstrate that (1) our covert channels achieve 12.87 Mb/s
and 14.16 Mb/s communication throughput, respectively, which is up to 4.91$\times$ and 5.41$\times$ faster than the state-of-the-art main
memory-based covert channels~\cite{pessl2016drama}, and (2) our side-channel attack allows the attacker to leak secrets with a low error rate. To avoid such covert and side channels in emerging PIM systems, we propose and evaluate three \omix{defense techniques} that \omix{attempt to} fundamentally eliminate the timing channel. 
For more detail on the IMPACT work, we refer the reader to our preprint on~\cite{kanellopoulos2024amplifying}.
}

Second, the PIM paradigm enables new opportunities to increase the security and privacy of computations and data, and thus entire computing systems. If data and computation stay within one chip, then the exposure of such data and computation to many attacks will likely be minimized. By eliminating data movement between memory and processor, the PIM paradigm takes a large step towards getting rid of one of the most attacker-exposed type of  data movement \omv{within a computing node}, i.e., data movement over the main memory bus. Enabling the secure and private execution of computations in PIM systems can therefore potentially enable fundamentally more secure computing systems. This requires providing support for such secure computation, as we discussed earlier in Section~\ref{sec:puf-trng}. 
For example, our afore-described DRAM latency PUF~\cite{kim.hpca18}, \juanrrr{DRAM latency True Random Number Generator~\cite{kim.hpca19}, and QUAC True Random Number Generator~\cite{olgun2021quactrng}} are notable examples of novel in-DRAM security primitives that take advantage of \gf{PUM} that were  briefly discussed in Section~\ref{sec:puf-trng}. 
We envision future works on PIM will provide many other security primitives, applications, use cases \omv{as well as end-to-end execution solutions~\cite{olgun2022pidram, bostanci2022dr}.}

%% file: sections/09-other-resources.tex
\section{Other Resources on PIM}
\label{sec:otherresources}

\juanr{As we have shown in previous sections, PIM is \omv{currently} capturing a lot of attention from \juanrr{both} industry and academia. 
It represents a key topic in \juanrr{our} advanced Computer Architecture courses~\gfv{\cite{CompArch_fall2021, CompArch_fall2022,CompArch_fall2023, CompArch_fall2024, Seminar_fall2021,Seminar_fall2022, Seminar_fall2023, Seminar_fall2024}}. 
\juanrr{We have} recently established PIM courses~\gf{\cite{PIM_fall2021, PIM_spring2022, PIM_spring2023, PIM_fall2022}}, where} \juanrr{we comprehensively cover many concepts, ideas, issues in PIM with a focus on both research and practical aspects.} 
\juanrrr{We have also held 
\gf{(1)}~a special session on PIM~\cite{isvlsi2022.pim} (at the ISVLSI 2022 conference~\cite{isvlsi2022}) with \juanrrri{nine} talks about recent research on tools \& methodologies for PIM~\cite{oliveira2022methodologies, olgun2022pidram, olgun2021pidram}, applications (ML~\cite{gomez2022machine, gomez2022experimental}, databases~\cite{oliveira2022heterogeneous}, \juanrrri{neural networks~\cite{oliveira2022heterogeneous}}, genomics~\cite{mansouri2022genstore.arxiv, mansouri2022genstore}, SpMV~\cite{giannoula2022sigmetrics, giannoula2022sparsep, giannoula2022towards}, time series \juanrrri{analysis}~\cite{fernandez2020natsa}, etc.), and challenges \& opportunities~\cite{ghose2022road}\gf{;
(2)~real-world PIM tutorials at several top venues, including HPCA 2023~\cite{pimtutorialhpca}, ASPLOS 2023~\cite{pimtutorialasplos}, ISCA 2023~\cite{pimtutorialisca}, and MICRO 2023~\cite{pimtutorialmicro}; and}}
\gfv{(3)~more recent and broader memory-centric computing system tutorials at several top venues, including HEART 2024~\cite{pimtutorialheart24}, ISCA 2024~\cite{pimtutorialisca24}, and MICRO 2024~\cite{pimtutorialmicro24}.}
\omvi{New versions of the memory-centric computing tutorial and future workshop are coming up soon at PPoPP 2025~\cite{pimtutorialppopp25} and ASPLOS 2025~\cite{pimworkshop}. 
We also point the reader to two major interviews that cover \omvii{progress in} memory-centric computing~\cite{hipeacinterview24,hipeacinterview18}. }

\juanr{These courses \juanrrr{and talks}, together with the growing amount of open-source infrastructures for PIM (e.g., simulators~\gfv{\cite{damov2021_repo, ramulator-pim,safari2023ramulator2,safariplutogithub}}, prototyping platforms~\gfv{\cite{olgun2021pidram_repo, softmc.github,safari-drambender}}, benchmark suites~\cite{damov2021_repo, gomezluna2021repo}, \juanrr{application implementations~\gfv{\cite{giannoula2022repo,giannoula2022sigmetrics,giannoula2022sparsep,giannoula2022towards,diab2022high, gomez2022machine,transpimlib2023repo,diab2023repo}}}), can greatly contribute to disseminating knowledge about PIM and foster more developments and innovations.} 
\juanrr{We hope that the future of memory-centric computing can be shaped in a way that can greatly improve our computing systems in a widespread manner.}

%% file: sections/10-conclusion.tex
\glsresetall

\section{Conclusion and Future Outlook}
\label{sec:conclusion}

Data movement is a major performance and energy bottleneck plaguing
modern computing systems. A large fraction of system energy is spent
on moving data across the memory hierarchy into the processors (and
accelerators), the only place where computation is performed in a
modern system. Fundamentally, the large amounts of data movement \sgii{are}
caused by the processor-centric design paradigm of modern computing systems:
processing of data is performed only in the processors (and
accelerators), which are far away from the data, and as a result, data
moves a lot in the system, to facilitate computation on it.

In this work, we argue for a \omv{fundamental} paradigm shift in the design of computing
systems\omv{:} a data-centric design paradigm that enables computation
capability in places where data resides \omv{or is generated} and thus performs computation
with minimal data movement.  \omv{Processing-In-Memory (PIM)} is
  a fundamentally data-centric design approach for computing systems
  that enables the ability to perform operations in or near memory \omv{structures}.
  Recent advances in modern memory architectures have enabled us to
  extensively explore two \omv{types of} approaches to designing PIM
  architectures: \gls{PUM} and \gls{PNM}.  First, we show that \gls{PUM} exploits the existing \omv{memory (e.g., DRAM)} architecture and the operational principles of the DRAM circuitry, enabling a number of important and widely-used operations (e.g., memory copy, data initialization, \omv{functionally complete} bulk\omv{-}bitwise operations,
  data reorganization\omv{, true random number generation}) within DRAM, with \omv{small} changes to DRAM chips. 
  Similar \gls{PUM} approaches are also applicable to other types of memory chips, and all yield large performance and energy benefits.  
  Second, we demonstrate that \gls{PNM} can embed computation capability \gf{\omv{into} conventional memory chips \omvi{(next to each bank or subarray)}, memory modules\omv{, memory controllers,} or the logic layer(s) of 3D-stacked memory technologies} 
  in a variety of ways to provide significant
  performance improvements and energy savings, across a large range of
  application domains and computing platforms. 
  Similar \gls{PNM} approaches are applicable to different types of memories.

\sg{Despite the extensive design space that we have studied so far, a
  number of key challenges remain to enable the widespread adoption of
  PIM in future computing systems~\omvi{\cite{ghose.bookchapter19,
    ghose.bookchapter19.arxiv,mutlu2019,ghose2019processing}}.}  \sg{Important challenges include
  developing  
  \gfvi{(1)~}easy-to-use programming models for PIM (e.g., PIM
  application interfaces, \gfvi{data structures,} compilers\gfvi{,} and libraries designed to abstract
  away PIM architecture details from programmers), 
  \gfvi{(2)~}extensive
  runtime support for PIM (e.g., scheduling PIM operations, sharing
  PIM logic among CPU threads, cache coherence, virtual memory
  support)\omv{,
  \gfvi{(3)~}robust and secure hardware/software design for PIM systems}\gfvi{, and
  (4)~benchmarks and simulation \omvii{infrastructures} that \omvii{enable accurate
assessment of} the benefits and shortcomings of PIM}.}  We \sg{hope} that providing the community with (1)~a
large set of memory-intensive benchmarks that can potentially benefit
from PIM, (2)~rigorous \omv{methodologies} to identify PIM-suitable parts
within an application, and (3)~accurate simulation \omv{and prototyping} infrastructures for
estimating the benefits and overheads of PIM will empower researchers
to address \sg{remaining} challenges for the adoption of PIM.
\juanrrri{Real PIM hardware that starts to become a reality is also key to investigate adoption-related research, as it \omv{enables rapid software development and} represents a \omvi{real and useful} baseline for future research.}

\juan{We firmly believe that it is time to design principled system
  architectures to solve the data movement problem of modern computing
  systems, which is caused by the rigid dichotomy and imbalance
  between the computing unit (CPUs and accelerators) and the
  memory/storage unit. Fundamentally solving the data movement problem
  requires a paradigm shift to a more data-centric computing system
  design, where computation happens where data resides \omv{or \omvi{where data} is generated} (i.e., in or near memory/storage \omv{and sensors}), with
  minimal movement of data.  Such a paradigm shift can greatly push
  the boundaries of future computing systems, leading to orders of
  magnitude improvements in energy and performance (as we \omvi{have} demonstrated
  with \omv{many} examples in this \omvi{article}), \omv{while} potentially \omv{also} enabling new
  applications and computing \sgii{platforms.}}

  \omvi{We believe that memory should be designed, used, and programmed not as an inactive storage substrate, which is business as usual in modern systems, but instead as a combined computation and storage substrate where both computational capability and storage density are key goals. Although many challenges remain to enable widespread adoption of processing-in-memory, we believe the mindset and infrastructure shift necessary to enable such a combined computation-storage paradigm remains to be the largest challenge. Overcoming this mindset and infrastructure shift can unleash a fundamentally energy-efficient, high-performance, and sustainable way of designing, using, and programming computing systems.}
  \omv{We therefore believe the future of \gfvi{processing-in-memory} is very bright and promising, yet there needs to be many exciting challenges to be solved \omvii{across the computing stack} to facilitate widespread \omvi{and easy} adoption. }
